\documentclass[11pt,twoside]{mitthesis_alternative}
\usepackage{lgrind}
\pdfoutput=1

\usepackage{color}
\usepackage{graphicx}
\usepackage{amsmath}
\usepackage{amssymb}
\usepackage{xspace}
\usepackage[small]{subfigure}
\usepackage[hyperfootnotes=false]{hyperref}
\usepackage[numbers,sort&compress]{natbib}

\usepackage[countmax]{subfloat}

\usepackage{cancel}

\usepackage{braket}
\usepackage{graphicx}
\usepackage{multirow}
\usepackage{verbatim}
\usepackage{amsthm}
\usepackage{slashed}
\usepackage{wasysym}
\usepackage{simplewick}
\usepackage{mathtools}
\usepackage{soul}

\newcommand{\eq}[1]{Eq.~\eqref{eq:#1}}
\newcommand{\eqs}[2]{Eqs.~\eqref{eq:#1} and \eqref{eq:#2}}

\newcommand{\app}[1]{App.~\ref{app:#1}}
\renewcommand{\sec}[1]{Sec.~\ref{sec:#1}}
\newcommand{\secs}[2]{Secs.~\ref{sec:#1} and \ref{sec:#2}}

\newcommand{\fig}[1]{Fig.~\ref{fig:#1}}

\DeclareRobustCommand{\Sec}[1]{Sec.~\ref{#1}}
\DeclareRobustCommand{\Secs}[2]{Secs.~\ref{#1} and \ref{#2}}
\DeclareRobustCommand{\App}[1]{App.~\ref{#1}}
\DeclareRobustCommand{\Tab}[1]{Table~\ref{#1}}

\DeclareRobustCommand{\Fig}[1]{Fig.~\ref{#1}}

\DeclareRobustCommand{\Eq}[1]{Eq.~(\ref{#1})}
\DeclareRobustCommand{\Eqs}[2]{Eqs.~(\ref{#1}) and (\ref{#2})}
\DeclareRobustCommand{\Ref}[1]{Ref.~\cite{#1}}

\newcommand{\tab}[1]{Tab.~\ref{tab:#1}}
\newcommand{\head}[1]{
	\vspace{0.3cm}
	\noindent {\bf \underline {#1}}
	\vspace{0.1cm}
}

\newcommand{\mi}{{\mu}}
\newcommand{\as}{\alpha_s}

\newcommand{\ord}[1]{\mathcal{O}(#1)}

\newcommand{\mae}[3]{\langle#1\rvert#2\rvert#3\rangle}
\newcommand{\Mae}[3]{\bigl\langle#1\bigr\rvert#2\bigr\rvert#3\bigr\rangle}

\newcommand{\ang}[1]{\langle #1 \rangle}
\newcommand{\msbar}{\overline{\textrm{MS}}}

\newcommand{\df}{\mathrm{d}}

\newcommand{\sdt}{\!\cdot\!}
\newcommand{\tr}{\textrm{tr}}

\newcommand{\al}{\alpha}
\newcommand{\bt}{\beta}
\newcommand{\ga}{\gamma}

\newcommand{\ve}{\varepsilon}
\newcommand{\la}{\lambda}

\newcommand{\w}{\omega}

\newcommand\bn{{\bar n}}

\newcommand{\e}{\epsilon}

\newcommand{\balpha}{{\bar \alpha}}
\newcommand{\bbeta}{{\bar \beta}}

\newcommand{\cB}{{\mathcal B}}

\newcommand{\cL}{{\mathcal L}}

\newcommand{\cP}{{\mathcal P}}

\newcommand{\cY}{{\mathcal Y}}

\newcommand{\bnP}{\overline {\mathcal P}}

\newcommand{\nslash}{\slashed{n}}
\newcommand{\bnslash}{\slashed{\bar{n}}}
\newcommand{\pslash}{\slashed{p}}

\newcommand{\qslash}{\slashed{q}}

\newcommand{\nn}{\nonumber}

\newcommand{\lqcd}{\Lambda_\mathrm{QCD}}

\newcommand{\hard}{\mathrm{hard}}
\newcommand{\dyn}{\mathrm{dyn}}

\newcommand{\BPS}{\mathrm{BPS}}

\newcommand{\lp}{ {\tilde{p}} }

\newcommand{\SCETa}{\ensuremath{{\rm SCET}_{\rm I}}\xspace}

\newcommand{\SCETi}{\mbox{${\rm SCET}_{\rm I}$}\xspace}
\newcommand{\SCETii}{\mbox{${\rm SCET}_{\rm II}$}\xspace}

\allowdisplaybreaks[2]

\definecolor{mygreen}{rgb}{0,0.65,0}
\definecolor{myblue}{rgb}{0,0,1}
\definecolor{myorange}{rgb}{1,0.5,0}
\definecolor{mydarkblue}{rgb}{0,0,0.6}
\definecolor{myred}{rgb}{1,0,0}

\def\cB{\mathcal{B}}

\def\cL{\mathcal{L}}

\def\cO{\mathcal{O}}

\def\cP{\mathcal{P}}

\def\cW{\mathcal{W}}

\def\cY{\mathcal{Y}}

\def\tr{{\rm tr}}

\newcommand{\V}{V}

\def\nn{{\nonumber}}

\DeclareRobustCommand{\Sec}[1]{Sec.~\ref{#1}}
\DeclareRobustCommand{\Secs}[2]{Secs.~\ref{#1} and \ref{#2}}
\DeclareRobustCommand{\App}[1]{App.~\ref{#1}}
\DeclareRobustCommand{\Tab}[1]{Table~\ref{#1}}

\DeclareRobustCommand{\Fig}[1]{Fig.~\ref{#1}}

\DeclareRobustCommand{\Eq}[1]{Eq.~(\ref{#1})}
\DeclareRobustCommand{\Eqs}[2]{Eqs.~(\ref{#1}) and (\ref{#2})}
\DeclareRobustCommand{\Ref}[1]{Ref.~\cite{#1}}

\def\be{\begin{equation}}
\def\ee{\end{equation}}

\def\l{\langle}
\def\r{\rangle}

\def\bt{\beta}

	\newcommand{\Sl}[1]{\slashed{#1}}

	\allowdisplaybreaks[3]
	
	\renewcommand{\sec}[1]{Sec.~\ref{sec:#1}}

	\newcommand{\vT}{\bar{T}}
	
	\newcommand{\vC}{\vec{C}}
	\newcommand{\vO}{\vec O}
	
	\newcommand{\ldel}{\tilde \delta} 

	\renewcommand{\P}{\mathrm{P}}       

\def\cL{\mathcal{L}}

\def\cO{\mathcal{O}}

\def\cP{\mathcal{P}}

\def\tr{{\rm tr}}

\def\bt{{\bf t}}
\def\vC{\vec C}

\begin{document}

\title{Accuracy and Precision in Collider Event Shapes}

\author{Daniel W.~Kolodrubetz}

\department{Department of Physics}

\degree{Doctor of Philosophy in Physics}

\degreemonth{June}
\degreeyear{2016}
\thesisdate{May 16, 2016}

\supervisor{Iain W.~Stewart}{Professor of Physics}

\chairman{Nergis Mavalvala}{Associate Department Head for Education}

\maketitle

\cleardoublepage

 \pagestyle{empty}
\setcounter{savepage}{\thepage}
\begin{abstractpage}

In order to gain a deeper understanding of the Standard Model of particle physics and test its limitations, it is necessary to carry out accurate calculations to compare with experimental results. Event shapes provide a convenient way for compressing the extremely complicated data from each collider event into one number. Using effective theories and studying the appropriate limits, it is possible to probe the underlying physics to a high enough precision to extract interesting information from the experimental results.

In the initial sections of this work, we use a particular event shape, C-parameter, in order to make a precise measurement of the strong coupling constant, $\alpha_s$. First, we compute the $e^+ e^-$ \mbox{C-parameter} distribution using the Soft-Collinear Effective Theory (SCET) with a resummation to N${}^3$LL$^\prime$ accuracy of the most singular partonic terms. Our result holds for $C$ in the peak, tail, and far-tail regions. We treat hadronization effects using a field theoretic nonperturbative soft function, with moments $\Omega_n$, and perform a renormalon subtraction while simultaneously including hadron mass effects. 

We then present a global fit for $\alpha_s(m_Z)$, analyzing the available C-parameter data in the resummation region, including center-of-mass energies between $Q=35$ and $207$\,GeV. We simultaneously also fit for the dominant hadronic parameter, $\Omega_1$. The experimental data is compared to our theoretical prediction, which has a perturbative uncertainty for the cross section of $\simeq 2.5\%$ at $Q=m_Z$ in the relevant fit region for $\alpha_s(m_Z)$ and $\Omega_1$. We find $\alpha_s(m_Z)=0.1123\pm 0.0015$ and $\Omega_1=0.421\pm 0.063\,{\rm GeV}$ with $\chi^2/\rm{dof}=0.988$ for $404$ bins of data. These results agree with the prediction of  universality for $\Omega_1$ between thrust and C-parameter within 1-$\sigma$.

The latter parts of this study are dedicated to taking SCET beyond leading power in order to further increase the possible precision of calculations. On-shell helicity methods provide powerful tools for determining scattering amplitudes, which have a one-to-one correspondence with leading power helicity operators in SCET away from singular regions of phase space. We show that helicity based operators are also useful for enumerating power suppressed SCET operators, which encode subleading amplitude information about singular limits. In particular, we present a complete set of scalar helicity building blocks that are valid for constructing operators at any order in the SCET power expansion. We also describe an interesting angular momentum selection rule that restricts how these building blocks can be assembled.

\end{abstractpage}

\cleardoublepage

\pagestyle{plain}

\tableofcontents
\newpage
\listoffigures
\newpage
\listoftables


\chapter{Introduction}
\label{ch:intro}

\section{Effective Theories} \label{sec:EFTphys}

Physics lives at many different scales. There are hierarchies in energy, distance, time and numerous other measurables. We have to apply very different intuition and reasoning at each scale that we study. By focusing only on the degrees of freedom that are important for the physical system we want to study, we can simplify the physics to its core components. When looking at gravity in the galaxy, we can treat the stars and planets as mere pointlike objects. In the physics of our everyday life, we do not care about the individual motion of atoms inside a baseball. As we zoom in further, atoms and even nuclei can no longer be viewed as having simple pointlike interactions with each other. Our comprehension for the physics that governs everyday scales is not appropriate for understanding the interactions that take place between fundamental particles on the order of $10^{-15}$ meters. At this scale, physics is governed by field theory, specifically the standard model, made up of matter and the forces of Quantum Chromodynamics (QCD), Quantum Electrodynamics (QED), and the weak force.

The field theories that directly govern the Standard Model can be computationally intractable. Due to this, we turn to effective theories to isolate the most important physical aspects of a system. A key component of an effective field theory is a definite power counting: the ability to expand in a small power counting parameter that allows us to control the size of corrections to our calculation. Often times, this power counting parameter is related to an energy scale. If we are studying dynamics at a particular energy scale $\mu$, we do not need the detailed behavior of the system at a much higher energy $Q \gg \mu$ and can expand our theory in $\mu / Q$. By doing this expansion, we pick out only the most important degrees of freedom for the process we are studying, and capture the leading (and most crucial) part of the calculation.

Particle colliders are our most powerful tool for probing physics at the high energy (or short distance) scales that are governed by the strong force, QCD. In addition to a plethora of data from older colliders such as LEP and the Tevatron, there are large amounts of data coming out of the LHC every day. In many cases, the precision of the experimental measurement exceeds the accuracy of a theoretical prediction for that measurement. Effective theories provide a controlled way of approaching calculations to higher precision. Additionaly, every collision has a variety of scales associated with it. These include the center of mass energy (usually denoted $Q$), the masses $m_i$ of the particles involved in the collision (both incoming and outgoing), as well as many different measurements of the separation between particles, which can be converted into effective energies for the groups of particles known as jets. These hierarchies of scales lend themselves naturally to an effective field theory approach. In the rest of this section, we will introduce a specific type of measurement at colliders and discuss Soft-Collinear Effective Theory (SCET), an effective theory that has proved fruitful for making high precision calculations for collider measurements.

These tools are the basis for the work done in the remainder of this thesis. The broad goal is to use and improve SCET in order to push the precision boundary for theoretical predictions of collider physics. First, we will apply SCET to increase precision on the calculation of the C-parameter cross section, with the result of improving the accuracy of the measurement of the strong coupling. Then, we will develop a formalism using helicity operators which will simplify applying the SCET power expansion to cross sections at subleading power in a well controlled way.

\section{Event Shapes at Particle Colliders} \label{sec:eventShapesColliders}

\begin{figure}[t!]
	\begin{center}
		\includegraphics[width=0.6\columnwidth]{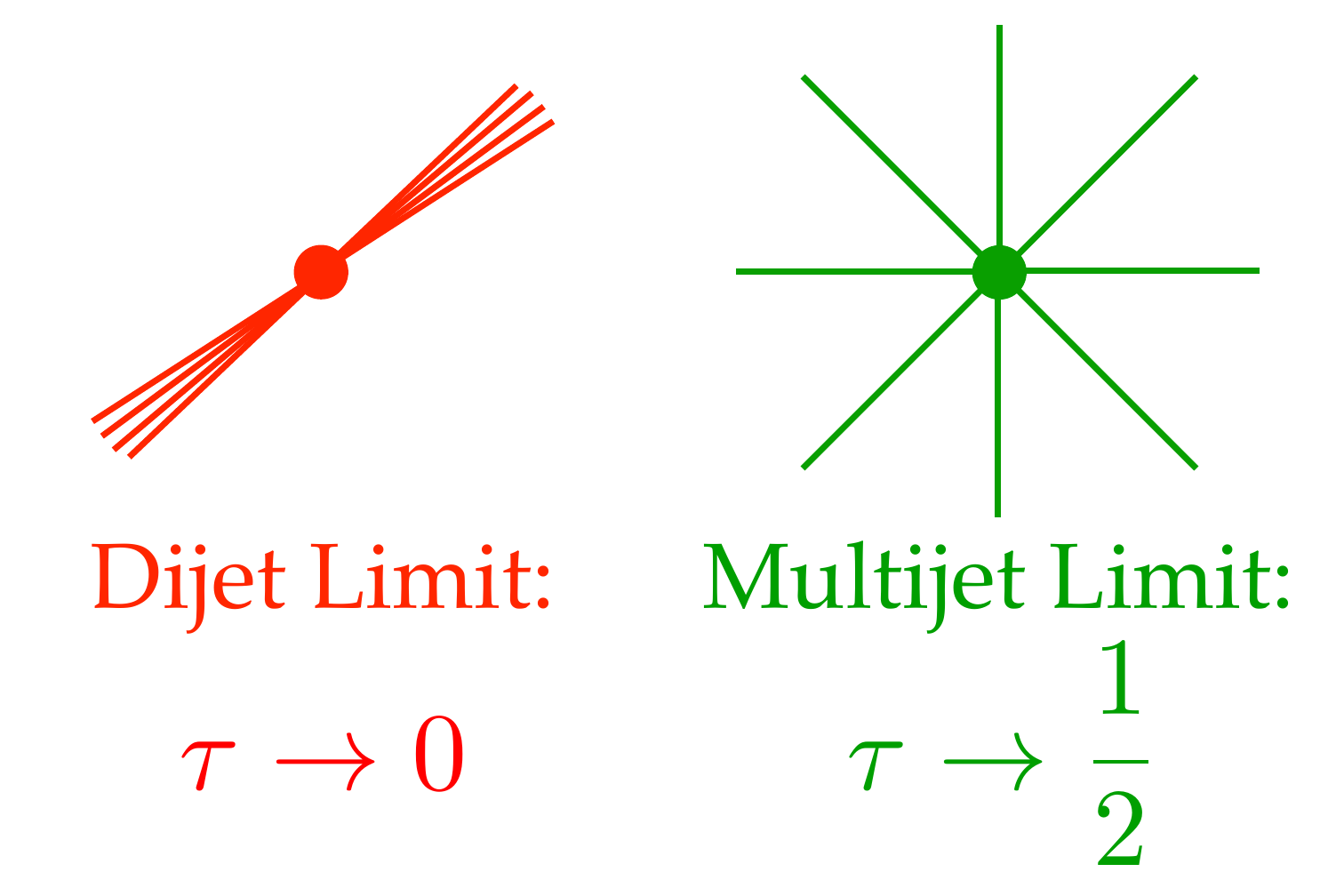}
		\caption[Kinematics of thrust regions]{Schematic drawing of particle flow in two limits of thrust. As $\tau \to 0$, the particles are all closely aligned in two back-to-back jets. As $\tau \to 1/2$, the particles cover the full region of phase space in the multijet region}\label{fig:dijet_figure} 
	\end{center}
\end{figure}

Particle colliders provide detailed particle tracking data on a scale that is not feasible to deal with individually. Rather than computing theoretical results for individual particle tracks and motion, it makes sense to combine this information into one number for each collision, called an event shape. These observables are designed to measure the geometrical properties of momentum flow for a collision event. The classic example of an event shape is thrust in $e^+ e^-$ colliders \cite{Farhi:1977sg}, which we will take to be
\begin{equation} \label{eq:thrustdef}
\tau \,=\, 1-T \,=\,
\min_{\vec{n}}\! \left( 1 -\, \frac{ \sum_i|\vec{n} \cdot \vec{p}_i|}{\sum_j |\vec{p}_j|} \right),
\end{equation}
where the sum is over all of the particles in an event and $\vec{n}$ is called the thrust axis. It follows from the above equation that $0 \le \tau \le 1/2$ and we can see that these two limits give very different kinematic structures for the event. As $\tau \to 0$, all of the particles are aligned with the thrust axis and the event will be two thin back-to-back dijets. As $\tau \to 1/2$, the particles are spread over all available phase space, creating a spherical event. The different structure of the particle flow in the two extremes is illustrated in \fig{dijet_figure}. While the exact limits are unique to thrust, the idea behind all event shapes is to separate such geometric regimes using one number. Additionally, we will focus in this work on event shapes that go to zero in the dijet limit.

\begin{figure}[t!]
	\begin{center}
		\includegraphics[width=0.6\columnwidth]{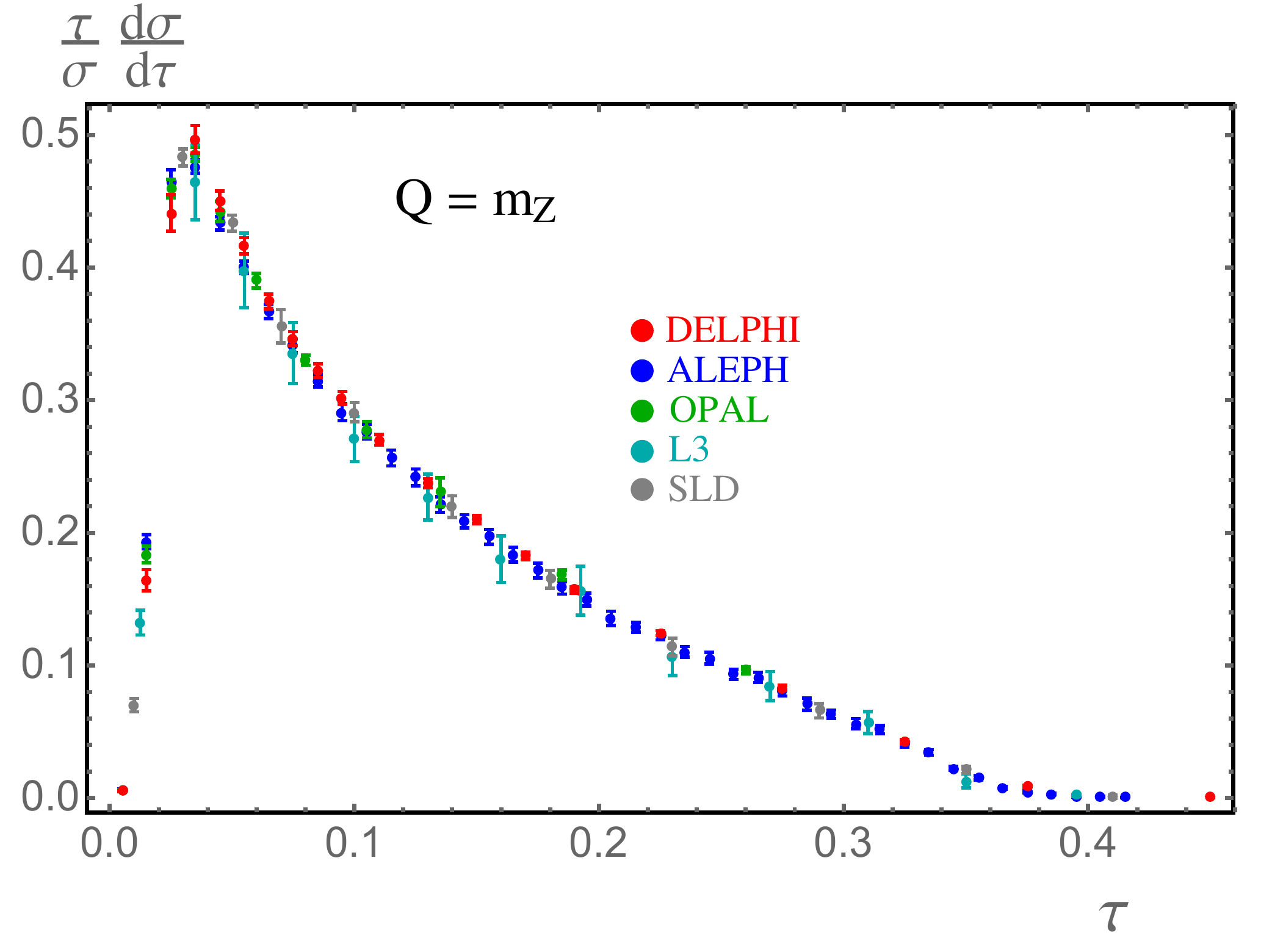}
		\caption[Plot of thrust data]{Plot of thrust data from various $e^+ e^-$ collider experiments. The colors indicate different experiments and the error bars are a combined statistical and systematic uncertainty. }\label{fig:thrust-data} 
	\end{center}
\end{figure}

Experimentally, we can measure the differential cross section as a function of this event shape,  $\dfrac{\df \sigma}{\df \tau}$,  counting how many events occur for a given range of thrust values. An example of this data is show in \fig{thrust-data}, with the experimental values taken from  SLD ~\cite{Abe:1994mf}, L3~\cite{Achard:2004sv}, DELPHI \cite{Wicke:1999zz}, OPAL
~\cite{Abbiendi:2004qz} and ALEPH~\cite{Heister:2003aj}. In this plot we can see the typical form that event shape cross sections take at $e^+ e^-$ colliders. For small values of $\tau$, we are in the peak region, where the physics is dominated by nonperturbative effects. This region can be roughly defined by $Q \tau \gtrsim \Lambda_\text{QCD}$. To the right of this, we have the tail region, where $\tau$ is still small, but is out of the regime where nonperturbative physics dominates. Finally, there is the far-tail, where $\tau \sim 1/2$. 

Like many event shapes, it is interesting to study the limit where thrust goes to zero, which forces us into the dijet configuration. In this limit, all of the particles in the event are either collinear, meaning that they lie extremely close to one of the two jets, or soft, meaning that their energy is low enough that they contribute only negligibly to the value of thrust. If we perform a fixed order calculation of the cross section, it will contain logs of thrust, which come suppressed by the strong coupling constant as $\dfrac{\df \sigma}{\df \tau} \sim \alpha_s^{n} \dfrac{1}{\tau} \log^{2n-1} \tau$ for various integers $n>0$. These logs capture the singular behavior of QCD in the soft and collinear limits. As we move to smaller $\tau$, theses logs become more and more important. At some point, the value of $\log \tau$ will be large enough that $\alpha_s \log \tau \sim 1$. Once this happens, a fixed order expansion in $\alpha_s$ is no longer enough to capture the leading behavior of the distribution, as if we truncate all terms of $\cO(\alpha_s^{n})$ we are potentially missing large pieces of the form $\alpha_s^{n} \dfrac{1}{\tau} \log^{2n -1} \tau$. To appropriately deal with these pieces, we need to sum the entire tower of logs, using a process known as resummation. In order to understand the various orders of resummation, we can look at a schematic expansion of the cross section. The best way to separate the structure of the large logs is to take the cumulant of the cross section,
\begin{align}
\Sigma (\tau) \equiv \int_0^\tau \df \tau' \frac{1}{\sigma_0} \frac{\df \sigma}{\df \tau'}.
\end{align}
The towers of large logs are most easily seen in the expansion of the log of this cumulant, which is schematically given by
\begin{align} \label{eq:log-towers}
\log \Sigma(\tau) \sim \alpha_s (& \log^2 \tau + \log \tau + 1)
\\
+ \alpha_s^2 (& \log^3 \tau + \log^2 \tau + \log \tau + 1)
\nn \\
+ \alpha_s^3 (& \log^4 \tau + \log^3 \tau + \log^2 \tau + \log \tau + 1)
\nn \\
+ \cdots \nn\,.
\end{align}
For small $\tau$, all of the terms that scale as $\alpha_s^n \log^{n+1} \tau$ are of the same order and adding all logs of this form is called Leading Log (LL) resummation. Similarly, if we can resum all of the terms of the form $\alpha_s^{n+1} \log^{n+1} \tau$, this is known as Next-to-Leading Log (NLL) resummation. Generically, we will have that N$^i$LL resummation will capture all terms of the form $\alpha_s^{n+i} \log^{n+1} \tau$. Additionally, it often makes sense to include one further fixed order piece, particularly when we look at a region in $\tau$ where we are transitioning between the regime with large logs and the regime where only the fixed order expansion is important. We denote the inclusion of that extra piece by an additional prime. So, for example, N$^2$LL$'$ will contain all of the pieces that scale up to $\alpha_s^{n+2} \log^{n+1} \tau$ and will additionally have the $\alpha_s^2$ non-logarithmic piece. As we will discuss in the next section, SCET is an ideal framework for performing this resummation.

One important application of event shapes has been the measurement of the strong coupling constant, $\alpha_s$. As this coupling runs with the energy scale, it is standard to report a determination of $\alpha_s(m_Z)$, measured at the mass of the Z-boson. Event shapes, particularly at $e^+ e^-$ colliders, have several nice properties that make them a good choice for performing the coupling extraction. Since we generally look at event shapes that sum over all of the particles in an event, they have  global nature, which gives them nice theoretical properties and makes accurate QCD predictions possible. Additionally, their leading order term comes with an $\alpha_s$, which gives them high sensitivity to the coupling. In contrast, an inclusive cross section (such as $e^+ e^- \to$ hadrons) will only contain $\alpha_s$ in the correction terms. Thanks to these benefits, there is a long history of event shape determinations of $\alpha_s$ (see the review~\cite{Kluth:2006bw} and the workshop proceedings~\cite{Bethke:2012jm}), including recent analyses which include higher-order resummation and corrections up to ${\cal O}(\alpha_s^3)$~\cite{GehrmannDeRidder:2007bj,Davison:2008vx,Becher:2008cf,Dissertori:2009qa,GehrmannDeRidder:2009dp,Chien:2010kc,Abbate:2010xh,Abbate:2012jh,Gehrmann:2012sc,Hoang:2014wka}.

The procedure for extracting the strong coupling from an event shape cross section is deceptively simple. First, calculate the cross section as a function of $\alpha_s$. Then, compare that calculation with the available data and perform a $\chi^2$ analysis to determine which value of $\alpha_s$ gives you the best fit. Of course, constructing a theoretical cross section with uncertainties that are similar to the experimental error is extremely difficult. As we will see in Secs. \ref{ch:Cparam-theory} and \ref{ch:Cparam-fit}, one must work to high order in both the fixed order calculation and the resummation to get a precise prediction. There is also the key issue of nonperturbative effects which must be addressed and included.

It is important to compare with other methods of determining the strength of the strong coupling. See the PDG~\cite{Agashe:2014kda} for a thorough review, which calculates a world average of $\alpha_s(m_Z)=0.1185\pm 0.0006$. This is dominated by the lattice QCD determination~\cite{Chakraborty:2014aca}. Recent higher-order event-shape analyses~\cite{Davison:2008vx,Abbate:2010xh,Gehrmann:2012sc,Abbate:2012jh,Gehrmann:2012sc} have found values of $\alpha_s(m_Z)$ significantly lower than this number.  This includes the determination carried out for thrust at N$^3$LL$^\prime\,$+$\,{\cal O}(\alpha_s^3)$ in Ref.\,\cite{Abbate:2010xh}\footnote{Note that results at N$^3$LL require the currently unknown QCD \mbox{four-loop} cusp anomalous, but conservative estimates show that this has a negligible impact on the perturbative uncertainties. Results at N$^3$LL$^\prime$ also technically require the unknown 3-loop non-logarithmic constants for the jet and soft functions which are also varied when determining our uncertainties, but these parameters turn out to only impact the peak region which is outside the range of their $\alpha_s(m_Z)$ fits in the resummation region.}, which is also consistent with analyses at N$^2$LL\,+\,${\cal O}(\alpha_s^3)$~\cite{Gehrmann:2009eh,Gehrmann:2012sc} which consider the resummation of logs at one lower order. In Ref.~\cite{Banfi:2014sua} a framework for a numerical code with N$^2$LL precision for many $e^+e^-$ event shapes was found, which could also be utilized for $\alpha_s$ fits in the future. In Ref.~\cite{Abbate:2010xh} it was pointed out that including a proper fit to power corrections for thrust causes a significant negative shift to the value obtained for $\alpha_s(m_Z)$, and this was also confirmed by subsequent analyses~\cite{Gehrmann:2012sc}. Recent results for $\alpha_s(m_Z)$ from $\tau$ decays~\cite{Boito:2014sta}, DIS data~\cite{Alekhin:2012ig}, the static potential for quarks~\cite{Bazavov:2014soa}, as well as global PDF fits~\cite{Ball:2011us,Martin:2009bu} also find values below the world average. This tension in the accepted value of the strong coupling motivates further analysis of event shapes to high precision in order to either confirm or refute the earlier results. This will be the goal of Chapters \ref{ch:Cparam-theory} and \ref{ch:Cparam-fit}.

\section{Review of SCET} \label{sec:SCET}
\begin{figure}[t!]
	\begin{center}
		\includegraphics[width=0.6\columnwidth]{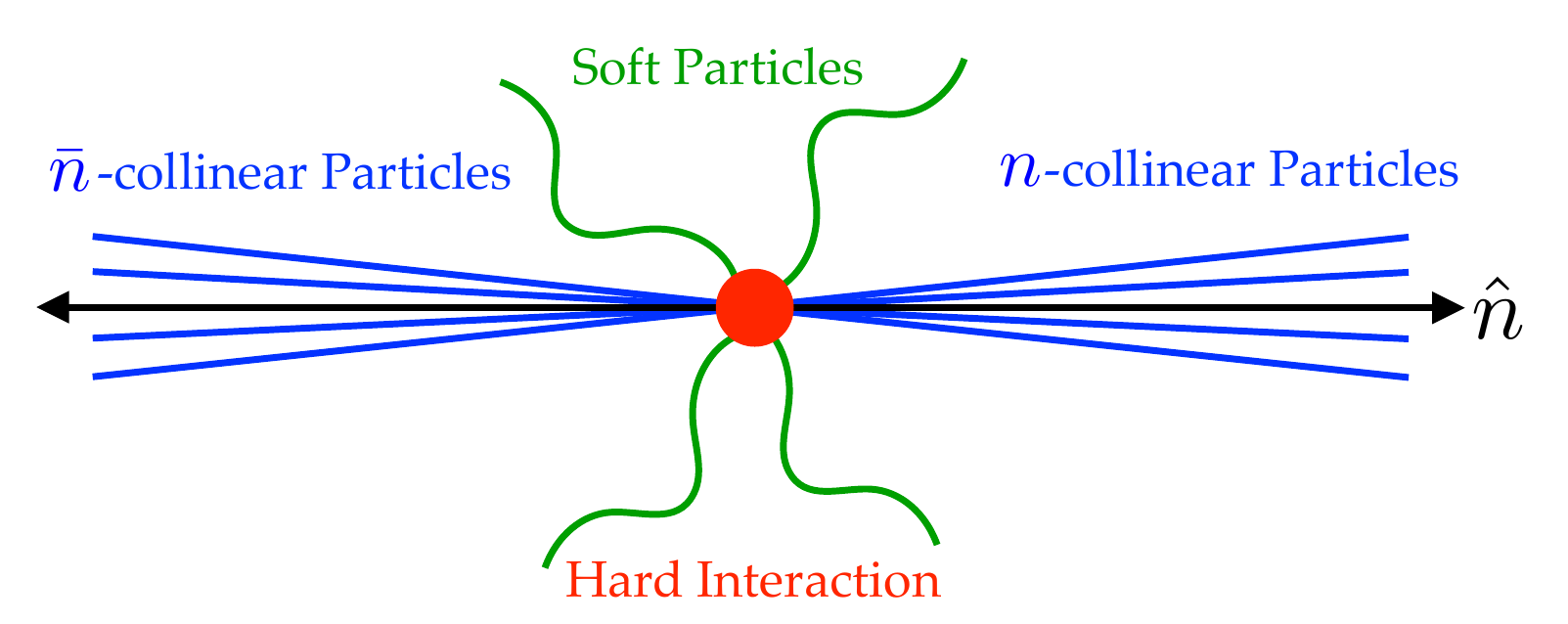}
		\caption[Kinematics of soft and collinear particles in SCET]{Kinematics of soft particles (green) and collinear particles (blue) in SCET. The red dot indicates the off-shell hard modes that are integrated out in the effective theory. }\label{fig:scet-kinematics} 
	\end{center}
\end{figure}

The collider events that we most often study  restrict particles to be in collimated jets or have low energy. The Soft-Collinear Effective Theory (SCET) is an effective field theory of QCD describing the interactions of collinear and soft particles in the presence of a hard interaction \cite{Bauer:2000ew, Bauer:2000yr, Bauer:2001ct, Bauer:2001yt, Bauer:2002nz}. A physical picture of a typical dijet event that would be well described by this theory is shown in \fig{scet-kinematics}. 
Since SCET describes collinear particles (which are characterized by a large momentum along a particular light-like direction), as well as soft particles, it is convenient to use light-cone coordinates. For each jet direction we define two light-like reference vectors $n_i^\mu$ and $\bn_i^\mu$ such that $n_i^2 = \bn_i^2 = 0$ and $n_i\sdt\bn_i = 2$. One typical choice for these quantities is
\begin{equation}
	n_i^\mu = (1, \vec{n}_i)
	\,,\qquad
	\bn_i^\mu = (1, -\vec{n}_i)
	\,,\end{equation}
where $\vec{n}_i$ is a unit three-vector. Given a choice for $n_i^\mu$ and $\bn_i^\mu$, any four-momentum $p$ can then be decomposed in light-cone coordinates as
\begin{equation} \label{eq:lightcone_dec}
	p^\mu =   n_i\sdt p\,\frac{\bn_i^\mu}{2} + \bn_i\sdt p\,\frac{n_i^\mu}{2} + p^\mu_{n_i\perp} \equiv (n_i \cdot p,\bn_i \cdot p, p_{n_i\,\perp}) \equiv (p^+,p^-,p_\perp)
	\,.\end{equation}
An ``$n_i$-collinear'' particle has momentum $p$ close to the $\vec{n}_i$ direction, so that the components of $p$ scale as $(n_i\!\cdot\! p, \bn_i \!\cdot\! p, p_{n_i\perp}) \sim \bn_i\!\cdot\! p$ $\,(\la^2,1,\la)$, where $\la \ll 1$ is a small parameter determined by the form of the measurement or kinematic restrictions. To ensure that $n_i$ and $n_j$ refer to distinct collinear directions, they have to be well separated, which corresponds to the condition
\begin{equation} \label{eq:nijsep}
	n_i\sdt n_j  \gg \la^2 \qquad\text{for}\qquad i\neq j
	\,.\end{equation}
Two different reference vectors, $n_i$ and $n_i'$, with $n_i\cdot n_i' \sim \ord{\lambda^2}$ both describe the same jet and corresponding collinear physics. Thus, each collinear sector can be labelled by any member of a set of equivalent vectors, $\{n_i\}$. This freedom is manifest as a symmetry of the effective theory known as reparametrization invariance (RPI) \cite{Manohar:2002fd,Chay:2002vy}. Specifically, the three classes of RPI transformations are
\begin{alignat}{3}\label{eq:RPI_def}
	&\text{RPI-I} &\qquad &  \text{RPI-II}   &\qquad &  \text{RPI-III} \nn \\
	&n_{i \mu} \to n_{i \mu} +\Delta_\mu^\perp &\qquad &  n_{i \mu} \to n_{i \mu}   &\qquad & n_{i \mu} \to e^\alpha n_{i \mu} \nn \\
	&\bar n_{i \mu} \to \bar n_{i \mu}  &\qquad &  \bar n_{i \mu} \to \bar n_{i \mu} +\epsilon_\mu^\perp  &\qquad & \bar n_{i \mu} \to e^{-\alpha} \bar n_{i \mu}\,.
\end{alignat}
Here, we have $\Delta^\perp \sim \lambda$, $\epsilon^\perp \sim \lambda^0$, and $\alpha\sim \lambda^0$. The parameters $\Delta^\perp$ and $\epsilon^\perp$ are infinitesimal, and satisfy $n_i\cdot \Delta^\perp=\bar n_i\cdot \Delta^\perp=n_i \cdot \epsilon^\perp=\bar n_i \cdot \epsilon^\perp=0$. RPI can be exploited to simplify the structure of operator bases within SCET.

The effective theory is constructed by expanding momenta into label and residual components
\begin{equation} \label{eq:label_dec}
	p^\mu = \lp^\mu + k^\mu = \bn_i \sdt\lp\, \frac{n_i^\mu}{2} + \lp_{n_i\perp}^\mu + k^\mu\,.
	\,\end{equation}
Here, $\bn_i \cdot\lp \sim Q$ and $\lp_{n_i\perp} \sim \la Q$ are the large label momentum components, where $Q$ is the scale of the hard interaction, while $k\sim \la^2 Q$ is a small residual momentum. A multipole expansion is then performed to obtain fields with momenta of definite scaling, namely collinear quark and gluon fields for each collinear direction, as well as soft quark and gluon fields. Independent gauge symmetries are enforced for each set of fields.

The SCET fields for $n_i$-collinear quarks and gluons, $\xi_{n_i,\lp}(x)$ and $A_{n_i,\lp}(x)$, are labeled by their collinear direction $n_i$ and their large momentum $\lp$. They are written in position space with respect to the residual momentum and in momentum space with respect to the large momentum components. Derivatives acting on the fields pick out the residual momentum dependence, $i \partial^\mu \sim k \sim \la^2 Q$. The large label momentum is obtained from the label momentum operator $\cP_{n_i}^\mu$, e.g. $\cP_{n_i}^\mu\, \xi_{n_i,\lp} = \lp^\mu\, \xi_{n_i,\lp}$. When acting on a product of fields, $\cP_{n_i}$ returns the sum of the label momenta of all $n_i$-collinear fields. For convenience, we define $\bnP_{n_i} = \bn\sdt\cP_{n_i}$, which picks out the large momentum component.  Frequently, we will only keep the label ${n_i}$ denoting the collinear direction, while the momentum labels are summed over (subject to momentum conservation) and suppressed.

The soft degrees of freedom separate SCET into two different effective theories. The first is \SCETi, where the low energy particles are ultrasofts, which have momenta that scale as $k^\mu \sim \lambda^2 Q$ for all components. The second is \SCETii, where the low energy particles are called softs and have momenta that scale as $k^\mu \sim \lambda Q$. For a given physical process, whether we are in \SCETi or \SCETii is determined by a measurement that constrains our low energy radiation to a particular scaling. For the C-parameter applications in the paper, we are in the \SCETi theory\,\footnote{Due to the fact that we are exclusively in the \SCETi theory, in Chapters \ref{ch:Cparam-theory} and \ref{ch:Cparam-fit}, we will simply call ultrasoft particles soft.}. When extending SCET beyond leading power using helicity formalism, we will include the tools needed for working in \SCETii, but our main focus will be \SCETi.

For SCET$_\text{I}$, the ultrasoft modes do not carry label momenta, but have residual momentum dependence with $i \partial^\mu \sim \la^2Q$. They are therefore described by fields $q_{us}(x)$ and $A_{us}(x)$ without label momenta. The ultrasoft degrees of freedom are able to exchange residual momenta between the jets in different collinear sectors. Particles that exchange large momentum of $\ord{Q}$ between different jets are off-shell by $\ord{n_i\cdot n_j Q^2}$, and are integrated out by matching QCD onto SCET.  Before and after the hard interaction the jets described by the different collinear sectors evolve independently from each other, with only ultrasoft radiation between the jets.

\begin{figure}[t!]
	\begin{center}
		\includegraphics[width=0.6\columnwidth]{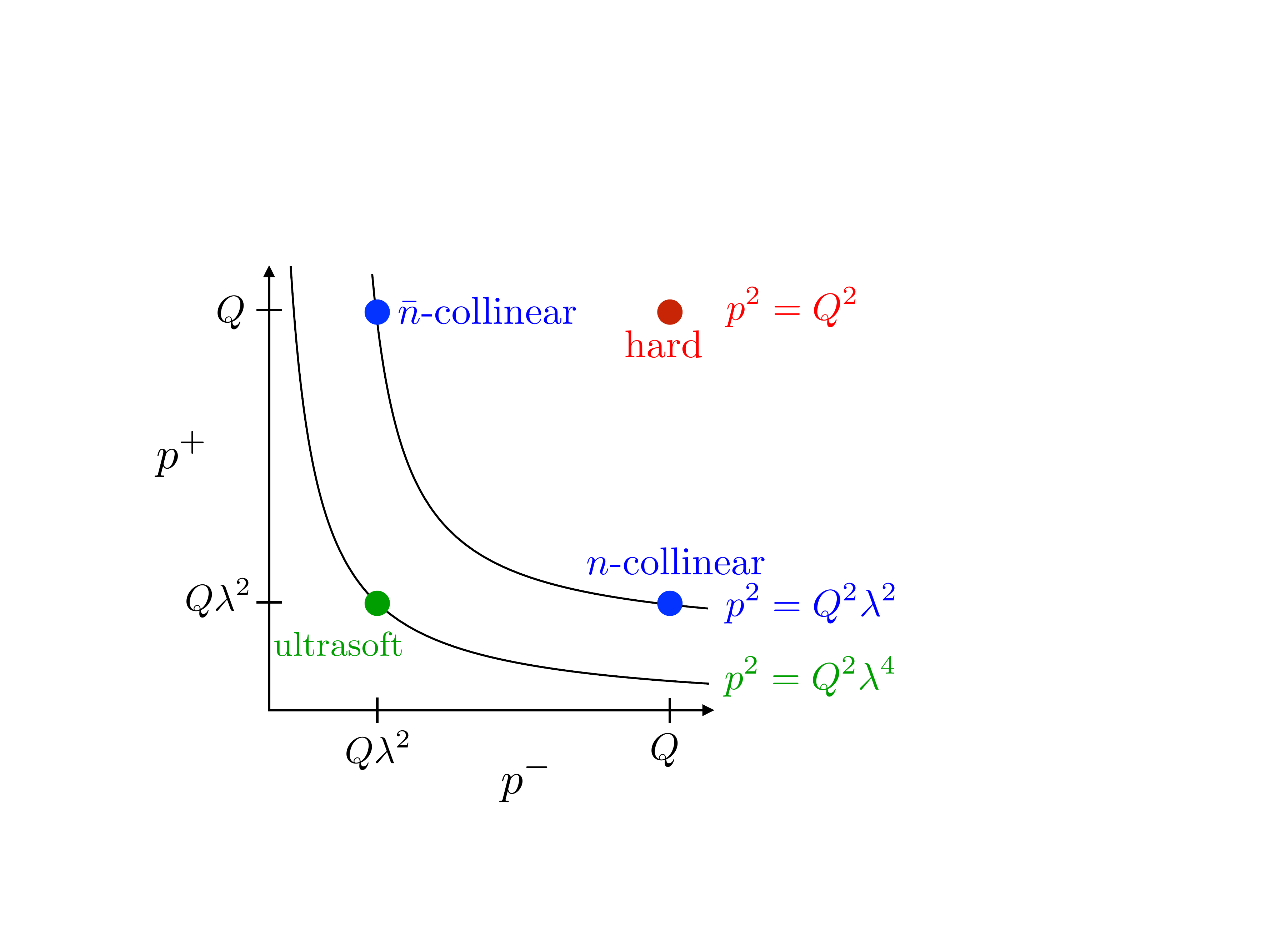}
		\caption[Modes in the \SCETa theory]{Location of the various modes that contribute in \SCETa. The axes indicate the projection of momentum along the $n$ and $\bn$ directions and each hyperbola lives at a different overall momentum scale, $p^2$.}\label{fig:scalehyperbola} 
	\end{center}
\end{figure}

The various modes for our \SCETa theory are summarized diagrammatically in \fig{scalehyperbola}. Each hyperbola lives at a different overall momentum scale, and our effective theory will allow us to separate the dynamics at these different scales. Any exchange of momentum between modes must occur in the component that sits at a common scale (e.g. the $p^-$ momentum for ultrasoft and $\bn$-collinear modes).

SCET is formulated as an expansion in powers of $\la$, constructed so that manifest power counting is maintained at all stages of a calculation. As a consequence of the multipole expansion, all fields acquire a definite power counting \cite{Bauer:2001ct}, shown in \Tab{tab:PC}. The SCET Lagrangian is also expanded as a power series in $\lambda$
\begin{align} \label{eq:SCETLagExpand}
	\cL_{\text{SCET}}=\cL_\hard+\cL_\dyn= \sum_{i\geq0} \cL_\hard^{(i)}+\sum_{i\geq0} \cL^{(i)} \,,
\end{align}
where $(i)$ denotes objects at ${\cal O}(\lambda^i)$ in the power counting. The Lagrangians $ \cL_\hard^{(i)}$ contain the hard scattering operators $O^{(i)}$, whose structure is determined by the matching process to QCD. The $\cL^{(i)}$ describe the dynamics of ultrasoft and collinear modes in the effective theory.  Expressions for the leading power Lagrangian $\cL^{(0)}$ can be found in~\cite{Bauer:2001yt}, and expressions for ${\cal L}^{(1)}$, and ${\cal L}^{(2)}$ can be found in~\cite{Bauer:2003mga} (see also~\cite{Pirjol:2002km,Beneke:2002ni,Chay:2002vy,Manohar:2002fd,Beneke:2002ph}). 

Much of the power of SCET comes in the ability to separate the hard, collinear and soft scales when making calculations for particle colliders. Specifically,
factorization theorems used in jet physics are typically derived at leading power in $\lambda$. In this case, interactions involving hard processes in QCD are matched to a basis of leading power SCET hard scattering operators $O^{(0)}$, the dynamics in the effective theory are described by the leading power Lagrangian, $\cL^{(0)}$, and the measurement function, which defines the action of the observable, is expanded to leading power. Higher power terms in the $\lambda$ expansion, known as power corrections, arise from three sources: subleading power hard scattering operators $O^{(i)}$, subleading Lagrangian insertions, and subleading terms in the expansion of the measurement functions which act on soft and collinear radiation. The first two sources are independent of the form of the particular measurement, while the third depends on its precise definition.

\begin{table}
	\begin{center}
		\begin{tabular}{| l | c | c |c |c|c| r| }
			\hline                       
			Operator & $\cB_{n_i\perp}^\mu$ & $\chi_{n_i}$& $\cP_\perp^\mu$&$q_{us}$&$D_{us}^\mu$ \\
			Power Counting & $\lambda$ &  $\lambda$& $\lambda$& $\lambda^3$& $\lambda^2$ \\
			\hline  
		\end{tabular}
	\end{center}
	\caption{
		Power counting for building block operators in $\text{SCET}_\text{I}$.
	}
	\label{tab:PC}
\end{table}

In SCET, collinear operators are constructed out of products of fields and Wilson lines that are invariant under collinear gauge transformations~\cite{Bauer:2000yr,Bauer:2001ct}.  The smallest building blocks are collinearly gauge-invariant quark and gluon fields, defined as
\begin{align} \label{eq:chiB}
	\chi_{{n_i},\w}(x) &= \Bigl[\delta(\w - \bnP_{n_i})\, W_{n_i}^\dagger(x)\, \xi_{n_i}(x) \Bigr]
	\,,\\
	\cB_{{n_i}\perp,\w}^\mu(x)
	&= \frac{1}{g}\Bigl[\delta(\w + \bnP_{n_i})\, W_{n_i}^\dagger(x)\,i  D_{{n_i}\perp}^\mu W_{n_i}(x)\Bigr]
	\,. \nn
\end{align}
By considering the case of no emissions from the Wilson lines, we can see that with this definition of $\chi_{{n_i},\w}$ we have $\w > 0$ for an incoming quark and $\w < 0$ for an outgoing antiquark. For $\cB_{{n_i},\w\perp}$, $\w > 0$ ($\w < 0$) corresponds to outgoing (incoming) gluons. In \eq{chiB},
\begin{equation}
	i  D_{{n_i}\perp}^\mu = \cP^\mu_{{n_i}\perp} + g A^\mu_{{n_i}\perp}\,,
\end{equation}
is the collinear covariant derivative and
\begin{equation} \label{eq:Wn}
	W_{n_i}(x) = \biggl[~\sum_\text{perms} \exp\Bigl(-\frac{g}{\bnP_{n_i}}\,\bn\sdt A_{n_i}(x)\Bigr)~\biggr]\,,
\end{equation}
is a Wilson line of ${n_i}$-collinear gluons in label momentum space. In general the structure of Wilson lines must be derived by a matching calculation from QCD. These Wilson lines sum up arbitrary emissions of ${n_i}$-collinear gluons off of particles from other sectors, which, due to the power expansion, always appear in the ${\bar{n}_i}$ direction. The emissions summed in the Wilson lines are $\ord{\lambda^0}$ in the power counting. The label operators in \eqs{chiB}{Wn} only act inside the square brackets. Since $W_{n_i}(x)$ is localized with respect to the residual position $x$, we can treat
$\chi_{{n_i},\w}(x)$ and $\cB_{{n_i},\w}^\mu(x)$ as local quark and gluon fields from the perspective of ultrasoft derivatives $\partial^\mu$ that act on $x$. 

The complete set of collinear and ultrasoft building blocks for constructing hard scattering operators or subleading Lagrangians at any order in the power counting is given in \Tab{tab:PC}. All other field and derivative combinations can be reduced to this set by the use of equations of motion and operator relations~\cite{Marcantonini:2008qn}. Since these building blocks carry vector or spinor Lorentz indices they must be contracted to form scalar operators, which also involves the use of objects like $\{n_i^\mu, \bn_i^\mu, \gamma^\mu, g^{\mu\nu}, \epsilon^{\mu\nu\sigma\tau}\}$. As we will see in Ch. \ref{ch:Subleading} a key advantage of the helicity operator approach discussed below is that this is no longer the case; all the building blocks will be scalars. 

As shown in \Tab{tab:PC}, both the collinear quark and collinear gluon building block fields scale as ${\cal O}(\lambda)$. For the majority of jet processes there is a single collinear field operator for each collinear sector at leading power.  (For fully exclusive processes that directly produce hadrons there will be multiple building blocks from the same sector in the leading power operators since they form color singlets in each sector.) Also, since $\cP_\perp\sim \lambda$, this operator will not typically be present at leading power (exceptions could occur, for example, in processes picking out P-wave quantum numbers). At subleading power, operators for all processes can involve multiple collinear fields in the same collinear sector, as well as $\cP_\perp$ operator insertions. The power counting for an operator is obtained by adding up the powers for the building blocks it contains. To ensure consistency under renormalization group evolution the operator basis in SCET must be complete, namely all operators consistent with the symmetries of the problem must be included.

Dependence on the ultrasoft degrees of freedom enters the operators through the ultrasoft quark field $q_{us}$, and the ultrasoft covariant derivative $D_{us}$, defined as 
\begin{equation}
	i  D_{us}^\mu = i  \partial^\mu + g A_{us}^\mu\,,
\end{equation}
from which we can construct other operators including the ultrasoft gluon field strength. All operators in the theory must be invariant under ultrasoft gauge transformations. Collinear fields transform under ultrasoft gauge transformations as background fields of the appropriate representation. The power counting for these operators is shown in \Tab{tab:PC}. Since they are suppressed relative to collinear fields, ultrasoft fields typically do not enter factorization theorems in jet physics at leading power. An example where ultrasoft fields enter at leading power is $B \to X_s \gamma$ in the photon endpoint region, which is described at leading power by a single collinear sector, and an ultrasoft quark field for the b quark. 

One of the most powerful aspects of SCET is the ability to fully decouple the ultrasoft and collinear degrees of freedom. Specifically, this is done through BPS field redefinition defined by \cite{Bauer:2002nz}
\be \label{eq:BPSfieldredefinition}
\cB^{a\mu}_{n\perp}\to \cY_n^{ab} \cB^{b\mu}_{n\perp} , \qquad \chi_n^\alpha \to Y_n^{\alpha \beta} \chi_n^\beta,
\ee
and is performed in each collinear sector. Here $Y_n$, $\cY_n$ are fundamental and adjoint ultrasoft Wilson lines, respectively, and we note that

\be \label{eq:adjointtofundamental}
Y_n T^a Y_n^\dagger =  T^b {\cal Y}_n^{ba}\,.
\ee
For a general representation, r, the ultrasoft Wilson line is defined by
\be
Y^{(r)}_n(x)=\bold{P} \exp \left [ -ig \int\limits_{0}^{\infty} ds\, n\cdot A^a_{us}(x+sn)  T_{(r)}^{a}\right]\,,
\ee
where $\bold P$ denotes path ordering.  The BPS field redefinition has the effect of decoupling the ultrasoft degrees of freedom from the leading power collinear Lagrangian \cite{Bauer:2002nz}. When this is done consistently for S-matrix elements it accounts for the full physical path of ultrasoft Wilson lines~\cite{Chay:2004zn,Arnesen:2005nk}.
After the BPS field redefinition, the collinear fields $\cB_{n\perp}$, and $\chi_n$ are ultrasoft gauge singlets, but still carry a global color index. As we will see in Ch. \ref{ch:Subleading}, we can use the BPS field redefinition to define ultrasoft quark and gluon fields that are ultrasoft gauge invariant.   

Post BPS field redefinition, the separation of the hard, collinear and soft modes in SCET leads us naturally into a factorization framework. Using event shapes as an example, we focus on a certain class of $e^+ e^- \to$ dijet observables, determined by specifying a function $f_e$ according to
\begin{align} \label{eq:general_event_shape}
e(X) = \frac{1}{Q} \sum_{i \in X} f_e (\eta_i) | \mathbf{p}_i^T |\,,
\end{align}
where the sum is over final state particles, $\eta_i$ gives the rapidity of each with respect to the beam axis and $|\mathbf{p}_i^T|$ gives the magnitude of the transverse momentum of each. If we have this particular decomposition, we can write the cross section as a factorized product of SCET functions \cite{Bauer:2008dt},

\begin{align} \label{eq:facorizationdef}
\frac{1}{\sigma_0} \frac{\df \sigma}{\df e} = H(\mu) \int \df e_1 \df e_2 J_1(e_1, \mu) J_2(e_2, \mu) S(e, \mu) \delta(e - e_1 - e_2)\,.
\end{align}

Here $H$ is the hard function, which is related to the Wilson coefficients of the matching between SCET and QCD and is perturbatively calculable. $J_1$ and $J_2$ are jet functions, which incorporate the collinear radiation from each of the final state jets and are given by the expectation of collinear quark fields at leading power. The soft function $S$ is given by a matrix element of ultrasoft Wilson lines and encodes the behavior of the ultrasoft radiation. Additionally, as we will see later, the leading nonpertubative physics is included in this function.

The equation above ignores power suppressed terms, which scale with our power counting parameter $\lambda$. For a \SCETa type observable, we have the relationship $e \sim \lambda^2$. From this power counting, we see that each of these factorized objects (which incorporates separated modes from our theory) has a different natural energy scale. We usually consider the hard scale to be $\mu_H \sim Q$, the center of mass energy of the collision. The jet functions live at the scale $\mu_J \sim Q \sqrt{e}$ and the soft function lives at the lowest scale, $\mu_S \sim Qe$. Another way to see this is to observe the logs that show up in the perturbative calculation of each function. The hard function will contain $\log^n (\mu/Q)$, the jet function will have $\log^n (\mu/(Q \sqrt{e}))$ and the soft function will carry $\log^n (\mu / (Qe))$.

\begin{figure}[t!]
 	\begin{center}
 		\includegraphics[width=0.45\columnwidth]{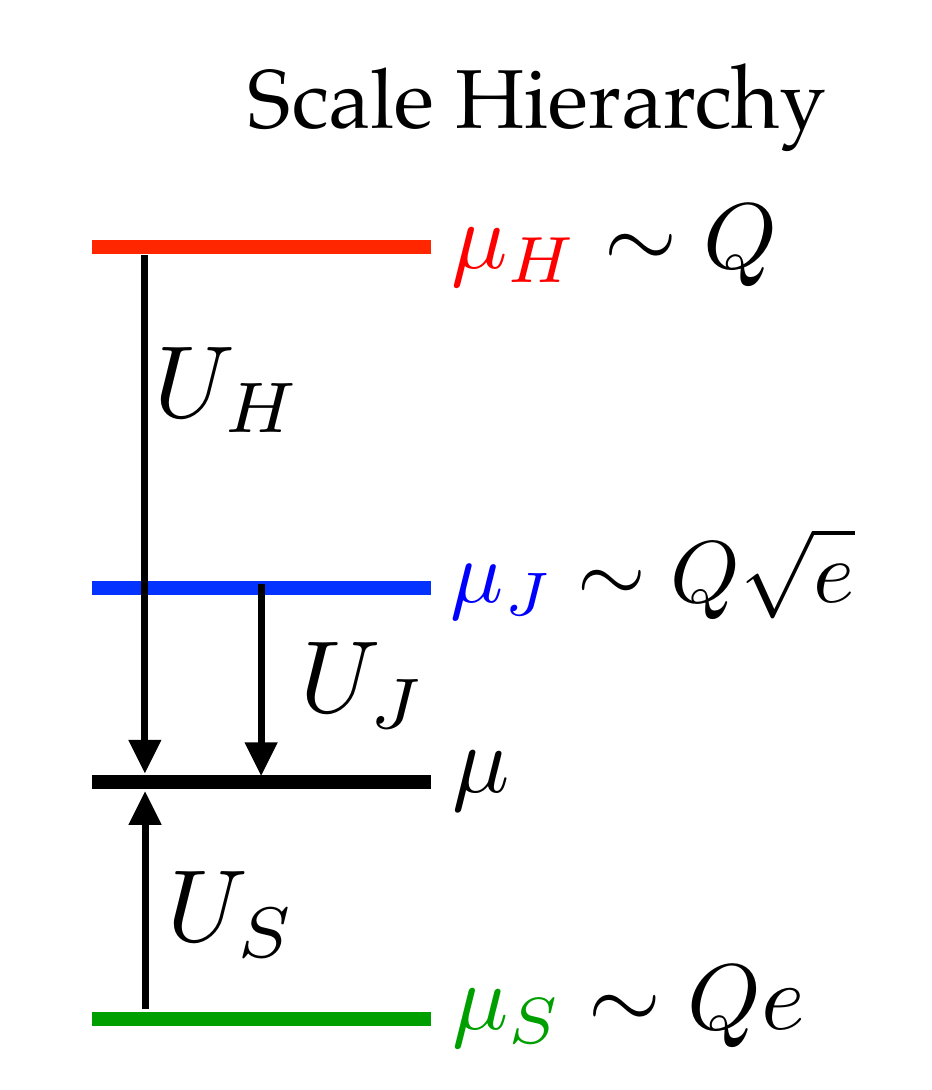}
 		\caption[Hierarchy of scales for SCET resummation]{Schematic drawing of of the hierarchy of scales for $e^+ e^- \to$ dijets resummation in SCET. The $U$ factors resum the large logs between different scales. The exact location of the common scale $\mu$ is arbitrary.}\label{fig:scalehierarchy} 
 	\end{center}
 \end{figure}

In order to have a well behaved cross section, we want the logs in each of these functions to be small. Obviously, this is not possible with a single choice of the scale $\mu$. In order to set each function to its own natural scale and minimize the logs, we must calculate how each object runs with the resolution scale, $\mu$. By calculating and solving the renormalization group equations for each function to a given order, we resum a specific tower of the large logs shown in \eq{log-towers}. After solving these equations, the resummed factorization formula can be written as
 \begin{align} \label{eq:facorizationresummed}
 \frac{1}{\sigma_0} \frac{\df \sigma}{\df e} = H(\mu_H) U_H(\mu_H,\mu) \int \df e_1 \df e_2 &\bigl(J_1(e_1, \mu_J) \otimes U_{J_1}(\mu_J,\mu) \bigr) \bigl(J_2(e_2, \mu) \otimes U_{J_2}(\mu_J,\mu) \bigr)
 \nn \\
 & \times  \bigl(S(e, \mu_S) \otimes U_{S}( \mu_S,\mu) \bigr)\delta(e - e_1 - e_2)\,,
 \end{align}
 where each $U$ is simply the appropriate function that solves the running equations in order to take the hard, jet or soft functions from their natural scale to the common scale $\mu$. As is shown schematically by the $\otimes$ symbol, these evolution kernels are convolved with the associated function evaluated at the appropriate scale. This running scheme is summarized in \fig{scalehierarchy}. The state of the art in precision calculations does this resummation at Next-to-next-to-next-to leading log (N$^3$LL), meaning that it includes all logs that scale as $\alpha_s^{n+2} \log^{n} e$ in the log of the cumulant of our distribution.

 From both a formal and a phenomenological perspective, we can see that SCET is a powerful tool. The fields involved capture only the most important soft and collinear degrees of freedom. Due to this, it is extremely useful for doing precise calculations in situations (such as $e^+ e^- \to$ dijets), where these degrees of freedom dominate. We will now turn to the main purpose of this work: using and expanding this tool for further precision in the realm of  collider physics.

\section{Outline} \label{sec:Outline}

Chapter \ref{ch:Cparam-theory} gives the theoretical results necessary for calculation the N${}^3$LL$'$ resummed C-parameter cross section. We begin by explaining the properties and kinematics related to C-parameter. Next, we give a detailed explanation of the factorization and resummation. The full range of effects that are included to give the most accurate result for the cross section are then described, including the nonsingular terms, the extraction of 2 loop soft parameters, the power corrections (including hadron mass effects) and the profile functions required for proper resummation. The final part of this chapter enumerates the theoretical results for the precise calculation of the C-parameter cross section, including plots that show the convergence and predictions, at the highest order of resummation. Formula that are required to calculate the cross section are given in App. \ref{ap:Cformula} and a general calculation of the one loop soft function for an event shape of the form given in \eq{general_event_shape} is given in App. \ref{ap:softoneloop}. This work has been published in \cite{Hoang:2014wka}.

In chapter \ref{ch:Cparam-fit}, we present the extraction of $\alpha_s(M_Z)$ from experimental data. First, we explain the data used in the measurement. Then, we go through the fit procedure. Finally, we present results for the values of $\alpha_s(M_Z)$ and the first hadronic moment $\Omega_1$. These results are compared with earlier thrust measurements to show universality in event shapes. Additionally, in App. \ref{app:fit-compare} we compare our final fits to those done with alternate renormalon subtractions as well as with older thrust profiles. This determination of $\alpha_s(M_Z)$ was first published in \cite{Hoang:2015hka}.

The content of chapter \ref{ch:Subleading} is a discussion of subleading helicity building blocks for SCET. First, we review the use of helicity fields at leading power in SCET. We then give the complete set of helicity based building blocks to construct operators at subleading power. Following, we discuss the angular momentum constraints that can arise at subleading power and how they limit the allowed helicity field content in each collinear sector. We conclude this chapter with an example of a process with two collinear directions, which illustrates the improvement that these helicity building blocks give in organizing and constructing operator bases. Associated with this chapter are our helicity conventions and several useful identities in App. \ref{app:helicity}. This work has been submitted for publicaiton in \cite{Kolodrubetz:2016uim}. We conclude in chapter \ref{ch:conclusion}.

\chapter{C-Parameter at N${}^3$LL$'$}
\label{ch:Cparam-theory}

In this chapter we give the calculation of the cross section for C-parameter at an $e^+ e^-$ collider, resumming the large logs to Next-to-next-to-next-to-leading order (N${}^3$LL$'$). This calculation was first presented in \cite{Hoang:2014wka}. After motivating this study in \sec{motivation}, we define the C-parameter and enumerate some of its important properties in \sec{definition}. Then, in \secs{kinematics}{factorization-singular} we develop a factorization formula in the dijet limit. Sec. \ref{sec:nonsingular} is devoted to understanding the kinematically nonsingular terms while \sec{softtwoloop} contains an extraction of the two loop pieces of the soft function. Other effects, including power corrections, the renormalon free scheme and hadron mass effects are discussed in \secs{power}{hadronmass}. The profile functions used to set the renormalization scales are presented in \sec{profiles}. Finally, we present the full set of results of this calculation in \sec{theory-results}.

There are seven sections of appendices associated with this chapter: Appendix~\ref{ap:formulae} provides all needed formulae for the singular cross section beyond those in the main body of the chapter. In App.~\ref{ap:MC-comparison} we present a comparison of our SCET prediction, expanded at fixed order, with the numerical results in full QCD at $\mathcal{O}(\alpha_s^2)$ and $\mathcal{O}(\alpha_s^3)$ from EVENT2 and EERAD3, respectively. In App.~\ref{ap:Gijcoefficients} we give analytic expressions for the $B_i$ and $G_{ij}$ coefficients of the fixed-order singular logs up to $\mathcal{O}(\alpha_s^3)$ according to the exponential formula of Sec.~\ref{subsec:Gij}. The R-evolution of the renormalon-free gap parameter is described in App.~\ref{ap:hadronmassR}. Appendix~\ref{ap:ArcTan} is devoted to a discussion of how the Rgap scheme is handled in the shoulder region above $C_{\rm shoulder}=3/4$. In App.~\ref{ap:subtractionchoice} we show results for the perturbative gap subtraction series based on the C-parameter soft function.  Finally, in App.~\ref{ap:softoneloop} we give a general formula for the one-loop soft function, valid for any event shape which is not recoil sensitive. 

\section{Motivation and Background} \label{sec:motivation}

As discussed in \ref{sec:eventShapesColliders}, SCET has been used for high accuracy calculations of several event shapes, most notably thrust, which provided a high precision extraction of $\alpha_s(m_Z)$ \cite{Abbate:2010xh}. Our main motivations for studying the C-parameter distribution are to:
\begin{itemize}
	\item[a)] Extend the theoretical precision of the logarithmic resummation for C-parameter from next-to-leading log (NLL)$\,+\,{\cal O}(\alpha_s^3)$ to N$^3$LL\,+\,${\cal O}(\alpha_s^3)$.
	\item [b)] Implement the leading power correction $\Omega_1$ using only field theory and with sufficient theoretical precision to provide a serious test of universality between C-parameter and thrust.
	\item [c)] Determine $\alpha_s(m_Z)$ using N$^3$LL$^\prime$\,+\,${\cal O}(\alpha_s^3)+\Omega_1$ theoretical precision for $C$, to make this independent extraction competitive with the thrust analysis carried out at this level in Refs.~\cite{Abbate:2010xh,Abbate:2012jh}.
\end{itemize}
In this chapter we present the theoretical calculation and analysis that yields a N$^3$LL$^\prime$\,+\,${\cal O}(\alpha_s^3)+\Omega_1$ cross section for C-parameter, and we analyze its convergence and perturbative uncertainties. The next chapter will be devoted to a numerical analysis that obtains $\alpha_s(m_Z)$ from a fit to a global \mbox{C-parameter} dataset and investigates the power correction universality.

A nice property of C-parameter is that its definition does not involve any minimization procedure, unlike thrust. This makes its determination from data or Monte Carlo simulations computationally inexpensive.
Unfortunately, this does not translate into a simplification of perturbative theoretical computations, which are similar to those for thrust.

The resummation of singular logarithms in \mbox{C-parameter} was first studied by Catani and Webber in Ref.~\cite{Catani:1998sf} using the coherent branching formalism~\cite{Catani:1992ua}, where NLL accuracy was achieved. Making use of SCET, we achieve a resummation at N$^3$LL order. The relation between thrust and \mbox{C-parameter} in SCET discussed here has been used in the Monte Carlo event generator GENEVA \cite{Alioli:2012fc}, where a next-to-next-to-leading log primed (N$^2$LL$'$) C-parameter result was presented. Nonperturbative effects for the C-parameter distribution have been studied by a number of authors: Gardi and Magnea \cite{Gardi:2003iv}, in the context of the dressed gluon approximation; Korchemsky and Tafat \cite{Korchemsky:2000kp}, in the context of a shape function; and Dokshitzer and Webber \cite{Dokshitzer:1995zt}, in the context of the dispersive model.

Catani and Webber\cite{Catani:1998sf} showed that up to NLL the cross sections for thrust and the reduced C-parameter
\begin{align}
\widetilde C = \frac{C}{6} \,,
\end{align}
are identical. Gardi and Magnea \cite{Gardi:2003iv} showed that this relation breaks down beyond NLL due to soft radiation
at large angles. Using SCET we confirm and extend these observations by demonstrating that the hard and jet functions, along
with all anomalous dimensions, are identical for thrust and $\widetilde{C}$ to all orders in perturbation theory. At any order in perturbation theory, the perturbative non-universality of the singular terms appears only through fixed-order terms in the soft function, which differ starting at $\mathcal{O}(\alpha_s)$.

There is also a universality between the leading power corrections for thrust and C-parameter which has been widely
discussed~\cite{Dokshitzer:1995zt,Akhoury:1995sp,Korchemsky:1994is,Lee:2006nr}. This universality has been proven nonperturbatively in Ref.~\cite{Lee:2006nr} using the field theory definition of the leading power correction with massless particle kinematics. In our notation this relation is
\begin{align} \label{eq:O1c}
\Omega_1^C = \frac{3\pi}{2}\, \Omega_1^\tau  \,.
\end{align}
Here $\Omega_1^e$ is the first moment of the nonperturbative soft functions for the event shape $e$ and, in the tail of the
distribution acts to shift the event shape variable 
\begin{align} \label{eq:shift}
\hat\sigma(e)\to \hat\sigma(e - \Omega_1^e/Q)\,,
\end{align}
at leading power.
The exact equality in \eq{O1c}
can be spoiled by hadron-mass effects~\cite{Salam:2001bd}, which have been formulated using a field theoretic definition
of the $\Omega_1^e$ parameters in Ref.~\cite{Mateu:2012nk}.  Even though nonzero hadron masses can yield quite large effects
for some event shapes, the universality breaking corrections between thrust and \mbox{C-parameter} are at the $2.5\%$ level and hence for our purposes are small relative to other uncertainties related to determining $\Omega_1$. Since relations like Eq.~(\ref{eq:O1c}) do not hold for higher moments $\Omega_{n>1}^e$ of the nonperturative soft functions, these are generically different for thrust and C-parameter.

\begin{figure}[t!]
	\begin{center}
		\includegraphics[width=0.65\columnwidth]{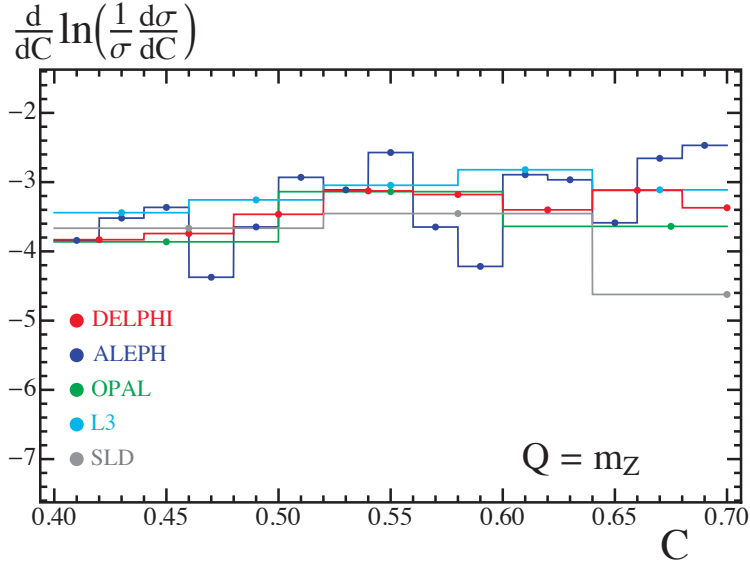}
		\caption[Plot of $h'(C)/h(C) $ from data]{Plot of $\df/\df C\, \ln[ (1/\sigma)\df\sigma/\df C]\simeq h'(C)/h(C)$,  computed from experimental data at \mbox{$Q=m_Z$}. The derivative is calculated using the central difference with neighboring points. }\label{fig:Numerical-derivative} 
	\end{center}
\end{figure}

Following Ref.~\cite{Abbate:2010xh}, a rough estimate of the
impact of power corrections can be obtained from the
experimental data with very little theoretical input.
We write $(1/\sigma)\, \df\sigma/\df C \simeq  h(C-\Omega_1^C/Q) = h(C) - h^\prime(C)\, \Omega_1^C/Q +\ldots$ for the tail region,
and assume the perturbative function $h(C)$ is proportional to $\alpha_s$. Then one can easily derive that if a value $\alpha_s$
is extracted from data by setting $\Omega_1^C=0$ then the change in the extracted value $\delta\alpha_s$ when $\Omega_1^C$
is present will be
\begin{align}
\frac{\delta\alpha_s}{\alpha_s} \simeq \frac{\Omega_1^C}{Q}\: \frac{h^\prime(C)}{h(C)} \,,
\end{align}
where the slope factor $h'(C)/h(C)$ should be constant at the level of these approximations.  By looking at the experimental results at the Z-pole shown in
Fig.~\ref{fig:Numerical-derivative}, we see that this is true at the level expected from these approximations, finding \mbox{$h'(C)/h(C) \simeq -\,3.3\pm 0.8$}.  This same analysis
for thrust \mbox{$T=1-\tau$} involves a different function $h$ and yields $[h'(\tau)/h(\tau)]_\tau \simeq -\,14\pm 4$~\cite{Abbate:2010xh}.  It is interesting to note that even this very simple analysis
gives a value $[\,h'(\tau)/h(\tau)\,]_\tau/[\,h'(C)/ h(C)\,]_C \simeq 4.2$ that is very close to the universality prediction of $3\pi/2 = 4.7$.
In the context of Eq.~(\ref{eq:O1c}), this already implies that in a C-parameter analysis we can anticipate the impact of the power correction in the extraction of the strong coupling to be quite similar to that in
the thrust analysis~\cite{Abbate:2010xh}, where $\delta\alpha_s/\alpha_s \simeq -\,9\%$.

In Ref.~\cite{Catani:1997xc} it was shown that within perturbation theory the C-parameter distribution reaches an infinite
value at a point in the physical spectrum $0 < C < 1$,  despite being an infrared and collinear safe observable.  This happens
for the configuration that distinguishes planar and non-planar partonic events and first occurs at $\mathcal{O}(\alpha_s^2)$ where one has enough partons to create a non-planar event at the value $C_{\rm shoulder} = 3/4$. However, this singularity is
integrable and related to the fact that at $\mathcal{O}(\alpha_s)$ the cross section does not vanish at the three-parton
endpoint $C_{\rm shoulder}$. In Ref.~\cite{Catani:1997xc} this deficiency was cured by performing soft gluon resummation
at $C_{\rm shoulder}$ to achieve a smooth distribution at leading-log (LL) order.
Since $C_{\rm shoulder}$ is far away from the dijet limit (in fact, it is a pure three-jet configuration), we will not include this resummation. In our analysis the shoulder effect is included in the non-singular contributions in fixed-order, and when the partonic distribution is convolved with a nonperturbative shape function, the shoulder effect is smoothed out, providing a continuous cross section across the entire $C$ range.

\section{Definition and properties of C-parameter}
\label{sec:definition}
\mbox{C-parameter} is defined in terms of the linearized momentum tensor
\cite{Parisi:1978eg,Donoghue:1979vi},
\begin{align}\label{eq:lintensor}
\Theta^{\alpha\beta}\,=\,\frac{1}{\sum_i |{\vec p}_i|}
\sum_i\frac{p_i^\alpha p_i^\beta}{|{\vec p}_i|}\,,
\end{align}
where $\alpha=1,2,3$ are spacial indices and $i$ sums over all final state particles. Since $\Theta$ is a symmetric positive semi-definite matrix\,\footnote{This property follows trivially from Eq.~(\ref{eq:lintensor}), since for any three-vector $\vec q$ one has that
	$q_\alpha q_\beta\Theta^{\alpha\beta}\propto \sum_i({\vec q}\cdot {\vec p}_i)^2/|{\vec p}_i|\ge 0$.},
its eigenvalues are real and non-negative. Let us denote them by
$\lambda_i$, \mbox{$i=1,2,3$}. As $\Theta$ has unit trace, $\sum_i \lambda_i = 1$, which implies that the eigenvalues are bounded $0\le \lambda_i\le 1$.
Without loss of generality we can assume $1\ge \lambda_1 \ge \lambda_2 \ge \lambda_3 \ge 0$. The characteristic polynomial for the eigenvalues of $\Theta$ is:
\begin{align}
x^3 -x^2 + (\lambda_1 \lambda_2 + \lambda_1 \lambda_3 + \lambda_2 \lambda_3)\,x 
- \lambda_1 \lambda_2 \lambda_3 = 0\,.
\end{align}
\mbox{C-parameter} is defined to be proportional to the coefficient of the term linear in $x$:
\begin{align}\label{eq:C-lambda}
C & = 3\, (\lambda_1 \lambda_2 + \lambda_1 \lambda_3 + \lambda_2 \lambda_3)\\
& = 3\, [\,(\lambda_1+\lambda_2)(1-\lambda_1) - \lambda_2^2\,]\,,\nonumber
\end{align}
where we have used the unit trace property to write $\lambda_3$ in terms of $\lambda_1$
and $\lambda_2$ in order to get the second line. Similarly one defines \mbox{D-parameter}
as $D = 27\,\lambda_1 \lambda_2 \lambda_3$, proportional to the $x$-independent term
in the characteristic equation. Trivially one also finds that
\begin{align}
{\rm Tr}\,[\,\Theta^2\,] = \sum_i \lambda_i^2 = 1 \,-\, \frac{2}{3}\,C\,,
\end{align}
where again we have used $\lambda_3 = 1\,-\,\lambda_1\,-\,\lambda_2$. We can easily compute
${\rm Tr}\,[\,\Theta^2\,]$ using Eq.~(\ref{eq:lintensor})
\begin{align}
{\rm Tr}\,[\,\Theta^2\,] & = \frac{1}{(\sum_i |{\vec p}_i|)^2}
\sum_{ij}\frac{({\vec p}_i\cdot {\vec p}_j)^2}{|{\vec p}_i||{\vec p}_j|}\\\nonumber
& = \frac{1}{(\sum_i |{\vec p}_i|)^2}
\sum_{ij}|{\vec p}_i| |{\vec p}_j| \cos^2 \theta_{ij} \\\nonumber
& = 1 \,-\, \frac{1}{(\sum_i |{\vec p}_i|)^2}
\sum_{ij}|{\vec p}_i| |{\vec p}_j| \sin^2 \theta_{ij}\,.
\end{align}
From the last relation one gets the familiar expression:
\begin{equation} \label{eq:Ckinematicdef}
C=\frac{3}{2} \, \frac{\sum_{i,j} | \vec{p}_i | | \vec{p}_j | \sin^2 \theta_{ij}}
{\left( \sum_i | \vec{p}_i | \right)^2}\,.
\end{equation}
From Eq.~(\ref{eq:Ckinematicdef}) and the properties of $\lambda_i$, it follows that $C \ge 0$, and from the second line of
Eq.~(\ref{eq:C-lambda}), one finds that $C\le 1$, and the maximum value is achieved for the
symmetric configuration \mbox{$\lambda_1 = \lambda_2 = \lambda_3 = 1/3$}. Hence, \mbox{$0\le C\le 1$}.
Planar events have $\lambda_3 = 0$.  To see this simply consider that the planar event defines
the $x-y$ plane, and then any vector in the $z$ direction is an eigenstate of $\Theta$ with zero eigenvalue.
Hence, planar events have $D=0$ and $C=3\,\lambda_1\,(1-\lambda_1)$, which gives a maximum value for
$\lambda_1 = \lambda_2 = 1/2$, and
one has $0\le C_{\rm planar} \le 3/4$. Thus $C>3/4$ needs at least four particles in the final state.
\mbox{C-parameter} is related to the first non-trivial Fox--Wolfram parameter~\cite{Fox:1978vu}. The Fox--Wolfram event
shapes are defined as follows:
\begin{align}\label{eq:Fox-Wolfram}
H_\ell = \sum_{ij}\frac{| \vec{p}_i | | \vec{p}_j |}{Q^2}\,P_\ell (\cos \theta_{ij})\,.
\end{align}
One has $H_0=1$, $H_1=0$, and
\begin{align}
H_2=1-\frac{3}{2}\,\frac{1}{Q^2}\sum_{ij}| \vec{p}_i | | \vec{p}_j |\sin^2 \theta_{ij}\,,
\end{align}
which is similar to Eq.~(\ref{eq:Ckinematicdef}). It turns out that for massless partonic particles
they are related in a simple way: \mbox{$H_2 \,=\, 1 \,-\, C_{\rm part}$}. As a closing remark, we note that for
massless particles \mbox{C-parameter} can be easily expressed in terms of scalar products with four vectors:
\begin{align}
C_{\rm part} = 3-\frac{3}{2}\,\frac{1}{Q^2}\sum_{ij}\frac{(p_i\cdot p_j)^2}{E_iE_j}\,.
\end{align}
\section{C-Parameter Kinematics in the Dijet Limit}
\label{sec:kinematics}
We now show that in a dijet configuration with only soft\,\footnote{In this chapter, for simplicity we denote as soft particles what are called ultrasoft particles in SCET$_{\rm I}$.}, $n$-collinear, and $\bar n$-collinear particles, the value of \mbox{C-parameter} can, up to corrections of higher power in the SCET counting parameter $\lambda$, be written as the sum of contributions from these three kinds of particles:
\begin{align}\label{eq:C-split}
C_{\rm dijet} = C_n + C_{\bar n} + C_{\rm s} + \mathcal{O}(\lambda^4)\,.
\end{align}
To that end we define
\begin{align}\label{eq:C-decomposition}
C_{\rm dijet} & = C_{n,n} + C_{\bar n,\bar n} + 2\,C_{n, \bar n} +
2\,C_{n, \rm s} + 2\,C_{\bar n, \rm s} + C_{\rm s,s}\,,\nonumber\\
C_{a,b} & \equiv \frac{3}{2}\, \frac{1}{\left( \sum_i | \vec{p}_i | \right)^2}\!\!
\sum_{i\in a,j\in b} \! | \vec{p}_i || \vec{p}_j | \sin^2 \theta_{ij}\,.
\end{align}
The various factors of $2$ take into account that for $a\neq b$ one has to add the symmetric
term $a \leftrightarrow b$.

The $\rm SCET_I$ power counting rules imply the following scaling for momenta:
$p_{\rm s}^\mu \sim Q\, \lambda^2$,
\mbox{$p_n\sim Q\,(\lambda^2,1,\lambda)$}, \mbox{$p_{\bar n}\sim Q\,(1,\lambda^2,\lambda)$}, where we use the
light-cone components $p^\mu = (p^+,p^-,p_\perp)$. Each one of the terms in \eq{C-decomposition}, as well
as $C_{\rm dijet}$ itself, can be expanded in powers of $\lambda$:
\begin{align}
C_{a,b} = \sum_{i=0} C_{a,b}^{(i)}\,,\quad C_{a,b}^{(i)} \sim \mathcal{O}(\lambda^{i+2})\,.
\end{align}
The power counting implies that $C_{\rm dijet}$ starts at $\mathcal{O}(\lambda^2)$ and
$C_{\rm s, \rm s}$ is a power correction since $C_{s,s}^{(0)}=0$, while $C_{s,s}^{(2)}\ne 0$. All $n-$ and ${\bar n}$-collinear particles together will be denoted as the collinear particles with $c = n \cup \bar n$. The collinear particles have masses much smaller than $Q\,\lambda$ and can be taken as massless at leading power. For soft particles we have perturbative components that can be treated as massless when $Q\lambda^2 \gg \Lambda_{\rm QCD}$, and nonperturbative components that always should be treated as massive. Also at leading order one can use $\sum_i | \vec{p}_i | = \sum_{i \in \rm c} | \vec{p}_i | + \mathcal{O}(\lambda^2) = Q + \mathcal{O}(\lambda^2)$ 
and $| \vec{p}_{i \in \rm c} | = E_{i \in \rm c} + \mathcal{O}(\lambda^2)$.

\vspace*{0.2cm}
Defining $C_s \equiv 2\,C_{n, \rm s} + 2\,C_{\bar n, \rm s}$ we find,
\begin{align} \label{eq:Cs}
C_{s} & \,=\, \frac{3}{Q^2}\!\!\! \sum_{i\in {\rm c},j\in {\rm s}}\!\!\!
| \vec{p}_i | | \vec{p}_j | \sin^2 \theta_{ij}\\\nonumber
& = \frac{3}{Q^2}\!\sum_{i \in \rm c} |\vec{p}_i|
\sum_{j \in \rm s} | \vec{p}_j | \big[\sin^2 \theta_{j} + O(\lambda^2)\,\big]\\\nonumber
& = \frac{3}{Q}\!\sum_{j \in \rm s} \frac{(p_j^\perp)^2}{| \vec{p}_j |} + O(\lambda^4)
= \frac{3}{Q}\sum_{j \in \rm s} \frac{p_j^\perp}{\cosh \eta_j} + O(\lambda^4)\,,
\end{align}
where the last displayed term will be denoted as $C_s^{(0)}$. Here $\theta_j$ is the angle between the three-momenta of a particle and the thrust axis and hence is directly related to the pseudorapidity $\eta_j$. Also $p_j^\perp \equiv | \vec{p}^{\,\perp}_{j} |$ is the magnitude of the three-momentum projection normal to the thrust axis. To get to the second line, we have used that $\sin \theta_{ij} = \sin \theta_{j} \,+\, O(\lambda^2)$, and to get the last line, we have used $\sin \theta_j = p^\perp_j/| \vec{p}_j\, |=1/\cosh\eta_j$. In order to compute the partonic soft function, it is useful to consider $C_s^{(0)}$ for the case of massless particles:
\begin{align}\label{eq:C-soft-massless}
C_s^{(0)}\Big|_{\rm m = 0} = \frac{6}{Q}\sum_{j \in \rm s} \frac{p_j^+ p_j^-}{p_j^+ + p_j^-}\,.
\end{align}

Let us next consider $C_{n,n}$ and $C_{\bar n, \bar n}$. Using energy conservation and momentum conservation in the thrust direction one can show that, up to $\mathcal{O}(\lambda^2)$, \mbox{$E_n = E_{\bar n} = Q/2$}. All $n$-collinear particles are in the plus-hemisphere, and all the $\bar n$-collinear particles are in the minus-hemisphere. Here the plus- and minus-hemispheres are defined by the  thrust axis. For later convenience we define \mbox{$\mathbb{P}_a^\mu = \sum_{i\in a} p_i^\mu$} and $E_a = \mathbb{P}_a^0$ with $a \in \{n, {\bar n}\}$ denoting the set of collinear particles in each hemisphere. We also define $s_a = \mathbb{P}_a^2$.

For $C_{n,n}$ one finds
\begin{align}\label{eq:Cnn}
& C_{n,n} = \frac{3}{2Q^2} \!\!\sum_{i,j\in n}| \vec{p}_i | | \vec{p}_j |
(1-\cos\theta_{ij})(1+\cos\theta_{ij})\\\nonumber
&= [\,1 + O(\lambda^2)\,]\,\frac{3}{Q^2}\!\!\sum_{i,j\in n}p_i\cdot p_j =
\frac{3}{Q^2} \bigg(\sum_{i\in n}p_i\bigg)^{\!\!2} + O(\lambda^4)\\\nonumber
&= \frac{3}{Q^2}\,s_n + O(\lambda^4) = \frac{3}{Q}\sum_{i\in n}p_i^+ + O(\lambda^4) \\\nonumber
&=
\frac{3}{Q}\,\mathbb{P}_n^+ + O(\lambda^4)\,,
\end{align}
and we can identify $C_{n,n}^{(0)}=3\, \mathbb{P}_n^+ /Q$. To get to the second line, we have used that for collinear particles in the same direction $\cos\theta_{ij} = 1 \,+\, \mathcal{O}(\lambda^2)$. In the third line, we use the property that the total perpendicular momenta of each hemisphere is exactly zero and that $\vec 0 = \sum_{i\in +}\vec p_i^{\,\perp} = \sum_{i\in n}\vec p_i^{\,\perp} + \mathcal{O}(\lambda^2)$ and $s_n = Q \, \mathbb{P}_n^+$. In a completely analogous way, we get
\begin{align}\label{eq:Cnbarnbar}
C_{\bar n,\bar n}^{(0)} \, 
=\, \frac{3}{Q}\sum_{i\in \bar n}p_i^- = \frac{3}{Q}\,\mathbb{P}_{\bar n}^-\,.
\end{align}
The last configuration to consider is $C_{n,\bar n}$:
\begin{align}\label{eq:Cnnbar}
2\,C_{n,\bar n}  
&= \frac{3}{Q^2} \!\!\!\sum_{i,j\in n, \bar n}\!\!\! | \vec{p}_i | | \vec{p}_j |
(1-\cos\theta_{ij})(1+\cos\theta_{ij})\\\nonumber
&=\frac{6}{Q^2}\!\!\sum_{i,j\in n}(2E_i E_j - p_i\cdot p_j) 
[\,1 + O(\lambda^2)\,]\,
\\\nonumber
&= \frac{3}{Q^2}\, (\,\mathbb{P}_n^+\,\mathbb{P}_{\bar n}^+ +
\mathbb{P}_n^-\,\mathbb{P}_{\bar n}^-
-2\,\mathbb{P}_n^\perp\cdot \mathbb{P}_{\bar n}^\perp\,)+ O(\lambda^4)\\\nonumber
&= \frac{3}{Q}\,(\,\mathbb{P}_n^+ + \mathbb{P}_{\bar n}^-\,) + O(\lambda^4)
= C_{n,n}^{(0)} + C_{\bar n,\bar n}^{(0)}+ O(\lambda^4). \nn
\end{align}
In the second equality, we have used that for collinear particles in opposite directions $\cos\theta_{ij} = -\, 1 + \mathcal{O}(\lambda^2)$; in the third equality, we have written $2E = p^+ + p^-$; and in the fourth equality, we have discarded the scalar product of perpendicular momenta since it is $\mathcal{O}(\lambda^4)$ and also used that at leading order $\mathbb{P}_n^- = \mathbb{P}_{\bar n}^+ = Q + \mathcal{O}(\lambda^2)$ and $\mathbb{P}_n^- \sim \lambda^2$, $\mathbb{P}_{\bar n}^+\sim \lambda^2$. For the final equality, we use the results obtained in Eqs.~(\ref{eq:Cnn}) and (\ref{eq:Cnbarnbar}). Because the final result in \Eq{eq:Cnnbar} just doubles those from \Eqs{eq:Cnn}{eq:Cnbarnbar}, we can define
\begin{align} \label{eq:Cndef}
C_n^{(0)} &\equiv 2\, C_{n,n}^{(0)} \,,\quad
 C_{\bar n}^{(0)} \equiv 2\, C_{\bar n,\bar n}^{(0)} \,.
\end{align}
Using Eqs.~(\ref{eq:Cnn}), (\ref{eq:Cnbarnbar}), and (\ref{eq:Cnnbar}), we then have
\begin{align}\label{eq:C-col}
\!C_n^{(0)} \,=\, \frac{6}{Q}\, & \mathbb{P}_n^+\,, \quad
C_{\bar n}^{(0)} \,=\, \frac{6}{Q}\, \mathbb{P}_{\bar n}^-\,.
\end{align}
Equation~(\ref{eq:Cs}) together with \Eq{eq:C-col} finalize the proof of Eq.~(\ref{eq:C-split}). As a final comment, we note that one can express $p^{\pm} = p^\perp \exp(\mp\, \eta)$, and since for $n$-collinear particles $2\cosh \eta = \exp(\eta) [\,1\,+\,{\cal O}(\lambda^2)\,]$ whereas for $\bar n$-collinear particles $2\cosh \eta = \exp(-\,\eta)[\,1\,+\,{\cal O}(\lambda^2)\,]$, one can also write
\begin{equation}
C_n^{(0)} = \frac{3}{Q} \sum_{i \in n} \frac{p_i^\perp}{\cosh \eta_i}\,,\quad
C^{(0)}_{\bar{n}} = \frac{3}{Q}\sum_{i \in \bar{n}} \frac{p_i^\perp}{\cosh \eta_i}\,,
\end{equation}
such that the same master formula applies for soft and collinear particles in the dijet limit, and we can write 
\begin{align}
C^{(0)}_{\rm dijet}= \frac{3}{Q} \sum_{i} \frac{p_i^\perp}{\cosh \eta_i} \,.
\end{align}

\section{Factorization and Resummation}
\label{sec:factorization-singular}
The result in Eq.~(\ref{eq:C-split}) leads to a factorization in terms of hard, jet, and soft functions. The dominant nonperturbative corrections at the order at which we are working come from the soft function and can be factorized with the following formula in the $\overline{\rm MS}$ scheme for the power corrections~\cite{Korchemsky:2000kp,Hoang:2007vb,Ligeti:2008ac}:
\begin{align}\label{eq:singular-nonperturbative}
\frac{1}{\sigma_0}\frac{\df \sigma}{\df C} &= \!\int \!\df p \,\frac{1}{\sigma_0}
\frac{\df \hat\sigma}{\df C}\Big(C-\frac{p}{Q}\Big)F_C(p)\,,\\
\frac{\df \hat\sigma}{\df C} & \,=\, \frac{\df {\hat\sigma}_{\rm s}}{\df C} \,+\,
\frac{\df {\hat\sigma}_{\rm ns}}{\df C}\,.\nn
\end{align}
Here $F_C$ is a shape function describing hadronic effects, and whose first moment $\Omega_1^C$ is the
leading nonperturbative power correction in the tail of the distribution. $\Omega_1^C$ and $\Omega_1^\tau$
are related to each other, as will be discussed further along with other aspects of power corrections in Sec.~\ref{sec:power}. The terms $\df \hat\sigma/\df C$, $\df \hat\sigma_{\rm s}/\df C$, and $\df \hat\sigma_{\rm ns}/\df C$ are the total partonic cross section and the singular and nonsingular contributions, respectively. The latter will be discussed in Sec.~\ref{sec:nonsingular}.

After having shown Eq.~(\ref{eq:C-split}), we can use the general results of Ref.~\cite{Bauer:2008dt} for the factorization theorem for the singular terms of the partonic cross section that splits into a sum of soft and collinear components. One finds
\begin{align}\label{eq:factorization-partonic-singular}
\frac{1}{\sigma_0}\frac{\df \hat\sigma_{\rm s}}{\df C}
&=\frac{1}{6} \frac{1}{\sigma_0}\frac{\df \hat\sigma_{\rm s}}{\df \widetilde C}
\\
&=\frac{Q}{6} H(Q,\mu)\!
\int\! \df s\, J_\tau(s,\mu) \hat S_{\widetilde C}\Big( Q \widetilde C- \frac{s}{Q},\mu\Big)\,,
\nn
\end{align}
where in order to make the connection to thrust more explicit we have switched to the variable $\widetilde C = C/6$. Here $J_\tau$ is the thrust jet function which is obtained by the convolution of the two hemisphere jet functions and where our definition for $J_\tau$ coincides with that of Ref.~\cite{Abbate:2010xh}. It describes the collinear radiation in the direction of the two jets. Expressions up to $\mathcal{O}(\alpha_s^2)$ and the logarithmic terms determined by its anomalous dimension at three loops are summarized in \App{ap:formulae}.

The hard factor $H$ contains short-distance QCD effects and is obtained from the Wilson
coefficient of the SCET to QCD matching for the vector and axial vector currents. The hard function is the same
for all event shapes, and its expression up to $\mathcal{O}(\alpha_s^3)$
is summarized in \App{ap:formulae}, together with the full anomalous dimension for $H$  at three loops.

The soft function $S_{\widetilde C}$ describes wide-angle soft radiation between the
two jets. It is defined as
\begin{align}
S_{C}(\ell,\mu) & = \frac{1}{N_c} \big\langle \, 0\,\big|\, {\rm tr}\: \overline{Y}_{\bar n}^T Y_n
\delta(\ell- Q{\widehat C}) Y_n^\dagger \overline{Y}_{\bar n}^*\, \big|\, 0\,\big\rangle\,,\\
S_{\widetilde C}(\ell,\mu) 
& = \frac{1}{N_c}  \big\langle \, 0\,\big|\, {\rm tr}\: \overline{Y}_{\bar n}^T Y_n
\delta\Big(\ell- \frac{Q{\widehat C}}{6}\Big) Y_n^\dagger \overline{Y}_{\bar n}^*\, \big|\, 0\,\big\rangle
\nn\\
& = 6\,S_{C}(6\,\ell,\mu) 
, \nn
\end{align}
where $Y_n^\dagger$ is a Wilson line in the fundamental representation from $0$ to $\infty$ and $\overline Y_{\bar n}^\dagger$ is a Wilson lines in the anti-fundamental representation from $0$ to $\infty$. Here $\widehat C$ is an operator whose eigenvalues on physical states correspond to the value of \mbox{C-parameter} for that state:
$\widehat C \,|X\rangle = C(X)\, |X\rangle$. Since the hard and jet functions are the same as for
thrust, the anomalous dimension of the C-parameter soft function has to coincide with the anomalous
dimension of the thrust soft function to all orders in $\alpha_s$ by consistency of the RGE. This allows us to determine all logarithmic terms of $S_C$ up to ${\cal O}(\alpha_s^3)$.  Hence one only needs to determine the
non-logarithmic terms of $S_C$. We compute it analytically at one loop and use EVENT2 to numerically determine the two-loop constant, $s_2^{\widetilde C}$. The three-loop constant $s_3^{\widetilde C}$ is currently not known and we estimate it with a Pad\'e, assigning a very conservative error. We vary this constant in our theoretical uncertainty analysis, but it only has a noticeable impact in the peak region.

In Eq.~(\ref{eq:factorization-partonic-singular}) the hard, jet, and
soft functions are evaluated at a common scale $\mu$. There is no
choice that simultaneously minimizes the logarithms of these three
matrix elements. One can use the renormalization group equations to evolve to $\mu$ from
the scales $\mu_H\sim Q$, $\mu_J\sim Q \sqrt{\widetilde C}$, and
$\mu_S\sim Q \widetilde C$ at which logs are minimized in each piece. In this way large
logs of ratios of scales are summed up in the renormalization group
factors:
\begin{align}\label{eq:singular-resummation}
\frac{1}{\sigma_0}\frac{\df \hat\sigma_{\rm s}}{\df  C} =
\frac{Q}{6} H(Q,\mu_H)\,U_H(Q,\mu_H,\mu)\! \int\! & \df s\, \df s^\prime\, \df k\!\,
J_\tau(s,\mu_J)\,U_J^\tau(s-s^\prime,\mu,\mu_J)\,U_S^\tau(k,\mu,\mu_S) 
\nn \\
&\times\!
e^{-\,3\pi\frac{\delta(R,\mu_{\!s})}{Q}\frac{\partial}{\partial C}}
\hat S_{\widetilde C}\bigg(\!\frac{Q C-3\pi\bar\Delta(R,\mu_S)}{6}- \frac{s}{Q}-k,\mu_S\!\bigg).
\end{align}
The terms $\delta$ and $\bar \Delta$ are related to the definition of the leading power correction in a
renormalon-free scheme, as explained in Sec.~\ref{sec:power} below.

\section{Nonsingular Terms}
\label{sec:nonsingular}

%
We include the kinematically power suppressed terms in the C-parameter distribution using the nonsingular
partonic distribution, $\mathrm{d} \hat{\sigma}_{\rm ns} / \mathrm{d}C$. We calculate the nonsingular distribution using
\begin{align} \label{eq:nssubt}
\frac{\df \hat{\sigma}_{\rm ns}}{\df C}(Q,\mu_{\rm ns}) \,=\, \frac{\df \hat{\sigma}_{\text{full}}^{\text{FO}}}{\df C}(Q,\mu_{\rm ns}) \,-\,
\frac{\df \hat{\sigma}_{s}^{\text{FO}}}{\df C}(Q,\mu_{\rm ns})\,.
\end{align}
Here $\df \hat{\sigma}_{s}^{\text{FO}}/\df C$ is obtained by using Eq.~(\ref{eq:singular-resummation}) with $\mu = \mu_H = \mu_J = \mu_S = \mu_{\rm ns}$.
This nonsingular distribution is independent of the scale $\mu_{\rm ns}$ order by order in perturbation theory as an expansion in $\alpha_s(\mu_{\rm ns})$.
We can identify the nontrivial ingredients in the nonsingular distribution by choosing $\mu_{\rm ns} = Q$ to give
\begin{align} \label{eq:nonsingular-expansion}
\frac{1}{\sigma_0} \frac{\df \hat{\sigma}_{\rm ns}}{\df C} & \,=\, \frac{\alpha_s(Q)}{2\pi}\, f_1(C)\,+\,\left(\frac{\alpha_s(Q)}{2\pi}\right)^{\!\!2}\!\! f_2(C)
+\,\left(\frac{\alpha_s(Q)}{2\pi}\right)^{\!\!3}\!\! f_3(C) \,+\,\ldots
\end{align}
We can calculate each $f_i(C)$ using an order-by-order subtraction of the fixed-order singular distribution from the full fixed-order distribution as displayed in \Eq{eq:nssubt}.
\begin{figure}
	\begin{center}
		\includegraphics[width=0.6\columnwidth]{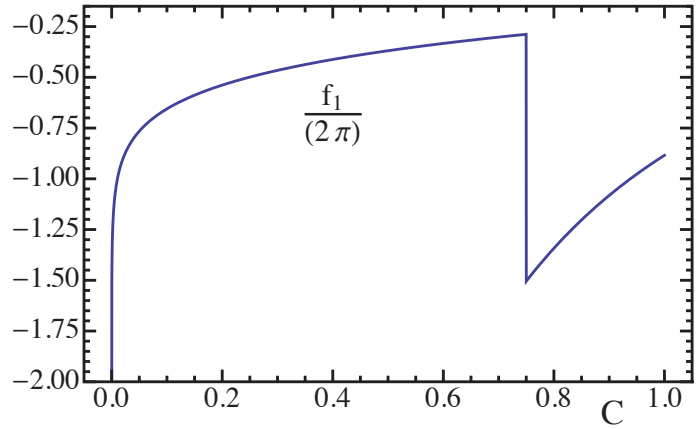}
		\caption[$\ord {\alpha_s}$ nonsingular C-parameter distribution]{$\ord {\alpha_s}$ nonsingular C-parameter distribution, corresponding to Eq.~(\ref{eq:1NS}).}
		\label{fig:1-loopNS}
	\end{center}
\end{figure}
\begin{figure}
	\begin{center}
		\includegraphics[width=0.6\columnwidth]{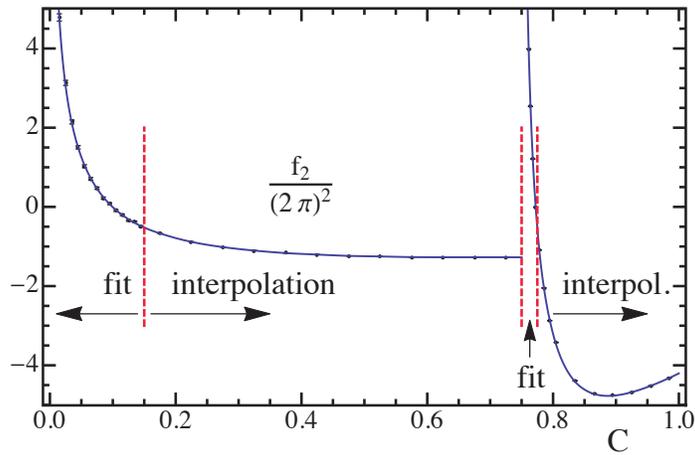}
		\caption[$\ord {\alpha_s^2}$ nonsingular C-parameter distribution]{$\ord {\alpha_s^2}$ nonsingular C-parameter distribution. The solid line shows our reconstruction, whereas dots with error bars correspond to the EVENT2 output with the singular terms subtracted. Our reconstruction consists of fit functions to the left of the red dashed line at $C = 0.15$ and between the two red dashed lines at $C = 0.75$ and $C = 0.8$ and interpolation functions elsewhere.}
		\label{fig:2-loopNS}
	\end{center}
\end{figure}
\begin{figure}
	\begin{center}
		\includegraphics[width=0.6\columnwidth]{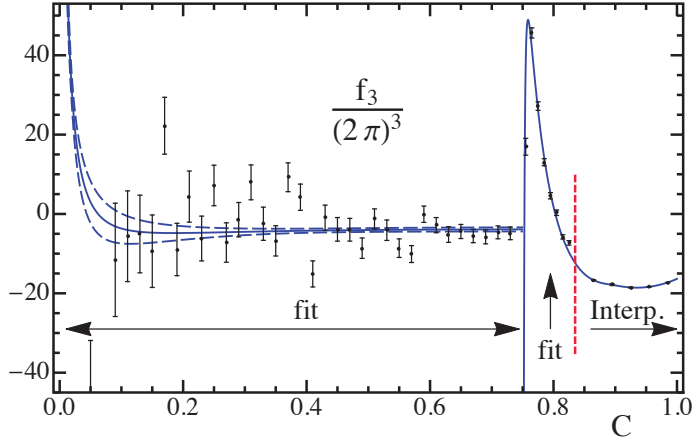}
		\caption[$\ord {\alpha_s^3}$ nonsingular C-parameter distribution]{$\ord {\alpha_s^3}$ nonsingular C-parameter distribution. The solid line shows our reconstruction, whereas dots with error bars correspond to the EERAD3 output with the singular terms subtracted. Our reconstruction consists of two fit functions, one for $C < 0.75$ and another one for $0.75 < C < 0.835$, and an interpolation for $C>0.835$ (to the right of the red dashed vertical line).}
		\label{fig:3-loopNS}
	\end{center}
\end{figure}
At one loop, we can write down the exact form of the full distribution as a two-dimensional integral 
\cite{Ellis:1980nc}
\begin{align}\label{eq:1loop-FO}
\frac{1}{\sigma_0} \bigg|\frac{\df \hat{\sigma}}{\df C}\bigg|^{\rm 1-loop} \,= &\frac{\alpha_s}{2 \pi}\, C_F \!\!\int_0^1\! \df x_1\! \int_0^1\! \df x_2 \, \theta( x_1 + x_2 -1 ) 
\,\frac{x_1^2+x_2^2}{(1-x_1)(1-x_2)} \\
&\times \delta\! \left( \!C - \frac{6\,(1-x_1)(1-x_2)(x_1+x_2-1)}{x_1 x_2 (2-x_1-x_2)}\right),\nonumber
\end{align}
which has support for $0<C<3/4$ and jumps to zero for $C> 3/4$. After resolving the delta function, it becomes a one-dimensional integral that can be easily evaluated numerically. After subtracting off the one-loop singular piece discussed in Sec.~\ref{sec:factorization-singular}, we obtain the result for $f_1$ shown in Fig.~\ref{fig:1-loopNS}. For $C>3/4$ the nonsingular distribution at this order is simply given by the negative of the singular, and for practical purposes one can find a parametrization for $f_1$ for $C < 3/4$, so we use
\begin{align}\label{eq:1NS}
 f_1(C<0.75) \,=&\, -\,2.25168 + 0.506802\, C \,+\, 0.184585\, C^2 +\, 0.121051\, C^3 \nonumber\\
 & \,+ (0.890476 - 0.544484 \,C -\, 0.252937 \,C^2 - 0.0327797 \,C^3) \ln(C)\,,\nn\\
 f_1(C>0.75) \,=&\, \frac{4}{3C}\bigg[3+4 \ln\bigg(\frac{C}{6}\bigg)\bigg].
\end{align}
For an average over $C$, this result for $f_1(C)$ is accurate to $10^{-7}$ and at worst for a particular $C$ is accurate at $10^{-5}$. An exact closed form in terms of elliptic functions for the integral in Eq.~(\ref{eq:1loop-FO}) has
been found in Ref.~\cite{Gardi:2003iv}.

The full $\ord{\alpha_s^2}$ and $\ord{\alpha_s^3}$ fixed-order distributions can be obtained numerically
from the Fortran programs EVENT2 \cite{Catani:1996jh, Catani:1996vz} and EERAD3
\cite{GehrmannDeRidder:2009dp,Ridder:2014wza}, respectively. At $\ord{\alpha_s^2}$ we use log-binning EVENT2 results
for $C<0.2$ and linear-binning (with bin size of 0.02) results for $0.2<C<0.75$. We then have additional
log binning from $0.75<C<0.775$ (using bins in $\ln[C-0.75]$) before returning to linear binning
for $C>0.775$. We used runs with a total of $3 \times 10^{11}$ events and an infrared cutoff $y_0=10^{-8}$. In the regions
of linear binning, the statistical uncertainties are quite low and we can use a numerical interpolation for $f_2(C)$.
For $C<0.15$ we use the ansatz, $f_2(C)=\sum_{i=0}^3 a_i \ln^i C + a_4\, C \ln C$  and fit the coefficients from EVENT2
output, including the constraint that the total fixed-order cross section gives the known $\ord{\alpha_s^2}$ coefficient for
the total cross section.  The resulting values for the $a_i$ are given as functions of $s_2^{\widetilde C}$,
the non-logarithmic coefficient in the partonic soft function. Details on the determination of this fit function and the determination of $s_2^{\widetilde C}$ can be found below in Sec.~\ref{sec:softtwoloop}. We  find 
\begin{align} \label{eq:s2result}
s_2^{\widetilde C}=-\,43.2\,\pm\,1.0 \,, 
\end{align}
whose central value is used in Figs.~\ref{fig:2-loopNS}, \ref{fig:3-loopNS}, \ref{fig:component-plot}, and \ref{fig:component-plot-sum}, and whose uncertainty is included in our uncertainty analysis. For $0.75<C<0.8$ we employ another ansatz, \mbox{$f_2(C)\,=\,\sum_{i=0}^1 \sum_{k=0}^2 b_{i k}\, (C - 0.75)^i \ln^k [\,8\,(C-0.75)/3]$}. We use the values calculated in Ref.~\cite{Catani:1997xc}, \mbox{$b_{01}=59.8728$}, $b_{02}=43.3122$, and $b_{12} = -\,115.499$ and fit the rest of the coefficients to EVENT2 output. The final result for the two-loop nonsingular cross section coefficient then has the form
\begin{align}
f_2(C)\,+\,\epsilon^\text{low}_{2}\, \delta^\text{low}_{2}(C) \,
+\,\epsilon^\text{high}_{2}\, \delta^\text{high}_{2}(C) \,. 
\end{align}
Here $f_2(C)$ gives the best fit in all regions, and $\delta_2^\text{low}$ and  $\delta_2^\text{high}$ give the $1$-$\sigma$ error functions for the lower fit ($C<0.75$) and upper fit ($C>0.75$), respectively. The two variables $\epsilon^\text{low}_{2}$ and $\epsilon^\text{high}_{2}$ are varied during our theory scans in order to account for the error in the nonsingular function. In Fig.~\ref{fig:2-loopNS}, we show the EVENT2 data as dots and the best-fit nonsingular function as a solid blue line. The uncertainties are almost invisible on the scale of this plot.

In order to determine the $\ord{\alpha_s^3}$ nonsingular cross section $f_3(C)$, we follow a similar procedure. The EERAD3 numerical output is based on an infrared cutoff $y_0=10^{-5}$ and calculated with $6 \times 10^7$ events for the three leading color structures and $10^7$ events for the three subleading color structures. The results are linearly binned with a bin size of 0.02 for $C<0.835$ and a bin size of $0.01$ for $C>0.835$. As the three-loop numerical results have larger uncertainties than the two-loop results, we employ a fit for all $C<0.835$ and use interpolation only above that value. The fit is split into two parts for $C$ below and above $0.75$. For the lower fit, we use the ansatz $f_3(C)=\sum_{i=1}^5 a_i \ln^i (C)$. The results for the $a_i$ depend on the $\ord{\alpha_s^2}$ partonic soft function coefficient $s_2^{\widetilde C}$ and a combination of the three-loop coefficients in the partonic soft and jet function, $s_3^{\widetilde C}\,+\,2\, j_3$. Due to the amount of numerical uncertainty in the EERAD3 results, it is not feasible to fit for this combination, so each of these parameters is left as a variable that is separately varied in our theory scans. Above $C=0.75$ we carry out a second fit, using the fit form \mbox{$f_3(C)=\sum_{i=0}^4 b_i \ln^i(C-0.75)$}. We use the value \mbox{$b_4 = 122.718$} predicted by exponentiation in Ref.~\cite{Catani:1997xc}. The rest of the $b$'s depend on $s_2^{\widetilde C}$. The final result for the three-loop nonsingular cross section coefficient can once again be written in the form
\begin{align}
f_3(C)+\epsilon^\text{low}_{3} \delta_3^\text{low}(C) +\epsilon^\text{high}_{3} \delta^\text{high}_{3}(C) \,,
\end{align}
where $f_3(C)$ is the best-fit function and the $\delta_3$'s give the $1$-$\sigma$ error function for the low ($C<0.75$) and high ($C>0.75$) fits. Exactly like for the $\ord{\alpha_s^2}$ case, the $\epsilon_3$'s are varied in the final error analysis. In Fig.~\ref{fig:3-loopNS}, we plot the EERAD3 data as dots, the best-fit function $f_3$ as a solid line, and the nonsingular results with $\epsilon^\text{low}_{3}=\epsilon^\text{high}_{3}=\pm\,1$ as dashed lines. In this plot we take $s_2^{\widetilde C}$ to its best-fit value and $j_3=s_3^{\widetilde C}=0$. 

In the final error analysis, we vary the nonsingular parameters encoding the numerical extraction uncertainty $\epsilon_i$'s, as well as the profile parameter $\mu_{\rm ns}$. The uncertainties in our nonsingular fitting are obtained by taking $\epsilon^{\rm low}_2,\,\epsilon^\text{high}_{2},\,\epsilon^\text{low}_{3},$ and $\epsilon^\text{high}_3$ to be $-\,1$, $0$, and $1$ independently. The effects of $\epsilon^\text{low}_{2}$, $\epsilon^\text{high}_{2}$ and $\epsilon^\text{high}_{3}$ are essentially negligible in the tail region. Due to the high noise in the EERAD3 results, the variation of $\epsilon^\text{high}_{3}$ is not negligible in the tail region, but because it comes in at ${\cal O}(\alpha_s^3)$, it is still small. We vary the nonsingular renormalization scale $\mu_{\rm ns}$ in a way described in Sec. \ref{sec:profiles}.

In order to have an idea about the size of the nonsingular distributions with respect to the singular terms, we quote numbers
for the average value of the one-, two-, and three-loop distributions between $C = 0.2$ and $C = 0.6$:
$5, 21, 76$ (singular at one, two, and three loops); $-\,0.4,-\,1,-\,4$ (nonsingular at one, two, and three loops). Hence, in the region to be used for fitting $\alpha_s(m_Z)$, the singular distribution is $12$ (at 1-loop) to $20$ (at two and three loops) times larger than the nonsingular one and has the opposite sign. Plots comparing the singular and nonsingular cross sections for all $C$ values are given below in \Fig{fig:component-plot}.

\section{Determination of Two-Loop Soft Function Parameters}
\label{sec:softtwoloop}
In this section we will expand on the procedure used to extract the  $\ord{\alpha_s^2}$ non-logarithmic coefficient in the soft function using EVENT2\footnote{After this work was originally published, the exact two loop C-parameter soft function constant was calculated in \cite{Bell:2015lsf}. In our notation, they find $s_2^{\widetilde{C}}=-42.921$, which agrees with the result we give in \eq{s2result}.}. For a general event shape, we can separate the partonic cross section into a singular part  where the cross section involves $\delta(e)$ or $\ln^k(e)/e$,\,\footnote{The numerical outcome of a parton-level Monte Carlo such as EVENT2 contains only the power of logs, but the complete distributions can be obtained from knowledge of the SCET soft and jet functions.} and a nonsingular part with integrable functions, that diverge at most as $\ln^k(e)$. Of course, when these are added together and integrated over the whole spectrum of the event shape distribution, we get the correct fixed-order normalization:
\begin{equation} \label{eq:s2extractionintegral}
\hat\sigma_{\rm had} = \int_0^{e_{\rm max}} \!\df e \left( \frac{\df \hat\sigma_{\rm s}}{\df e} +
\frac{\df \hat\sigma_{\rm ns}}{\df e} \right).
\end{equation}
Here $e_\text{max}$ is the maximum value for the given event shape (for C-parameter, $e_\text{max}=1$).
Using SCET, we can calculate the singular cross section at $\mathcal{O}(\alpha_s^2)$, having the form
\begin{equation} \label{eq:ADs}
\dfrac{1}{\sigma_0} \frac{\df\hat\sigma_{\rm s}^{(2)}}{\df e}=A_{\delta}\,\delta(e)+\sum_{n=0}^3\,D_n\bigg[\frac{\ln^n e}{e}\bigg]_+\,.
\end{equation}
To define $\df\hat\sigma_{\rm s}^{(2)}/\df e$ we factor out $\alpha_s^2/16\pi^2$ and set $\mu=Q$. The only unknown term at ${\cal O} (\alpha_s^2)$ for the \mbox{C-parameter} distribution is the two-loop constant $s_{2}^{\widetilde C}$ in the soft function, which contributes to $A_\delta$. The explicit result for the terms in Eq.~(\ref{eq:ADs}) can be obtained from $H(Q,\mu)\,P(Q,QC/6,\mu)$ which are given in \App{ap:formulae}. This allows us to write the singular integral 
in (\ref{eq:s2extractionintegral}) as a function of $s_{2}^{\widetilde C}$ and known constants.

We extract the two-loop nonsingular portion of the cross section from EVENT2 data. Looking now at the specific case of
C-parameter, we use both log-binning (in the small $C$ region, which is then described with high accuracy)
and linear binning (for the rest). By default we use logarithmic binning for $C < C_{\rm fit} = 0.2$, but this boundary
is changed between $0.15$ and $0.25$ in order to estimate systematic uncertainties of our method. In the logarithmically
binned region we use a fit function to extrapolate for the full behavior of the nonsingular cross section.
In order to determine the coefficients of the fit function we use data between $C_{\rm cut}$ and $C_{\rm fit}$. By
default we take the value $C_{\rm cut} = 10^{-4}$, but we also explore different values between $5\times 10^{-5}$ and
$7.625\times 10^{-4}$ to estimate systematic uncertainties. We employ the following functional form, motivated by the expected nonsingular logarithms
\begin{align}
\dfrac{1}{\sigma_0}\frac{\df\hat\sigma^{\rm ns}_{\rm fit}}{\df C}=\sum_{i=0}^{3}\,a_i\,\ln^{i}C +
a_4\,C\ln^n C\,,
\end{align}
taking the value $n=1$ as default and exploring values \mbox{$0\le n \le 3$} as an additional source of systematic
uncertainty.

For the region with linear binning, we can simply calculate the relevant integrals by summing over the
bins. One can also sum bins that contain the shoulder region as its singular behavior is integrable.
These various pieces all combine into a final formula that can be used to extract the two-loop constant piece
of the soft function:
\begin{align} \label{eq:s2extractionfinalintegral}
\hat\sigma_{\rm had}^{(2)} = \!\int_0^{1}\! \df C\, \frac{\df \hat\sigma^{(2)}_{\rm s}}{\df C}
\,+ \!\int_0^{C_{\rm fit}}\!\! \df C \,\frac{\df \hat\sigma^{\rm ns}_{\rm fit}}{\df C} \,+
\!\int_{C_{\rm fit}}^{1}\!\! \df C\, \frac{\df \hat\sigma^{\text{ns}}_{\rm int}}{\df C}  .
\end{align}
Using \eq{s2extractionfinalintegral} one can extract $s_2^{\widetilde{C}}$, which can be decomposed into its various
color components as
\begin{align}
s_2^{\widetilde{C}} = C_F^2\, s_{2}^{[C_F^2]} +  C_F\, C_A s_{2}^{[C_FC_A]} +  C_F\, n_f T_F s_{2}^{[n_f]} \,.
\end{align} 
The results of this extraction are
\begin{align} \label{s2results}
s_{2}^{[C_F^2]}  = -0.46 \pm 0.75\,,\,\, 
s_{2}^{[C_FC_A]}  = - 29.08 \pm 0.13\,, \,\,
s_{2}^{[n_f]}  = 21.87 \pm 0.03\,. 
\end{align}
The quoted uncertainties include a statistical component coming from the fitting procedure and a systematical component coming from the parameter variations explained above, added in quadrature. Note that the value for $s_{2}^{[C_F^2]}$ is  consistent with zero, as expected from exponentiation \cite{Hoang:2008fs}. For our analysis we will always take $s_{2}^{[C_F^2]}=0$. We have cross-checked that, when a similar extraction is repeated for the  case of thrust, the extracted values are consistent with those calculated analytically in Refs.~\cite{Kelley:2011ng,Monni:2011gb}. This indicates a high level of accuracy in the fitting procedure.  We have also confirmed that following the alternate fit procedure of Ref.~\cite{Hoang:2008fs} gives compatible results, as shown in App.~\ref{ap:MC-comparison}.

\section{Power Corrections and Renormalon-Free Scheme}
\label{sec:power}
The expressions for the theoretical prediction of the C-parameter distribution in the dijet region shown in  Eqs.~(\ref{eq:singular-nonperturbative}) and (\ref{eq:factorization-partonic-singular}) incorporate that the full soft function can be written as a convolution of the partonic soft function $\hat S_C$  and the nonperturbative shape function $F_C$\,\cite{Hoang:2007vb}\,\footnote{Here we use the relations $\hat S_C(\ell,\mu) = \hat S_{\widetilde C}(\ell/6,\mu)/6$ and $F_C(\ell) = F_{\widetilde C}(\ell/6)/6$. }:
\begin{align}\label{eq:soft-nonperturbative}
S_C(k,\mu) = \!\int \!\df k' \,\hat S_C(k-k',\mu) F_C(k',\Delta_{C})\,.
\end{align}
Here, the partonic soft function $\hat S_C$ is defined in fixed order in $\overline{\text{MS}}$. The shape function $F_C$ allows a smooth transition between the peak and tail regions, where different kinematic expansions are valid, and $\Delta_{C}$ is a parameter of the shape function that represents an offset from zero momentum and that will be discussed further below. By definition, the shape function satisfies the relations $F_C(k, \Delta_C) = F_C(k - 2\Delta_C)$ and $F_C(k <0) = 0$. In the tail region, where $QC/6 \gg \Lambda_{\text{QCD}}$, this soft function can be expanded to give
\begin{align}\label{eq:soft-OPE}
S_C(k,\mu) = \hat S_C(k) - \frac{\df \hat S_C(k)}{\df k}\, \overline{\Omega}_1^C
+ {\cal O}\Big( \frac{\alpha_s \Lambda_{\rm QCD}}{QC},\frac{\Lambda_{\rm QCD}^2}{Q^2 C^2}\Big) ,
\end{align}
where $ \overline{\Omega}_1^C$ is the leading nonperturbative power correction in $\overline{\rm MS}$ which effectively introduces a shift of the distribution in the tail region \cite{Abbate:2010xh}. The $\overline \Omega_1^C$ power correction 
\begin{align}
\overline \Omega_1^C(\mu) & = \frac{1}{N_c} \big\langle 0 \big| {\rm tr}\: \overline Y_{\bar n}^T Y_{n}  (Q \widehat C) Y_n^\dagger  \overline Y_{\bar n}^*  \big| 0 \big\rangle \,,
\end{align}
is related to $\overline{\Omega}_1^\tau$, the first moment of the thrust shape function, as given in \Eq{eq:O1c}. In addition to the normalization difference that involves a factor of $3\pi/2$, their relation is further affected by hadron-mass effects which cause an additional deviation at the 2.5\% level (computed in \Sec{sec:hadronmass}). The dominant contributions of the ${\cal O}(\alpha_s \Lambda_{\rm QCD}/Q C)$ corrections indicated in \Eq{eq:soft-OPE} are log enhanced and will be captured once we include the $\mu$-anomalous dimension for $\Omega_1^C$ that is induced by hadron-mass effects~\cite{Mateu:2012nk}. There are additional ${\cal O}(\alpha_s \Lambda_{\rm QCD}/Q C)$ corrections, which we neglect, that do not induce a shift. We consider hadron-mass effects in detail in \Sec{sec:hadronmass}. 

From \Eq{eq:soft-nonperturbative} and the OPE of \Eq{eq:soft-OPE}, we can immediately read off the relations
\begin{align} \label{eq:gapmomentrelation}
&\!\int\! \df k' k' F_C(k',\Delta_{C}) = 2\,\Delta_{C} + \! \int \! \df k' k' F_C(k') = \overline{\Omega}_1^C\,, \qquad
\!\int\! \df k' F_C(k') = 1\,,
\end{align}
which state that the first moment of the shape function provides the leading power
correction and that the shape function is normalized. In the peak, it is no longer sufficient to keep only the first
moment, as there is no OPE when $QC/6 \sim \Lambda_{\text{QCD}}$ and we must keep the full dependence on the model
function in \Eq{eq:soft-nonperturbative}.

The partonic soft function in $\overline{\text{MS}}$ has an $\mathcal{O}(\Lambda_{\rm QCD})$ renormalon, an ambiguity which is related to a linear sensitivity in its perturbative series.
This renormalon ambiguity is in turn inherited to the numerical values for $\overline \Omega_1^C$ obtained in fits to the experimental data.
It is possible to avoid this renormalon issue by switching to
a different scheme for $\Omega_1^C$, which involves subtractions in the partonic soft function that remove this type of infrared sensitivity.
Following the results of \Ref{Hoang:2007vb}, we write  $\Delta_{C}$ as
\begin{align} \label{eq:deltasplitting}
\Delta_{C}  & =  \frac{3\pi}{2}[\,\bar{\Delta}(R,\mu) + \delta(R,\mu)\,]\,.
\end{align}
The term $\delta(R,\mu)$ is a perturbative series in $\alpha_s(\mu)$ which has the same
renormalon behavior as $\overline{\Omega}_1^C$. In the factorization formula, it is grouped into the partonic soft function $\hat S_C$ through the exponential factor involving $\delta(R,\mu)$ shown in \Eq{eq:singular-resummation}. 
Upon simultaneous perturbative expansion of the exponential together with $\hat S_C$, the $\mathcal{O}(\Lambda_{\rm QCD})$ renormalon is subtracted.
The term $\bar{\Delta}(R,\mu)$ then becomes a nonperturbative
parameter which is free of the $\mathcal{O}(\Lambda_{\rm QCD})$ renormalon. Its dependence on the subtraction scale $R$ and on $\mu$ is dictated by $\delta(R,\mu)$ since $\Delta_{C}$ is $R$ and $\mu$ independent.
The subtraction scale $R$ encodes the momentum scale associated with
the removal of the linearly infrared-sensitive fluctuations. 
The factor $3\pi/2$ is a normalization coefficient that relates the ${\cal O}(\Lambda_{\rm QCD})$ 
renormalon ambiguity of the $C$ soft function $S_{C}$ to the one for the thrust soft function.
Taking into account this normalization we can use for $\delta(R,\mu)$ the scheme for the thrust 
soft function already defined in \Ref{Abbate:2010xh},
\begin{equation} \label{eq:delta-scheme}
\!\!\!\!\delta(R,\mu) = \frac{R}{2} \, e^{\gamma_E} \frac{\df}{\df \ln(ix)}
\big[ \ln S^{\text{part}}_{\tau}(x,\mu) \big]_{x = (i R e^{\gamma_E})^{-1}},
\end{equation}
where $S^{\text{part}}_{\tau}(x,\mu)$ is the position-space thrust partonic soft function. 
From this, we find that the perturbative series
for the subtraction is
\begin{align} \label{eq:deltaseries}
\delta(R,\mu) =
R \,e^{\gamma_E} \!\sum_{i=1}^\infty \alpha_s^i(\mu)\, \delta^i(R,\mu)\,.
\end{align}
Here the $\delta_{i\ge 2}$ depend on both the adjoint Casimir
\mbox{$C_A=3$} and the number of light flavors in combinations that are unrelated to the QCD beta
function. Using five light flavors the first three coefficients have been calculated in \Ref{Hoang:2008fs} as
\begin{align} \label{eq:d123}
\frac{\pi}{2}\, \delta^1(R,\mu) &= -\,1.33333\, L_R \,, \nn \\
\frac{\pi}{2}\, \delta^2(R,\mu) &= -\,0.245482 - 0.732981\, L_R - 0.813459\, L_R^2 \,, \nn \\
\frac{\pi}{2}\, \delta^3(R,\mu) &= -\,0.868628\, - 0.977769\, L_R
\nn\\&\quad -1.22085\, L_R^2 - 0.661715\, L_R^3 \,,
\end{align}
where $L_R = \ln(\mu/R)$. Using these $\delta$'s, we can make a scheme change on the first moment to what we
call the Rgap scheme:
\begin{align} \label{eq:omegaschemechange}
\Omega_1^{C}(R,\mu) &= \overline{\Omega}_1^C(\mu) - 3\pi\,\delta (R,\mu)\,.
\end{align}
In contrast to the $\overline{\text{MS}}$ scheme $\overline{\Omega}_1^C(\mu)$, the Rgap scheme $\Omega_1^{C}(R,\mu)$  
is free of the $\Lambda_{\rm QCD}$ renormalon. From \Eq{eq:gapmomentrelation} it is then easy to see that the first moment of the shape function becomes
\begin{equation}
\int \!\df k \,k\,F_C\big(k - 3\pi\, \overline\Delta(R,\mu)\big) = \Omega_1^C(R,\mu)  \,.
\end{equation}
The factorization in \Eq{eq:soft-nonperturbative} can now be written as
\begin{align} \label{eq:soft-nonperturbative-subtract}
\!\!\!S_C(k,\mu) & = \!\int \!\df k' \,e^{-\,3\pi\delta(R,\mu) \frac{\partial}{\partial k}}
\hat S_C(k-k',\mu) \, F_C\big(k' - 3\pi \, \overline{\Delta}(R,\mu)\big)\,.
\end{align}
The logs in \Eq{eq:d123} can become large when $\mu$ and $R$ are far apart. This imposes a constraint that $R \sim \mu$,
which will require the subtraction scale to depend on $C$ in a way similar to $\mu$. On the other hand, we also must consider the power
counting $\overline{\Omega}^C \sim \Lambda_\text{QCD}$, which leads us to desire using $R \simeq 1$\,GeV. In order to satisfy both of these constraints in the tail region, where $\mu \sim QC/6 \gg 1$\,GeV, we (i) employ $R \sim \mu$ for the subtractions in $\delta(R,\mu)$ that are part of the Rgap partonic soft function and (ii) use the \mbox{R-evolution} to relate the gap parameter $\bar\Delta(R,\mu)$ to the reference gap parameter $\bar\Delta(R_\Delta,\mu_\Delta)$ with $R_\Delta \sim \mu_\Delta \sim {\cal O}(1\,\text{GeV})$ where the $\Lambda_{\rm QCD}$ counting applies~\cite{Hoang:2008yj, Hoang:2009yr,Hoang:2008fs}. The formulae for the \mbox{$R$-RGE} and $\mu$-RGE are
\begin{align} \label{eq:RandmuRGE}
R \frac{\df}{\df R}  \bar{\Delta} (R,R) &= -\,R \sum_{n=0}^\infty \gamma_n^R \left( \frac{\alpha_s(R)}{4 \pi} \right)^{\!\!n+1}  , \nn \\
\mu \frac{\df}{\df\mu}  \bar{\Delta} (R,\mu) &= 2\, R\, e^{\gamma_E} \sum_{n=0}^\infty \Gamma_n^\text{cusp} \left( \frac{\alpha_s(\mu)}{4 \pi} \right)^{\!\!n+1}  ,
\end{align}
where for five flavors the $\Gamma_n^\text{cusp}$ is given in App.~\ref{ap:formulae} and the $\gamma^R$ coefficients are given by
\begin{align} \label{eq:gammaR}
\gamma_0^R &= 0, \; \gamma_1^R = -\,43.954260\,,
\;\gamma_2^R = -\,606.523329\,.
\end{align}
The solution to \Eq{eq:RandmuRGE} is given, at N$^k$LL, by
\begin{align} \label{eq:DeltaRevolution}
\bar{\Delta}(R,\mu) &= \bar{\Delta}(R_\Delta,\mu_\Delta) + R\, e^{\gamma_E} \omega\, [\, \Gamma^\text{cusp},\mu,R\,] + R_\Delta e^{\gamma_E} \omega\,[\,\Gamma^\text{cusp},R_\Delta,\mu_\Delta\,] \nn \\
&\qquad+ \Lambda_{\text{QCD}}^{(k)} \,\sum_{j=0}^{k} (-1)^j S_j e^{i \pi \hat{b}_1} \big[\,\Gamma ( -\,\hat{b}_1 -j, t_1) - \Gamma(-\,\hat{b}_1  -j,t_0) \,\big]\nn\\[0.2cm]
&\equiv \bar{\Delta}(R_\Delta,\mu_\Delta) + \Delta^{\rm diff}(R_\Delta,R, \mu_\Delta,\mu)\,.
\end{align}
For the convenience of the reader, the definition for $\omega$ is provided in \Eq{eq:w}, and the values for $\hat{b}_1$ and the $S_j$ are given in \Eq{eq:Sjnor}. In order to satisfy the power counting criterion for $R$, we specify the parameter $\bar{\Delta} (R_\Delta,\mu_\Delta)$ at the low reference scales $R_\Delta = \mu_\Delta = 2 \,\text{GeV}$. We then use \Eq{eq:DeltaRevolution} to evolve this parameter up to a scale $R(C)$, which is given in \Sec{sec:profiles} and satisfies the condition $R(C) \sim \mu_S(C)$ in order to avoid large logs. This \mbox{R-evolution} equation yields a similar equation for the running of $\Omega_1^C(R,\mu_S)$, which is easily found from \Eqs{eq:deltasplitting}{eq:omegaschemechange}.
\begin{figure}[t!]
	\begin{center}
		\includegraphics[width=0.65\columnwidth]{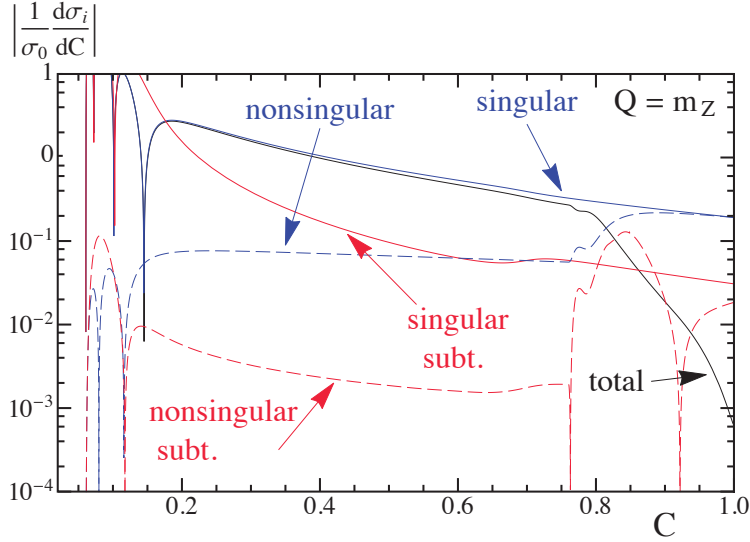}
		\caption[Singular and nonsingular components of the FO C-parameter cross section]{Singular and nonsingular components of the fixed-order C-parameter cross section, including up to ${\cal O}(\alpha_s^3)$ terms, with $\Omega_1=0.25\,$GeV and $\alpha_s(m_Z) = 0.1141$.}
		\label{fig:component-plot}
	\end{center}
\end{figure}

We also apply the Rgap scheme in the nonsingular part of the cross section by using the convolution
\begin{align} \label{eq:nonsingular-shape-convolution}
& \!\!\!\int  \!\df k'\, e^{-\,3 \pi \frac{\delta(R,\mu_{\!s})}{Q}\frac{\partial}{\partial C}}
\frac{\df \hat{\sigma}_{\text{ns}}}{\df C} \Big( C - \frac{k'}{Q}, \frac{\mu_{\text{ns}}}{Q} \Big) \times F_C\big(k' - 3\pi \,\bar{\Delta}(R,\mu_S)\big) \,.
\end{align}
By employing the Rgap scheme for both the singular and nonsingular pieces,
the sum correctly recombines in a smooth manner to the fixed-order result in the far-tail region. 

Note that by using \Eq{eq:deltasplitting} we have defined the renormalon-free moment parameter $\Omega_1^C(R,\mu)$
in a scheme directly related to the one used for the thrust analyses in Refs.~\cite{Abbate:2010xh,Abbate:2012jh}.
This is convenient as it allows for a direct comparison to the $\Omega_1$ fit results we obtained in both these analyses. However, many other renormalon-free schemes can be devised, and all these schemes are perturbatively related to each other through their relation to the $\overline{\text{MS}}$ scheme $\Delta_C$. As an alternative, we could have defined a renormalon-free scheme for $\Omega_1^C$ by determining the subtraction $\delta$ directly from the $\widetilde C$ soft function $S_{\widetilde C}$ using the analog to \Eq{eq:delta-scheme}. For future reference we quote the results for the resulting subtraction function $\delta_{\widetilde C}$ in App.~\ref{ap:subtractionchoice}. 

\begin{figure}[t!]
	\begin{center}
		\includegraphics[width=0.65\columnwidth]{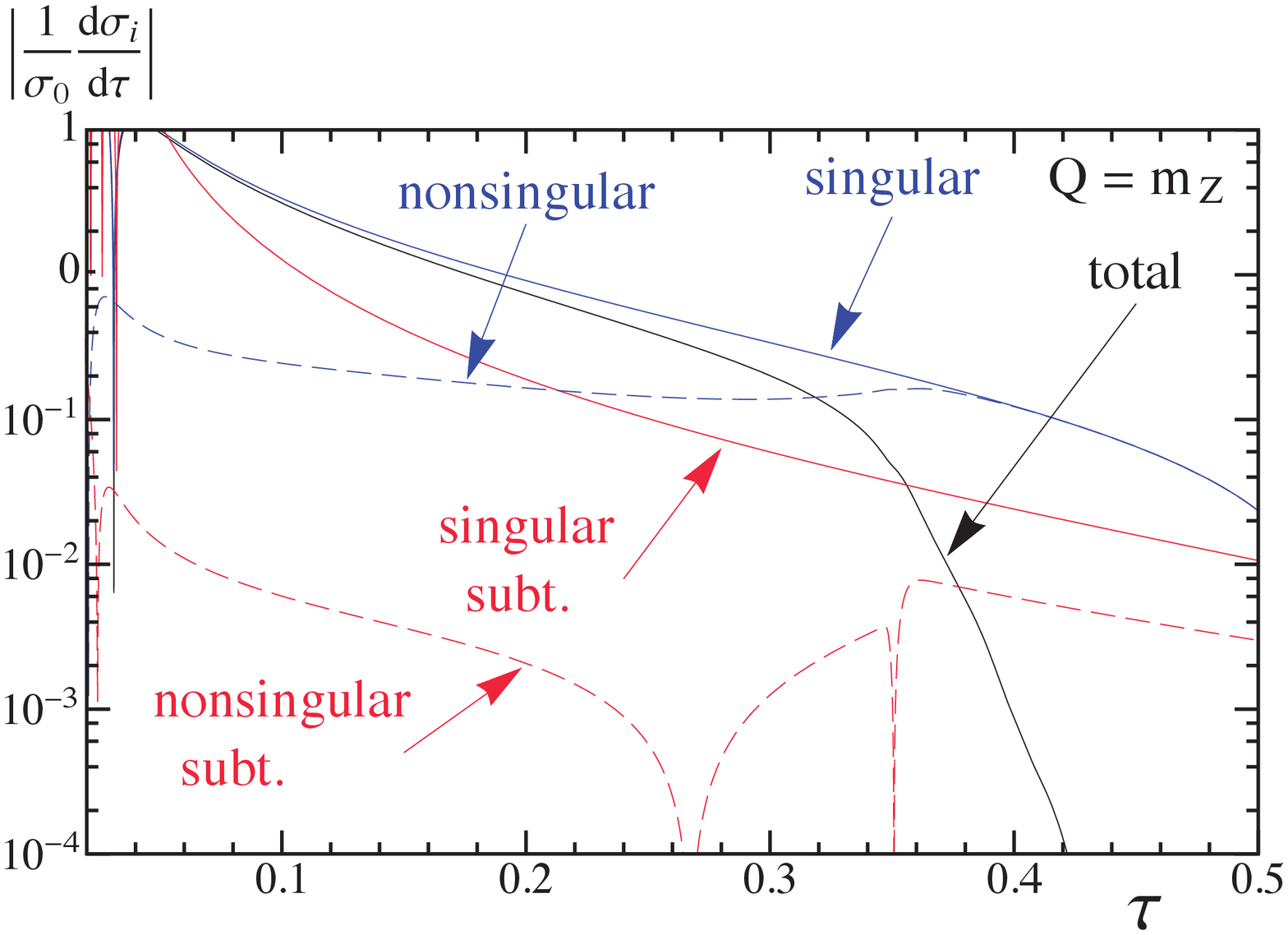}
		\caption[Singular and nonsingular components of the FO thrust cross section]{Singular and nonsingular components of the fixed-order thrust cross section, including up to ${\cal O}(\alpha_s^3)$ terms, with \mbox{$\Omega_1=0.25\,$GeV} and $\alpha_s(m_Z) = 0.1141$.}
		\label{fig:component-plot-tau}
	\end{center}
\end{figure}
In close analogy to Ref.~\cite{Abbate:2010xh}, we parametrize the shape function $F_C$ in terms of the basis functions introduced in Ref.~\cite{Ligeti:2008ac}. In this expansion the shape function has the form
\begin{equation} \label{eq:shapebasis}
F_C(k,\lambda,\{c_i\}) = \frac{1}{\lambda} \left[ \sum_{n=0}^N c_n f_n\left(\frac{k}{\lambda}\right)\right]^2,
\end{equation}
where the $f_n$ are given by
\begin{align} \label{eq:basisfunctions}
f_n(z) = 8 \sqrt{\frac{2z^3(2n+1)}{3}} e^{-2z} P_n [\,g(z)\,], \nn \\
g(z)=\frac{2}{3} \left[3 - e^{-4z} (3 + 12 z + 24 z^2 + 32 z^3) \right] - 1\,,
\end{align}
and $P_n$ denote the Legendre polynomials. The additional parameter $\lambda$ is irrelevant when $N \to \infty$. For finite $N$ it is strongly correlated with the first moment $\Omega_1^C$ (and with $c_1$). The normalization of the shape function requires that $\sum_{n=0}^N c_n^2 = 1$. When plotting and fitting in the tail region, where the first moment of the shape
function $\Omega_1^C$ is the only important parameter, it suffices to take $c_0=1$ and all $c_{i>0}=0$. In this case
the parameter $\lambda$ directly specifies our $\Omega_1^C$ according to
\begin{align} \label{eq:omega-from-lambda}
\Omega_1^C(R_\Delta,\mu_\Delta) 
= \lambda + 3 \pi\, \bar{\Delta}(R_\Delta,\mu_\Delta) 
\,.
\end{align}
In the tail region where one fits for $\alpha_s(m_Z)$, there is not separate dependence on the nonperturbative parameters $\lambda$ and $\bar{\Delta}(R_\Delta,\mu_\Delta)$; they only appear together through the parameter $\Omega_1^C(R_\Delta,\mu_\Delta)$. In the peak region, one should keep more $c_i$'s in order to
correctly parametrize the nonperturbative behavior.

In Fig.~\ref{fig:component-plot} we plot the absolute value of the four components of the partonic fixed-order $C$ distribution at $\mathcal{O}(\alpha_s^3)$ in the Rgap scheme at $Q=m_Z$. Resummation has been turned off. The cross section components include the singular terms (solid blue), nonsingular terms (dashed blue), and separately the contributions from terms that involve the subtraction coefficients $\delta_i$, for both singular subtractions (solid red) and nonsingular subtractions (dashed red). The sum of these four components gives the total cross section (solid black line). One can observe that the nonsingular terms are significantly smaller than the singular ones in the tail region below the shoulder, i.e.\ for $C<0.7$. Hence the tail region is completely dominated by the part of the cross section described by the SCET factorization theorem, where resummation matters most. Above the shoulder the singular and nonsingular $C$ results have comparable sizes. An analogous plot for the thrust cross section is shown in Fig.~\ref{fig:component-plot-tau}.  We see that the portion of the C-parameter distribution where the logarithmic resummation in the singular terms is important, is substantially larger compared to the thrust distribution.

\subsection{Hadron Mass Effects}
\label{sec:hadronmass}

Following the analysis in \Ref{Mateu:2012nk}, we include the effects of hadron masses by including 
the dependence of  $\Omega_1^C$ on the distributions of transverse velocities,
\begin{equation} \label{eq:def-trans-vel}
r \equiv \frac{p_\perp}{\sqrt{p_\perp^2 + m_H^2}} \,,
\end{equation}
where $m_H$ is the nonzero hadron mass and $p_\perp$ is the transverse velocity with respect to the thrust axis.
For the massless case, one has $r=1$. However, when the hadron masses are nonzero, $r$ can take any value
in the range $0$ to $1$. The additional effects of the finite hadron masses cause non-trivial modifications in the form of the first moment of the shape function,
\begin{equation} \label{eq:omega1functionofg}
\overline\Omega_1^e(\mu) \,=\, c_e\! \int_0^1\! \df r \, g_e(r)\, \overline\Omega_1(r,\mu)\,,
\end{equation}
where $e$ denotes the specific event shape that we are studying, $c_e$ is an event-shape-dependent constant, $g_e(r)$
is an event-shape-dependent function\,\footnote{As discussed in Ref.~\cite{Mateu:2012nk}, event shapes with
	a common $g_e$ function belong to the same universality class. This means that their leading power corrections are
	simply related by the $c_e$ factors. In this sense $g_e$ is universality-class dependent rather than event-shape
	dependent.} that encodes the dependence on the hadron-mass effects and $\overline\Omega_1(r)$ is a
universal $r$-dependent generalization of the first moment, described by a matrix element of the transverse
velocity operator. $\overline\Omega_1(r,\mu)$ is universal for all recoil-insensitive event shapes. Note that once hadron masses are included there is no limit of the hadronic parameters that reduces to the case where hadron masses are not accounted for.

For the cases of thrust and C-parameter, we have
\begin{align} \label{eq:thrust-C-r-constants}
c_C &= 3 \pi\, , && g_C(r)  = \frac{2 r^2}{\pi} K(1-r^2)\,, \\
c_\tau &= 2\, , && g_\tau(r) \,= 1 - E(1-r^2) + r^2 K(1-r^2)\,, \nn
\end{align}
where $E(x)$ and $K(x)$ are complete elliptic integrals, whose definition can be found in~\Ref{Salam:2001bd,Mateu:2012nk}.
Notice that $g_C(r)$ and $g_\tau(r)$ are within a few percent of
each other over the entire $r$ range, so we expect the relation 
\begin{align} \label{eq:O1univ}
2\,\overline\Omega_1^C(\mu) = 3 \pi\,\overline\Omega_1^\tau(\mu)
\big[ 1 +{\cal O}(2.5\%) \big]\,,
\end{align}
where the ${\cal O}(2.5\%)$ captures the breaking of universality due to the effects of hadron masses. We have determined the size of the breaking by 
\begin{align}
\frac{\int_0^1 \df r  \big[\,g_C(r) - g_\tau(r) \big] }{\int_0^1 \df r \big[\,g_C(r) + g_\tau(r)\big]/2} = 0.025 .
\end{align}
All $r\in[\,0,1]$ contribute roughly an equal amount to this deviation, which is therefore well captured by 
this integral.

As indicated in \Eq{eq:omega1functionofg} the moment $\overline\Omega_1(r,\mu)$ in the $\overline{\text{MS}}$ scheme is renormalization-scale dependent and at LL satisfies the RGE of the form~\cite{Mateu:2012nk}
\begin{equation} \label{eq:omega1-r-murunning}
\overline \Omega_1(r,\mu) = \overline \Omega_1(r,\mu_0)\!
\left[ \frac{\alpha_s(\mu)}{\alpha_s(\mu_0)} \right]^{\hat{\gamma}_1(r)}.
\end{equation}
In App.~\ref{ap:hadronmassR} we show how to extend this running to the Rgap scheme in order to remove the ${\cal O}(\Lambda_{\rm QCD})$ renormalon.
The result in the Rgap scheme is
\begin{align}  \label{eq:omega1R-r-murunning}
g_C(r)\,\Omega_1^C(R,\mu,r) =& g_C(r)\!
\left[ \frac{\alpha_s(\mu)}{\alpha_s(\mu_\Delta)} \right]^{\hat{\gamma}_1(r)}\!
\Omega_1^C(R_\Delta, \mu_\Delta,r) + R\, e^{\gamma_E} \!\left(\frac{\alpha_s(\mu)}{\alpha_s(R)} \right)^{\!\!\hat{\gamma}_1(r)}\!
\omega\, [\,\Gamma^{\text{cusp}},\mu,R\,] \nn \\
&+ R_\Delta e^{\gamma_E} \!\left(\frac{\alpha_s(\mu)}{\alpha_s(R_\Delta)} \right)^{\!\!\hat{\gamma}_1(r)}\!
\omega\, [\,\Gamma^{\text{cusp}},R_\Delta,\mu_\Delta]
\nn \\
&+ \Lambda_{\text{QCD}}^{(k)} \!\left( \frac{\beta_0 \alpha_s(\mu)}{2 \pi} \right)^{\!\!\hat{\gamma}_1(r)}\!
\sum_{j=0}^{k}S^r_j(r) (-1)^j e^{i \pi [\hat{b}_1 - \hat{\gamma}_1(r)]}
\nn \\
& \times
\big[\,\Gamma ( -\,\hat{b}_1 + \hat{\gamma}_1(r) -j, t_1) - 
\Gamma(-\,\hat{b}_1 + \hat{\gamma}_1(r) -j,t_0) \,\big] \nn\\
&\equiv g_C(r)\!\left[ \frac{\alpha_s(\mu)}{\alpha_s(\mu_\Delta)} \right]^{\hat{\gamma}_1(r)}\!
\Omega_1(R_\Delta, \mu_\Delta,r) \ \ +  \Delta^{\rm diff}(R_\Delta,R,\mu_\Delta,\mu,r)\,.
\end{align}
Here the formula is resummed to N$^k$LL, and $\Lambda_{\text{QCD}}^{(k)}$ is the familiar N$^k$LO perturbative expression for $\Lambda_{\text{QCD}}$. We always use $R_\Delta = \mu_\Delta = 2\,$GeV to define the initial hadronic parameter. The values for $\hat{b}_1$, $\gamma_1(r)$, $t_1$, $t_0$,and  the $S_j$ can all be found in App.~\ref{ap:hadronmassR} and the resummed $\omega$ is given in ~\Eq{eq:w}.

In order to implement this running, we pick an ansatz for the form of the moment at the low scales, $R_\Delta$
and $\mu_\Delta$, given by
\begin{align} \label{eq:omega1-ansatz}
\Omega_1(R_\Delta,\mu_\Delta,r) &= \big[\, a(R_\Delta,\mu_\Delta) f_a(r) + b(R_\Delta,\mu_\Delta) f_b(r) \,\big]^2, \nn \\
f_a(r) &= 3.510 \, e^{-\frac{r^2}{1-r^2}}, \\
f_b(r) &= 13.585 \, e^{-\frac{2\,r^2}{1-r^2}} - 21.687 \,\, e^{-\frac{4\,r^2}{1-r^2}}. \nn
\end{align}
The form of $\Omega_1(R_\Delta,\mu_\Delta,r)$ was chosen to always be positive and to smoothly go to zero at the endpoint $r=1$. In the Rgap scheme, $\Omega_1$ can be interpreted in a Wilsonian manner as a physical hadronic average momentum parameter, and hence it is natural to impose positivity. As $r\to 1$ we are asking about the vacuum-fluctuation-induced distribution of hadrons with large $p_\perp$ which is anticipated to fall off rapidly.
We also check other ans\"atze that satisfied these conditions, but choosing different positive definite functions has a minimal effect on the distribution. The functions $f_a$ and $f_b$ were chosen to satisfy
$\int_0^1 \df r g_C(r) f_a(r)^2 = \int_0^1 \df r g_C(r) f_b(r)^2 = 1$ and $\int_0^1 \df r g_C(r) f_a(r)\, f_b(r) = 0$.
This allows us to write
\begin{align} \label{eq:omegafunctionofaandb}
\Omega_1^C(R_\Delta,\mu_\Delta) 
&= c_C \!\!\int_0^1 \!\! \df r\:  g_C(r) \: \Omega_1(R_\Delta, \mu_\Delta ,r) 
\nn\\
&= 3\, \pi\, \big[\,a(R_\Delta,\mu_\Delta)^2 + b(R_\Delta,\mu_\Delta)^2\, \big]\,,
\end{align}
and to define an orthogonal variable,
\begin{equation} \label{eq:thetadef}
\theta(R_\Delta,\mu_\Delta) \equiv \arctan\left(\frac{b(R_\Delta,\mu_\Delta)}{a(R_\Delta,\mu_\Delta)}\right).
\end{equation}
The parameters $a$ and $b$ can therefore be swapped for $\Omega_1^C(R_\Delta,\mu_\Delta)$ and $\theta(R_\Delta,\mu_\Delta)$. This $\theta$ is defined as part of the model for the universal function $\Omega_1(R,\mu,r)$ and so should also exhibit universality between event shapes. In \Sec{subsec:hadronmass} below, we will demonstrate that $\theta$ has a small effect on the cross section for the C-parameter, and hence that $\Omega_1^C(R_\Delta,\mu_\Delta)$ is the most important hadronic parameter.

\section{Profile Functions}
\label{sec:profiles}

The ingredients required for cross section predictions at various resummed perturbative orders are given in \Tab{tab:ordercounting}. This includes the order for the cusp and non-cusp anomalous dimensions for $H$, $J_\tau$, and $S_C$; their perturbative matching order; the beta function $\beta[\alpha_s]$ for the running coupling: and the order for the nonsingular corrections discussed in \Sec{sec:nonsingular}.  It also includes the anomalous dimensions $\gamma_\Delta$ and subtractions $\delta$ discussed in this section. In our analysis we only use primed orders with the factorization theorem for the distribution. For the unprimed orders, only the formula for the cumulant cross section properly resums the logarithms, see Ref.~\cite{Almeida:2014uva}, but for the reasons discussed in Ref.~\cite{Abbate:2010xh}, we need to use the distribution cross section for our analysis. The primed order distribution factorization theorem properly resums the desired series of logarithms for $C$, and was also used in Refs.~\cite{Abbate:2010xh,Abbate:2012jh} to make predictions for thrust.

The factorization formula in Eq.~(\ref{eq:singular-resummation}) contains three characteristic renormalization scales, the hard scale $\mu_H$, the jet scale $\mu_J$, and the soft scale $\mu_S$. In order to avoid large logarithms, these scales
must satisfy certain constraints in the different $C$ regions:
\begin{align} \label{eq:profileconstraints}
& \text{1) nonperturbative:~} C \lesssim 3\pi\,\frac{\Lambda_\text{QCD}}{Q}\nn \\
&  \qquad  \mu_H  \sim Q,\  \mu_J \sim \sqrt{\Lambda_\text{QCD} Q},\  \mu_S \!\sim\! R \!\sim\!  \Lambda_\text{QCD} 
\,, \nn \\[5pt]
& \text{2) resummation:~} 3\pi\,\frac{\Lambda_\text{QCD}}{Q} \ll C < 0.75
\\
& \qquad  \mu_H \sim Q,\  \mu_J \sim Q \sqrt{\frac{C}{6}},\   \mu_S \! \sim \! R \!\sim\! \frac{QC}{6}  \gg \Lambda_{\rm QCD}
\,, \nn \\
& \text{3) fixed-order:~} C > 0.75
\nn \\
& \qquad \mu_H   = \mu_J  = \mu_S = R \sim Q\gg \Lambda_{\rm QCD}\nn
\,.\end{align}
In order to meet these constraints and have a continuous factorization formula, we make each scale a smooth function of $C$ using profile functions.

When one looks at the physical C-parameter cross-section, it is easy to identify the peak, tail, and far-tail as
distinct physical regions of the distribution. How much of the physical peak belongs to the nonperturbative vs
resummation region is in general a process-dependent statement, as is the location of the transition between the resummation
and fixed-order regions. For example, in $b\to s\,\gamma$ the entire peak is in the nonperturbative 
region~\cite{Ligeti:2008ac}, whereas for $pp\to H+1$ gluon initiated jet with $p_T\sim 400\,{\rm GeV}$, the entire peak
is in the resummation region~\cite{Jouttenus:2013hs}.  For thrust with $Q=m_Z$~\cite{Abbate:2010xh}, and similarly here
for C-parameter with $Q=m_Z$, the transition between the nonperturbative and resummation regions occurs near the maximum
of the physical peak.   Note that, despite the naming, in the nonperturbative region, where the full form of the shape function is needed, resummation is always important.
The tail for the thrust and C-parameter distributions is located in the resummation region, and the far-tail, which is dominated by events with three or more jets, exists in the fixed-order region.

For the renormalization scale in the hard function, we use
\begin{align}
\mu_H \,=\, e_H \, Q\,,
\end{align}
where $e_H$ is a parameter that we vary from 0.5 to 2.0 in order to account for theory uncertainties.
\begin{table}[t!]
	\centering
	\begin{tabular}{c|ccccccc}
		&  cusp &  non-cusp  &  matching  &  $\beta[\alpha_s]$  &  nonsingular  &  $\gamma_\Delta^{\mu,R,r}$  &  $\delta$  \\
		\hline
		LL  &  1 &  -  &  tree  &  1  &  -  &  -  &  - \\
		NLL  &  2 &  1  &  tree  &  2  &  -  &  1  &  - \\
		N${}^2$LL  &  3 &  2  &  1  &  3  &  1  &  2  &  1 \\
		N${}^3$LL  &  4${}^\text{pade}$ &  3  &  2  &  4  &  2  &  3  &  2 \\
		\hline
		NLL$^\prime$  &  2 &  1  &  1  &  2  &  1  &  1  &  1 \\
		N${}^2$LL$^\prime$  &  3 &  2  &  2  &  3  &  2  &  2  &  2 \\
		N${}^3$LL$^\prime$  &  4${}^\text{pade}$ &  3  &  3  &  4  &  3  &  3  &  3 \\
	\end{tabular}
	\caption[Loop corrections required for specified orders in log resummation.]{Loop corrections required for specified orders.}
	\label{tab:ordercounting}
\end{table}
\begin{figure}[t!]
	\begin{center}
		\includegraphics[width=0.65\columnwidth]{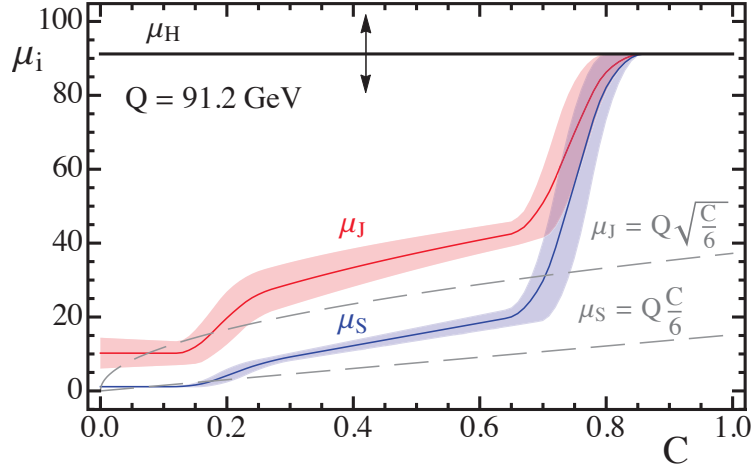}
		\caption[Profile variations for C-parameter]{
			Solid lines are the central results for the profile functions for the renormalization scales $\mu_H$, $\mu_J(C)$, $\mu_S(C)$ at $Q=m_Z$. The bands and up-down arrow indicate the results of varying the profile parameters. The result for $R(C)$ is identical to $\mu_S(C)$ at the resolution of this figure, differing only at small $C$. Above $C=t_s\simeq 0.8$ all the scales merge, $\mu_H=\mu_J=\mu_S=R$.   }
		\label{fig:profile-variation}
	\end{center}
\end{figure}

The profile function for the soft scale is more complicated, and we adopt the following form:
\begin{equation} \label{eq:muSprofile}
\!\mu_{S} = \left\{ \begin{tabular}{p{.5\columnwidth} l}
$\mu_0$                                             & $0 \le C < t_0$ \\
$\zeta(\mu_0,\,0,\,0,\,\frac{r_s\,\mu_H}{6},\,t_0,\,t_1,\,C)$  & $t_0 \le C < t_1$ \\
$r_s \,\mu_H \frac{C}{6}$         & $t_1 \le C < t_2$ \\
$\zeta(0,\,\frac{r_s\,\mu_H}{6},\,\mu_H,0,\,t_2,\,t_s,\,C)$ & $t_2 \le C < t_s$ \\
$\mu_H$     &           $t_s \le C < 1$
\end{tabular}
\right.\!.
\end{equation}
Here the 1st, 3rd, and 5th lines satisfy the three constraints in \Eq{eq:profileconstraints}. 
In particular, $\mu_0$ controls the intercept of the soft scale at $C=0$. The term $t_0$ controls the boundary of the purely nonperturbative region and the start of the transition to the resummation region, and $t_1$ represents the end of this 
transition. As the border between the nonperturbative and perturbative regions is $Q$ dependent, we actually use  $n_0 \equiv t_0 (Q/1$\,{\rm GeV}) and $n_1 \equiv t_1 (Q/1$\,{\rm GeV}) as the profile parameters. In the resummation region $t_1< C<t_2$, the parameter $r_s$ determines the linear slope with which $\mu_S$ rises.
The parameter $t_2$ controls the border and transition between the resummation and fixed-order
regions. Finally, the $t_s$ parameter sets the value of $C$ where the renormalization
scales all join. We require both $\mu_{S}$ and its first derivative to be continuous, and to
this end we have defined the function $\zeta(a_1,b_1,a_2,b_2,t_1,t_2,\,t)$ with $t_1 < t_2$,
which smoothly connects two straight lines of the form $l_1(t) = a_1 \,+\, b_1\,t$ for $t < t_1$
and $l_2(t) = a_2 \,+\, b_2\,t$ for $t > t_2$ at the meeting points $t_1$ and $t_2$. We find that
a convenient form for $\zeta$ is a piecewise function made out of two quadratic functions patched
together in a smooth way.
These two second-order polynomials join at the middle point $t_m=(t_1 + t_2)/2$:
\begin{align}
& \!\!\!\!\!\!\!
\zeta(a_1,b_1,a_2,b_2,t_1,t_2,t) 
\nn\\*
& = \left\{ \!\begin{tabular}{p{.5\columnwidth} l}
$\hat a_1 + b_1(t - t_1) + e_1(t - t_1)^2$  &  $~t_1 \le t \le t_m$  \nn \\
$\hat a_2 + b_2(t - t_2) + e_2(t - t_2)^2$  &  $~t_m \le t \le t_2$
\end{tabular}
\right.\!,\\[0.1cm]
\hat a_1 & = a_1 + b_1\,t_1\,,\qquad \hat a_2 = a_2 + b_2\,t_2\,,\nn\\
e_1 &=\frac{4\,(\hat a_2-\hat a_1)-(3\,b_1 + b_2)\,(t_2-t_1)}{2\,(t_2-t_1)^2}\,,\nn\\[0.1cm]
e_2 &=\frac{4\,(\hat a_1-\hat a_2)+(3\,b_2 + b_1)\,(t_2-t_1)}{2\,(t_2-t_1)^2}\,.
\end{align}
The soft scale profile in \Eq{eq:muSprofile} was also used in Ref.~\cite{Stewart:2014nna} for jet-mass distributions in $pp\to Z+1$-jet.
\begin{table}[t!]
	\begin{center}%
	\begin{tabular}{ccc}
		parameter\ & \ default value\ & \ range of values \ \\
		\hline
		$\mu_0$ & $1.1$\,GeV & - \\
		$R_0$ & $0.7$\,GeV &  - \\
		$n_0$ & $12$ & $10$ to $16$\\
		$n_1$ & $25$ & $22$ to $28$\\
		$t_2$ & $0.67$ & $0.64$ to $0.7$\\
		$t_s$ & $0.83$ & $0.8$ to $0.86$\\
		$r_s$ & $2$ & $1.78$ to $2.26$\\
		$e_J$ & $0$ & $-\,0.5$ to $0.5$\\
		$e_H$ & $1$ & $0.5$ to $2.0$\\
		$n_s$ & $0$ & $-\,1$, $0$, $1$\\
		\hline
		$\Gamma^{\rm cusp}_3$ & $1553.06$ & $-\,1553.06$ to $+\,4659.18$ \\
		$s_2^{\widetilde C}$ & $-\,43.2$ & $-\,44.2$ to $-\,42.2$ \\
		$j_3$ & $0$ & $-\,3000$ to $+\,3000$ \\
		$s_3^{\widetilde C}$ & $0$ & $-\,500$ to $+\,500$ \\
		\hline
		$\epsilon^\text{low}_{2}$ & $0$ & $-\,1$, $0$, $1$ \\
		$\epsilon^\text{high}_{2}$ & $0$ & $-\,1$, $0$, $1$ \\
		$\epsilon^\text{low}_{3}$ & $0$ & $-\,1$, $0$, $1$ \\
		$\epsilon^\text{high}_{3}$ & $0$ & $-\,1$, $0$, $1$ \\
	\end{tabular}
	\end{center}
	\caption[Variation of C-parameter theory parameters]{C-parameter theory parameters relevant for estimating the theory uncertainty, their
		default values, and range of values used for the scan for theory uncertainties.}
	\label{tab:theoryerr}
\end{table}
\begin{table}[tbh!]
	\begin{center}
	\begin{tabular}{ccc}
		parameter\ & \ default value\ & \ range of values \ \\
		\hline
		$\mu_0$ & $1.1$\,GeV & -\\
		$R_0$ & $0.7$\,GeV & -\\
		$n_0$ & $2$ & $1.5$ to $2.5$ \\
		$n_1$ & $10$ & $8.5$ to $11.5$\\
		$t_2$ & $0.25$ & $0.225$ to $0.275$\\
		$t_s$ & $0.4$ & $0.375$ to $0.425$\\
		$r_s$ & $2$ & $1.77$ to $2.26$\\
		$e_J$ & $0$ & $-\,1.5$ to $1.5$\\
		$e_H$ & $1$ & $0.5$ to $2.0$\\
		$n_s$ & $0$ & $-\,1$, $0$, $1$\\
		\hline
		$j_3$ & $0$ & $-\,3000$ to $+\,3000$ \\
		$s_3^{\tau}$ & $0$ & $-\,500$ to $+\,500$ \\
		\hline
		$\epsilon_{2}$ & $0$ & $-\,1$, $0$, $1$ \\
		$\epsilon_{3}$ & $0$ & $-\,1$, $0$, $1$ \\
	\end{tabular}
	\end{center}
	\caption[Variation of thrust theory parameters]{Thrust theory parameters relevant for estimating the theory uncertainty, their
		default values, and range of values used for the scan for theory uncertainties.}
	\label{tab:theoryerrthrust}
\end{table}

In Ref.~\cite{Abbate:2010xh} slightly different profiles were used. For instance there was no
region of constant soft scale. This can be reproduced from our new profiles by choosing
\mbox{$t_0 = 0$}. Moreover, in Ref.~\cite{Abbate:2010xh} there was only one quadratic form after
the linear term, and the slope was completely determined by other parameters. These new profiles have several
advantages. The most obvious is a variable slope, which allows us to balance the introduction of logs and the
smoothness of the profiles. Additionally, in the new set up, the parameters for different regions are more
independent. For example, the $n_0$ parameter will only affect the nonperturbative region in the new profiles,
while in the old profiles, changing $n_0$ would have an impact on the resummation region. This independence
makes analyzing the different regions more transparent.

For the jet scale, we introduce a ``trumpeting'' factor that modifies the natural relation to the hard and soft
scales in the following way:
\begin{equation} \label{eq:muJprofile}
\!\!\!\!\!\mu_J(C) = \left\{\!\! \begin{array}{lr}
\big[\,1 + e_J (C-t_s)^2\,\big] \sqrt{ \mu_H\, \mu_{S} (C)} & C \le t_s\\
\,\mu_H & C > t_s
\end{array}
\right.\!.
\end{equation}
The parameter $e_J$ is varied in our theory scans.

The subtraction scale $R(C)$ can be chosen to be the same as $\mu_S(C)$ in the resummation
region to avoid large logarithms in the subtractions for the soft function. In the
nonperturbative region we do not want the $\mathcal{O}(\alpha_s)$ subtraction piece
to vanish, see \Eq{eq:d123}, so we choose the form
\begin{equation} \label{eq:muRprofile}
\!\!\!\!\!R(C) = \left\{\!\! \begin{array}{ll}
R_0                             & 0 \le C < t_0 \\
\zeta(R_0,\,0,\,0,\,\frac{r_s\,\mu_H}{6},\,t_0,\,t_1,\,C) & t_0 \le C < t_1 \\
\mu_S(C)                        & t_1 \le C \le 1
\end{array}
\right.\!\!.
\end{equation}
The only free parameter in this equation, $R_0$, simply sets the value of $R$ at $C=0$.
The requirement of continuity at $t_1$ in both $R(C)$ and its first derivative are again
ensured by the $\zeta$ function.

In order to account for resummation effects in the nonsingular partonic cross section,
which we cannot treat coherently, we vary $\mu_{\rm ns}$. We use three possibilities:
\begin{equation} \label{eq:muNSprofile}
\mu_{\rm ns}(C) = \left\{\! \begin{array}{ll}
\frac{1}{2} \big[\,\mu_H(C) + \mu_J (C)\,\big] &~ n_s \,= ~~\,1 \\
\mu_H  & ~n_s \,= ~~\,0 \\
\frac{1}{2} \big[\,3\,\mu_H(C) - \mu_J (C)\,\big] &~ n_s \,= -\,1
\end{array}
\right.\!\!.
\end{equation}
Using these variations, as opposed to those in Ref.~\cite{Abbate:2010xh}, gives more symmetric uncertainty
bands for the nonsingular distribution.

The plot in Fig.~\ref{fig:profile-variation} shows the scales for the default parameters for the case $Q=m_Z$ (thick lines). Also shown (gray dashed lines) are plots of $QC/6$ and $Q\sqrt{C/6}$.  In the resummation region, these correspond fairly well with the profile functions, indicating that in this region our analysis will avoid large logarithms. Note that the soft and jet scales in the plot would exactly match the gray dashed lines in the region $0.25<C<0.67$ if we took $r_s=1$ as our default. For reasons discussed in \Sec{subsec:slope-C} we use $r_s=2$ as our default value. We also set as default values $\mu_0 = 1.1\, {\rm GeV}$, $R_0 = 0.7\, {\rm GeV}$, $e_H=1$, $e_J=0$, and $n_s=0$. Default central values for other profile parameters for $C$ are listed in \Tab{tab:theoryerr}.

Perturbative uncertainties are obtained by varying the profile parameters. We hold $\mu_0$ and $R_0$ fixed, which are the parameters relevant in the region impacted by the entire nonperturbative shape function. They influence the meaning of  the nonperturbative soft function parameters in $F_C$.  The difference of the two parameters is important for renormalon subtractions and hence should not be varied ($\mu_0-R_0 = 0.4\, {\rm GeV}$) to avoid changing the meaning of $F_C$.  Varying $\mu_0$ and $R_0$ keeping the difference fixed has a very small impact compared to variations from $F_C$ parameters, as well as other profile parameters, and hence is also kept constant.  We are then left with eight profile parameters to vary during the theory scan, whose central values and variation ranges used in our analysis are $r_s = 2\times 1.13^{\pm1}$, $n_0=12\,\pm\,2$, $n_1=25\,\pm\,3$, $t_2 = 0.67\,\pm\,0.03$, $t_s=0.83\,\pm\,0.03$, $e_J = 0\,\pm\, 0.5$,  $e_H = 2^{\pm1}$, and $n_s = 0\,\pm\,1$.  The resulting ranges are also listed in \Tab{tab:theoryerr}, and the effect of these variations on the scales is plotted in \Fig{fig:profile-variation}. Since  we have so many events in our EVENT2 runs, the effect of $\epsilon_2^{\rm low}$ is completely negligible in the theory uncertainty scan. Likewise, the effect of $\epsilon_2^{\rm high}$ is also tiny above the shoulder
region.

Due to the advantages of the new profile functions, we have implemented them for the thrust predictions from Refs.~\cite{Abbate:2010xh} as well. For thrust we redefine $r_s \to 6\, r_s$, which eliminates all four appearances of the factor of 1/6 in \Eqs{eq:muSprofile}{eq:muRprofile}. After making this substitution, we can specify the theory parameters for thrust, which are summarized in \Tab{tab:theoryerrthrust}.These choices create profiles and profile variations that are very similar to those used in Ref.~\cite{Abbate:2010xh}. The only noticeable difference is the flat $\mu_S$ in the nonperturbative region (which is relevant for a fit to the full shape function but is irrelevant for the $\alpha_s$ tail fit).

\begin{figure*}[t!]
	\subfigure[]{
		\includegraphics[width=0.3\textwidth]{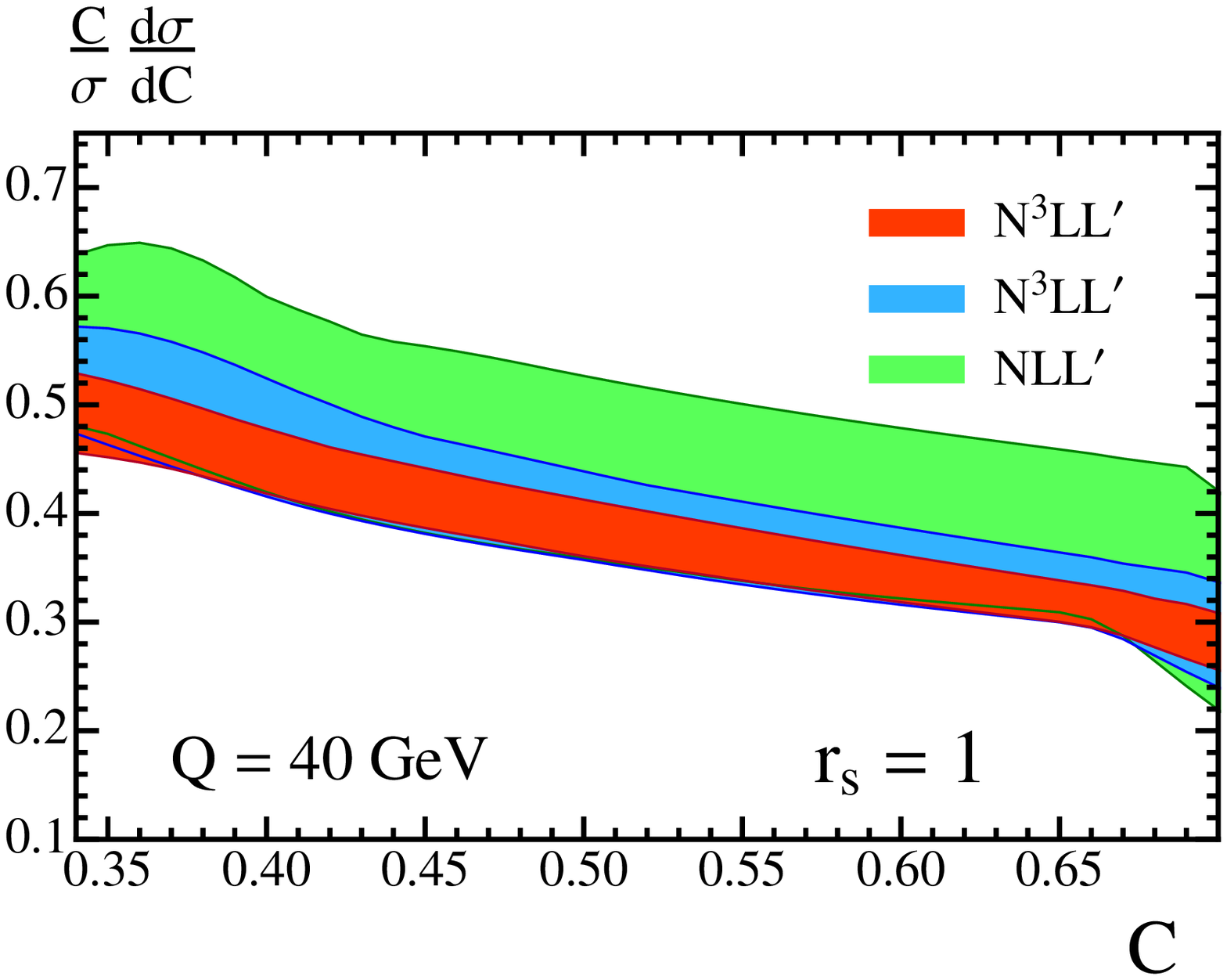}
		\label{fig:slope1-40}
	}
	\subfigure[]
	{
		\includegraphics[width=0.3\textwidth]{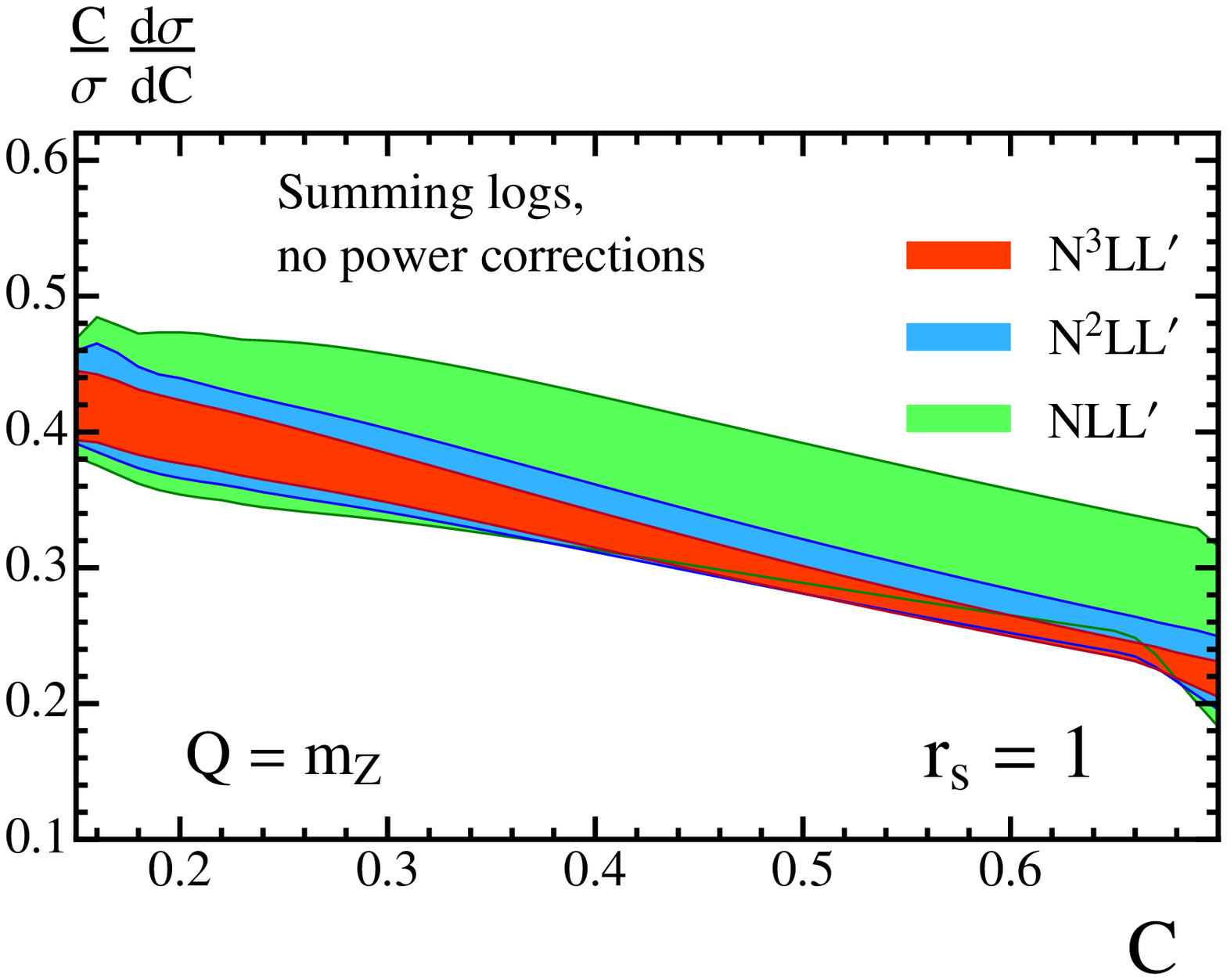}
		\label{fig:slope1-Mz}
	}
	\subfigure[]{
		\includegraphics[width=0.3\textwidth]{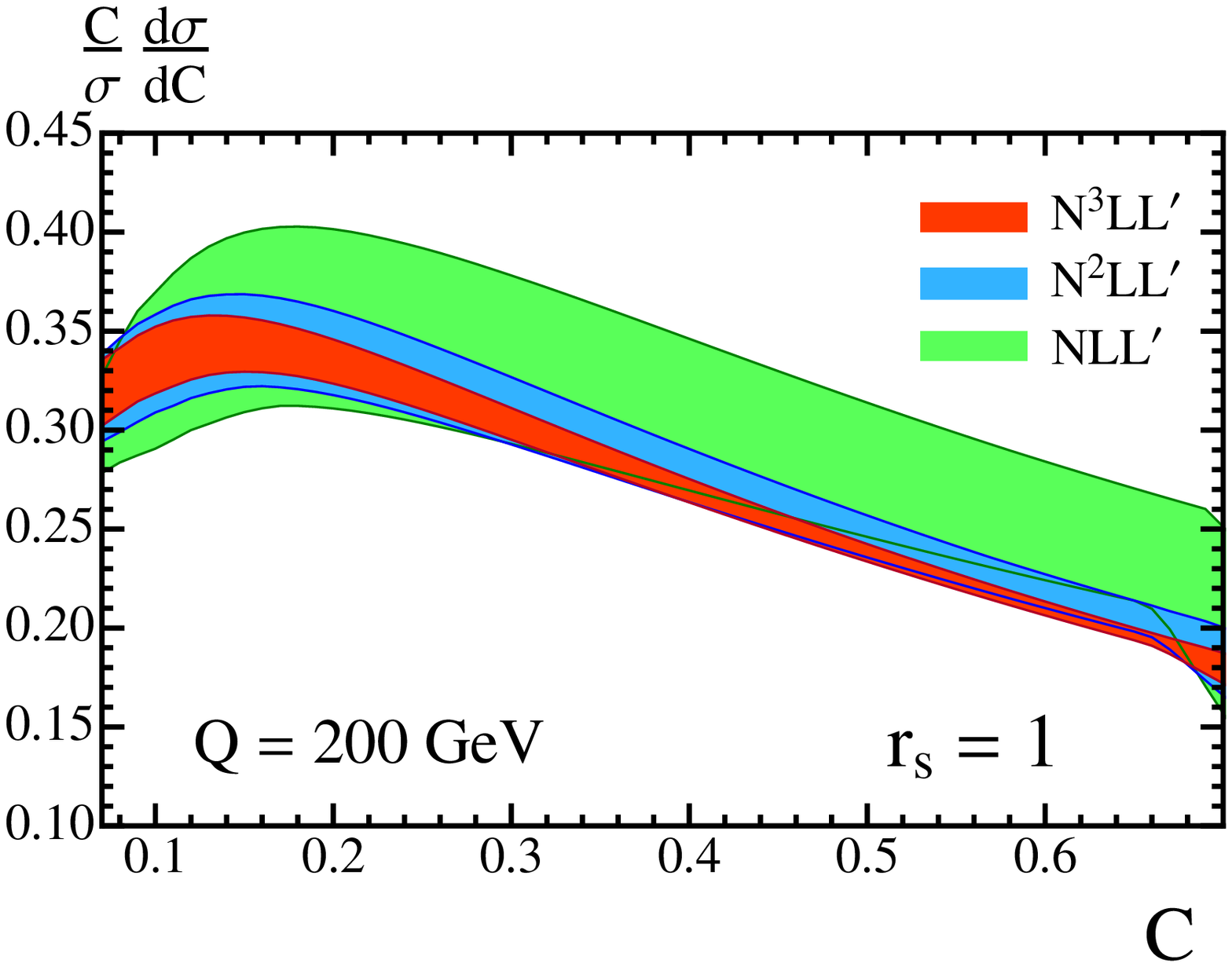}
		\label{fig:slope1-200}
	}
	\subfigure[]{
		\includegraphics[width=0.3\textwidth]{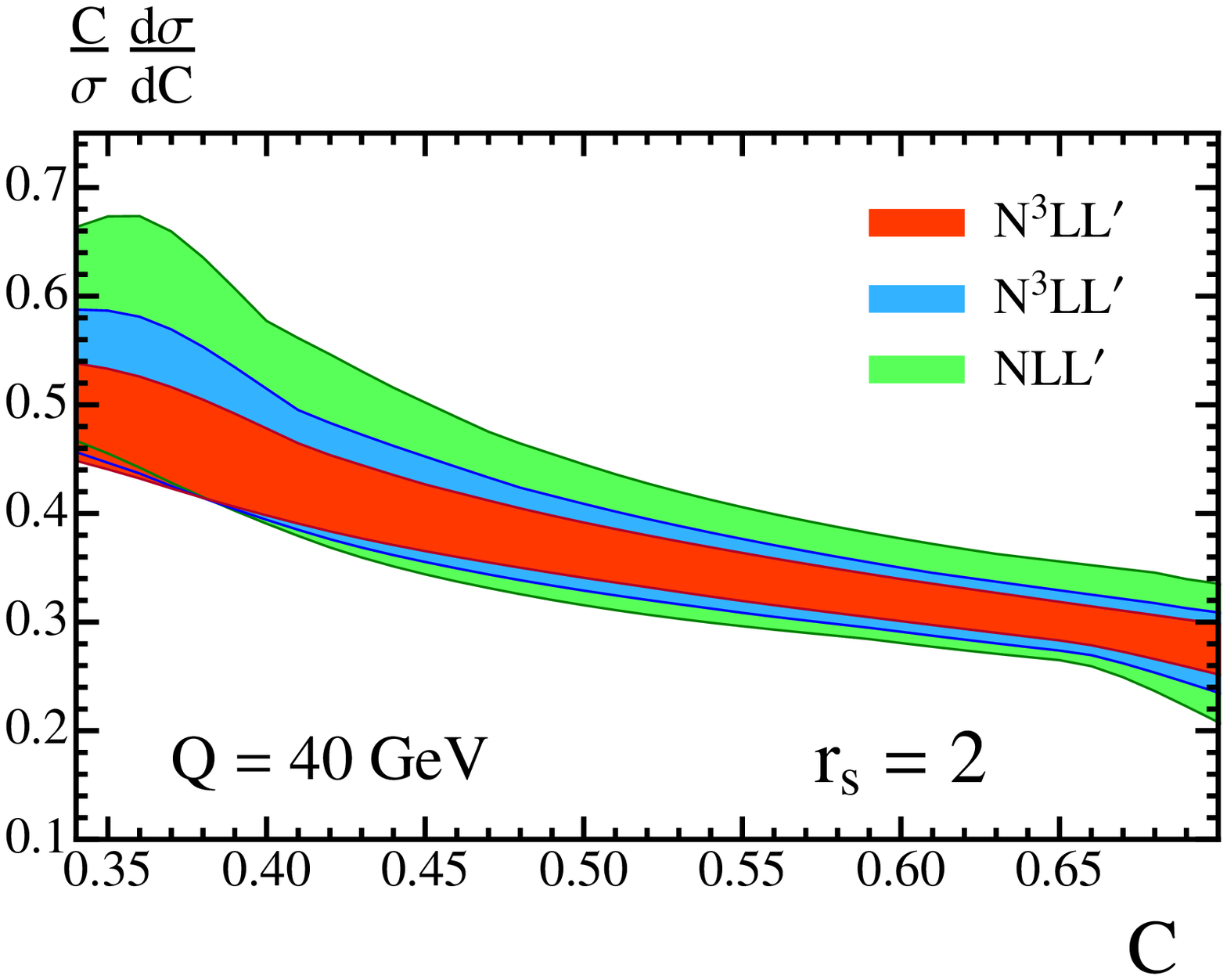}
		\label{fig:slope2-40}
	}
	\subfigure[]{
		\includegraphics[width=0.3\textwidth]{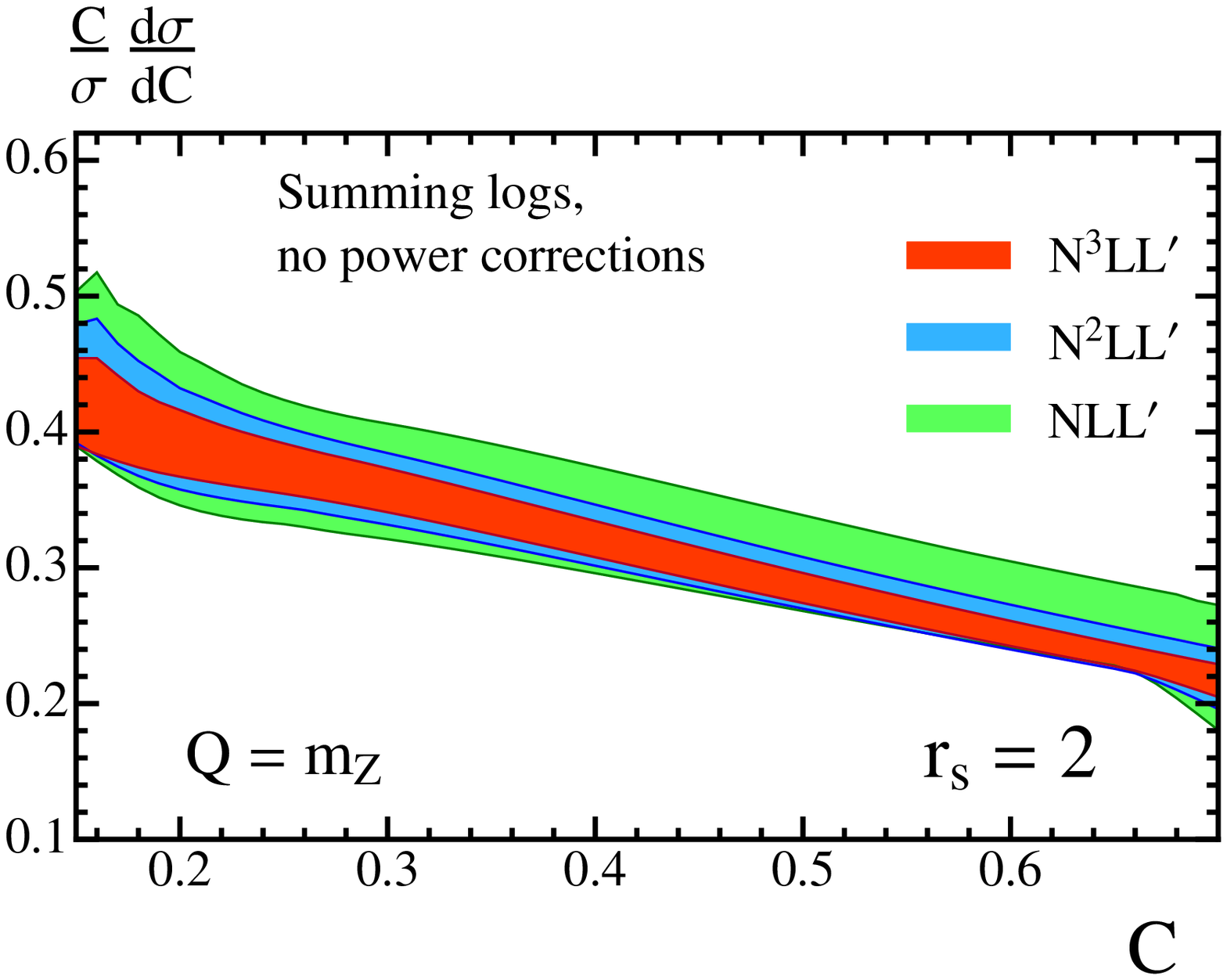}
		\label{fig:slope2-Mz}
	}
	\subfigure[]{
		\includegraphics[width=0.3\textwidth]{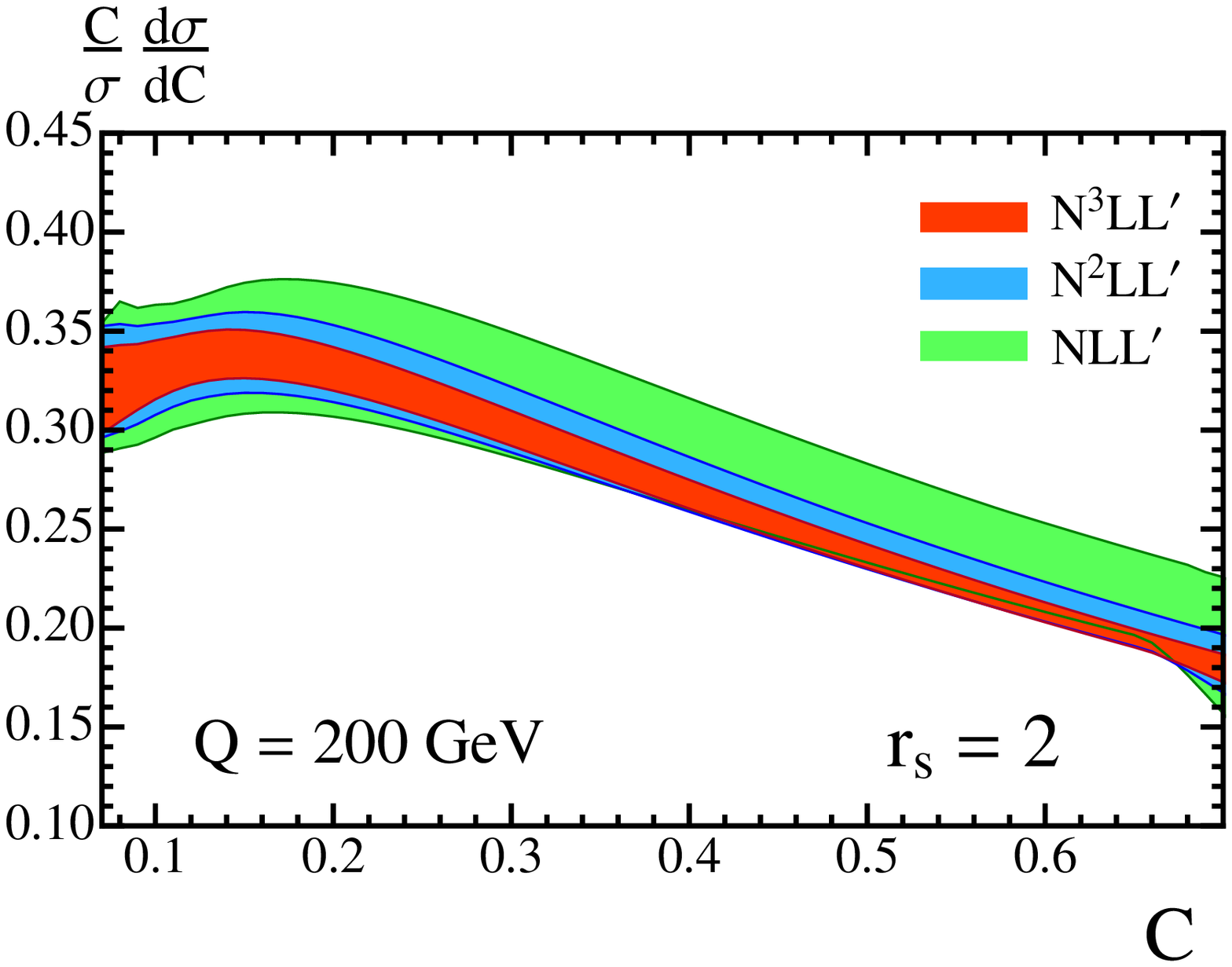}
		\label{fig:slope2-200}
	}
	\subfigure[]
	{
		\includegraphics[width=0.3\textwidth]{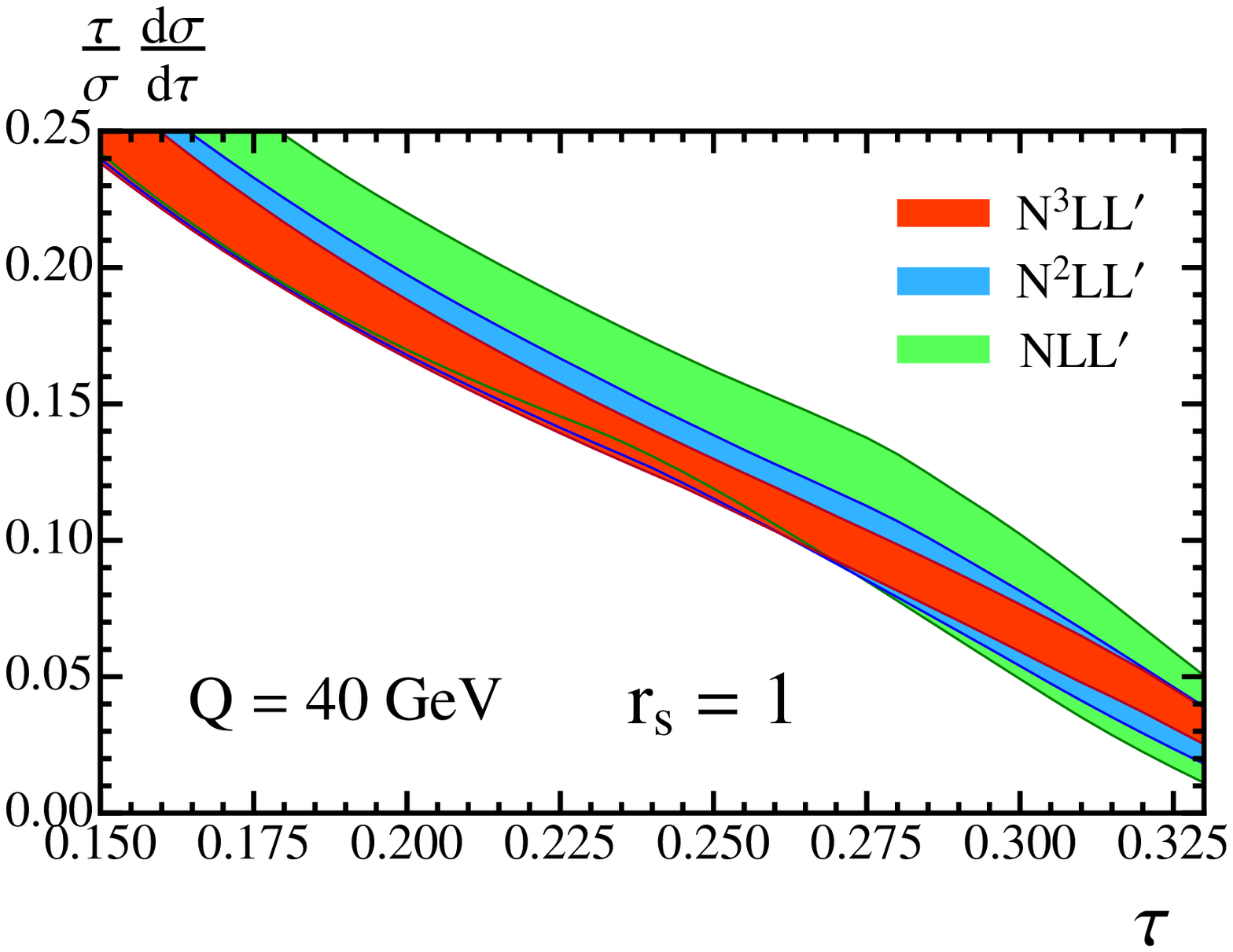}
		\label{fig:slope1-40-T}
	}
	\subfigure[]{
		\includegraphics[width=0.3\textwidth]{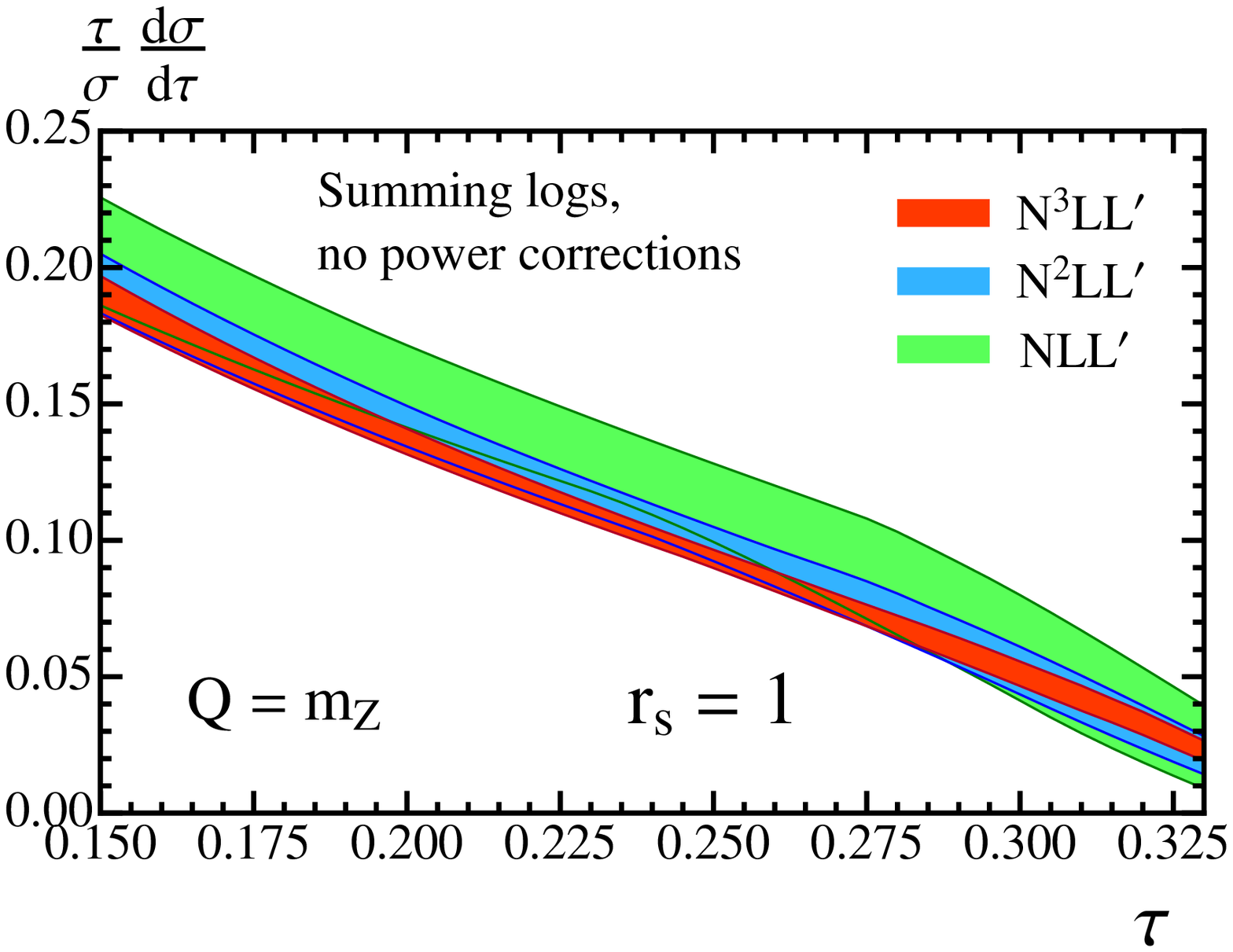}
		\label{fig:slope1-Mz-T}
	}
	\subfigure[]{
		\includegraphics[width=0.3\textwidth]{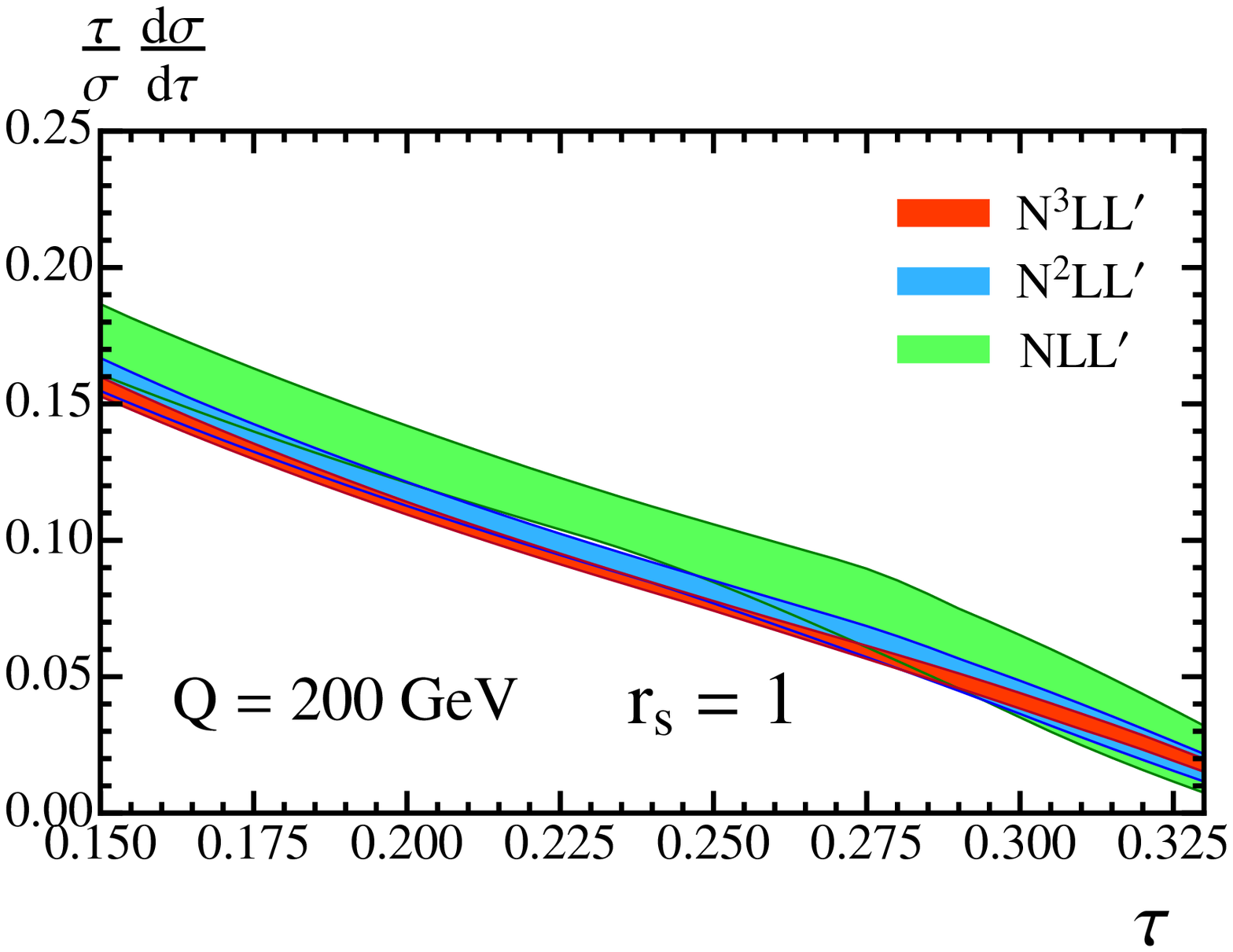}
		\label{fig:slope1-200-T}
	}
	\subfigure[]{
		\includegraphics[width=0.3\textwidth]{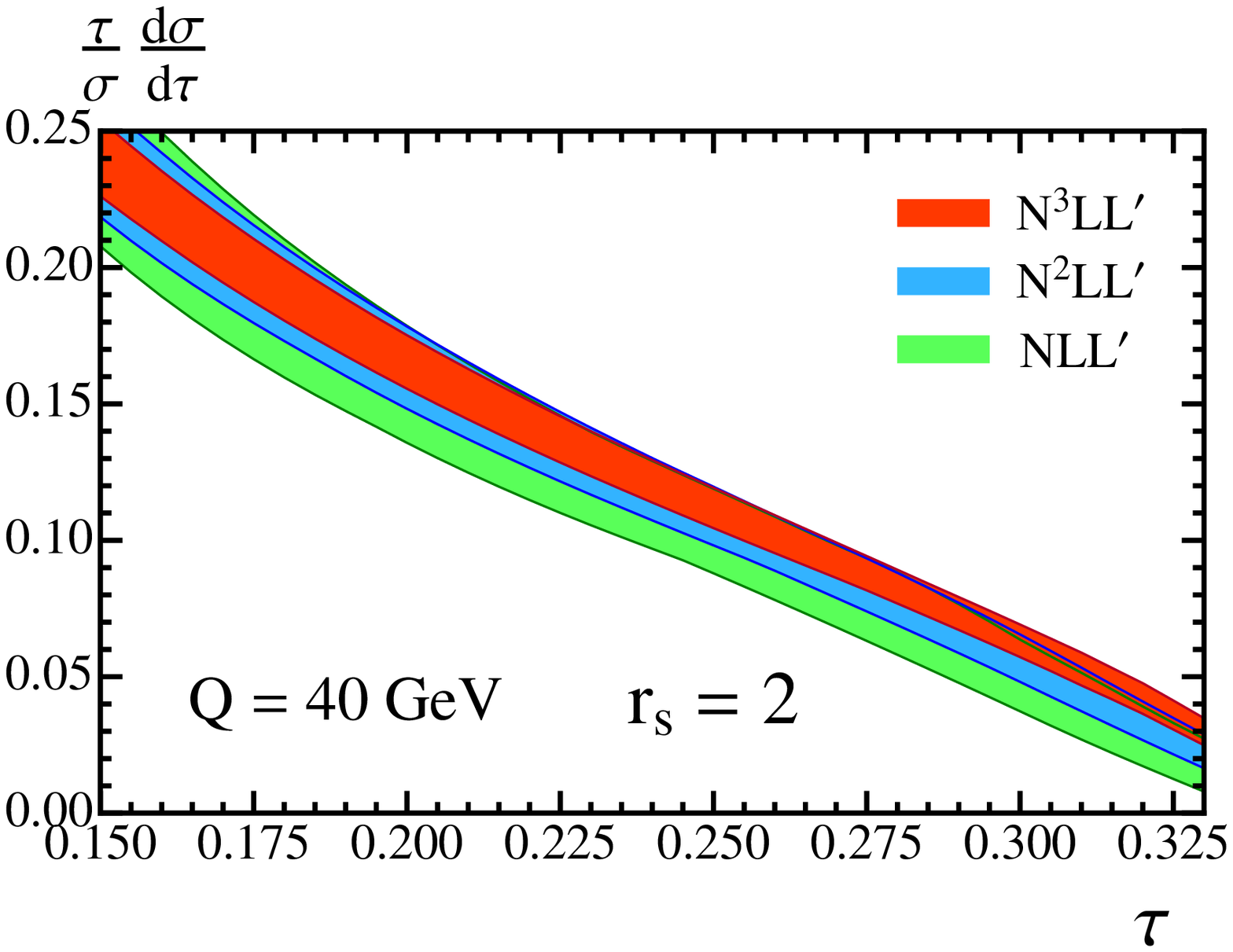}
		\label{fig:slope2-40-T}
	}
	\subfigure[]{
		\includegraphics[width=0.3\textwidth]{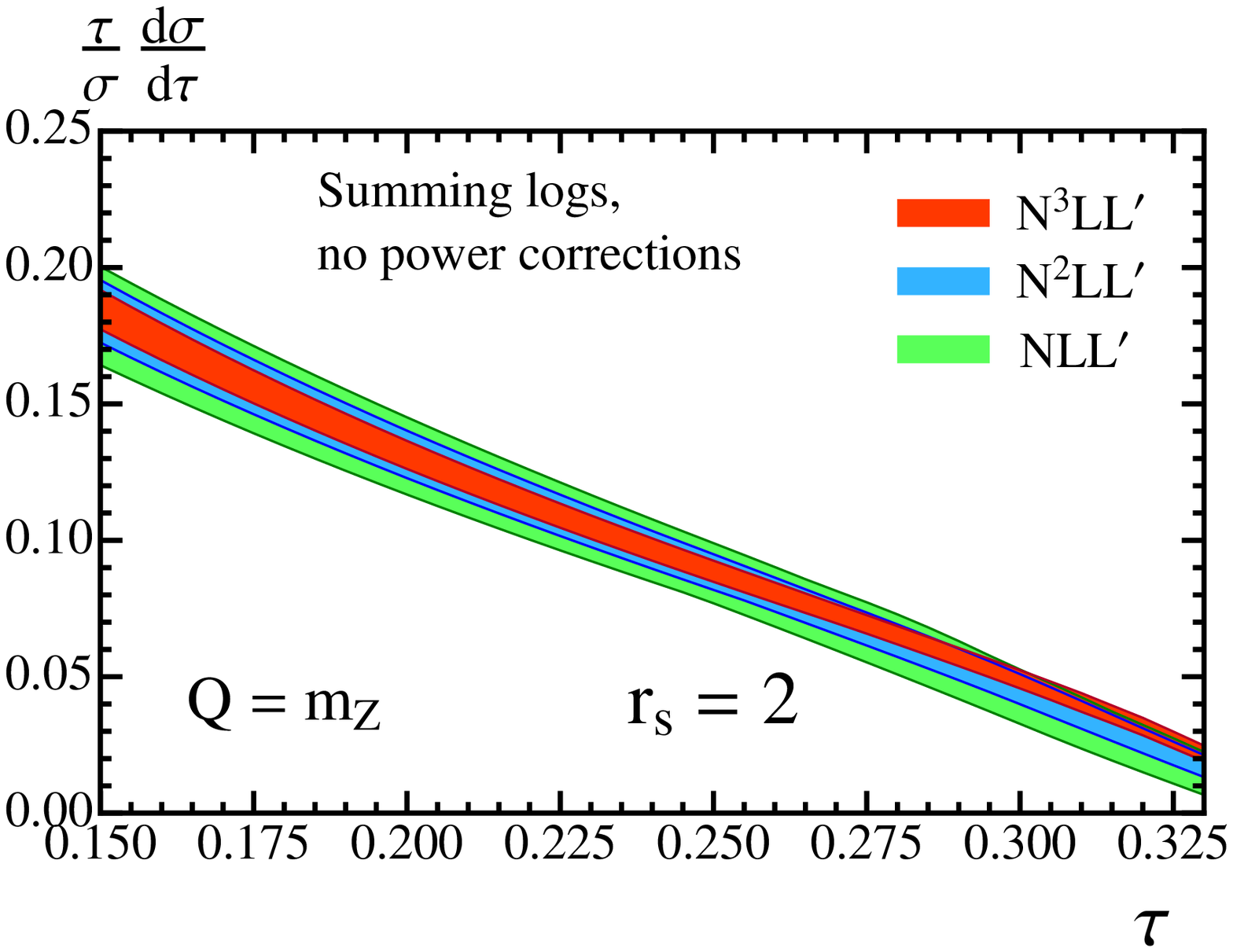}
		\label{fig:slope2-Mz-T}
	}
	\subfigure[]{
		\includegraphics[width=0.3\textwidth]{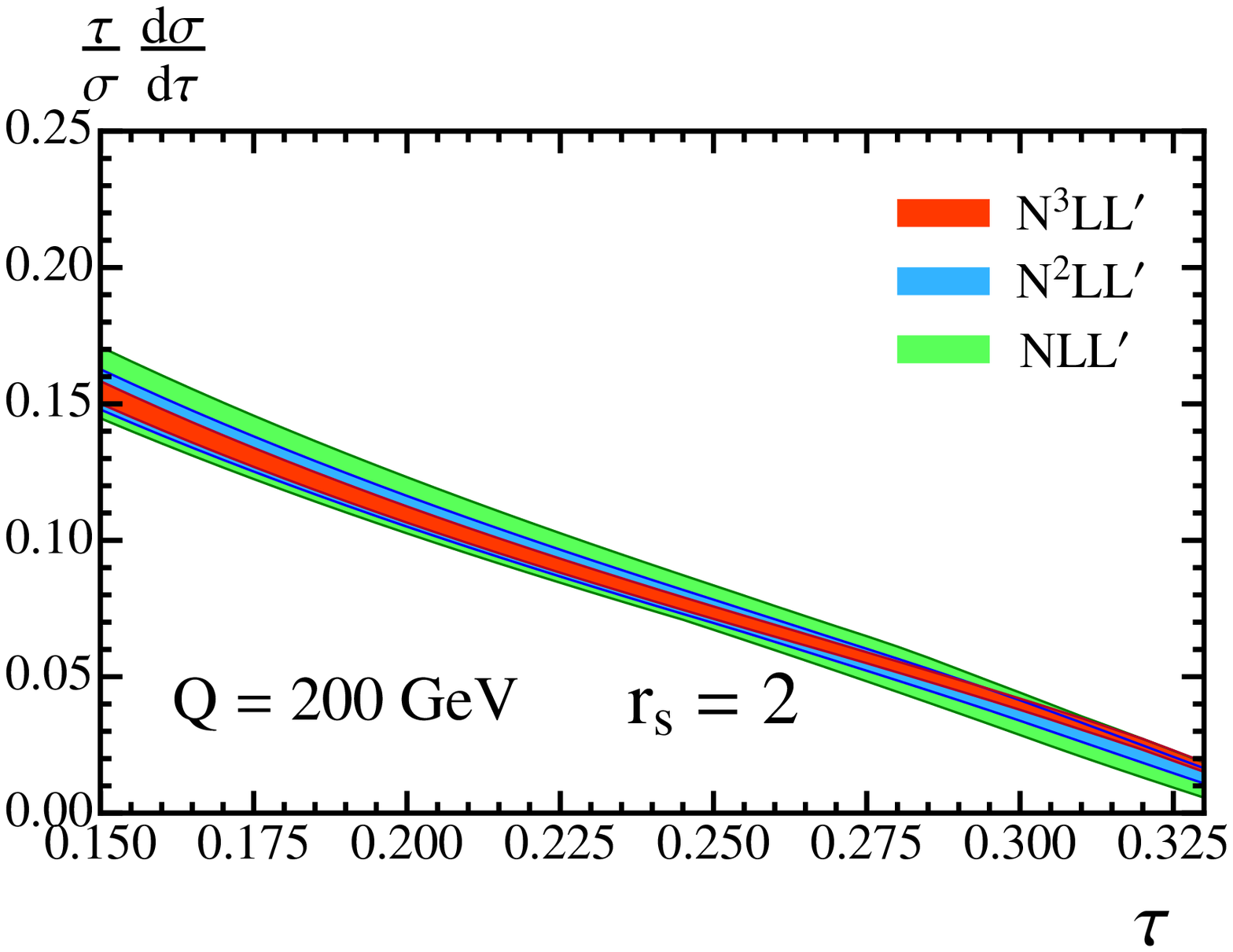}
		\label{fig:slope2-200-T}
	}
	\caption[Cross sections scans for C-parameter and thrust]{Theory scan for C-parameter [\,panels (a) to (f)\,] and Thrust [\,panels (g) to (l)\,] cross section uncertainties at
		various orders, for few center-of-mass energies $Q$ and slopes $r_s$. The theoretical predictions include log resummation and are purely perturbative. The respective upper and lower rows use $r_s = 1$ and $r_s = 2$, respectively. The left, center, and right columns correspond to $Q = 40$\,GeV, $91.2$\,GeV, and $200$\,GeV, respectively. Here we use $\alpha_s(m_Z)=0.1141$. }
	\label{fig:12-plots}%
\end{figure*}

\section{Results}
\label{sec:theory-results}
In this section we present our final results for the \mbox{C-parameter} cross section, comparing  to the thrust cross section  when appropriate.  We will use different levels of theoretical accuracy for these analyses. When indicating the perturbative precision, and whether or not the power correction $\Omega_1$ is included and at what level of precision, we follow the notation
\begin{align}
& {\cal O}(\alpha_s^k) 
& \phantom{x} & \text{fixed order up to ${\cal O}(\alpha_s^k)$} 
\nn\\ 
& \text{N}^k\text{LL}^\prime \!+\! {\cal O}(\alpha_s^k) 
& \phantom{x} & \text{perturbative resummation} 
\nn\\ 
& \text{N}^k\text{LL}^\prime \!+\! {\cal O}(\alpha_s^k)  \!+\! {\overline \Omega}_1
& \phantom{x} & \text{$\overline{\rm MS}$ scheme for $\Omega_1$} 
\nn\\ 
& \text{N}^k\text{LL}^\prime \!+\! {\cal O}(\alpha_s^k) 
\!+\!  {\Omega}_1(R,\mu)
& \phantom{x} & \text{Rgap scheme for $\Omega_1$} 
\nn\\ 
& \text{N}^k\text{LL}^\prime \!+\! 
{\cal O}(\alpha_s^k) \!+\! {\Omega}_1(R,\mu,r)
& \phantom{x} & \text{Rgap scheme with }
\nn\\
& & \phantom{x} & \  \text{hadron masses for $\Omega_1$} 
\,. \nn
\end{align}
In the first three subsections, we discuss the determination of higher-order perturbative coefficients in the cross section, the impact of the slope parameter $r_s$, and the order-by-order convergence and uncertainties. 
Since the effect of the additional running induced by the presence of hadron masses is a relatively small effect on the cross section, we leave their discussion to the final fourth subsection.

\begin{table}[t!]
	\begin{center}
	\begin{tabular}{c|c}
		Resummation Order&  Calculable $G_{ij}$'s and $B_i$'s \\
		\hline
		LL  &  $G_{i,\,i+1}$ \\
		NLL$^\prime$  &  $G_{i,\,i}$ and $B_1$\\
		N${}^2$LL$^\prime$  &  $G_{i,i-1}$ and $B_2$\\ 
		N${}^3$LL$^\prime$  &  $G_{i,i-2}$ and $B_3$\\ 
	\end{tabular}
	\end{center}
	\caption[Hierarchy of $G_{ij}$ coefficients]{By doing resummation to the given order, we can access results for the entire hierarchy of $G_{ij}$'s listed.}
	\label{tab:Gijorders}
\end{table}
\begin{figure*}[t!]
	\includegraphics[width=0.5\columnwidth]{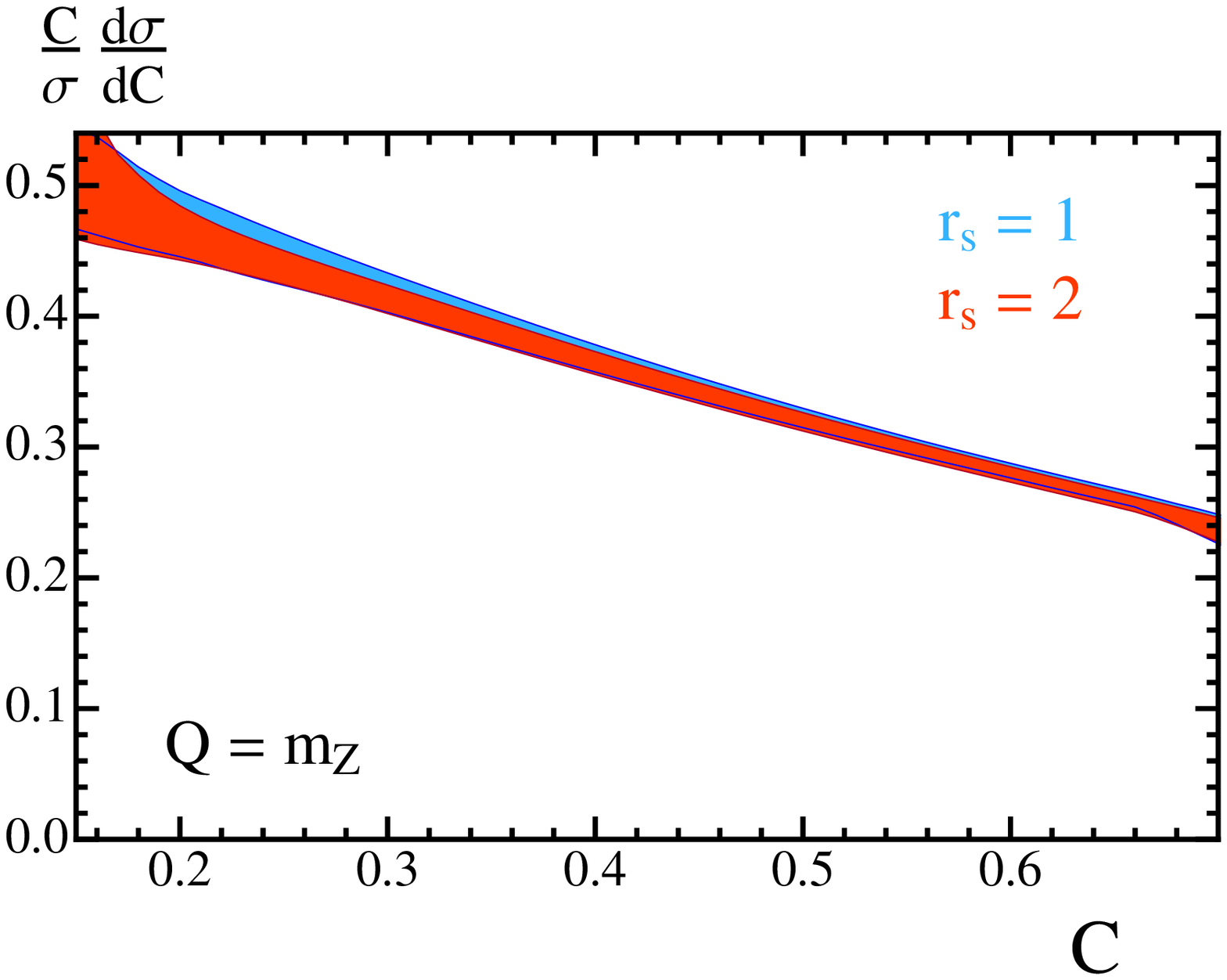}
	\includegraphics[width=0.5\columnwidth]{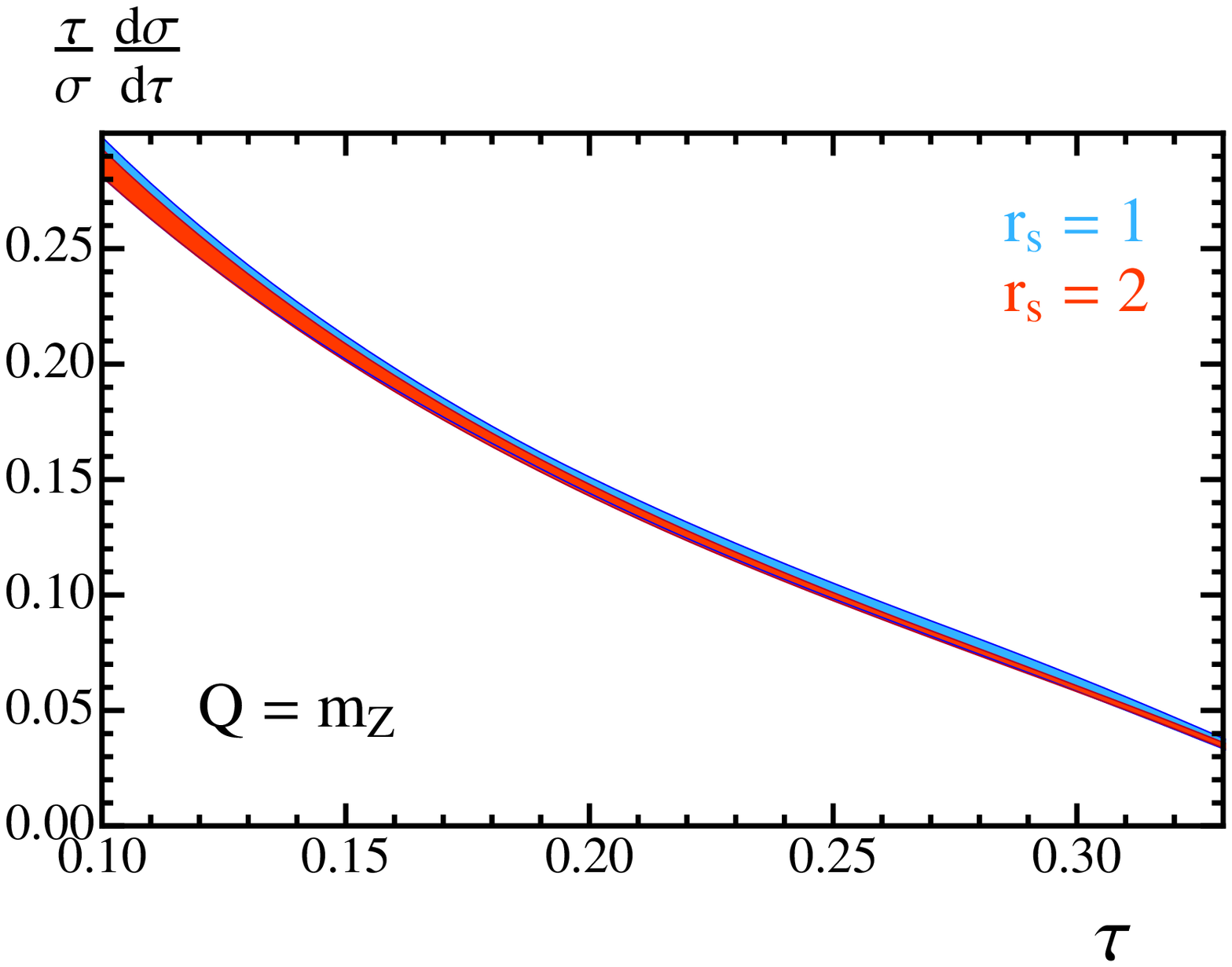}
	\caption[Slope comparison for C-parameter and thrust]{C-parameter (left panel) and thrust (right panel) cross section predictions at N$^3$LL$^\prime+{\cal O}(\alpha_s^3)$ + $\Omega_1(R,\mu)$ for $r_s = 1$ (blue) and $r_s = 2$ (red).
		For these plots we use our most complete setup, with power corrections in the renormalon-free Rgap scheme. Here we use $\alpha_s(m_Z)=0.1141$ and $\Omega_1(R_\Delta,\mu_\Delta)=0.33\,{\rm GeV}$.}
	\label{fig:highest-slope}
\end{figure*}
\subsection{$\mathbf{G_{ij}}$ expansion}
\label{subsec:Gij}
Our N$^3$LL$^\prime$ resummed predictions can be used to compute various coefficients of the most singular terms in the cross section. In this determination only perturbative results are used.  In order to exhibit the terms that are determined by the logarithmic resummation, one can take $\mu=Q$ and write the cumulant function as:
\begin{align}\label{eq:BG1}
\Sigma_0 (C) = & \,\dfrac{1}{\sigma_0}\!\int^C_0\!\!\! \df C^\prime\, \dfrac{\df \sigma}{\df C^\prime} =
\left(\! 1 + \sum_{m=1}^{\infty} B_{m}^{[0]} \!\left( \dfrac{\alpha_s(Q)}{2 \pi} \right)^{\!\!m} \right) \exp\! \left( \sum_{i=1}^{\infty} \sum_{j=1}^{i+1} G_{i j}\! \left( \dfrac{\alpha_s(Q)}{2 \pi} \right)^{\!\!i}
\!\ln^j\!\!\left( \dfrac{6}{C} \right) \!\right)\!,
\end{align}
where $\sigma_0$ is the tree-level total cross section.
A different normalization with respect to the total cross section including all QCD corrections is also used in the literature and reads
\begin{align}\label{eq:BG2}
\Sigma (C) = &\, \dfrac{1}{\sigma_\text{had}}\!\int^C_0 \!\df C^\prime \dfrac{\df \sigma}{\df C^\prime} =
\left(\! 1 + \sum_{m=1}^{\infty} B_m\! \left( \dfrac{\alpha_s(Q)}{2 \pi} \right)^{\!\!m} \right) \exp\! \left( \sum_{i=1}^{\infty} \sum_{j=1}^{i+1} G_{i j}\! \left( \dfrac{\alpha_s(Q)}{2 \pi} \right)^{\!\!i}
\!\ln^j\!\! \left( \dfrac{6}{C} \right)\! \right)\!.
\end{align}
Notice that the different normalizations do not affect the $G_{ij}$'s and only change the non-logarithmic pieces. 
Our $\text{N}^3 \text{LL}^\prime$ result allows us to calculate the $B_{m}^{[0]}$'s and $B_{m}$'s to third order and 
entire hierarchies of the $G_{ij}$ coefficients as illustrated in \Tab{tab:Gijorders}.

The results for these coefficients through $G_{34}$ are collected in App.~\ref{ap:Gijcoefficients}. Note that, due to the equivalence of the thrust and C-parameter distributions at NLL (when using $\widetilde{C}$), we know that the $G_{i,\,i+1}$ and $G_{i,\,i}$ series are equal for these two event shapes~\cite{Catani:1998sf}. From our higher-order resummation analysis, we find that the $G_{i,i-1}$ and $G_{i,i-2}$ coefficients differ for the C-parameter and thrust event shapes because the fixed-order $s_1^{\widetilde C,\tau}$ and $s_2^{\widetilde C,\tau}$ constants differ and enter into the resummed results at N$^2$LL$^\prime$ and N$^3$LL$^\prime$ respectively. The values of all the $B_i$ coefficients are also different. (If we had instead used the unprimed counting for logarithms, N$^k$LL, there is less precision obtained at a given order, and each index of a $B_i$ in \Tab{tab:Gijorders} would be lowered by $1$.)

The $G_{ij}$ and $B_m$ serve to illustrate the type of terms that are included by having resummation and fixed-order terms at a given order. They are not used explicitly for the resummed analyses in the following sections, which instead exploit the full resummed factorization theorem in \Eq{eq:singular-resummation}.

\begin{figure*}[t!]
	\includegraphics[width=.5\columnwidth]{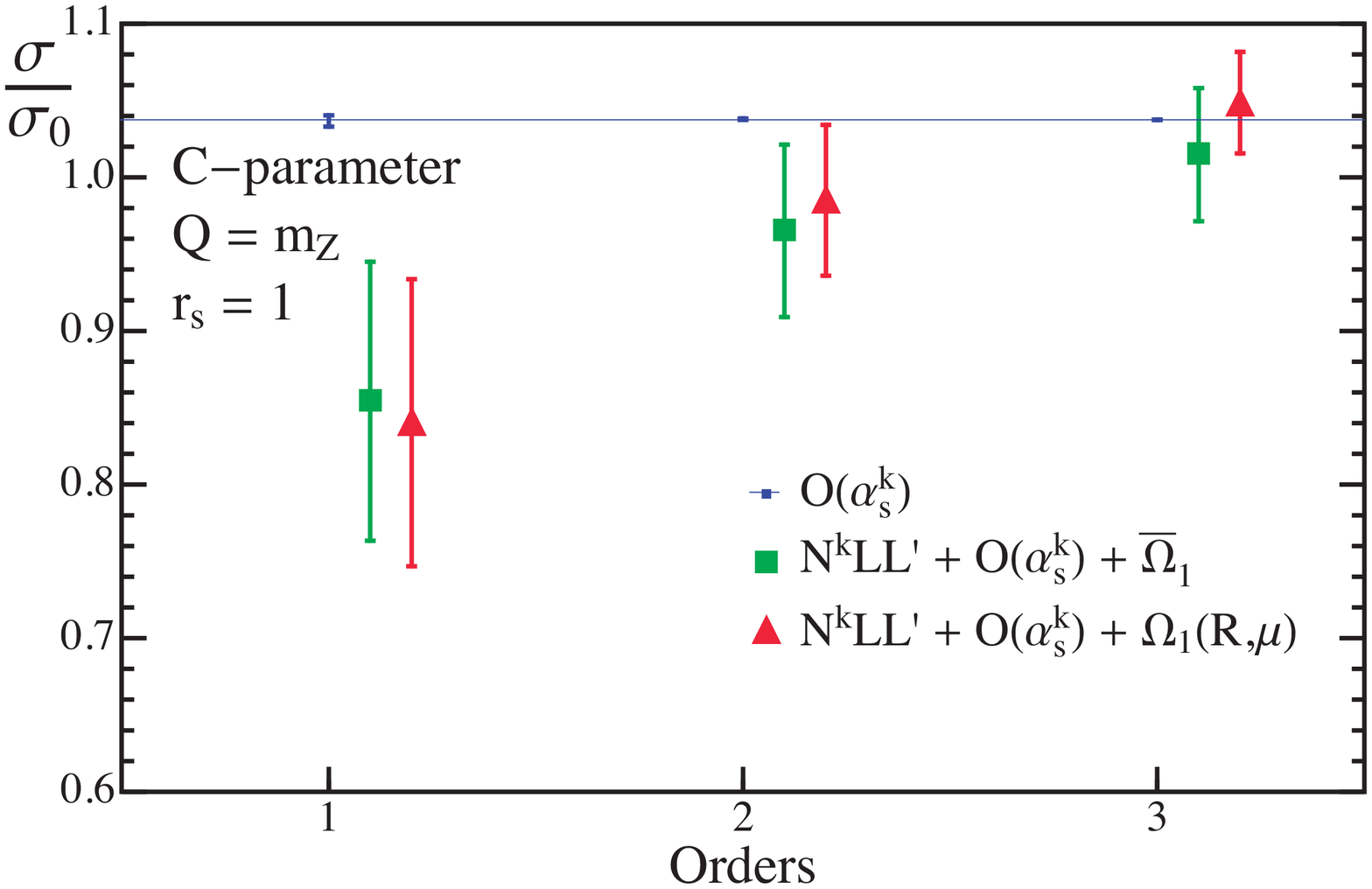}
	\includegraphics[width=.5\columnwidth]{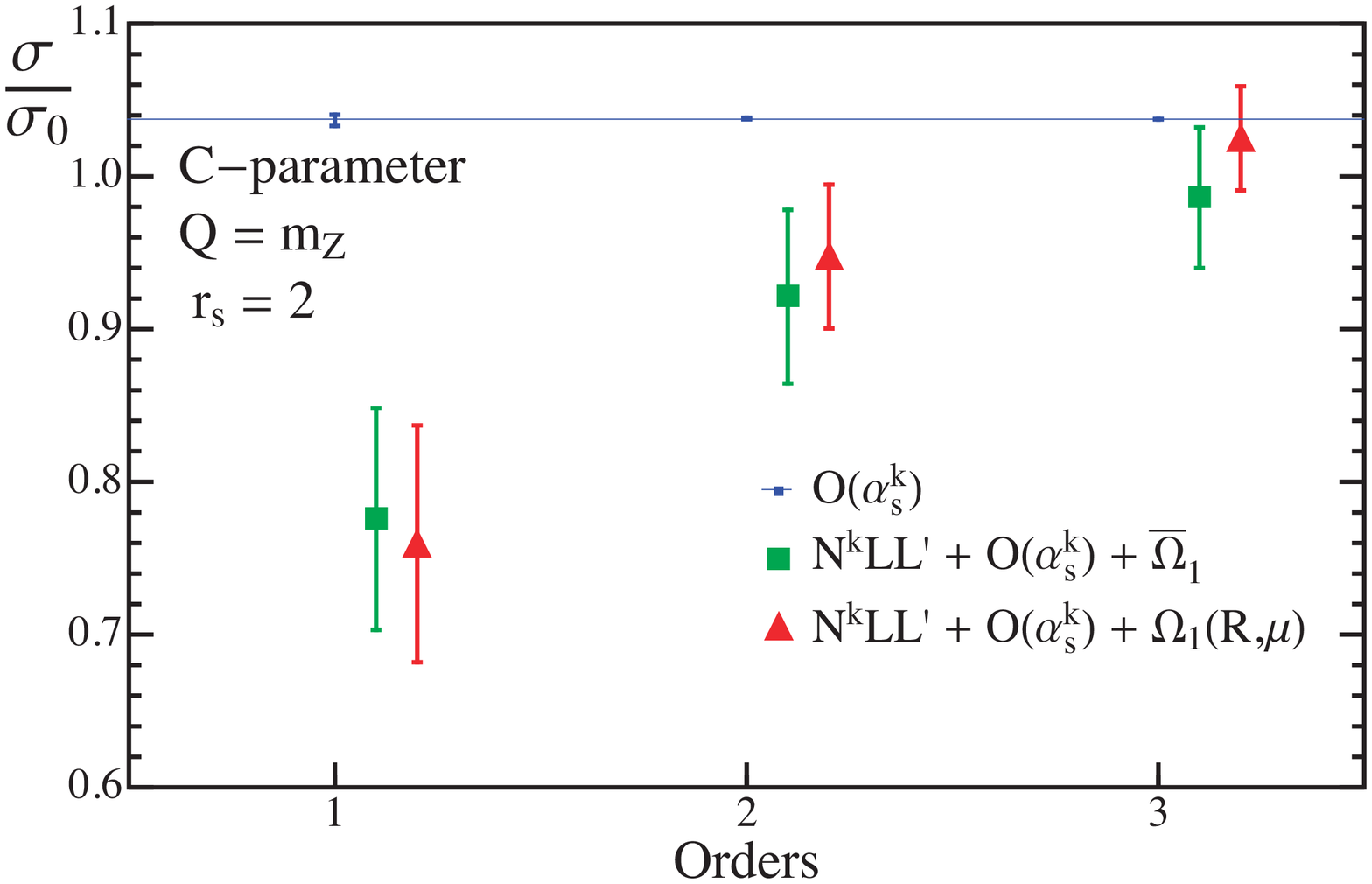}
	\includegraphics[width=.5\columnwidth]{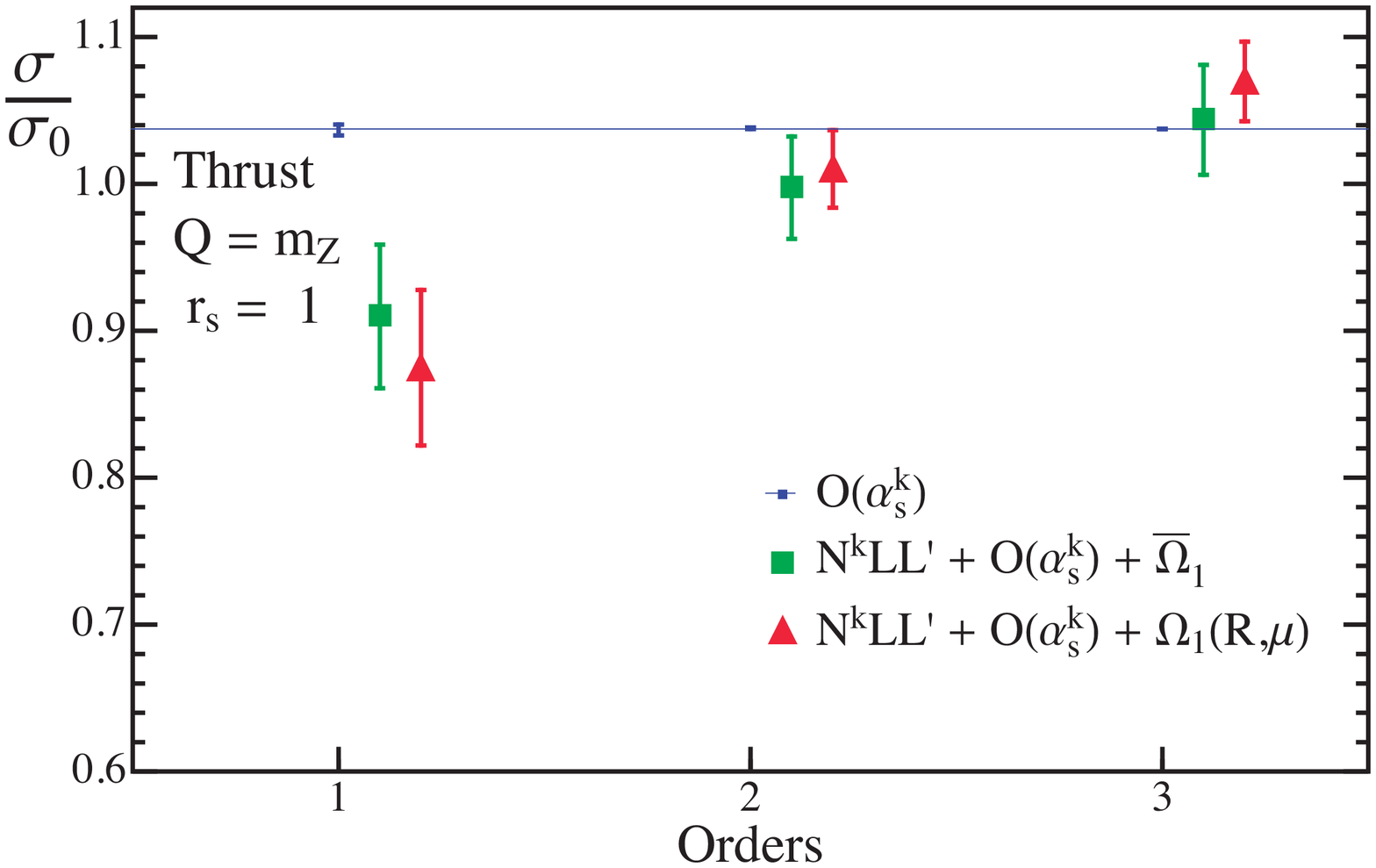}
	\includegraphics[width=.5\columnwidth]{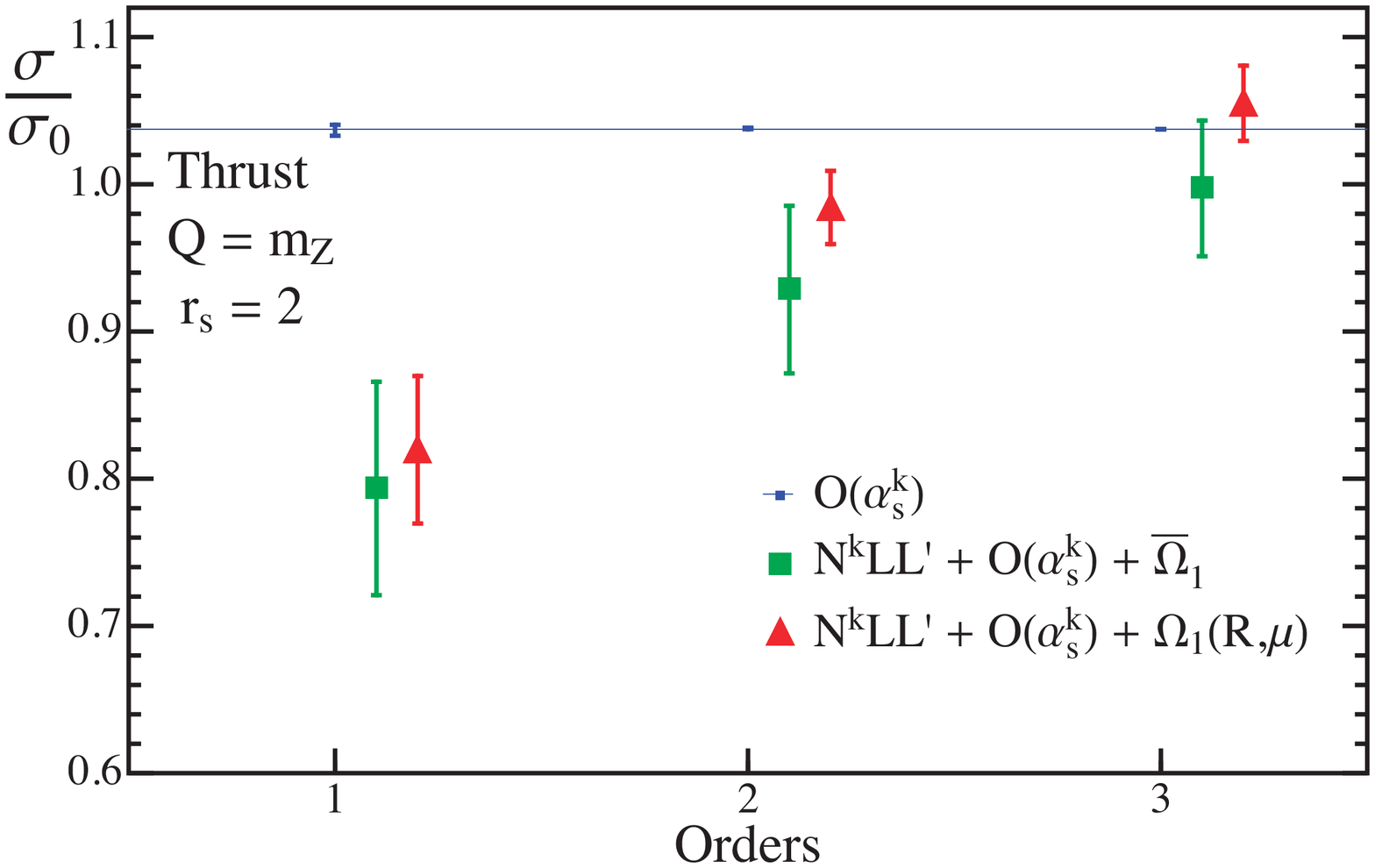}
	\caption[Total hadronic cross section for C-parameter]{\label{fig:norm-C}Total hadronic cross section obtained from integrating the resummed cross section. The top two panels show the prediction for $r_s = 1$ and $r_s = 2$ for C-parameter, respectively. Likewise, the bottom two panels  show the thrust results. Green squares correspond to the prediction with log resummation and the power correction in the $\msbar$ scheme, whereas red triangles have log resummation and the power correction in  the Rgap scheme. The blue points correspond to the fixed-order prediction, and the blue line shows the highest-order FO prediction. }
\end{figure*}
\begin{figure}[t!]
	\begin{center}
		\includegraphics[width=0.65\columnwidth]{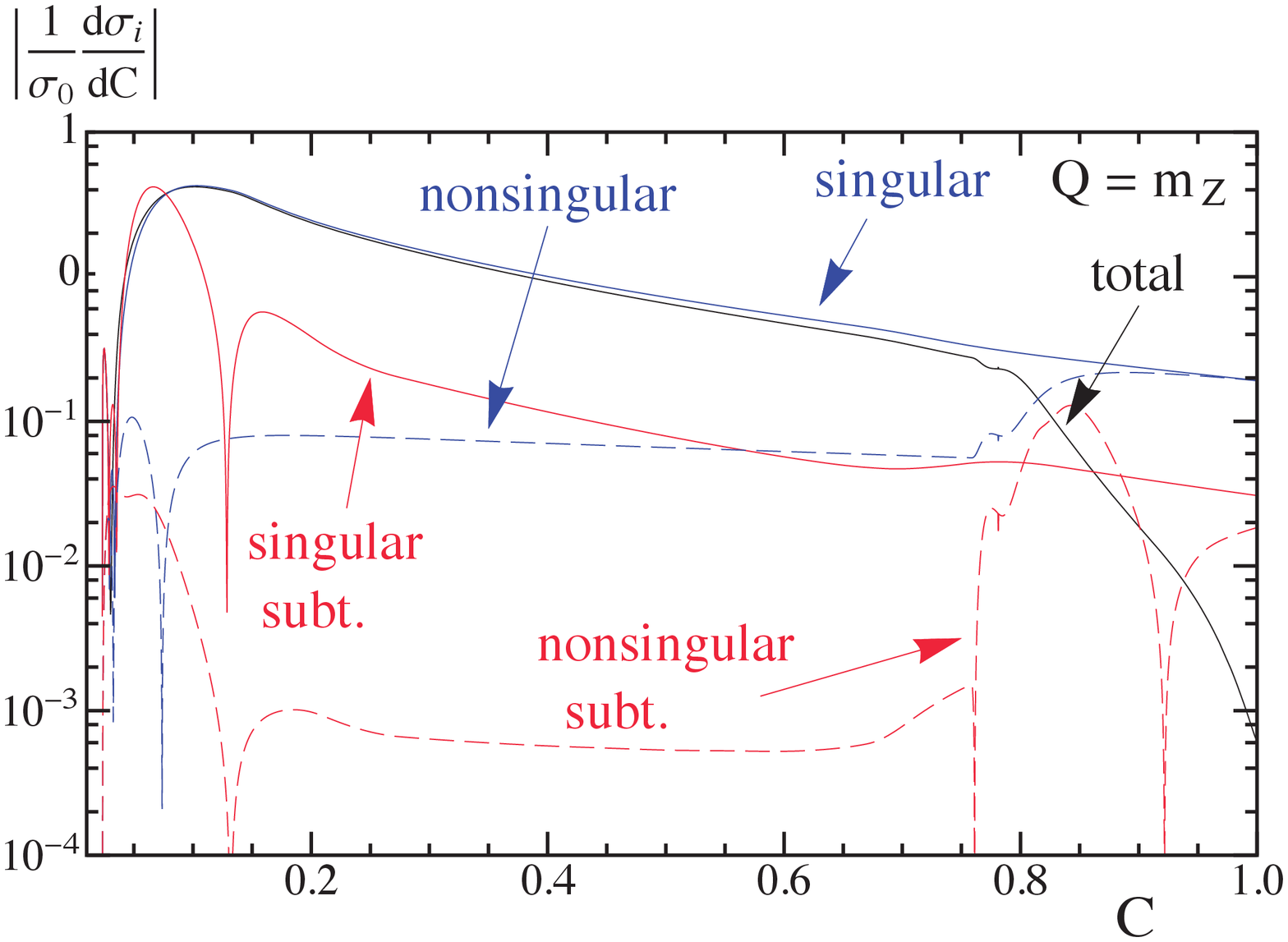}
		\caption[Components of resummed C-parameter cross section]{Components of the C-parameter cross section with resummation  at N$^3$LL$^\prime+{\cal O}(\alpha_s^3)+\Omega_1(R,\mu)$ with $\Omega_1(R_\Delta,\mu_\Delta)=0.33\,$GeV
			and $\alpha_s(m_Z) = 0.1141$.}
		\label{fig:component-plot-sum}
	\end{center}
\end{figure}

\subsection{The slope $\mathbf{r_s}$ for C-parameter and Thrust}
\label{subsec:slope-C}
In the profiles of \Sec{sec:profiles}, the parameter $r_s$ was defined as the dimensionless slope of $\mu_S$ in the resummation region. It would seem natural to pick $r_s=1$ to eliminate the powers of $\ln(6\, \mu_S/(QC))$ and $\ln(\mu_S/(Q\tau))$ that appear in the cross section formula for $C$ and $\tau$. However, having a slightly steeper rise may also yield benefits by smoothing out the profile. Using an $r_s$ that is larger than 1, such as $r_s=2$, will only shift small $\ln(r_s)$ factors between different orders of the resummed cross section. Just like other profile parameters the dependence on $r_s$ will decrease as we go to higher orders in perturbation theory, but the central value choice may improve the accuracy of lower-order predictions.

In order to determine whether $r_s=1$ or $r_s=2$ is a better choice for the slope parameter, we examine the convergence of the cross section between different orders of resummation. For this analysis we will compare the three perturbative orders N$^3$LL$^\prime +{\cal O}(\alpha_s^3)$, N$^2$LL$^\prime +{\cal O}(\alpha_s^2)$, and NLL$^\prime +{\cal O}(\alpha_s)$. We also fix $\alpha_s(m_Z)=0.1141$, which is the value favored by the QCD only thrust fits~\cite{Abbate:2010xh}. [Use of larger values of $\alpha_s(m_Z)$ leads to the same conclusions that we draw below.]  In \Fig{fig:12-plots} we show the perturbative C-parameter cross section (upper two rows) and thrust cross section (lower two rows) with a scan over theory parameters (without including $\Omega_1$ or the shape function) for both $r_s=1$ (first respective row) and $r_s=2$ (second respective row). Additionally, we plot with different values of $Q$, using $Q=40$\,GeV in the first column, $Q=91.2$\,GeV  in the second column, and $Q=200$\,GeV in the third column.  The bands here correspond to a theory parameter scan with $500$ random points taken from Tabs.~\ref{tab:theoryerr} and \ref{tab:theoryerrthrust}. We conclude from these plots that $r_s=2$ has better convergence between different orders than $r_s=1$. For all of the values of $Q$, we can see that in the slope 1 case the N${}^2$LL$^\prime$ band lies near the outside of the edge of the NLL$^\prime$ band, while in the slope 2 plots, the scan for N${}^2$LL$^\prime$ is entirely contained within the scan for NLL$^\prime$. A similar picture can be seen for the transition from N${}^2$LL$^\prime$ to N${}^3$LL$^\prime$. This leads us to the conclusion that the resummed cross section prefers $r_s=2$ profiles which we choose as our central value for the remainder of the analysis. (In the thrust analysis of \Ref{Abbate:2010xh}, the profiles did not include an independent slope parameter, but in the resummation region, their profiles are closer to taking $r_s = 2$ than $r_s=1$.)
\begin{figure*}[t!]
	\subfigure[]
	{
		\includegraphics[width=0.48\textwidth]{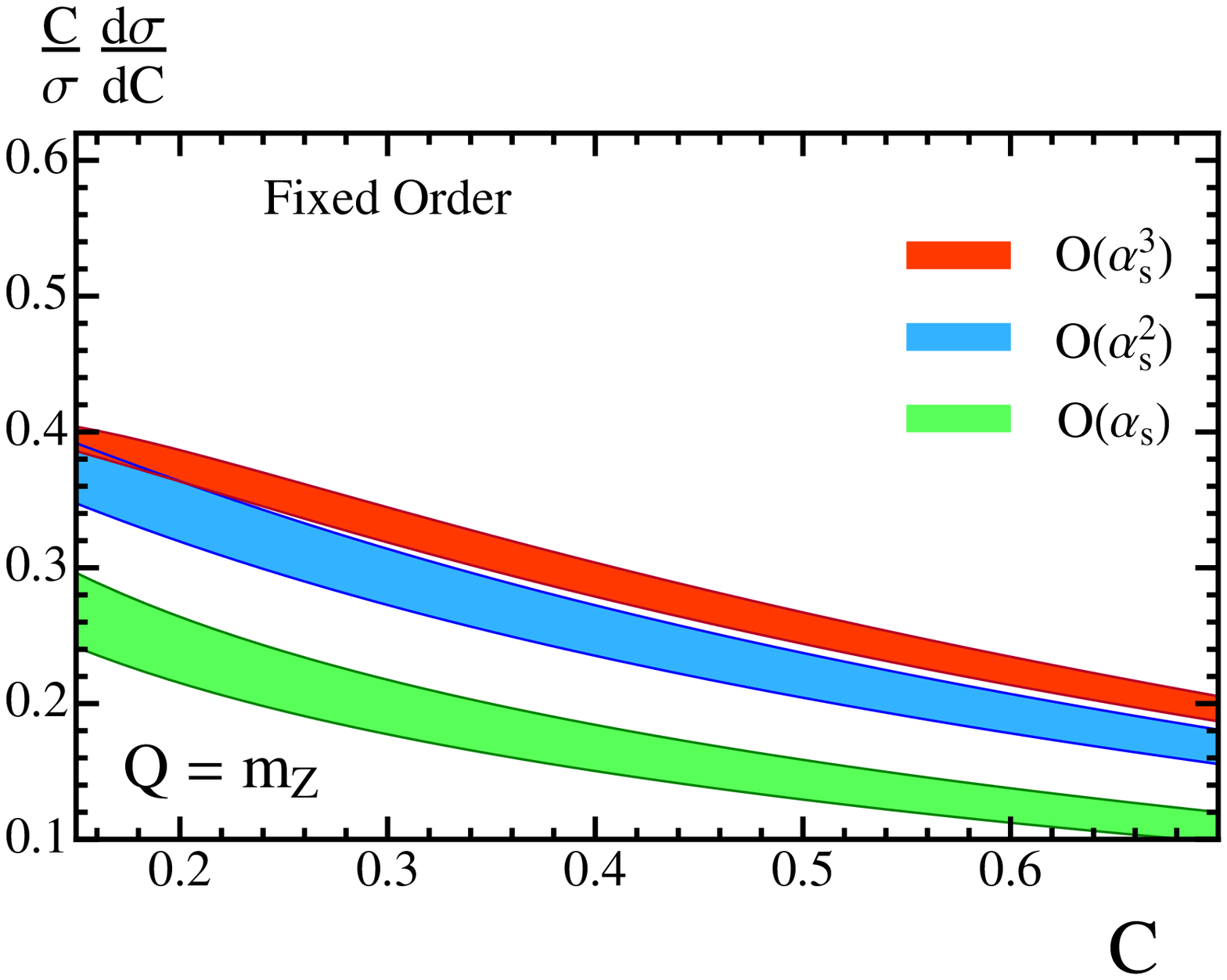}
		\label{fig:tailbandFO}
	}
	\subfigure[]{
		\includegraphics[width=0.48\textwidth]{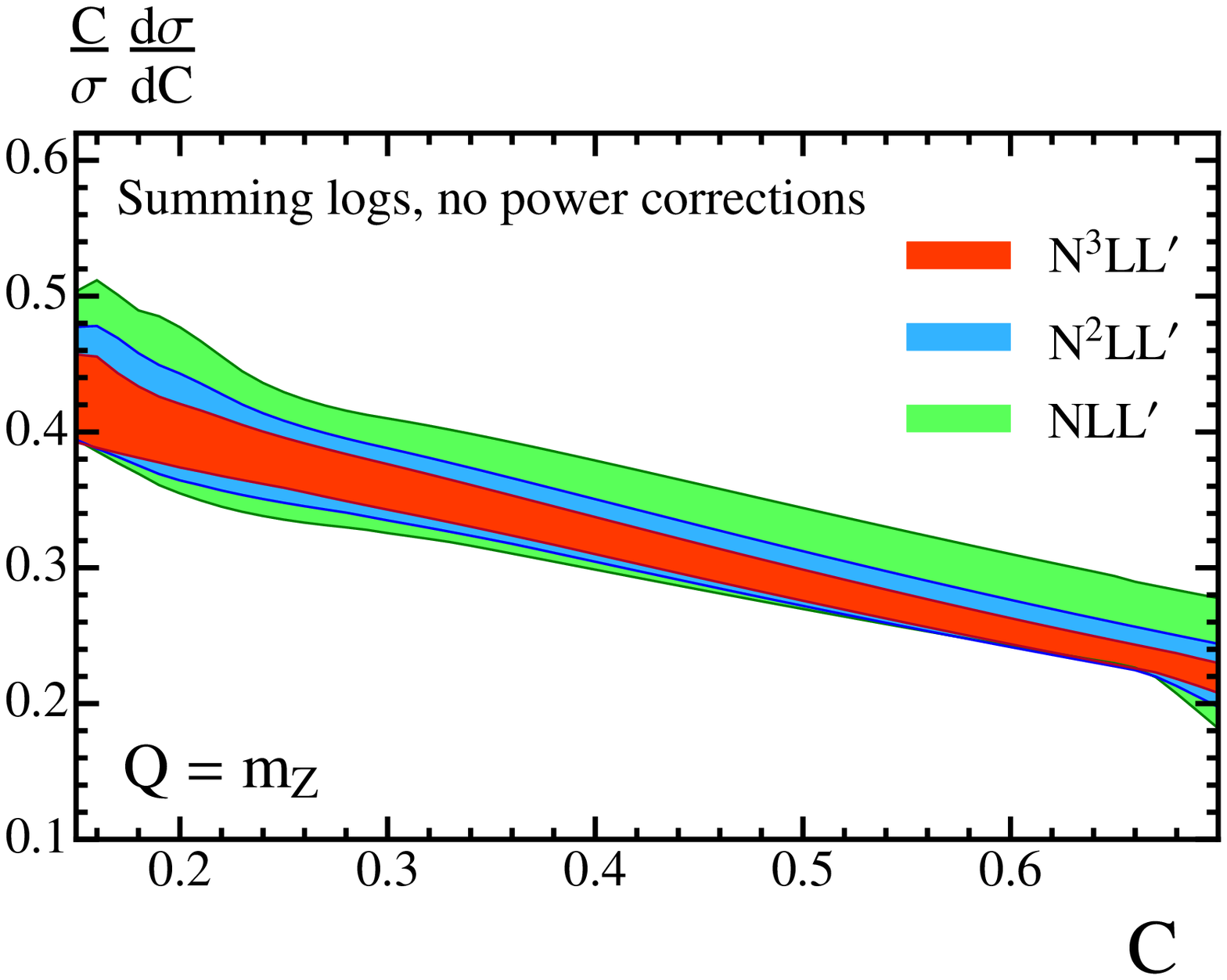}
		\label{fig:tailbandnoF}
	}
	\subfigure[]{
		\includegraphics[width=0.48\textwidth]{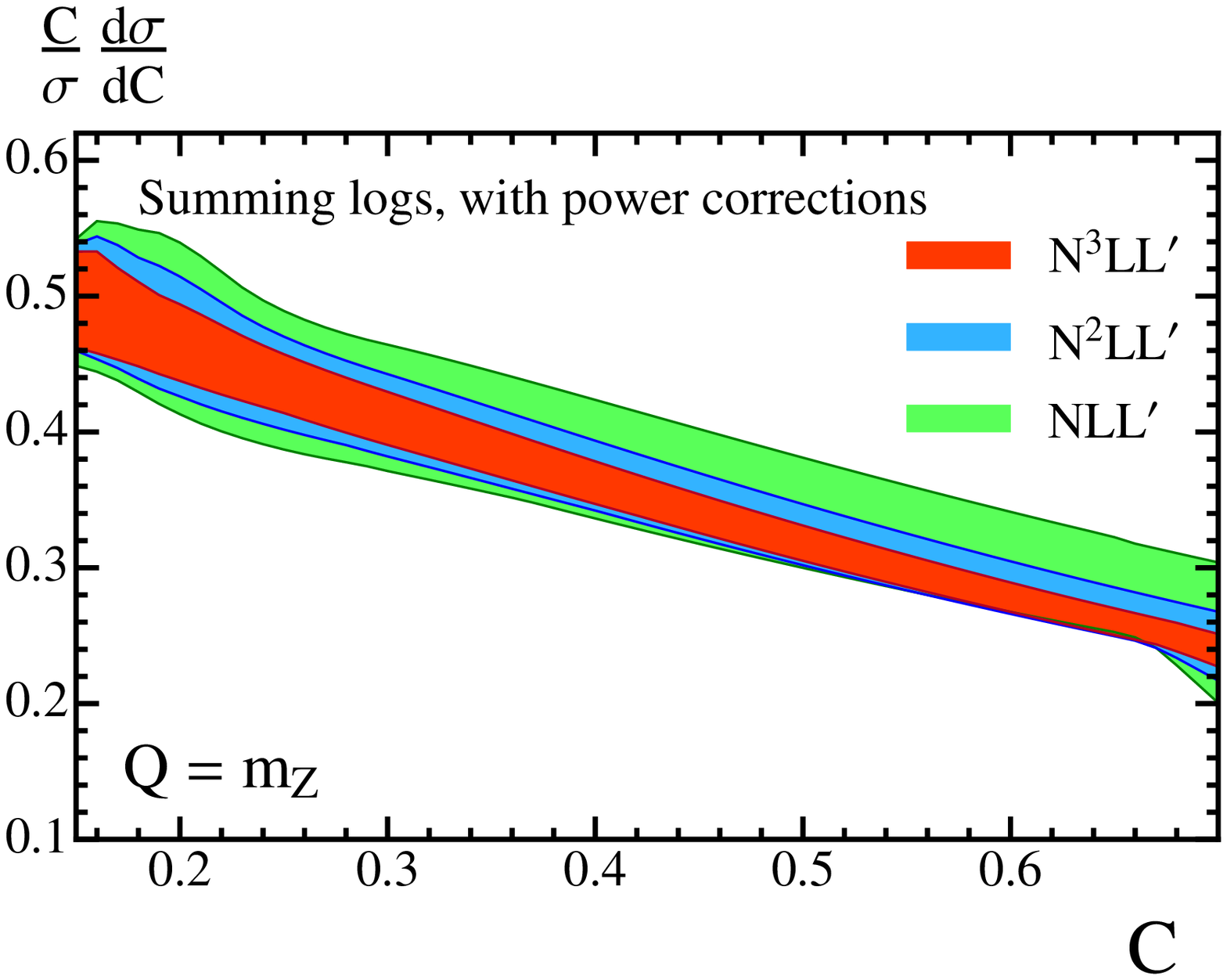}
		\label{fig:tailbandnogap}
	}
	\subfigure[]{
		\includegraphics[width=0.48\textwidth]{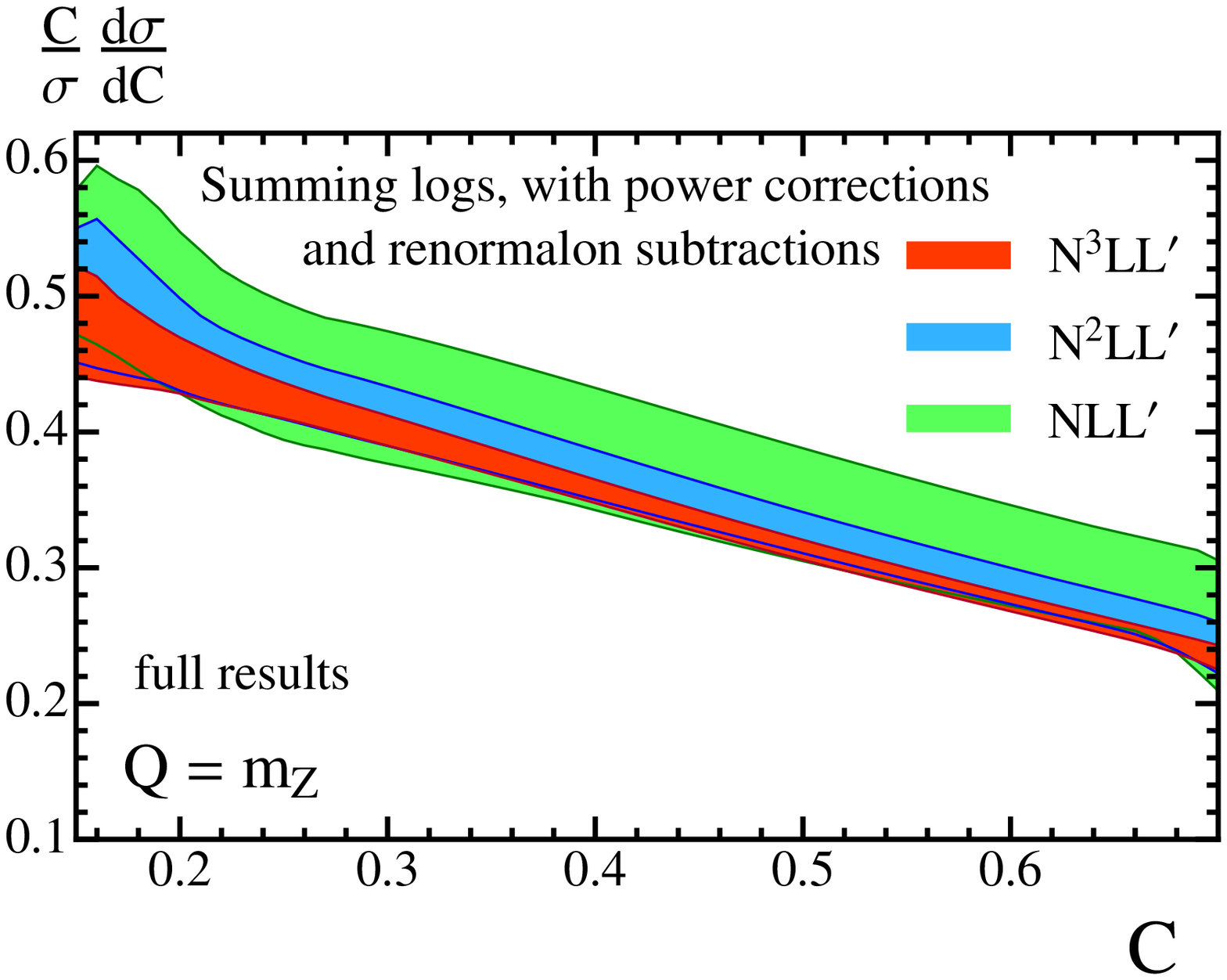}
		\label{fig:tailbandwithgap}
	}
	\caption[Theory scan for cross section uncertainties in C-parameter]{Theory scan for cross section uncertainties in C-parameter. The panels are (a) fixed order, (b) resummation with no nonperturbative function, (c) resummation with a nonperturbative function using the $\msbar$ scheme for $\overline\Omega_1^{C}$ without renormalon subtraction, and (d) resummation with a nonperturbative function using the Rgap scheme for $\Omega_1^{C}$ with renormalon subtraction.}
	\label{fig:4-plots}%
\end{figure*}

At the highest order, N${}^3$LL$^\prime+{\cal O}(\alpha_s^3)$, the choice of $r_s=1$ or $r_s=2$ has very little impact on the resulting cross section, both for C-parameter and thrust. Indeed, the difference between these two choices is smaller than the remaining (small) perturbative uncertainty at this order. This is illustrated in Fig.~\ref{fig:highest-slope}, which shows the complete N$^3$LL$^\prime+{\cal O}(\alpha_s^3)+\Omega_1(R,\mu)$ distributions for $r_s = 1$ (blue) and $r_s = 2$ (red) for C-parameter in the left panel, and for thrust in the right panel. Here we use $\Omega_1(R_\Delta,\mu_\Delta)=0.33\,{\rm GeV}$. Thus, the choice of $r_s$ is essentially irrelevant for our highest-order predictions, but has a bit of impact on conclusions drawn about the convergence of the lower to highest orders. Another thing that is clear from this figure is that the perturbative uncertainties for the \mbox{C-parameter} cross section at $Q=m_Z$, which are on average $\pm \,2.5\%$ in the region $0.25<C<0.65$, are a bit larger than those for thrust where we have on average $\pm\, 1.8\%$ in the region $0.1< \tau< 0.3$. This $\pm\, 1.8\%$, obtained with the profile and variations discussed here, agrees well with the $\pm\, 1.7\%$ quoted in Ref.~\cite{Abbate:2010xh}.

One can also look at the effect that the choice of $r_s$ has on the total integral over the C-parameter and thrust distributions at N$^k$LL$^\prime+{\cal O}(\alpha_s^k)+\Omega_1(R,\mu)$, which should reproduce the total hadronic cross section. For C-parameter, the outcome is shown in the first row of Fig.~\ref{fig:norm-C}, where green squares and red triangles represent resummed cross sections with power corrections in the $\msbar$ and Rgap schemes, respectively. In blue we display the fixed-order prediction. Comparing the predictions for $r_s = 1$ (left panel) and $r_s = 2$ (right panel), we observe that the former achieves a better description of the fixed-order prediction at N$^2$LL$^\prime$, in agreement with observations made in Ref.~\cite{Alioli:2012fc}.  For the case of thrust (second row of Fig.~\ref{fig:norm-C}), similar conclusions as for C-parameter can be drawn by observing the behavior of the total cross section. Again, at the highest order the result is  independent of the choice of the slope within uncertainties for both $C$ and thrust. Since our desired fit to determine $\alpha_s(m_Z)$ and $\Omega_1^C$ requires the best predictions for the shape of the normalized cross section, we do not use the better convergence for the normalization as a criteria for using $r_s=1$. Our results for cross section shapes are self-normalized using the central profile result.

In Fig.~\ref{fig:component-plot-sum} we present a plot analogous to Fig.~\ref{fig:component-plot} but including resummation at N$^3$LL$^\prime+\,{\cal O}(\alpha_s^3)\,+\,\Omega_1(R,\mu)$ with $r_s = 2$  (a similar plot for thrust can be found in Ref.~\cite{Abbate:2010xh}). The suppression of the dashed blue nonsingular curve relative to the solid upper blue singular curve is essentially the same as observed earlier in  Fig.~\ref{fig:component-plot}. The subtraction components are a small part of the cross section in the resummation region but have an impact at the level of precision obtained in our computation. Above the shoulder region, the singular and nonsingular terms appear with opposite signs and largely cancel. This is
clear from the figure where the individual singular and nonsingular lines are
larger than the total cross section in this region. The same cancellation
occurs for the singular subtraction and nonsingular subtraction terms.
The black curve labeled total in Fig.~\ref{fig:component-plot-sum} shows the central value for our full prediction. Note that the small dip in this black curve, visible at $C\simeq 0.75$, is what survives for the log-singular terms in the shoulder after the convolution with $F_C$.

\subsection{Convergence and Uncertainties: Impact of Resummation and Renormalon Subtractions}
\label{subsec:impact}

Results for the C-parameter cross sections at $Q=m_Z$ are shown at various levels of theoretical sophistication in Fig.~\ref{fig:4-plots}. The simplest setup is the purely perturbative fixed-order ${\cal O}(\alpha_s^k)$ QCD prediction (i.e.\ no resummation and no power corrections), shown in panel (a), which does not make use of the new perturbative results in this chapter. Not unexpectedly, the most salient feature at fixed order is the lack of overlap between the $\mathcal{O}(\alpha_s)$ (green), $\mathcal{O}(\alpha_s^2)$ (blue), and $\mathcal{O}(\alpha_s^3)$ (red) results. This problem is cured once the perturbative resummation is included with N$^k$LL$^\prime+{\cal O}(\alpha_s^k)$ predictions shown in panel (b): the NLL$^\prime$ (green), N$^2$LL$^\prime$ (blue), and N$^3$LL$^\prime$ (red) bands now nicely overlap. To achieve this convergence and overlap with our setup, it is important to normalize the cross section bands with the integrated norm at a given order using the default profiles. (The convergence for the normalization in \Fig{fig:norm-C} was slower. Further discussion of this can be found in Ref.~\cite{Abbate:2010xh}.)

The panel (b) results neglect power corrections. Including them in the $\msbar$ scheme, N$^k$LL$^\prime+{\cal O}(\alpha_s^k)+{\overline\Omega}_1$, as shown in panel (c), does not affect the convergence of the series but rather simply shifts the bands toward larger $C$ values. This was mentioned above in \Eq{eq:shift}. 

\begin{figure*}[t!]
	\begin{center}
		\includegraphics[width=0.5\columnwidth]{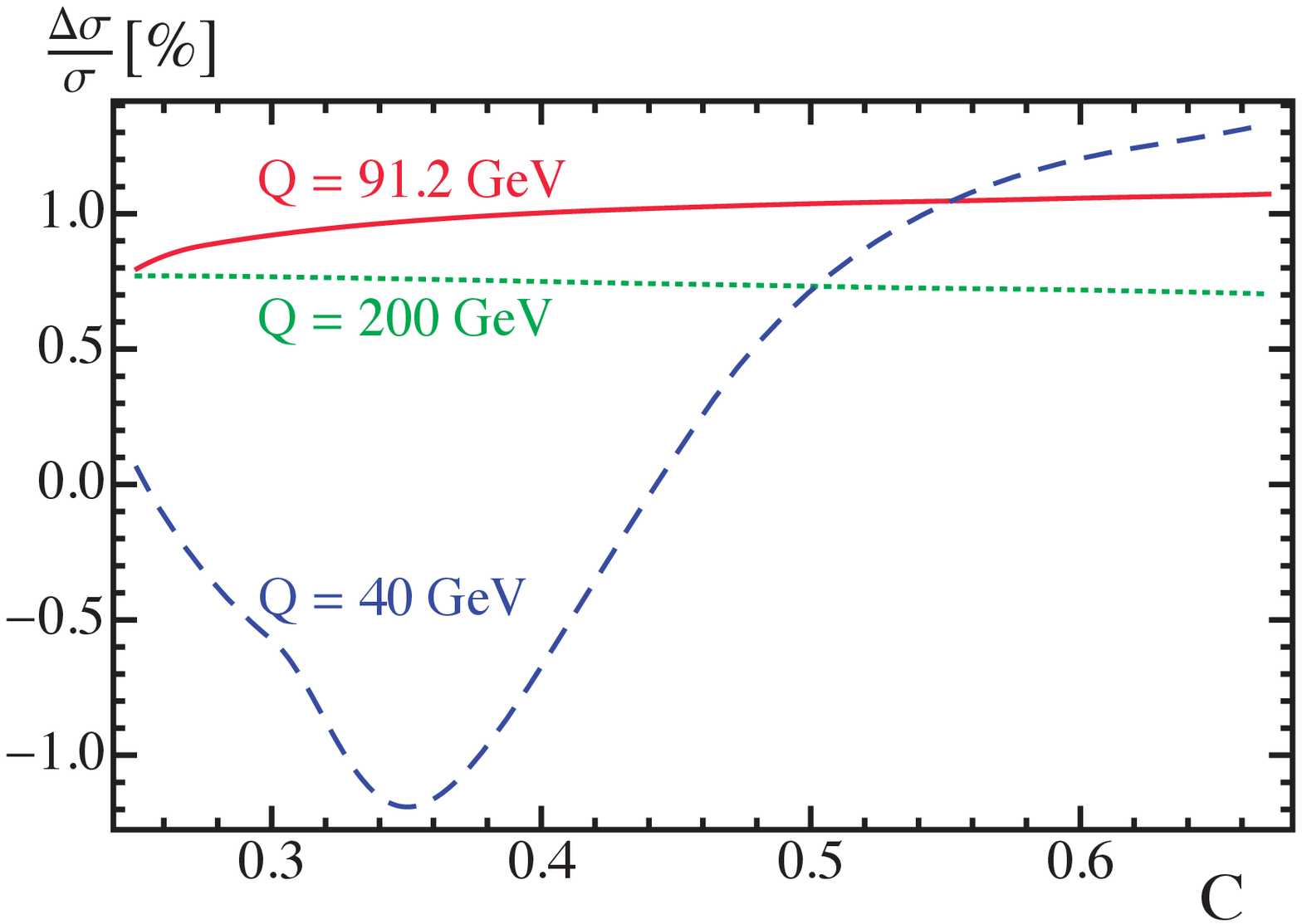}~~~~~
		\includegraphics[width=0.5\columnwidth]{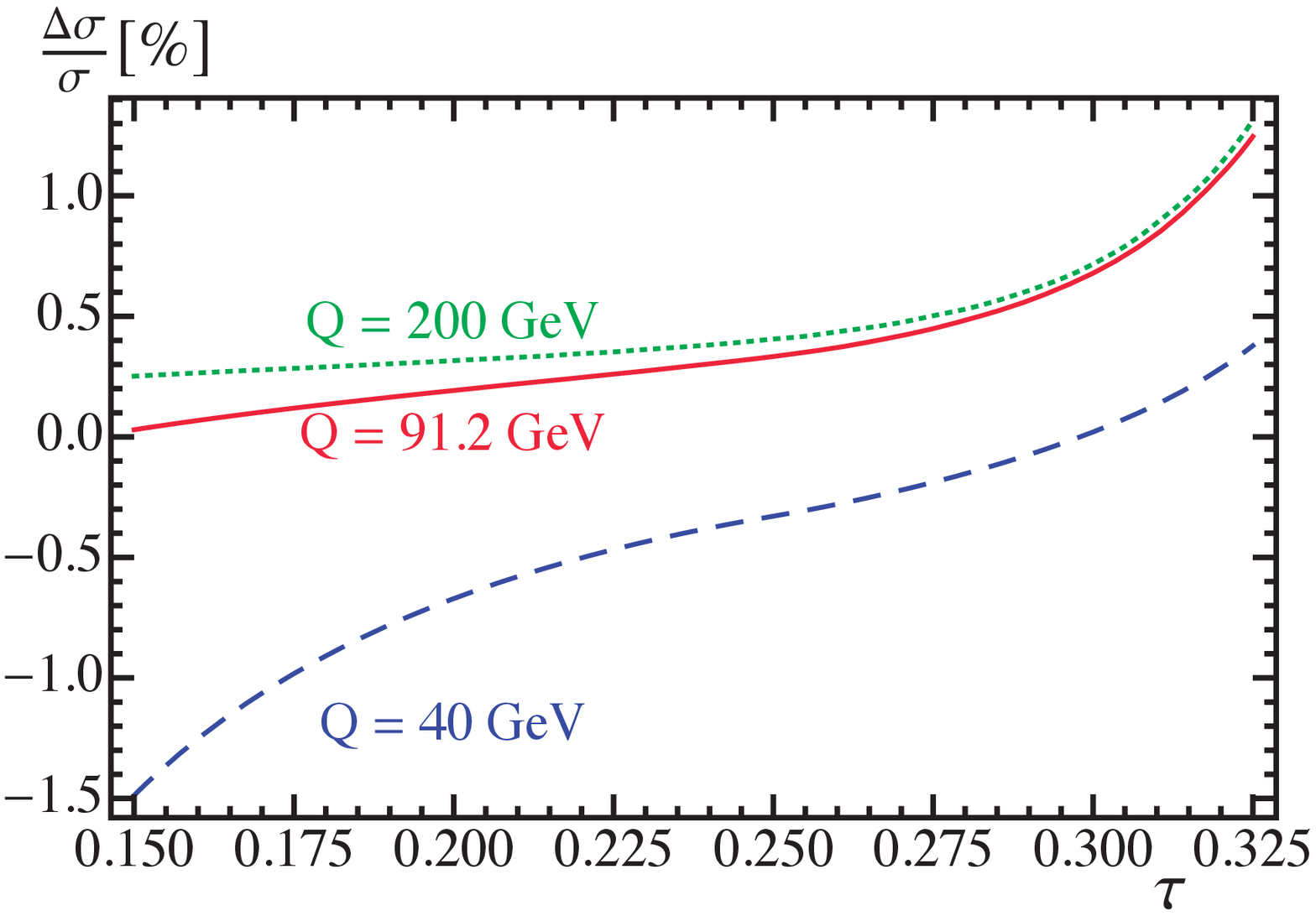}
		\caption[Effects of hadron masses on C-parameter and thrust differential cross sections]{Effects of hadron masses on the differential cross section for C-parameter (left) and thrust (right). Curves correspond to the percent difference between the cross section with and without hadron-mass effects, for three center-of-mass energies: $Q = 91.2$\,GeV (solid red), $200$\,GeV (dashed blue), and $40$\,GeV (dotted green). The cross section with hadron-mass effects uses $\theta(R_\Delta,\mu_\Delta) = 0$ and $\Omega_1(R_\Delta,\mu_\Delta) = 0.32\,(0.30)\,$GeV for C-parameter\,(thrust), while the cross section without hadron-mass effects has $\Omega_1(R_\Delta,\mu_\Delta) = 0.33\,$GeV. We set $\alpha_s(m_Z) = 0.1141$.}
		\label{fig:hadron-masses}
	\end{center}
\end{figure*}

In panel (d) we show our results, which use the Rgap scheme for the power correction, at N$^k$LL$^\prime+{\cal O}(\alpha_s^k)+{\Omega}_1(R,\mu)$. In this scheme a perturbative series is subtracted from the partonic soft function to remove its ${\cal O}(\Lambda_{\rm QCD})$ renormalon. This subtraction entails a corresponding scheme change for the parameter $\Omega_1$ which becomes a subtraction-scale-dependent quantity. In general the use of renormalon-free schemes stabilizes the perturbative behavior of cross sections. The main feature visible in panel (d) is the noticeable reduction of the perturbative uncertainty band at the two highest orders, with the bands still essentially contained inside lower-order ones.

We can see the improvement in convergence numerically by comparing the average percent uncertainty between different orders at $Q=m_Z$. If we first look at the results without the renormalon subtraction, at N$^k$LL$^\prime+{\cal O}(\alpha_s^k)+{\overline\Omega}_1$, we see that in the region of interest for $\alpha_s(m_Z)$ fits ($0.25<C<0.65$) the NLL$^\prime$ distribution has an average percent error of $\pm\,11.7\%$, the N$^2$LL$^\prime$ distribution has an average percent error of $\pm\,7.0\%$, and the highest-order N$^3$LL$^\prime$ distribution has an average percent error of only $\pm\,4.3\%$. Once we implement the Rgap scheme to remove the renormalon, giving N$^k$LL$^\prime+{\cal O}(\alpha_s^k)+{\Omega}_1(R,\mu)$, we see that the NLL$^\prime$ distribution has an average percent error of $\pm\,11.8\%$, the N$^2$LL$^\prime$ distribution has an average percent error of $\pm\,4.9\%$, and the most precise N$^3$LL$^\prime$ distribution has an average percent error of only $\pm\,2.5\%$. Although the renormalon subtractions for C-parameter induce a trend toward the lower edge of the perturbative band of the predictions at one lower order, the improved convergence of the perturbative series makes the use of these more accurate predictions desirable.
\begin{figure*}[t!]
	\begin{center}
		\includegraphics[width=0.5\columnwidth]{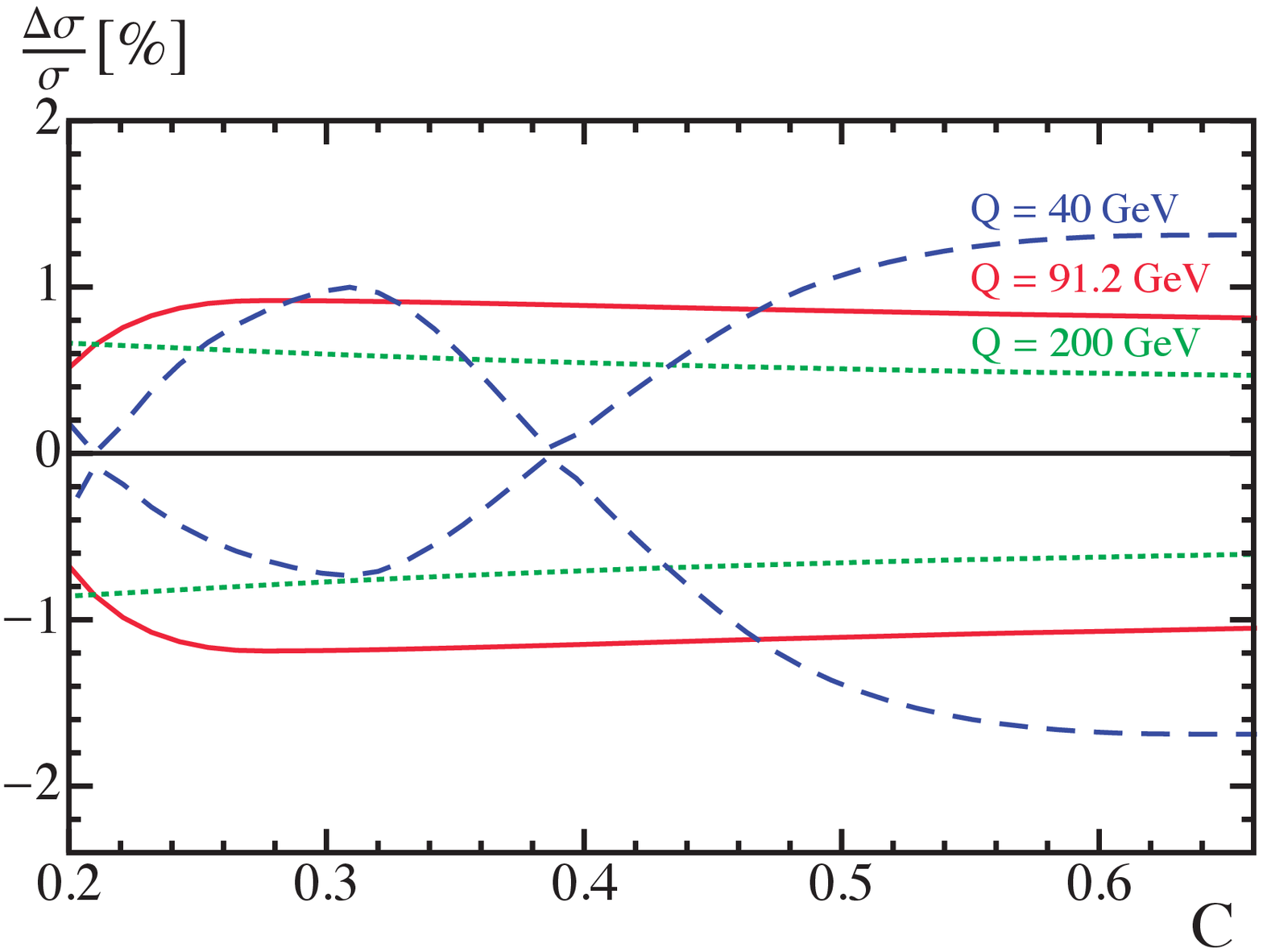}~~~~~
		\includegraphics[width=0.5\columnwidth]{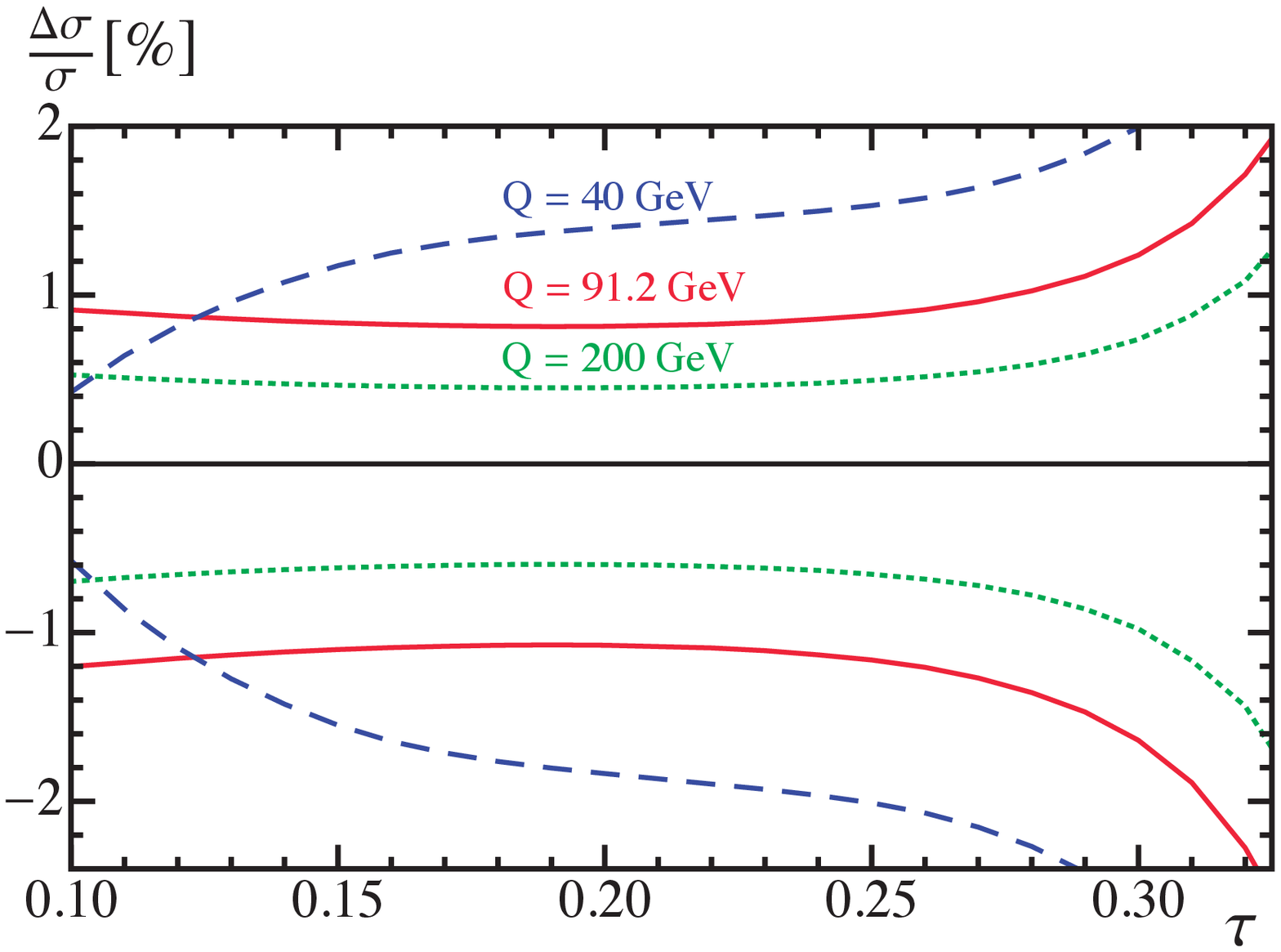}
		\caption[Effects of $\theta(R_\Delta,\mu_\Delta)$ on the cross section for C-parameter and thrust]{Effects of $\theta(R_\Delta,\mu_\Delta)$ on the cross section for C-parameter (left) and thrust (right). The lines correspond to the largest variation achievable by varying $\theta(R_\Delta,\mu_\Delta)$ in both directions (which happen for $\theta = 0.23\,\pi$ and $\theta = -\,0.27\,\pi$), with respect to the cross section with $\theta(R_\Delta,\mu_\Delta) = 0$. The solid (red), dashed (blue), and dotted (green) lines correspond to $Q = 91.2$, $200$, and $40$\,GeV, respectively. We use $\Omega_1(R_\Delta,\mu_\Delta) = 0.32\,(0.30)\,$GeV for C-parameter\,(thrust), and $\alpha_s(m_Z) = 0.1141$.}
		\label{fig:theta-effect}
	\end{center}
\end{figure*}
\subsection{Impact of Hadron Mass Effects}
\label{subsec:hadronmass}

In this section we discuss the impact of adding hadron-mass effects, which gives the orders denoted \mbox{N$^k$LL$^\prime+{\cal O}(\alpha_s^k)+{\Omega}_1(R,\mu,r)$}. Hadron masses induce an additional anomalous dimension for $\Omega_1$ and an associated series of logarithms of the form $\ln(Q C / \Lambda_{\rm QCD})$ that need to be resummed. They also impact the definition of $\Omega_1$ and its RGE equations in the $\overline{\rm MS}$ and Rgap schemes.  Since the overall effect of hadron masses on C-parameter and thrust are rather small, they do not change the perturbative convergence discussed in the previous section. Therefore, we study these effects here making use of only the highest-order perturbative results at N$^3$LL$^\prime+{\cal O}(\alpha_s^3)$.

In Fig.~\ref{fig:hadron-masses} we show the effect of hadron-mass running on the cross section. We compare differential cross sections with and without hadron masses, at the same center-of-mass energies. When the running effects from hadron masses are turned off, the value of $\Omega_1(R_\Delta,\mu_\Delta)$ preferred by the experimental data will attempt to average away these effects by absorbing them into the value of the initial parameter. (The running of $\Omega_1$ with and without hadron masses is shown below in Fig.~\ref{fig:Omega-run}.) Therefore, the specific values used in Fig.~\ref{fig:hadron-masses} are obtained by fixing $\alpha_s(m_Z)$ and then fitting for $\Omega_1(R_\Delta,\mu_\Delta)$ to minimize the difference between the cross section with and without hadron masses in the tail region. As the values for $\alpha_s(m_Z)$ and $\Omega_1$ in the case of no hadron-mass effects come from a fit to data (from Ref.~\cite{Abbate:2010xh}), choosing $\Omega_1(R_\Delta,\mu_\Delta)$ with the outlined procedure is similar to a full fit to data and will give results that allow comparison between the two cases. With hadron masses on, we fix $\theta(R_\Delta,\mu_\Delta)=0$, so the effects observed in Fig.~\ref{fig:hadron-masses} are related to the additional log resummation for $\Omega_1$. The effect is largest at $Q = 40\,$GeV where it varies between a $-\,1.25\%$ and $+\,1.25\%$ shift for C-parameter and between a $-\,1.5\%$ and $+\,0.5\%$ shift for thrust. For $Q=m_Z$ it amounts to a $1.0\%$ shift for C-parameter and a shift of $0\%$ to $1.3\%$ for thrust.

With hadron masses the additional hadronic parameter $\theta(R_\Delta,\mu_\Delta)$ encodes the fact that the extra resummation takes place in $r$ space, and therefore induces some dependence on the shape of the $\Omega_1(R,\mu,r)$ parameter. In contrast the dominant hadronic parameter $\Omega_1(R_\Delta,\mu_\Delta)$ is the normalization of this $r$-space hadronic function. In Fig.~\ref{fig:theta-effect} we show the effect that varying \mbox{$-\,\pi/2< \theta(R_\Delta,\mu_\Delta)<\pi/2$} has on the cross section for three center-of-mass energies, fixing \mbox{$\Omega_1(R_\Delta,\mu_\Delta) = 320$\,MeV} for C-parameter and $300$\,MeV for thrust.  The sensitivity of $\Delta\sigma$ here is proportional to $\Omega_1(R_\Delta,\mu_\Delta)$.  In these plots we pick the value of $\theta$ that gives the largest deviation from the $\theta = 0$ cross section (these values are listed in the figure caption). This maximum deviation is only $1.5\%$  for C-parameter and occurs at $Q = 40$\,GeV at larger $C$. For thrust, the largest deviation also occurs for $Q=40$\,GeV and for higher values of $\tau$ and is $\lesssim 2.0\%$ for $\tau\le 0.3$. For $Q=m_Z$ the effect is roughly $\pm 1.0\%$ for C-parameter and around $1.0\%$ for thrust when $\tau < 0.25$, growing to $2\%$ by $\tau=0.33$.

We conclude that the effect of hadron masses on log resummation should be included if one wishes to avoid an additional $\sim 1.5\%$ uncertainty on the cross section. Furthermore, one should consider fitting $\theta(R_\Delta,\mu_\Delta)$ as an additional parameter if one wants to avoid another $\sim 1\%$ uncertainty in the cross section that it induces. (Recall that Fig.~\ref{fig:theta-effect} shows the worst-case scenario.)


\chapter{Precision $\alpha_s$ Measurements from C-parameter}
\label{ch:Cparam-fit}

This chapter contains the extraction of $\alpha_S(m_Z)$ using the calculation of the C-parameter cross section given in Ch. \ref{ch:Cparam-theory} and was first presented in \cite{Hoang:2015hka}. In \sec{data} we begin by discussing the experimental data used in the extraction. The fit procedure is presented in \sec{fitprocedure}. In \sec{results}, we present the results of our analysis and discuss various sources of uncertainty. The final extraction gives $\alpha_s(m_Z)=0.1123 \pm 0.0002_{\rm exp} \pm 0.0007_{\rm hadr} \pm  0.0014_{\rm pert}$ and 
$\Omega_1(R_\Delta,\mu_\Delta)  = 
0.421 \pm 0.007_{\rm exp} \pm 0.019_{\rm \alpha_s(m_Z)} 
\pm  0.060_{\rm pert}$ GeV. In \sec{peak-tail} we discuss the predictions of our theory outside of the fit region and in \sec{universality} we compare our result with earlier results from thrust analyses and find universality between the two event shapes. In \App{app:fit-compare}, we compare with some alternate fit choices.

\section{Experimental Data}
\label{sec:data}
Data on the C-parameter cross section are given by several experiments for a 
range of $Q$ from $35$ to $207$ GeV.  We use data from ALEPH\,\footnote{The ALEPH dataset with $Q=91.2\,$GeV has  two systematic uncertainties for each bin. The second of
	these uncertainties is treated as correlated while the first one is treated as an uncorrelated uncertainty
	and simply added in quadrature to the statistical uncertainty.} with $Q = \{91.2$,
$133$, $161$, $172$, $183$, $189$, $200$, $206\}$ GeV \cite{Heister:2003aj},
DELPHI with $Q = \{45$, $66$, $76$, $89.5$, $91.2$, $93$, $133$, $161$, $172$, $183$, 
$189$, $192$, $196$, $200$, $202$, $205$, $207\}$ GeV \cite{Abdallah:2004xe, 
	Abreu:1996mk, Abreu:1999rc, Wicke:1999zz}, JADE with $Q=\{35$, $44\}$ GeV 
\cite{Biebel:1999zt}, L3 with $Q=\{91.2$, $130.1$, $136.1$, $161.3$, $172.3$, $182.8$, 
$188.6$, $194.4$, $200.2$, $206.2\}$ GeV \cite{Achard:2004sv, Adeva:1992gv}, OPAL with 
$Q=\{91$, $133$, $177$, $197\}$ GeV \cite{Abbiendi:2004qz}, and SLD with $Q=91.2$ GeV \cite{Abe:1994mf}. As each of these datasets is given in binned form we integrate the full cross section from Ch. \ref{ch:Cparam-theory} is integrated over each bin before being compared to the data.
The default range on $C$ used in fitting the data is $25\,\text{ GeV}/Q \le C \le 0.7$. A lower limit of $25\text{ GeV}/Q$  eliminates the peak region where higher nonperturbative moments $\Omega_{n>1}^C$  become important. The upper limit is  chosen to be 0.7 in order to avoid the far-tail region as well as the Sudakov shoulder at  $C=0.75$. Any bin that contains one of the end points of our range ($C= 25\, \text{GeV}/Q$  or 0.7) is included if more than half of that bin lies within the range. Using the default range and datasets gives a total of $404$ bins. As a further test of the stability of our analysis, both this C-parameter range and the selection of datasets is varied in the numerical analysis  contained in Sec.~\ref{sec:results}. 

\begin{figure}[t!]
	\begin{center}
		\subfigure[{}]{
			\includegraphics[width=0.5\columnwidth]{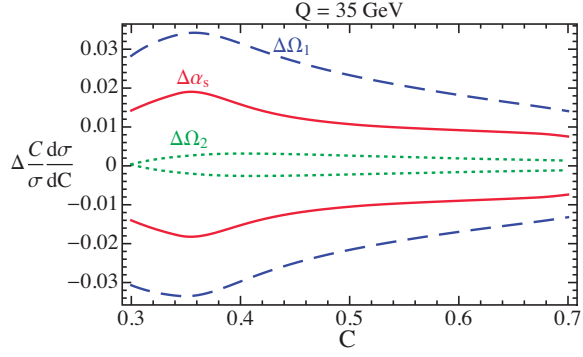}
		}
		
		\subfigure[{}]{
			\includegraphics[width=0.5\columnwidth]{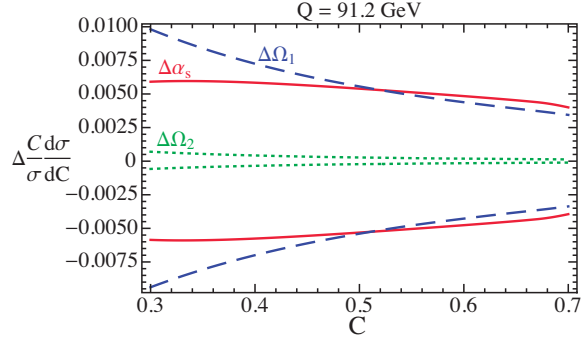}
		}
		\subfigure[{}]{
			\includegraphics[width=0.5\columnwidth]{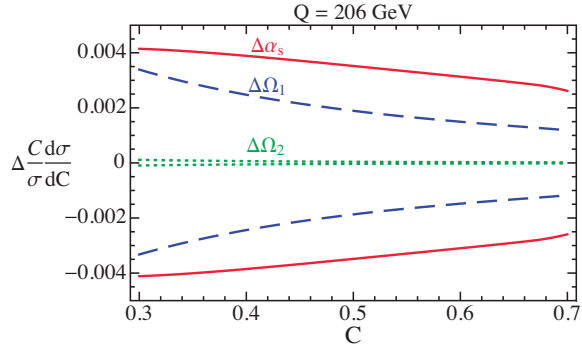}
		}
		\caption[Effect of varying only one parameter on C-parameter cross section]{Difference between the default cross section and the cross section varying only one parameter.
			We vary $\alpha_s (m_Z)$ by $\pm\, 0.001$ (solid red), $2\,\Omega_1$ by $\pm \,0.1$ (dashed blue) and
			$\Omega_2^C$ by $\pm\,0.5$ (dotted green). The three plots correspond to three different center
			of mass energies: (a)~$Q=35$\,GeV, (b)~$Q=91.2$\,GeV, (c)~$Q=206$\,GeV.}
		\label{fig:Degeneracy}
	\end{center}
\end{figure}
In our fitting procedure, we consider both the statistical and systematic experimental uncertainties. The statistical uncertainties can be treated as independent between bins. The systematic experimental uncertainties come from various sources and full documentation of their correlations are not available, so dealing with them in our $\chi^2$ analysis is more complicated, and we have to use a correlation model. For this purpose we follow the LEP QCD working group~\cite{Heister:2003aj,Abbiendi:2004qz} and use the minimal overlap model. Within one C-parameter dataset, which consists of various C-parameter bins at one $Q$ value for one experiment, we take for the bin $i$, bin $j$ off-diagonal entry of the experimental covariance matrix $[{\rm min}(\Delta_i^{\rm sys},\Delta_j^{\rm sys})]^2$. Here $\Delta_{i,j}^{\rm sys}$ are the quoted systematic uncertainties of the bins $i$ and $j$. Within each dataset, this model implies a positive correlation of systematic uncertainties. In addition to this default model choice, we also do the fits assuming uncorrelated systematic uncertainties, in order to test whether the minimal overlap model introduces any bias. See Sec.~\ref{sec:random} for more details on the correlation matrix.

\section{Fit Procedure} \label{sec:fitprocedure}

In order to accurately determine both $\alpha_s(m_Z)$ and the leading power correction in the same fit, it is important to perform a global analysis, that is, simultaneously fitting $C$-spectra for a wide range of center-of-mass energies $Q$. For each $Q$, effects on the cross sections induced by changes in $\alpha_s(m_Z)$ can be partly compensated by changes in $\Omega_1$, resulting in a fairly strong degeneracy. This is resolved by the global fit, just as in the thrust analysis of Ref.~\cite{Abbate:2010xh}. Fig.~\ref{fig:Degeneracy} shows the difference between the theoretical prediction for the cross section at three different values of $Q$, when
$\alpha_s(m_Z)$ or $\Omega_1$ are varied by $\pm\,0.001$ and $\pm\,0.05$ GeV, respectively. It is clear that the potential degeneracy in these parameters is broken by having data at multiple $Q$ values. In Fig.~\ref{fig:Degeneracy}  we also vary the higher-order power correction parameter $\Omega^C_2$, which clearly has a much smaller effect than the dominant power correction parameter $\Omega_1$.

To carry out a fit to the experimental data we fix the profile and theory parameters to the values shown in \tab{theoryerr}.  The default values are used for our primary theory cross section.  We integrate the resulting theoretical distribution over the same C-parameter bins as those available experimentally, and construct a $\chi^2$ function with the uncorrelated statistical experimental uncertainties and correlated systematic uncertainties. This $\chi^2$ is a function of $\alpha_s(m_Z)$ and $\Omega_1$, and is very accurately described by a quadratic near its global minimum, which therefore determines the central values and experimental uncertainties. The value of $\Omega_1$ and its associated uncertainties encode the dominant hadronization effect as well as the dominant residual uncertainty from hadronization.  

To obtain the perturbative theoretical uncertainty we consider the range of values shown for the theory parameters in \tab{theoryerr}. Treating each of these as a flat distribution, we randomly generate values for each of these parameters and then repeat the fit described above with the new $\chi^2$ function.  This random sampling and fit is then repeated 500 times. We then construct the minimum ellipse that fully contains all 500 of the central-fit values by first creating the convex envelope that contains all of these points within it. Then, we find the equation for the ellipse that best fits the points on the envelope, with the additional restrictions that all values lie within the ellipse and its center is the average of the maximum and minimum values in each direction. This ellipse determines the perturbative theoretical uncertainty, which turns out to be the dominant uncertainty in our fit results. In our final results the perturbative and experimental uncertainties are added in quadrature.  This procedure is similar to that discussed in the Appendix of Ref.~\cite{Abbate:2012jh}.

\section{Results}
\label{sec:results}
In this section we discuss the results from our global analysis. We split the presentation into several subsections. In Sec.~\ref{sec:resumpower} we discuss the impact that resummation and the inclusion of power corrections have on the fit results. 
In Sec.~\ref{sec:random} we present the analysis which yields the perturbative uncertainty in detail, cross-checking our method by analyzing the order-by-order convergence.  We also analyze the impact of removing the renormalon.  In Sec.~\ref{sec:exptfit} we discuss the experimental uncertainties obtained from the fit.  Section~\ref{sec:up-down} discusses the impact that varying the theory parameters one by one has on the best-fit points, allowing us to determine which parameters dominate the theoretical uncertainty. The impact of hadron-mass resummation is discussed in detail in \Sec{sec:hadmassresum}. We examine the effects of changing the default dataset in \Sec{sec:dataset}. The final fit results are collected in Sec.~\ref{sec:final}. When indicating the perturbative precision, and whether or not the power correction $\Omega_1$ is included and at what level of precision, we use the following notation:
\begin{align}
& {\cal O}(\alpha_s^k) 
& \phantom{x} & \text{fixed order up to ${\cal O}(\alpha_s^k)$} 
\nn\\ 
& \text{N}^k\text{LL}^\prime \!+\! {\cal O}(\alpha_s^k) 
& \phantom{x} & \text{perturbative resummation} 
\nn\\ 
& \text{N}^k\text{LL}^\prime \!+\! {\cal O}(\alpha_s^k)  \!+\! {\overline \Omega}_1
& \phantom{x} & \text{$\overline{\rm MS}$ scheme for $\Omega_1$} 
\nn\\ 
& \text{N}^k\text{LL}^\prime \!+\! {\cal O}(\alpha_s^k) 
\!+\!  {\Omega}_1(R,\mu)
& \phantom{x} & \text{Rgap scheme for $\Omega_1$} 
\nn\\ 
& \text{N}^k\text{LL}^\prime \!+\! 
{\cal O}(\alpha_s^k) \!+\! {\Omega}_1(R,\mu,r)
& \phantom{x} & \text{Rgap scheme with }
\nn\\
& & \phantom{x} & \  \text{hadron masses for $\Omega_1$} 
\,. \nn
\end{align}
\begin{figure*}[t!]
	\begin{center}
		\includegraphics[width=.75\columnwidth]{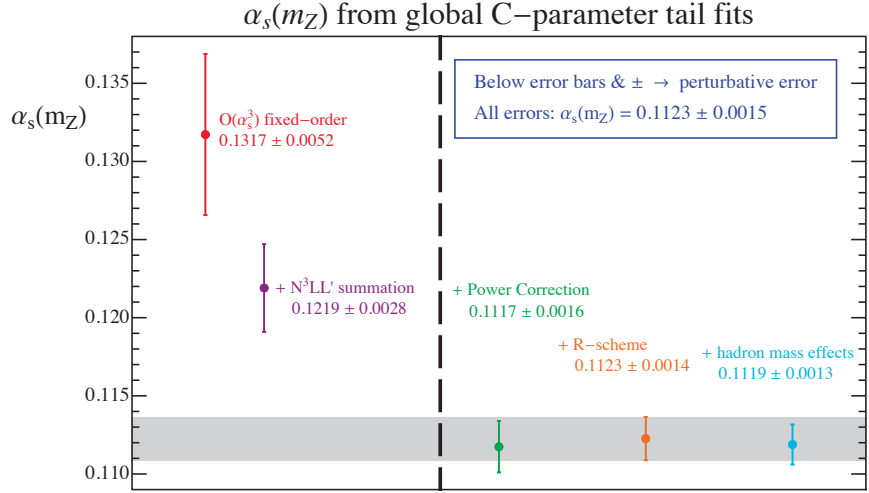}
		\caption[Evolution of the value of $\alpha_s(m_Z)$ adding components of C-parameter cross section calculation to fit]{The evolution of the value of $\alpha_s(m_Z)$ adding components of the calculation. An additional $\sim 8\%$ uncertainty from not including power corrections is not included in the two left points. }
		\label{fig:alpha-evolution-C}
	\end{center}
\end{figure*}

\subsection{Impact of Resummation and Power Corrections} 
\label{sec:resumpower}
In Fig.~\ref{fig:alpha-evolution-C} we show $\alpha_s(m_Z)$ extracted
from fits to the tail of the C-parameter distribution including sequential improvements to the treatment of perturbative and nonperturbative components of our code, using the highest perturbative accuracy at each stage. The sequence from left to right shows the fit results using:
${\cal O}(\alpha_s^3)$ fixed-order results only, adding N$^3$LL resummation, adding the $\overline\Omega_1$ power correction, adding renormalon
subtractions and using the Rgap power correction parameter $\Omega_1(R_\Delta,\mu_\Delta)$, and adding hadron-mass effects. These same results together with the corresponding $\chi^2/{\rm dof}$ are also collected in Tab.~\ref{tab:lesser}.  
The fit with only fixed-order ${\cal O}(\alpha_s^3)$ results has a relatively large $\chi^2/{\rm dof}$ and also its central value has the largest value of $\alpha_s(m_Z)$.  Including the resummation of large logarithms decreases the central $\alpha_s(m_Z)$ by 8\% and also decreases the perturbative uncertainty by $\sim 50\%$. Due to this smaller perturbative uncertainty it becomes clear that the theoretical cross section has a different slope than the data, which can be seen, for example, at $Q=m_Z$ for $0.27 < C< 0.35$. This leads to the increase in the $\chi^2/{\rm dof}$ for the ``N$^3$LL$^\prime$ no power corr.''~fit, and makes it quite obvious that power corrections are needed. When the power correction parameter $\Omega_1$ is included in the fit, shown by the third entry in Tab.~\ref{tab:lesser} and the result just to the right of the vertical dashed line in Fig.~\ref{fig:alpha-evolution-C}, the $\chi^2/{\rm dof}$ becomes $1.004$ and this issue is resolved. Furthermore, a reduction by $\sim 50\%$ is achieved for the perturbative uncertainty in $\alpha_s(m_Z)$. This reduction makes sense since some of the perturbative uncertainty of the cross section is now absorbed in $\Omega_1$, and a much better fit is achieved for any of the variations associated to estimating higher-order corrections. The addition of $\Omega_1$ also caused the fit value of $\alpha_s(m_Z)$ to drop by another 8\%, consistent with our expectations for the impact of power corrections and the estimate made in Ch. \ref{ch:Cparam-theory}. 
Note that the error bars of the first two purely perturbative determinations, shown at the left-hand side of the vertical thick dashed line in Fig.~\ref{fig:alpha-evolution-C} and in the last two entries in \tab{lesser}, do not include the $\sim 8\%$ uncertainties associated with the lack of power corrections.

The remaining corrections we consider are the  use of the R-scheme for  $\Omega_1$ which includes the renormalon subtractions, and the inclusion of the log-resummation effects associated to the hadron-mass effects.  Both of these corrections have a fairly small impact on the determination of $\alpha_s(m_Z)$, shifting the central value by $+0.5\%$ and $-\,0.3\%$ respectively. Since adding the $-\,0.3\%$ shift from the hadron mass corrections in quadrature with the $\simeq 1.2\%$ perturbative uncertainty does not change the overall uncertainty we will use the R-scheme determination for our main result. This avoids the need to fully discuss the extra fit parameter $\theta(R_\Delta,\mu_\Delta)$ that appears when hadron masses are included. Further discussion of the experimental uncertainties and the perturbative uncertainty from the random scan are given below in \Secs{sec:random}{sec:up-down}, and a more detailed discussion of the impact of hadron-mass resummation is given below in \Sec{sec:hadmassresum}.

The values of $\Omega_1$ obtained from the fits discussed above can be directly compared to the $\Omega_1$ power correction obtained from the thrust distribution. Values for $\Omega_1$ from the C-parameter fits are given below in \Secs{sec:random}{sec:up-down} and the comparison with thrust is considered in \Sec{sec:universality}.

\begin{table}[t!]
	\begin{center}
		\begin{tabular}{l|c c}
		& $\alpha_s(m_Z)$ & $\chi^2/{\rm dof}$ \\
		\hline
		N$^3$LL$^\prime$ + hadron                          & ~~$0.1119(13)(06)$ & $0.991$\\
		N$^3$LL$^\prime$ with $\Omega_1(R,\mu)$        & ~~$0.1123(14)(06)$ & $0.988$\\
		N$^3$LL$^\prime$ with $\overline\Omega_1$ & ~~$0.1117(16)(06)$ & $1.004$\\
		N$^3$LL$^\prime$ no power corr.\,\,                & ~~$0.1219(28)(02)$& $2.091$\\
		\!\!\parbox{20ex}{${\cal O}(\alpha_s^3)$ fixed order\\[2pt]
			no power corr.} & ~~$0.1317(52)(03)$ & $1.486$
	\end{tabular}
	\end{center}
	\caption[Comparison of C-parameter tail fit results for analyses adding various components of the theoretical result]{
		Comparison of C-parameter tail fit results for analyses when we add various components of the theoretical result (from the bottom to top). The first parentheses gives the theory uncertainty, and the second is the experimental and hadronic uncertainties added in quadrature for the first three rows, and experimental uncertainty for the last two rows.
		\label{tab:lesser}}
\end{table}

\subsection{Perturbative Uncertainty from the Scan}
\label{sec:random}
\begin{figure*}[t!]
	\subfigure[]{
		\includegraphics[width=0.45\textwidth]{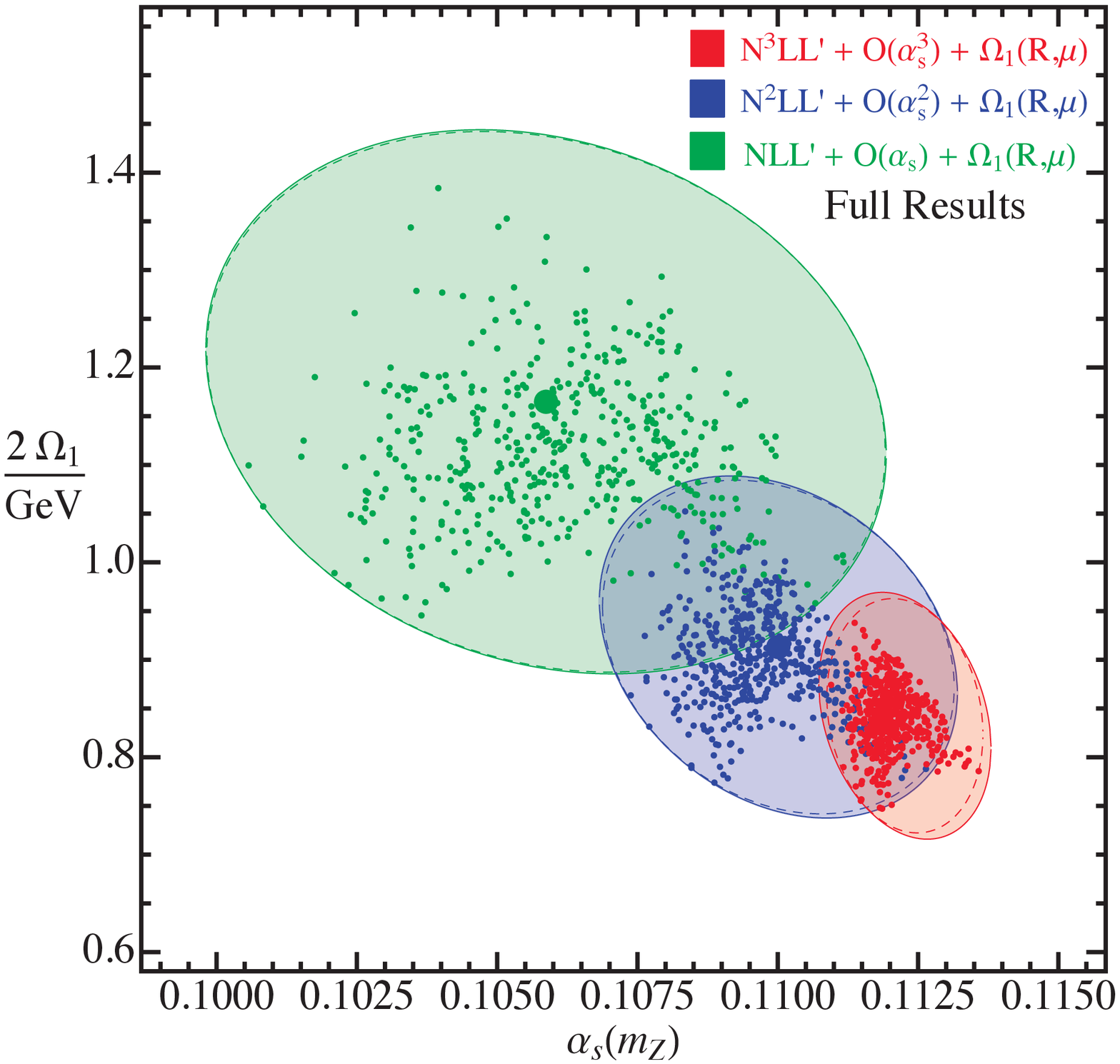}
		\label{fig:alphagap}
	}
	\subfigure[]{
		\includegraphics[width=0.45\textwidth]{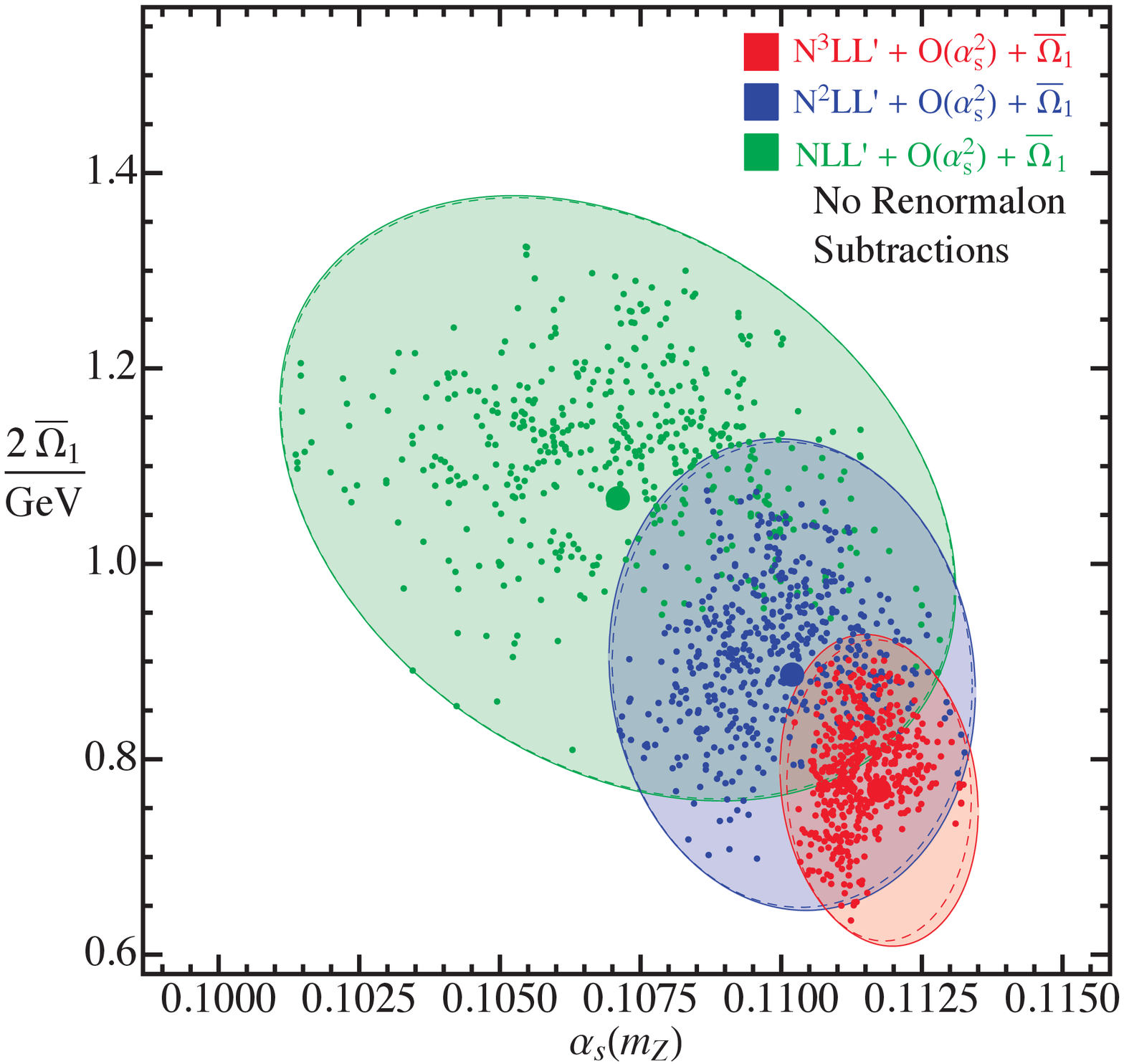}
		\label{fig:alphanogap}
	}
	\subfigure[]{
		\includegraphics[width=0.45\textwidth]{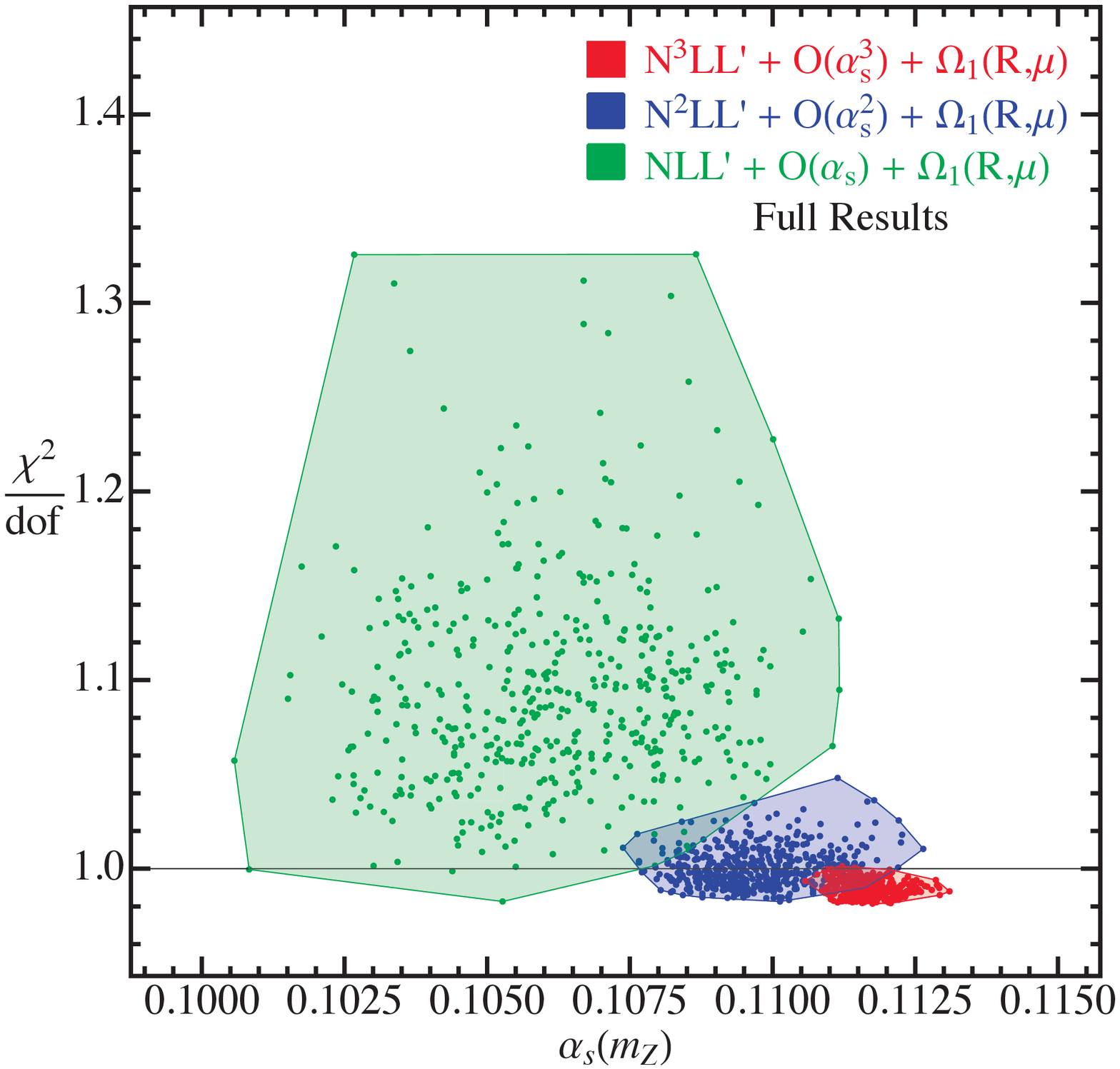}
		\label{fig:chi2alphagap}
	}
	\subfigure[]{
		\includegraphics[width=0.45\textwidth]{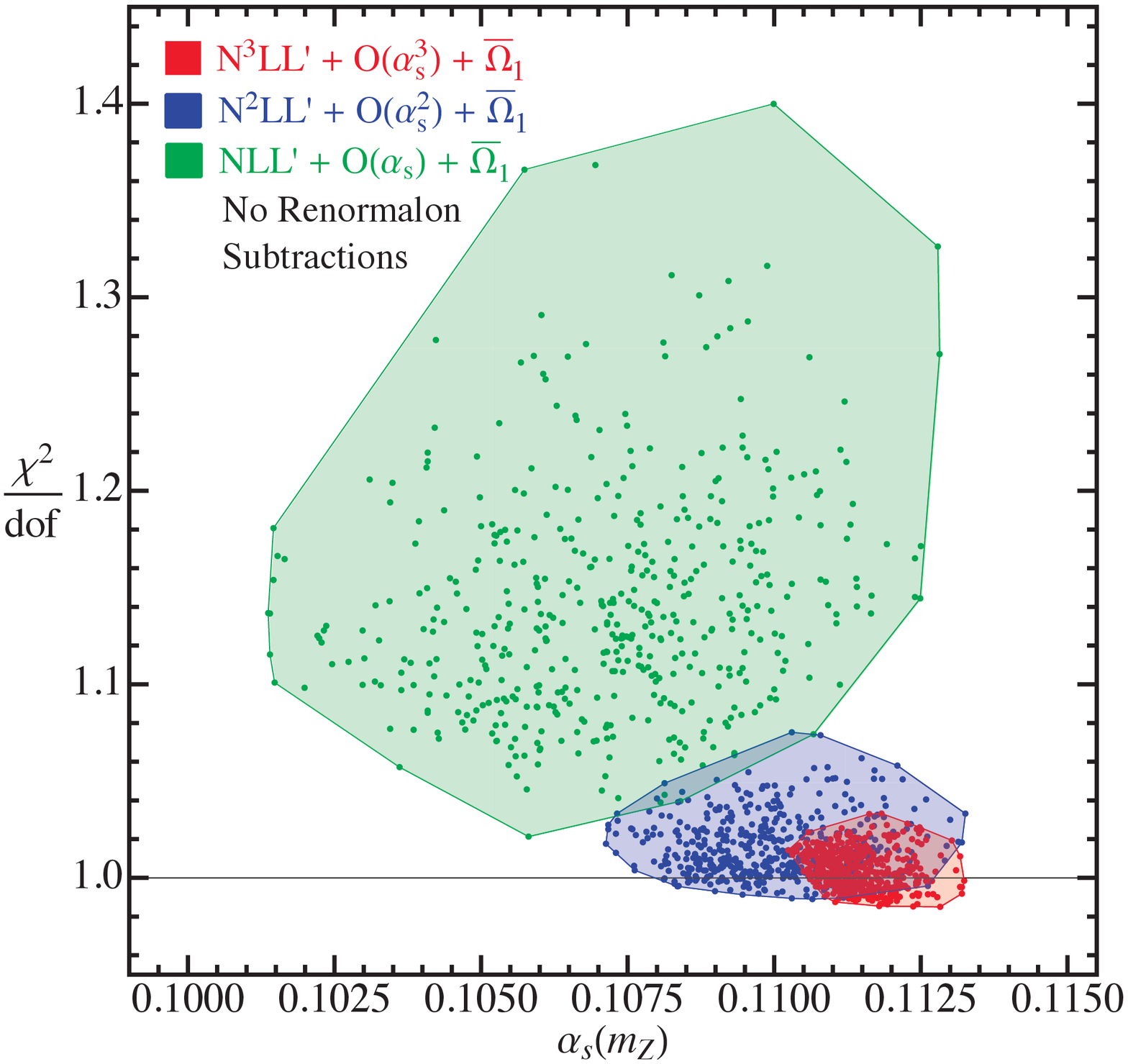}
		\label{fig:chi2alphanogap}
	}
	\vspace{-0.2cm}
	\caption[Best fit points in C-parameter $\alpha_s(m_Z)$-$\Omega_1$ fits]{The first two panels show the distribution of best-fit points in the $\alpha_s(m_Z)$-$2\Omega_1$ and
		$\alpha_s(m_Z)$-$2\overline\Omega_1$ planes. Panel~(a) shows results including perturbation theory, resummation of large
		logs, the soft nonperturbative function and $\Omega_1$ defined in the Rgap scheme with renormalon subtractions.
		Panel~(b) shows the results as in panel~(a), but with $\overline\Omega_1$ defined in the $\msbar$ scheme without renormalon
		subtractions. In both panels the dashed lines corresponds to an ellipse fit to the contour of the best-fit points to
		determine the theoretical uncertainty. The respective total (experimental\,+\,theoretical) 39\% CL standard uncertainty
		ellipses are displayed (solid lines), which correspond to $1$-$\sigma$ (68\% CL) for either one-dimensional projection.
		The big points represent the central values in the random scan for $\alpha_s(m_Z)$ and $2\,\Omega_1$.
		Likewise, the two panels at the bottom show the distribution of best-fit points in the $\alpha_s(m_Z)$-$\chi^2/{\rm dof}$
		plane. Panel~(c) shows the $\chi^2/{\rm dof}$ values of the points given in 
		panel~(a), whereas panel~(b) shows the $\chi^2/{\rm dof}$ values of the points given in panel~(b).
		\label{fig:alpha} }
\end{figure*}

To examine the robustness of our method of determining the perturbative uncertainty by the random scan, we consider the convergence and overlap of the results at different perturbative orders. Figure~\ref{fig:alpha} shows the spread of best-fit values at NLL$^\prime$, N$^2$LL$^\prime$ and N$^3$LL$^\prime$.
The upper left panel, Fig.~\ref{fig:alphagap}, shows results
from fits performed in the Rgap scheme, which implements a renormalon subtraction for $\Omega_1$,
and the upper right-panel, Fig.~\ref{fig:alphanogap}, shows results in the $\msbar$
scheme without renormalon subtractions. Each point in the plot represents the
outcome of a single fit, and different colors correspond to different orders in perturbation theory. Not unexpectedly, fits in the Rgap scheme show generally smaller theory uncertainties.

In order to estimate correlations induced by theoretical uncertainties, each ellipse in the $\alpha_s$-$2\Omega_1$ plane is constructed following the procedure discussed in \Sec{sec:fitprocedure}.
Each theory ellipse constructed in this manner is interpreted as an estimate for the \mbox{1-$\sigma$} theoretical uncertainty ellipse for each individual parameter (39\% confidence for the two parameters), and is represented by a dashed ellipse in Fig.~\ref{fig:alpha}.
The solid lines represent the combined (theoretical plus experimental) standard uncertainty ellipses at 39\% confidence for two parameters, obtained by adding the theoretical and experimental error matrices from the individual
ellipses, where the experimental ellipse corresponds to $\Delta \chi^2 = 1$.
Figure~\ref{fig:alpha} clearly shows a substantial reduction of the perturbative uncertainties when increasing the resummation accuracy, and given that they are 39\% confidence regions for two parameters, also show good overlap between the results at different orders. 

The results for $\alpha_s(m_Z)$ and $\Omega_1$ from the theory scan at different perturbative orders are collected in Tabs~\ref{tab:power-results} and \ref{tab:Omega1results}. Central values here are determined from the average of the maximal and minimal values of the theory scan, and are very close to the central values obtained when running with our default parameters.  The quoted perturbative uncertainties are one-parameter uncertainties.

\begin{table}[t!]
	\begin{center}
		\begin{tabular}{ccc}
		order &$\alpha_s(m_Z)$ (with $\overline\Omega_1$) & $\alpha_s(m_Z)$
		(with $\Omega_1(R_\Delta,\mu_\Delta)$)\\
		\hline
		NLL$^\prime$             & $0.1071(60)(05)$ & $0.1059(62)(05)$  \\
		N$^2$LL$^\prime$            & $0.1102(32)(06)$ & $0.1100(33)(06)$ \\
		N$^3$LL$^\prime$ (full)  & $0.1117(16)(06)$ & $\mathbf{0.1123(14)(06)}$
	\end{tabular}
	\end{center}
	\caption[Fit results for $\alpha_s(m_Z)$ at various 
	orders with theoretical, experimental and hadronic uncertainties]{Central values for $\alpha_s(m_Z)$ at various 
		orders with theory uncertainties from the parameter scan (first value in 
		parentheses), and experimental and hadronic uncertainty added in quadrature (second
		value in parentheses). The bold N$^3$LL$^\prime$ value is our final result.}
	\label{tab:power-results}
\end{table}
\begin{table}[t!]
	\begin{center}
		\begin{tabular}{ccc}
		order & \hspace{5mm}$\overline\Omega_1$ [GeV]\hspace{4mm} 
		& \hspace{1mm}$\Omega_1(R_\Delta,\mu_\Delta)$ [GeV]\hspace{1mm} \\
		\hline
		NLL$^\prime$             & $0.533(154)(18)$ & $0.582(134)(16)$ \\
		N$^2$LL$^\prime$            & $0.443(119)(19)$  & $0.457(83)(19)$ \\
		N$^3$LL$^\prime$ (full)  & $0.384(91)(20)$  & $\mathbf{0.421(60)(20)}$ \\
	\end{tabular}
	\end{center}
	\caption[Fit results for $\Omega_1$ at various 
	orders with theoretical, experimental and hadronic uncertainties]{Central values for  $\Omega_1$ at the 
		reference scales $R_\Delta=\mu_\Delta=2$\,GeV and for $\overline\Omega_1$ and at various
		orders. The parentheses show first the theory uncertainties from the parameter scan,
		and second the experimental plus the uncertainty due to the imprecise determination of $\alpha_s$ (added in quadrature). The bold N$^3$LL$^\prime$ value is our final result.}
	\label{tab:Omega1results}
\end{table}

In Tab.~\ref{tab:lesser} above we also present $\alpha_s(m_Z)$ results with no power corrections and either using resummation or fixed-order perturbative results. Without power corrections there is no fit for $\Omega_1$, so we take the central value to be the average of the maximum and minimum value of $\alpha_s(m_Z)$ that comes from our parameter scan. Our estimate of the uncertainty is given by the difference between our result and the maximum fit value. For the fixed-order case, since there is only one renormalization scale, we know that the uncertainties from our parameter variation for $e_H$, $s_2^{\widetilde C}$, $\epsilon_2^{\rm low}$ and $\epsilon_3^{\rm low}$ are uncorrelated. So, we take the fit value for $\alpha_s(m_Z)$ with the default parameters as our result and add the uncertainties from variations of these parameter in quadrature to give the total uncertainty.

An additional attractive result of our fits is that the experimental data is better described when increasing the order of the resummation and fixed-order terms. This can be seen by looking at the minimal $\chi^2/$dof values for the best-fit points, which are shown in Fig.~\ref{fig:alpha}. In Figs.~\ref{fig:chi2alphagap} and \ref{fig:chi2alphanogap} we show the distribution of $\chi^2_{\rm min}/{\rm dof}$ values for the various $\alpha_s(m_Z)$ best-fit points. 
Figure~\ref{fig:chi2alphagap}
displays the results in the Rgap scheme, whereas Fig.~\ref{fig:chi2alphanogap} shows the results in the $\msbar$ scheme. In both cases we find that the $\chi^2_{\rm min}$ values
systematically decrease with increasing perturbative order. The highest-order analysis in the
$\msbar$ scheme leads to $\chi^2_{\rm min}/{\rm dof}$ values around unity and thus provides
an adequate description of the whole dataset, however one also observes that accounting for the
renormalon subtraction in the Rgap scheme leads to a substantially improved theoretical
description having $\chi^2_{\rm min}/{\rm dof}$ values below unity essentially for all points
in the random scan. Computing the average of the $\chi^2_{\rm min}$ values we find at  N$^3$LL$^\prime$ order for the Rgap and $\msbar$ schemes $0.988$ and $1.004$, respectively (where the spread of values is smaller in the Rgap scheme). Likewise for N$^2$LL$^\prime$ we find $1.00$ and $1.02$, and for NLL$^\prime$ we find $1.09$ and $1.14$.
These results show the excellent description of the experimental data for
various center-of-mass energies. They also validate the smaller theoretical uncertainties
obtained for $\alpha_s$ and $\Omega_1$ at N$^2$LL$^\prime$ and N$^3$LL$^\prime$ orders in the Rgap scheme.

\subsection{Experimental Fit Uncertainty} \label{sec:exptfit}
\begin{figure}[t]
	\vspace{0pt}
	\begin{center}
	\includegraphics[width=.65\linewidth]{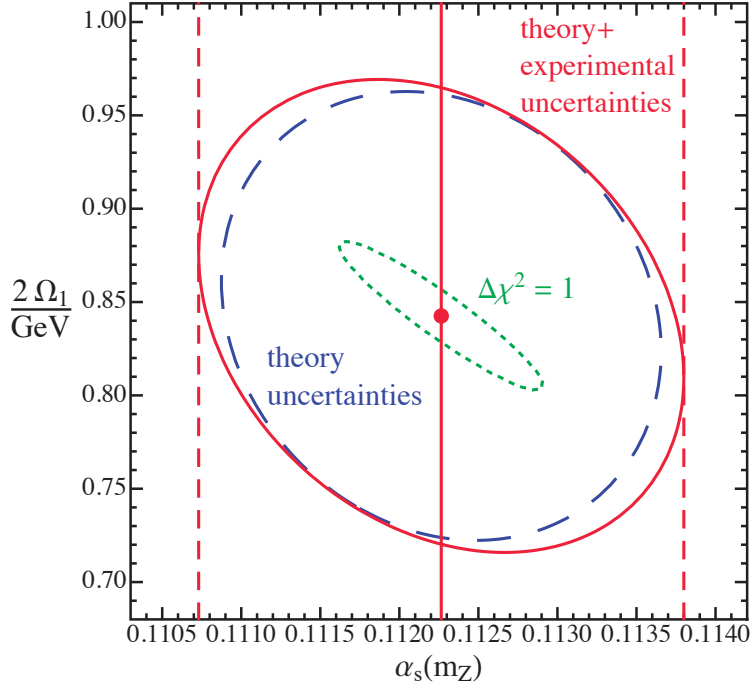}
	\end{center}
	\caption[Uncertainty ellipses in $\alpha_s$\,-\,$2\Omega_1$ plane]{Experimental $\Delta\chi^2=1$ standard uncertainty ellipse (dotted green) at
		N${}^3$LL$^\prime$ accuracy with renormalon subtractions, in the
		$\alpha_s$\,-\,$2\Omega_1$ plane.  The dashed blue ellipse represents the
		theory uncertainty which is obtained by fitting an ellipse to the contour of the
		distribution of the best-fit points. This ellipse should be interpreted as
		the $1$-$\sigma$ theory uncertainty for one parameter (39\% confidence for
		two parameters). The solid red ellipse represents the total (combined
		experimental and perturbative) uncertainty ellipse.}
	\label{fig:ellipses}
\end{figure}
Next we discuss in more detail the experimental uncertainty in $\alpha_s(m_Z)$ and the hadronization parameter $\Omega_1$ as well as the combination with the perturbative uncertainty done to obtain the total uncertainty. 

Results are depicted in Fig.~\ref{fig:ellipses} for our highest order fit including resummation, power corrections and renormalon subtractions. The inner green dotted ellipse, blue dashed ellipse, and solid red ellipse represent the  $\Delta \chi^2=1$ uncertainty ellipses for the experimental, theoretical, and combined theoretical and experimental uncertainties respectively. These ellipses correspond to the \mbox{one-dimensional} projection of the
uncertainties onto either $\alpha_s(m_Z)$ or $\Omega_1$ (39\% confidence ellipse for two parameters). The correlation matrix of the experimental,
theory, and total uncertainty ellipses are (for $i,j=\alpha_s, 2\,\Omega_1$),
\begin{align} \label{eq:Vijresult}
V_{ij}&  =\, 
\left( \begin{array}{cc}
\sigma_{\alpha_s}^2
& \,\, 2 \sigma_{\alpha_s} \sigma_{\Omega_1}\rho_{\alpha\Omega}\\
2\sigma_{\alpha_s} \sigma_{\Omega_1}\rho_{\alpha\Omega} 
& \,\,4 \sigma_{\Omega_1}^2
\end{array}\right) ,
\\
V^{\rm exp}_{ij}& =
\left( \begin{array}{cr}
4.18(52)\cdot 10^{-7}  & \,\, -\,0.24(5)\cdot 10^{-4}\,\mbox{GeV}\\ 
-\,0.24(5)\cdot 10^{-4}\,\mbox{GeV} & \,\, 1.60(47)\cdot 10^{-3}\,\mbox{GeV}^2
\end{array}\right) \! , \nn\\
V^{\rm theo}_{ij}& =
\left( \begin{array}{cr}
1.93\cdot 10^{-6}  & \,\, -\,0.27\cdot 10^{-4}~\mbox{GeV}\\
-\,0.27\cdot 10^{-4}~\mbox{GeV} & \,\, 1.45\cdot 10^{-2}~\mbox{GeV}^2
\end{array}\right), \nn\\
V^{\rm tot}_{ij}& =
\left( \begin{array}{cr}
2.35(5)\cdot 10^{-6}  & \,\, -\,0.51(5)\cdot 10^{-4}\,\mbox{GeV}\\ 
-\,0.51(5)\cdot 10^{-4}\,\mbox{GeV} & \,\, 1.61(5)\cdot 10^{-2}\,\mbox{GeV}^2
\end{array}\right)\! . \nn
\end{align}
Note that the theoretical uncertainties dominate by a significant amount. 
The experimental correlation coefficient is significant and
reads
\begin{align} \label{eq:rhoaO}
\rho^{\rm exp}_{\alpha\Omega}\,=\,-\,0.93(15) \,.
\end{align}
The theory correlation coefficient is small, $\rho^{\rm theo}_{\alpha\Omega}\,=\,-\,0.16$, and since these uncertainties dominate 
it reduces the correlation coefficient for the total uncertainty to
\begin{align}\label{eq:rhoaOtot}
\rho_{\alpha\Omega}^{\rm total} \,=\, -\,0.26(2)\,.
\end{align}
In both \eqs{rhoaO}{rhoaOtot} the numbers in parentheses indicate a $\pm$ range that captures all values obtained from the theory scan.  The correlation exhibited by the green dotted experimental uncertainty ellipse in
Fig.~\ref{fig:ellipses} is given by the line describing the semimajor axis
\begin{align}
\frac{\Omega_1}{30.84\,{\rm GeV}} = 0.1257 - \alpha_s(m_Z) \,.
\end{align}
Note that extrapolating this correlation to the extreme case where we neglect
the nonperturbative corrections ($\Omega_1=0$) gives $\alpha_s(m_Z)\to 0.1257$ which is consistent with the $0.1219 \pm 0.0028$ result of our fit without power corrections in \tab{lesser}.

\begin{figure*}[t!]
	\begin{center}
		\includegraphics[width=0.48\textwidth]{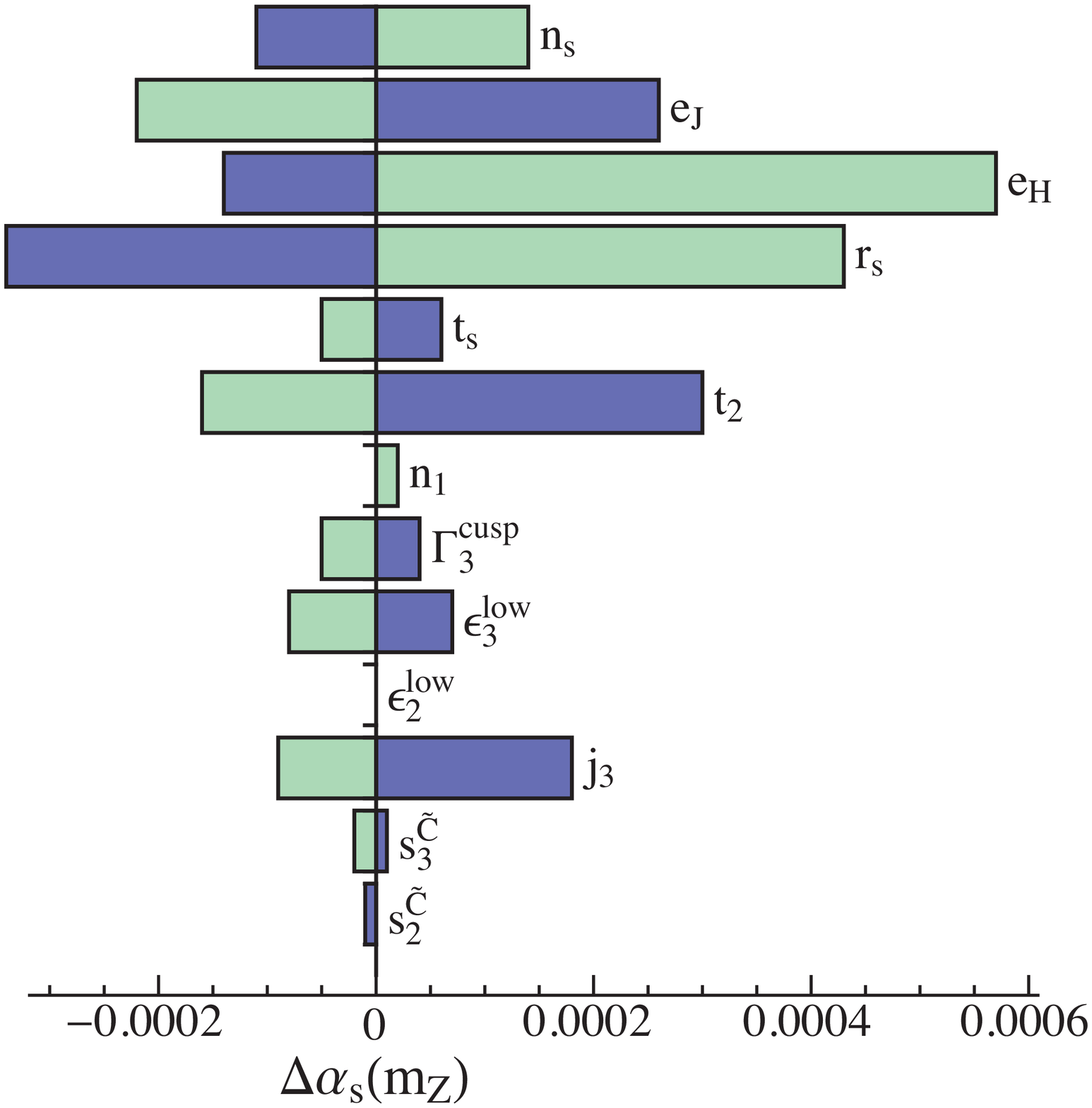}~~~
		\includegraphics[width=0.48\textwidth]{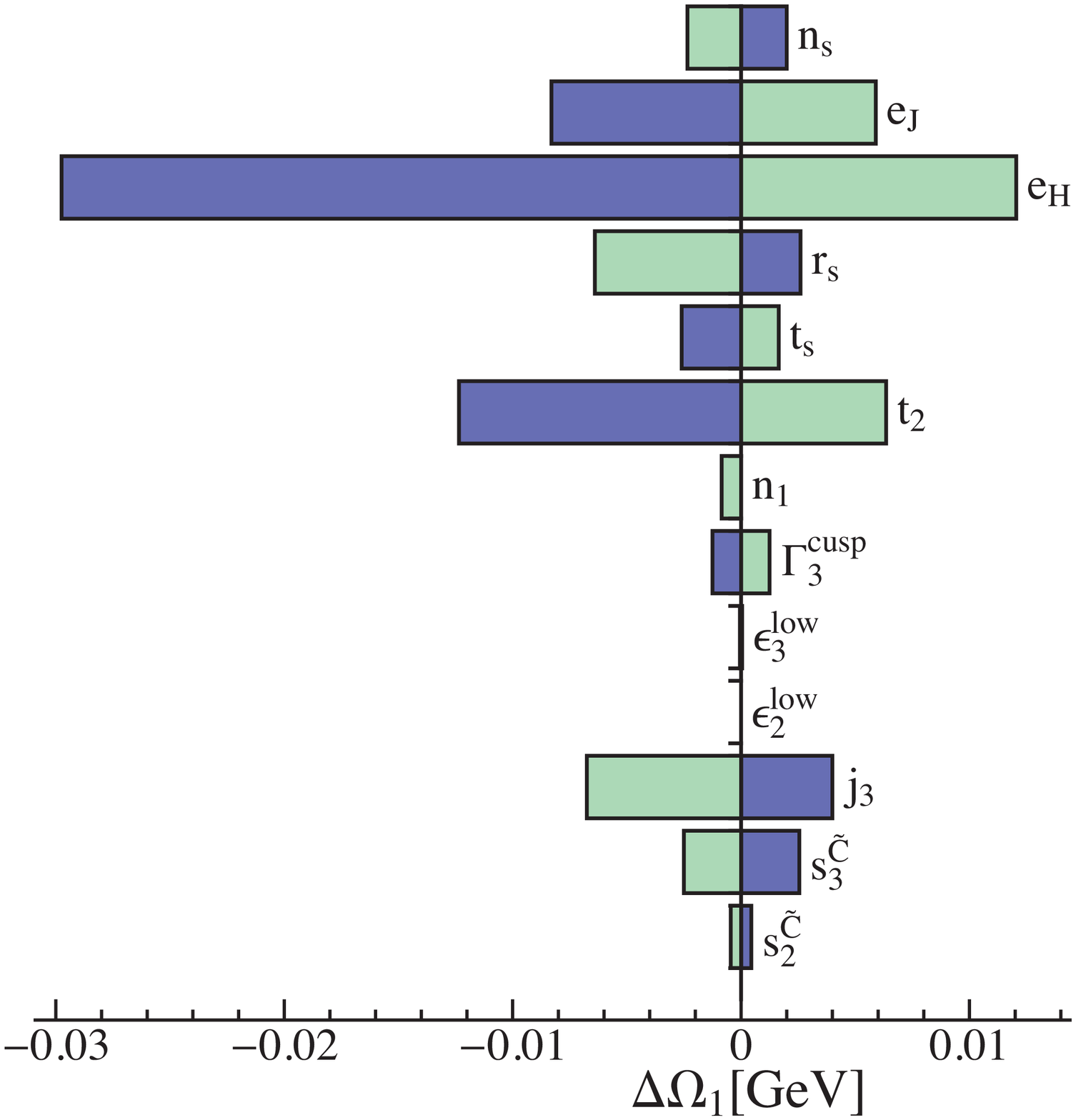}
		\caption[Variations of the best-fit values for $\alpha_s(m_Z)$ and $\Omega_1$ from up and down 
		variations for the theory parameters]{Variations of the best-fit values for $\alpha_s(m_Z)$ and $\Omega_1$ from up (dark blue) and down (light green) 
			variations for the theory parameters according to Tab.~\ref{tab:theoryerr}. We do not display those parameters 
			which do not affect the fit region ($\epsilon_2^{\rm high}$, $\epsilon_3^{\rm high}$, $\mu_0$, $R_0$, $n_0$).}
		\label{fig:UpDown}
	\end{center}
\end{figure*}

From $V_{ij}^{\rm exp}$ in Eq.~(\ref{eq:Vijresult}) it is possible
to extract the experimental uncertainty for $\alpha_s(m_Z)$ and the uncertainty due to the imprecise determination of $\Omega_1$,
\begin{align}
\sigma_{\alpha_s}^{\rm exp} 
& = \,\sigma_{\alpha_s}\sqrt{1-\rho^2_{\alpha\Omega}}
=  \,0.0002 \,,
\nonumber\\
\sigma_{\alpha_s}^{\rm \Omega_1} 
& = \,\sigma_{\alpha_s}\, |\rho_{\alpha\Omega}|\,
=  \,0.0006 \,,
\end{align}
and to extract the experimental uncertainty for $\Omega_1$ and its uncertainty due to the imprecise determination of $\alpha_s(m_Z)$,
\begin{align}
\sigma_{\Omega_1}^{\rm exp} 
& = \,\sigma_{\Omega_1}\sqrt{1-\rho^2_{\alpha\Omega}}
=  \,0.014~\mbox{GeV} 
\,, \nn \\
\sigma_{\Omega_1}^{\rm \alpha_s} 
& = \,\sigma_{\Omega_1}\, |\rho_{\alpha\Omega}|\,
=  \,0.037~\mbox{GeV}
\,.
\end{align}

The projections of the outer solid ellipse in Fig.~\ref{fig:ellipses}\,  show the total  uncertainty in our final one-parameter results obtained from $V_{ij}^{\rm tot}$, which are quoted below in
\eq{asfinal}.

\subsection{Individual Theory Scan Errors}
\label{sec:up-down}

To gain further insight into our theoretical precision and in order to estimate the dominant source for theory uncertainty from missing higher-order terms, we look at the size of the theory uncertainties caused by the individual variation of each one of the theory parameters included in our random scan. In Fig.~\ref{fig:UpDown}
two bar charts are shown with these results for $\alpha_s(m_Z)$ (left panel) and
$\Omega_1(R_\Delta,\mu_\Delta)$ (right panel) for fits corresponding to our best theoretical setup
(with N$^3$LL$^\prime$ accuracy and in the Rgap scheme). The dark blue bars correspond to the result of the fit with an upward variation of the given parameter from \tab{theoryerr}, while the light green bars correspond to the fit result from the downward variation in \tab{theoryerr}.
Here we vary a single parameter keeping the rest fixed at
their default values. 
We do not show parameters that have a negligibly small impact in the fit region, e.g.\ $\epsilon_2^{\rm high}$ and $\epsilon_3^{\rm high}$, which only
have an effect on the cross section to the right of the shoulder, or $n_0$, which only affects the cross section in the nonperturbative region.

We see that the dominant theory uncertainties are related to variations of the profile functions
($e_H, r_s, e_J, t_2$), where $e_H$ is the largest source of uncertainty, and is particularly dominant for $\Omega_1$.
The second most important uncertainty comes from $r_s$ for $\alpha_s$ and $t_2$ for $\Omega_1$, and $e_J$ also has a significant effect on both parameters. 

As expected, the parameters associated to the transitions on the sides of our fit region, $n_1$ and $t_s$, hardly matter. The
renormalization scale parameter $n_s$ for the nonsingular partonic distribution
${\rm d}\hat\sigma_{\rm ns}/{\rm d}C$ also causes a very small uncertainty since the nonsingular terms are always dominated by the singular terms in our fit region. The uncertainties
related to the numerical uncertainties of the perturbative constants ($s_2^{\widetilde C}$, $s_3^{\widetilde C}$, $j_3$) as well as the numerical uncertainties in the extraction of the nonsingular distribution for small $C$ values,
($\epsilon_2^{\rm low}$, $\epsilon_3^{\rm low}$) are -- with the possible exception of $j_3$
-- much smaller and do not play an important role. The uncertainty related to the unknown
$4$-loop contribution to the cusp anomalous dimension is always negligible. Adding quadratically
the symmetrized individual uncertainties shown in Fig.~\ref{fig:UpDown}, we find $0.0007$ for $\alpha_s$
and $0.05$ GeV for $\Omega_1$. This is about one half of the theoretical uncertainty
we have obtained by the theory parameter scan for $\alpha_s$ (or five sixths for $\Omega_1$), demonstrating that incorporating correlated variations through the theory parameter scan represents a more realistic method to estimate the theory uncertainty.

\subsection{Effects of $\mathbf{\Omega_1}$ hadron-mass resummation}
\label{sec:hadmassresum}

\begin{figure}[t]
	\vspace{0pt}
	\begin{center}
	\includegraphics[width=.65\linewidth]{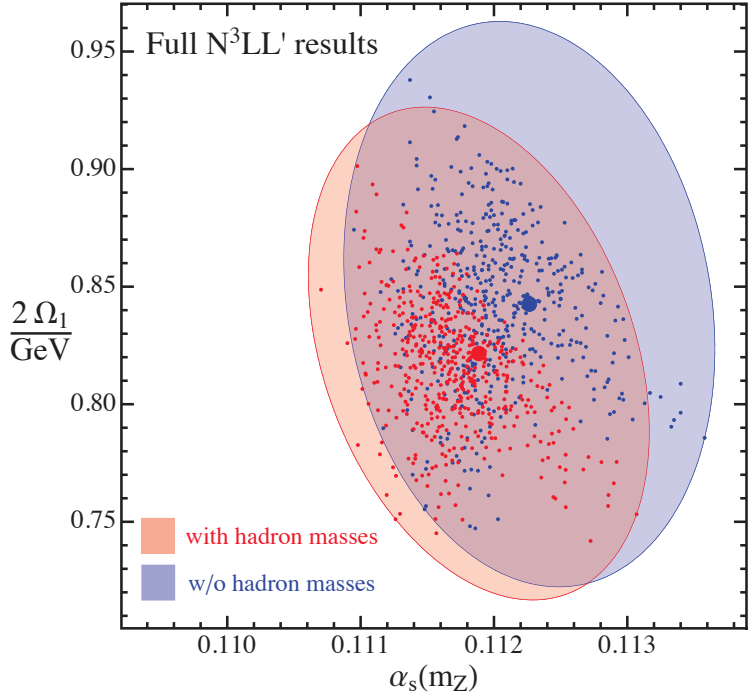}
	\end{center}
	\caption[Comparison of C-parameter tail fits with and without hadron mass effects]{Comparison of fits to the C-parameter tail distribution with theory prediction which include/ignore
		hadron-mass effects (in red/blue). Although a direct comparison of $\alpha_s$ values is possible, one has
		to keep in mind that $\Omega_1(\mu_\Delta, R_\Delta)$ has a different meaning once hadron mass running effects are included.}
	\label{fig:hadron}
\end{figure}

The fit results presented in the previous two sections ignored the small hadron-mass effects. These effects are analyzed in greater detail in this section. We again perform $500$ fits for a theory setup which includes N$^3$LL$^\prime$ accuracy and a power correction in the Rgap scheme, but this time it also includes hadron-mass-induced running.

Since the impact of hadron-mass effects is small, one finds that the experimental data in the tail of the distribution is not accurate enough to fit for $\theta(R_\Delta,\mu_\Delta)$ in \Eq{eq:thetadef}, in addition to $\alpha_s(m_Z)$ and $\Omega_1(R_\Delta,\mu_\Delta)$. This is especially true because it enters as a small modification to the power correction, which by itself is not the dominant term. Indeed, fitting for $a(R_\Delta,\mu_\Delta)$ and $b(R_\Delta,\mu_\Delta)$ as defined in Eq.~(\ref{eq:omega1-ansatz}) gives a strongly correlated determination of these two parameters. The dominant hadronic parameter $\Omega_1^C(R_\Delta,\mu_\Delta)$, which governs the normalization, is still as accurately determined from data as the $\Omega_1$ in Tab.~\ref{tab:Omega1results}. However, the orthogonal parameter $\theta(R_\Delta,\mu_\Delta)$ is only determined with very large statistical uncertainties. As discussed in Ch. \ref{ch:Cparam-theory} \cite{Hoang:2014wka}, the specific value of $\theta(R_\Delta,\mu_\Delta)$ has a very small impact on the cross section, which is consistent with the inability to accurately fit for it.

The results of our fit including hadron-mass effects are
\begin{align}
&\alpha_s(m_Z) = 0.1119 \pm 0.0006_{\rm exp+had} \pm 0.0013_{\rm pert}\,,\\[1mm]
&\Omega_1(R_\Delta,\mu_\Delta) = 0.411 \pm 0.018_{\rm exp+\alpha_s} \pm 0.052_{\rm pert}\,{\rm GeV}\,.\nn
\end{align}
Note that the meaning of $\Omega_1(R_\Delta,\mu_\Delta)$ here is different from the case in which hadron-mass running effects are ignored because there are extra evolution effects needed to translate this value to that used in the cross section at a given value of $C$, compared to the no-hadron-mass case.

In Fig.~\ref{fig:hadron} we compare the outcome of the $500$ fits at N$^3$LL$^\prime$ in the Rgap scheme. Results with hadron-mass effects give the red ellipse on the left, and without hadron-mass effects give the blue ellipse on the right. (The latter ellipse is the same as the one discussed above in Sec.~\ref{sec:random}.) The effects of hadron masses on $\alpha_s(m_Z)$ are to decrease its central value by $0.3\%$ and reduce the percent perturbative uncertainty by $0.1\%$. Given that the total perturbative uncertainties are $1.2\%$, these effects are not statistically significant. When studying the effect on $\Omega_1$ one has to keep in mind that its meaning changes when hadron-mass effects are included. Ignoring this fact we observe that hadron masses shift the central value downwards by $2.4\%$, and reduce the percent theoretical uncertainty by $1.6\%$. Again, given that the perturbative uncertainty for $\Omega_1$ is $14\%$, this shift is not significant.

Since the theory uncertainties become slightly smaller when hadron-mass effects are incorporated, one could use this setup as our default. However we take a more conservative approach and consider the $0.3\%$ shift on the central value as an additional source of uncertainty, to be added in quadrature to the hadronization uncertainty already discussed in Sec.~\ref{sec:random}. This increases the value of the hadronization uncertainty from $0.0006$ to $0.0007$, and does not affect the total $\alpha_s$ uncertainty. The main reason we adopt this more conservative approach is that, while well motivated, the ansatz that we take in Eq.~(\ref{eq:omega1-ansatz}) is not model independent. We believe that this ansatz serves as a good estimate of what the numerical effect of hadron masses are, but should likely not be used for the central fit until further theoretical insight on the form of $\Omega_1(r)$ is gained. We do not add an additional uncertainty to $\Omega_1$ since hadron-mass effects change its meaning and uncertainties for $\Omega_1$ are large enough that these effects are negligible.

In \App{ap:subtractions} we also consider fits performed using the Rgap scheme with C-parameter gap subtractions, rather than our default Rgap scheme with thrust gap subtractions. The two results are fully compatible. As discussed in Ch. \ref{ch:Cparam-theory} the thrust gap subtractions give better perturbative convergence, and hence are used for our default cross section.

\subsection{Dataset dependence}
\label{sec:dataset}
\begin{figure}[t]
	\vspace{0pt}
	\begin{center}
	\includegraphics[width=.65\linewidth]{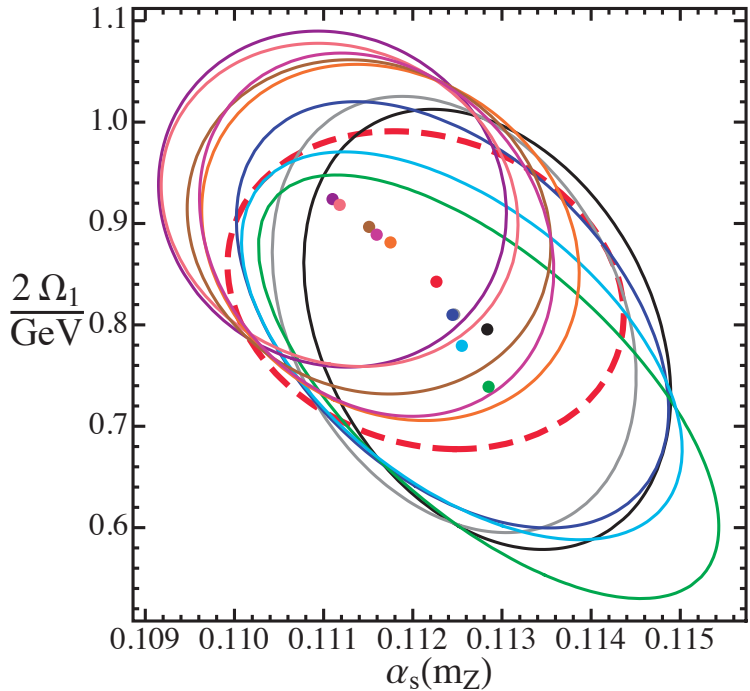}
\end{center}
	\caption[Fit results for different choices of C-parameter dataset]{Global fit results for different choices of dataset, using our best theory setup at N$^3$LL$^\prime$ with
		power corrections in the Rgap scheme. Considering the central values from left to right, the datasets read
		$[\,C_{\rm min}, C_{\rm max}\,]_{\rm \#~of~bins}$: $[\,29/Q, 0.7\,]_{371}$, $[\,22/Q, 0.75\,]_{453}$,
		$[\,23/Q, 0.7\,]_{417}$, $[\,0.24, 0.75\,]_{403}$, $[\,24/Q, 0.7\,]_{409}$, $[\,25/Q, 0.7\,]_{404}$ (default),
		$[\,25/Q, 0.6\,]_{322}$, $[\,25/Q, 0.75\,]_{430}$,
		$[\,27/Q, 0.7\,]_{386}$, $[\,25/Q, 0.65\,]_{349}$, $[\,22/Q, 0.7\,]_{427}$. We accept bins which are at least $50\%$ inside these
		fit regions. The ellipses correspond to total $1$-$\sigma$ uncertainties (experimental + theory) for two variables ($\alpha_s$ and
		$\Omega_1$), which are suitable for a direct comparison of the outcome of two-parameter fits. The center of the ellipses are
		also shown.}
	\label{fig:dataset}
\end{figure}
In this section we discuss how much our results depend on the dataset choice. Our default global dataset accounts for all experimental bins for $Q\ge 35\,$GeV in the intervals $[\,C_{\rm  min}, C_{\rm max}\,]=[\,25/Q,0.7\,]$, (more details are given in Sec.~\ref{sec:data}). The upper limit in this range is motivated by the fact that we do not want to include data too close to the shoulder, since we do not anticipate having the optimal theoretical description of this region. The lower limit avoids including data too close to the nonperturbative region, which is near the cross section peak for $Q=m_Z$,  since we by default only include the leading power correction $\Omega_1$ in the OPE of the shape function. To consider the impact of this dataset choice we can vary the upper and lower limits used to select the data.

In Fig.~\ref{fig:dataset} the best fits and the respective total experimental + theory $68\%$ CL uncertainty ellipses (for two parameters) are shown for global datasets based on different choices of data ranges. The result for our default global dataset is given in red, with a thicker, dashed ellipse. In the caption of Fig.~\ref{fig:dataset} the data ranges and the number of bins are specified for each one of the plotted ellipses.

Interestingly all uncertainty ellipses have very similar correlation and are lined up approximately along the line
\begin{align}
\frac{\Omega_1}{41.26 \,{\rm GeV}} = 0.1221 - \alpha_s(m_Z) \,.
\end{align}
As expected, the results of our fits depend only weakly on the $C$ range and the size of the global
datasets, as shown in Fig.~\ref{fig:dataset}. The size and tilt of the total uncertainty ellipses is very similar
for all datasets (with the exception of $[\,22/Q, 0.7\,]$, which clearly includes too much peak data).
Since the centers and the sizes of the uncertainty
ellipses are fully statistically compatible at the $1$-$\sigma$ level, this indicates that our theory uncertainty estimate at
N$^3$LL$^\prime$ really reflects the accuracy at which we are capable of describing the different regions of the spectrum.
Therefore a possible additional uncertainty that one could consider due to the arbitrariness of the dataset choice is actually already represented in our final uncertainty estimates.
\vspace*{-0.3cm}
\subsection{Final Results}
\label{sec:final}
\vspace*{-0.3cm}
As our final result for $\alpha_s(m_Z)$ and $\Omega_1$, obtained at N$^3$LL$^\prime$ order
in the Rgap scheme for $\Omega_1(R_\Delta,\mu_\Delta)$, we get
\begin{align} \label{eq:asfinal}
\alpha_s(m_Z) & \, = \,
0.1123 \,\pm\, 0.0002_{\rm exp} \,\pm\, 0.0007_{\rm hadr} \,\pm \, 0.0014_{\rm pert},
\\[2mm]
\Omega_1(R_\Delta,\mu_\Delta) & \, = \,
0.421 \,\pm\, 0.007_{\rm exp}        \,\pm\, 0.019_{\rm \alpha_s(m_Z)} 
\,\pm \, 0.060_{\rm pert}\,\mbox{GeV},
\nonumber
\end{align}
where $R_\Delta=\mu_\Delta=2$~GeV and we quote individual \mbox{$1$-$\sigma$} uncertainties for each parameter.
Here $\chi^2/\rm{dof}=0.99$.

Equation~(\ref{eq:asfinal}) accounts for the effect of hadron mass running through an additional (essentially negligible) uncertainty. Also, it neglects QED and finite bottom-mass corrections, which were found to be small effects in the corresponding thrust analysis in Ref.~\cite{Abbate:2010xh}.

\begin{figure}[t!]
	\begin{center}
		\includegraphics[width=0.65\columnwidth]{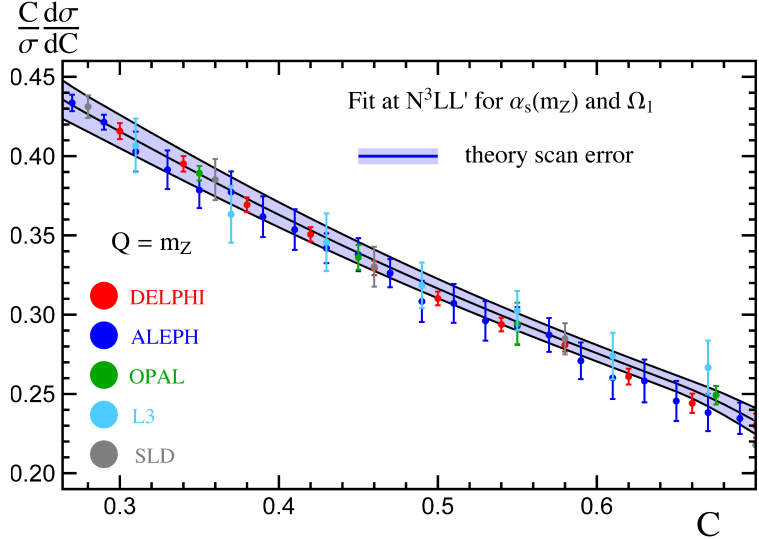}
		\caption[C-parameter distribution at N${}^3$LL$^\prime$ order for $Q=m_Z$]{C-parameter distribution at N${}^3$LL$^\prime$ order for $Q=m_Z$ showing the fit result for the values for $\alpha_s(m_Z)$ and
			$\Omega_1$. The blue band corresponds to the theory uncertainty as described in Sec.~\ref{sec:random}.
			Experimental data is also shown.}
		\label{fig:Theory-data-comparison}
	\end{center}
\end{figure}

Given that we treat the correlation of the systematic experimental uncertainties
in the minimal overlap model, it is useful to examine the
results obtained when assuming that all systematic experimental uncertainties are uncorrelated. At N$^3$LL$^\prime$ order in the Rgap scheme the results that are
analogous to Eq.~(\ref{eq:asfinal}) read
$\alpha_s(m_Z) = 0.1123 \pm 0.0002_{\rm exp} \pm 0.0007_{\rm hadr} \pm
0.0012_{\rm pert}$ and $\Omega_1(R_\Delta,\mu_\Delta) = 0.412 \,\pm\, 0.007_{\rm exp}
\pm 0.022_{\alpha_s} \pm 0.061_{\rm pert}$\,GeV
with a combined correlation coefficient of $\rho_{\alpha\Omega}^{\rm total}=-\,0.091$.
The results are compatible with Eq.~(\ref{eq:asfinal}), indicating that the ignorance of the precise
correlation of the systematic experimental uncertainties barely affects the outcome of the fit.

In Fig.~\ref{fig:Theory-data-comparison} we show the theoretical fit for the \mbox{C-parameter} distribution
in the tail region, at a center-of-mass energy corresponding to the $Z$-pole. We use the best-fit values given in 
Eq.~(\ref{eq:asfinal}). The band corresponds to the perturbative
uncertainty as determined by the scan. The fit result is shown in comparison with experimental
data from DELPHI, ALEPH, OPAL, L3 and SLD. Good agreement is observed for this spectrum, as well as for spectra at other center of mass values.

\begin{figure}
	\begin{center}
		\includegraphics[width=0.65\columnwidth]{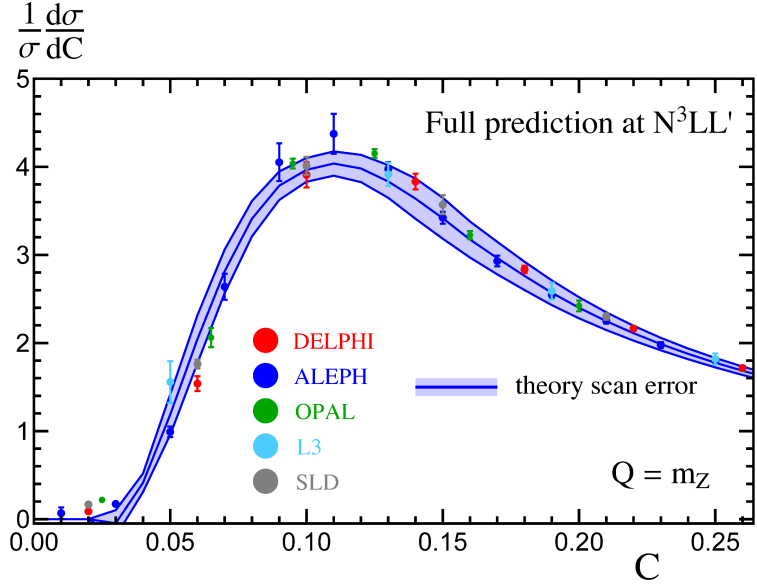}
		\caption[C-parameter distribution in the peak region]{C-parameter distribution below the fit region, shown at N${}^3$LL$^\prime$ order for $Q=m_Z$ using the best-fit values for $\alpha_s(m_Z)$ and
			$\Omega_1$. Again the blue band corresponds to the theory uncertainty and error bars are used for experimental data.}
		\label{fig:Peak-plot}
	\end{center}
\end{figure}
\begin{figure}[t!]
	\begin{center}
		\includegraphics[width=0.65\columnwidth]{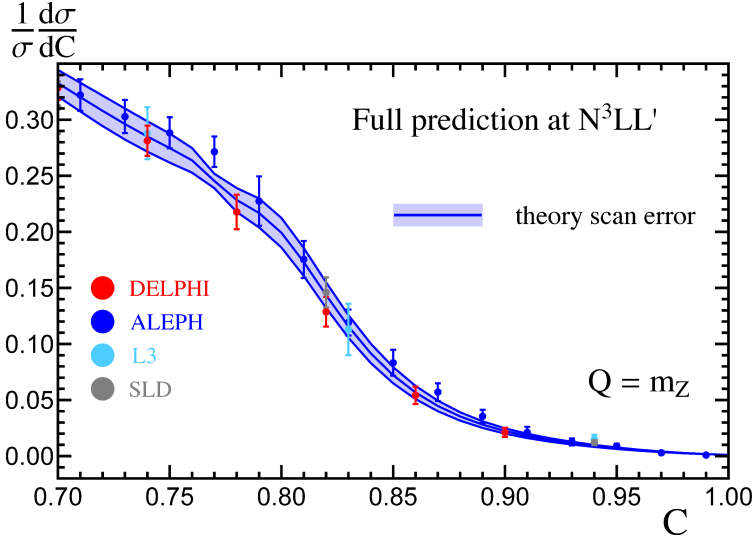}
		\caption[C-parameter distribution in the far-tail region]{C-parameter distribution above the fit range, shown at N${}^3$LL$^\prime$ order for $Q=m_Z$ using the best-fit values for $\alpha_s(m_Z)$ and
			$\Omega_1$. Again the blue band corresponds to the theory uncertainty and the error bars are used for experimental data. }
		\label{fig:Far-tail-plot}
	\end{center}
\end{figure}

\section{Peak and Far Tail Predictions}
\label{sec:peak-tail}

Even though our fits were performed in the resummation region which is dominated by tail data, our theoretical results also apply for the peak and far-tail regions. As an additional
validation for the results of our global analysis in the tail region, we use the best-fit values obtained for
$\alpha_s$ and $\Omega_1$ to make predictions in the peak and the far-tail regions where the corresponding data was not included in the fit.

Predictions from our full N$^3$LL$^\prime$ code in the Rgap scheme for the C-parameter cross section at the $Z$-pole in the peak
region are shown in Fig.~\ref{fig:Peak-plot}. The nice agreement within theoretical uncertainties (blue band) with the
precise data from DELPHI, ALEPH, OPAL, L3, and SLD indicates that the value of $\Omega_1$ obtained from the fit to the
tail region is the dominant nonperturbative effect in the peak. The small deviations between the theory band and the
experimental data can be explained due to the fact that the peak is also sensitive to higher-order power corrections
$\Omega_{k\ge 2}^C$, which have not been tuned to reproduce the peak data in our analysis.

In Fig.~\ref{fig:Far-tail-plot} we compare predictions from our full N$^3$LL$^\prime$ code in the Rgap scheme to the accurate
DELPHI, ALEPH, L3, and SLD data at $Q=m_Z$ in the far-tail region.\footnote{The OPAL data was excluded from the plot because its bins are
	rather coarse in this region, making it a bad approximation of the differential cross section.} We find excellent agreement with the data within the
theoretical uncertainties (blue band). The key feature of our theoretical prediction that matters most in the far-tail
region is the merging of the renormalization scales toward $\mu_S=\mu_J=\mu_H$ at $C\sim 0.75$ in the profile functions. This is a necessary condition for the cancellations between singular and nonsingular terms in the cross section to occur above the shoulder region.\footnote{It is worth mentioning that
	in the far-tail region we employ the $\overline{\rm MS}$ scheme for $\Omega_1$, since the subtractions implemented in the
	Rgap scheme clash with the partonic shoulder singularity, resulting in an unnatural behavior of the cross section around
	$C=0.75$. The transition between the Rgap and $\overline{\rm MS}$ schemes is performed smoothly, by means of a hybrid
	scheme which interpolates between the two in a continuous way. This hybrid scheme has been discussed at length in
	Ch. \ref{ch:Cparam-theory}.} At $Q=m_Z$ the theoretical cross section presented here 
obtains accurate predictions in the region both below and above the shoulder that agree with the data. Our analysis does not include the full ${\cal O}(\alpha_s^k \Lambda_{\rm QCD}/Q)$ power corrections (for $k < 4$), since they are not part of our master formula. Nevertheless, and in analogy with what was found in the case of thrust, agreement with the experimental data seems to indicate that these missing power corrections may be smaller than naively expected.

\section{Universality and Comparison to Thrust}
\label{sec:universality}

\begin{figure}[t!]
	\begin{center}
		\includegraphics[width=0.65\columnwidth]{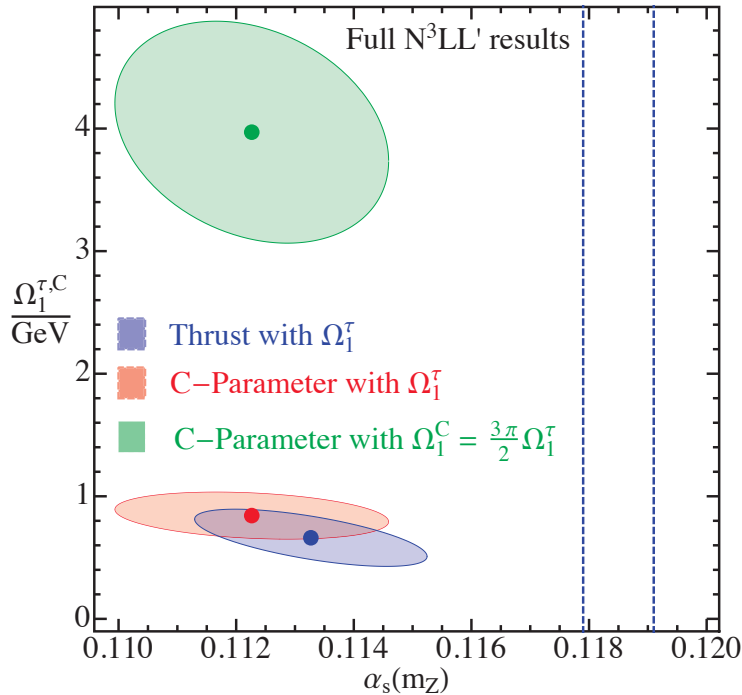}
		\caption[Comparison of determinations of $\alpha_s(m_Z)$ and $\Omega_1$ for different $\Omega_1$ normalizations]{Comparison of determinations of $\alpha_s(m_Z)$ and $\Omega_1$ with the corresponding total 1-$\sigma$
			uncertainty ellipses. As an illustration we display the determination of $\Omega_1^C$ obtained from fits
			to the C-parameter distribution (green), which is clearly different from $\Omega_1^\tau$ obtained from thrust fits (blue),
			and the determination of $\Omega_1^\tau$ as obtained from \mbox{C-parameter} distribution fits (red). All fits have been performed with N$^3$LL$^\prime$ theoretical predictions with power corrections and in the Rgap scheme. The dashed vertical lines indicate the PDG 2014~\cite{Agashe:2014kda} determination of $\alpha_s(m_Z)$. }
		\label{fig:Universality}
	\end{center}
\end{figure}

\begin{figure}[t!]
	\begin{center}
		\includegraphics[width=0.65\columnwidth]{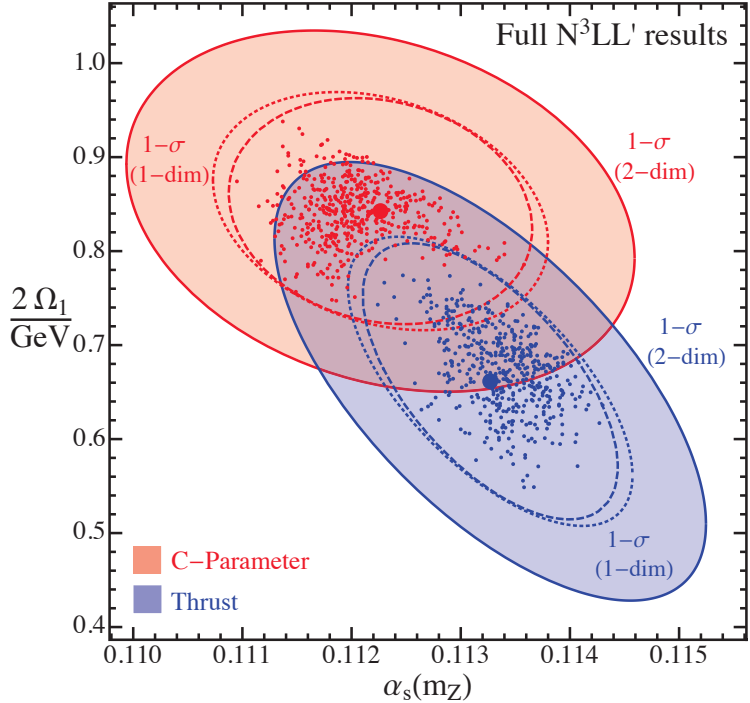}
		\caption[Comparison of determinations of $\alpha_s(m_Z)$ and $\Omega_1$ for C-parameter and thrust ]{Distribution of best-fit points in the \mbox{$\alpha_s(m_Z)$-$2\Omega_1$} plane for both thrust (blue) and
			C-parameter (red) at N${}^3$LL$^\prime+\mathcal{O}(\alpha_s^3) + \Omega_1(R,\mu)$. The outer solid ellipses show the $\Delta \chi^2 = 2.3$ variations, representing 1-$\sigma$ uncertainties for two variables. The inner dashed ellipses correspond to the 1-$\sigma$ theory uncertainties for each one of the fit parameters. The dotted ellipses correspond to $\Delta \chi^2 = 1$ variations of the total uncertainties. All fits have been performed with N$^3$LL$^\prime$ theoretical predictions with power corrections and in the Rgap scheme. This plot zooms in on the bottom two ellipses of Fig.~\ref{fig:Universality}.}
		\label{fig:Thrust-Cparam-comparison}
	\end{center}
\end{figure}

\begin{figure}[t!]
	\begin{center}
		\includegraphics[width=0.65\columnwidth]{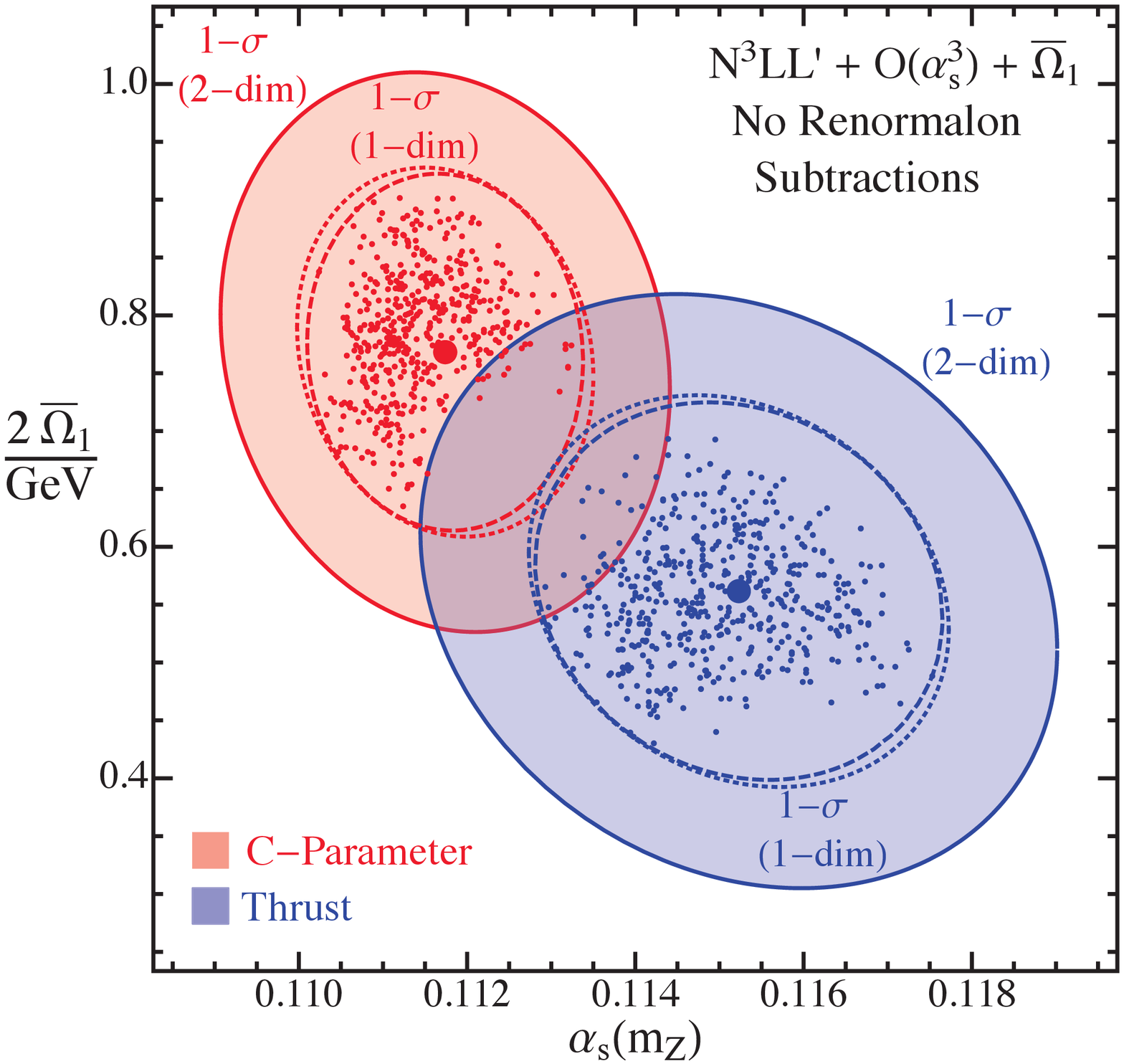}
		\caption[Comparison of determinations of $\alpha_s(m_Z)$ and $\Omega_1$ for C-parameter and thrust without renormalon subtraction]{Distribution of best-fit points in the \mbox{$\alpha_s(m_Z)$-$2\overline\Omega_1$} plane for both thrust (blue) and
			C-parameter (red) at N${}^3$LL$^\prime+\mathcal{O}(\alpha_s^3) + \overline\Omega_1$. The meaning of the different
			ellipses is the same as in Fig.~\ref{fig:Thrust-Cparam-comparison}.}
		\label{fig:Thrust-Cparam-NoGap}
	\end{center}
\end{figure}

An additional prediction of our theoretical formalism is the universality of $\Omega_1$ between the thrust and \mbox{C-parameter} event shapes. Therefore, a nontrivial test of our formalism can be made by comparing our result for $\Omega_1$ with the determination from the earlier fits of the thrust tail distributions in Ref.~\cite{Abbate:2010xh} and the first moment of the thrust distribution in Ref.~\cite{Abbate:2012jh}.

Since we have updated our profiles for thrust, it is expected that the outcome of the $\alpha_s$ and $\Omega_1$ determination is slightly (within theoretical uncertainties) different from that of Ref.~\cite{Abbate:2010xh}. We also have updated our code to match that of Ref.~\cite{Abbate:2012jh} (higher statistics for the two-loop nonsingular cross sections and using the exact result for the two-loop soft function non-logarithmic constant). In addition we have corrected the systematic uncertainty for the ALEPH data, $Q = 91.2$\,GeV of Ref.~\cite{Heister:2003aj}.\footnote{In Ref.~\cite{Abbate:2010xh}, they assumed that two quoted uncertainties where asymmetric uncertainties, but it turns out they are two sources of systematic uncertainties that need to be added in quadrature. This has no significant effect on the results of  Ref.~\cite{Abbate:2010xh}.} When we compare thrust and C-parameter we neglect bottom-mass and QED effects in both event shapes. In this setup, we find an updated result for thrust:
\begin{align} \label{eq:thrustQCD}
\alpha_s(m_Z) & \, =
0.1134 \pm 0.0002_{\rm exp} \pm 0.0005_{\rm hadr} \pm  0.0011_{\rm pert},
\\[2mm]
\Omega_1(R_\Delta,\mu_\Delta) & \, = 
0.329 \pm 0.009_{\rm exp}     \pm 0.021_{\rm \alpha_s(m_Z)} 
\pm  0.060_{\rm pert}\,\mbox{GeV}.
\nonumber
\end{align}
For completeness we also quote an updated thrust result when both QED and bottom-mass effects are taken into account:
\begin{align} \label{eq:thrustQED}
\alpha_s(m_Z) & \, =
0.1128 \pm 0.0002_{\rm exp}  \pm 0.0005_{\rm hadr} \pm 0.0011_{\rm pert},
\\[2mm]
\Omega_1(R_\Delta,\mu_\Delta) & \, =
0.322 \pm 0.009_{\rm exp}        \pm 0.021_{\rm \alpha_s(m_Z)} 
\pm  0.064_{\rm pert}\,\mbox{GeV}.
\nonumber
\end{align}
Both the results in Eqs.~(\ref{eq:thrustQCD}) and (\ref{eq:thrustQED}) are fully compatible at 1-$\sigma$ with those in Ref.~\cite{Abbate:2010xh}, as discussed in more detail in \App{ap:thrustresults}.

When testing for the universality of $\Omega_1$ between thrust and C-parameter, there is an important calculable numerical factor of $3\pi/2=4.7$ between $\Omega_1^\tau$ and $\Omega_1^C$ that must be accounted for; see  \Eq{eq:O1univ}. If we instead make a direct comparison of $\Omega_1^\tau$ and $\Omega_1^C$, as shown in Fig.~\ref{fig:Universality} (lowest blue ellipse vs uppermost green ellipse, respectively) then the results are $4.5$-$\sigma$ away from each other. Accounting for the $3\pi/2$ factor to convert from $\Omega_1^C$ to $\Omega_1^\tau$ the upper green ellipse becomes the centermost red ellipse, and the thrust and C-parameter determinations agree with one another within uncertainties. Due to our high-precision control of perturbative effects, the $\Omega_1$ parameters have only $\sim 15\%$ uncertainty, yielding a test of this universality at a higher level of precision than what has been previously achieved. 

A zoomed-in version of this universality plot is shown in Fig.~\ref{fig:Thrust-Cparam-comparison}. The upper red ellipse again shows the result from fits to the C-parameter distribution, while the lower blue 
ellipse shows the result from thrust tail fits. For both we show the theory uncertainty
(dashed lines) and combined theoretical and experimental (dotted lines) 39\% CL uncertainty ellipses,
as well as the solid ellipses which correspond to \mbox{$\Delta\chi^2=2.3$} which is the standard 1-$\sigma$ uncertainty for a two-parameter fit
(68\% CL). We see that
the two analyses are completely compatible at the 1-$\sigma$ level. An important  
ingredient to improve the overlap is the fact that we define the power corrections in the
renormalon-free Rgap scheme. This is shown by contrasting the Rgap result in Fig.~\ref{fig:Thrust-Cparam-comparison} with the overlap obtained when using the ${\overline {\rm MS}}$ scheme for $\Omega_1$, as shown in Fig.~\ref{fig:Thrust-Cparam-NoGap}.  
%

\chapter{Subleading Helicity Building Blocks}
\label{ch:Subleading}

\begin{figure}[h!]
	%
	%
	\begin{center}
		\includegraphics[width=0.23\columnwidth]{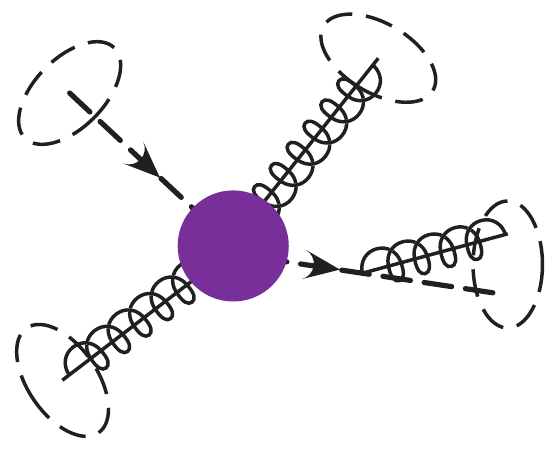} 
		\hspace{0.1cm}
		\includegraphics[width=0.23\columnwidth]{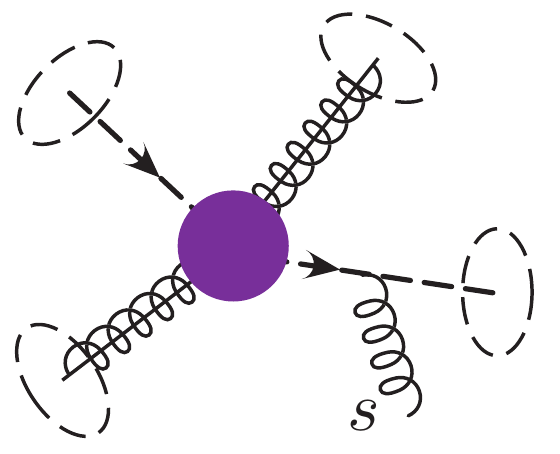} 
		\hspace{0.1cm}
		\includegraphics[width=0.23\columnwidth]{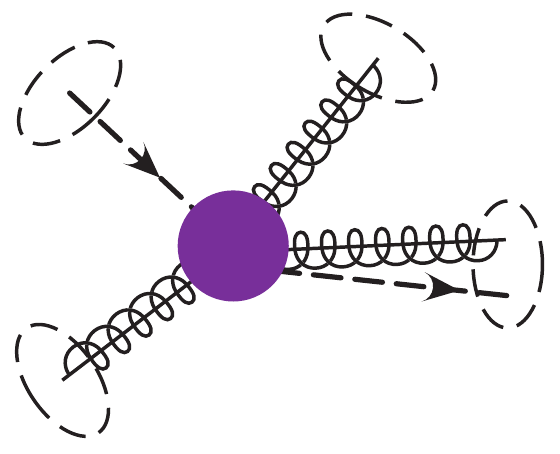} 
		\hspace{0.1cm}
		\includegraphics[width=0.23\columnwidth]{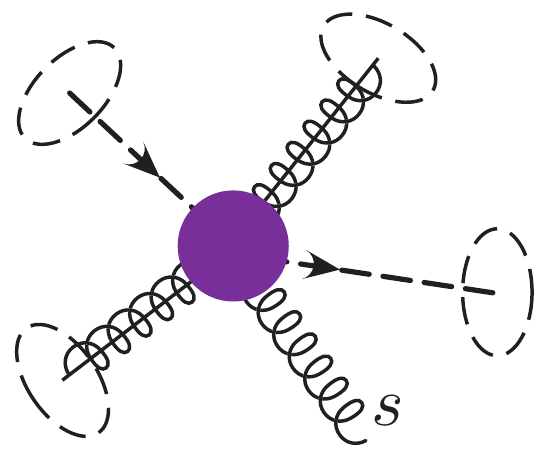} 
		\raisebox{0cm}{ \hspace{-0.2cm} 
			$a$)\hspace{3.4cm}
			$b$)\hspace{3.4cm} 
			$c$)\hspace{3.4cm}
			$d$)\hspace{4cm} } 
		\\[-25pt]
	\end{center}
	\vspace{-0.4cm}
	\caption{ 
		Example of scattering amplitudes with energetic particles in four distinct regions of phase space, at leading power in a) and b), and subleading power in c) and d).  There is an extra collinear gluon in a) from splitting, and in b) there is an extra gluon from soft emission. In c) the extra energetic gluon is collinear with the quark, but occurs without a nearly onshell parent propagator. Likewise in d) the extra soft emission amplitude is subleading. } 
	\label{fig:subleadingamp}
\end{figure}

This chapter contains an extension of the leading SCET helicity building block operators presented in \cite{Moult:2015aoa} to subleading power, and was submitted for publication in Ref. \cite{Kolodrubetz:2016uim}. 
Examples of the type of amplitudes that are described at leading and subleading power are shown in \fig{subleadingamp}. For leading power amplitudes with an extra collinear or soft gluon emission, such as those in \fig{subleadingamp}a,b, the extra gluon is accompanied by the enhancement from an additional nearly onshell propagator. In contrast, in the subleading amplitudes in \fig{subleadingamp}c,d we have an extra gluon emission without this enhancement. In order to enumerate the operators that can contribute to these amplitudes, it is more convenient to replace traditional SCET building blocks with helicity building blocks.

We begin this discussion with a review of leading SCET helicity fields in \sec{scethelicity}. The main result of this chapter is presented in \sec{helops}, where we show a complete set of helicity building blocks for constructing SCET operators at subleading power. In \sec{ang_cons} we give some angular momentum constraints that can be powerful tool for reducing the number of required subleading power operators. Finally, \sec{example} is devoted to a particular example in the case of dijets which illustrates the utility of this formalism. Our helicity conventions and some useful identities can be found in App. \ref{app:helicity}.

\section{Helicity Fields in SCET}\label{sec:scethelicity}

The use of on-shell helicity amplitudes has been fruitful for the study of scattering amplitudes in gauge theories and gravity (see e.g. \cite{Dixon:1996wi,Elvang:2013cua,Dixon:2013uaa,Henn:2014yza} for pedagogical reviews).  By focusing on amplitudes for external states with definite helicity and color configurations many simplifications arise. The helicity approach to SCET operators of \Ref{Moult:2015aoa} takes advantage of the fact that collinear SCET fields are themselves gauge invariant, and are each associated with a fixed external label direction with respect to which helicities can naturally be defined. Instead of considering operators formed from Lorentz and Dirac structures (each of which contributes to multiple states with different helicity combinations) helicity operators can be associated with external states of definite helicity. This approach greatly simplifies the construction of a minimal operator bases for processes with many active partons, and facilitates the matching to fixed order calculations which are often performed using spinor helicity techniques.

We now briefly summarize our spinor helicity conventions. Further identities, as well as our phase conventions, can be found in \app{helicity}. To simplify our discussion we take all momenta and polarization vectors as outgoing, and label all fields and operators by their outgoing helicity and momenta. We use the standard spinor helicity notation 
\begin{align} \label{eq:braket_def}
|p\rangle\equiv \ket{p+} &= \frac{1 + \ga_5}{2}\, u(p)
\,,
& |p] & \equiv \ket{p-} = \frac{1 - \ga_5}{2}\, u(p)
\,, \\
\bra{p} \equiv \bra{p-} &= \mathrm{sgn}(p^0)\, \bar{u}(p)\,\frac{1 + \ga_5}{2}
\,, 
& [p| & \equiv \bra{p+} = \mathrm{sgn}(p^0)\, \bar{u}(p)\,\frac{1 - \ga_5}{2}
\,, \nn 
\end{align}
with $p$ lightlike. The polarization vector of an outgoing gluon with momentum $p$ can be written
\begin{equation}
\ve_+^\mu(p,k) = \frac{\mae{p+}{\ga^\mu}{k+}}{\sqrt{2} \langle kp \rangle}
\,,\qquad
\ve_-^\mu(p,k) = - \frac{\mae{p-}{\ga^\mu}{k-}}{\sqrt{2} [kp]}
\,,\end{equation}
where $k\neq p$ is an arbitrary lightlike reference vector.

Since the building blocks in \tab{PC} carry vector or spinor Lorentz indices they must be contracted to form scalar operators, which involves the use of objects like $\{n_i^\mu, \bn_i^\mu, \gamma^\mu, g^{\mu\nu},$ $\epsilon^{\mu\nu\sigma\tau}\}$.  For operators describing many jet directions or for operators at subleading power, constructing a minimal basis in this manner becomes difficult. Rather than dealing with contractions of vector and spinor indices, one can exploit a decomposition into operators with definite helicity, and work with building blocks that are scalars.\footnote{Generically when we say scalar building blocks, we are not accounting for their transformations under parity. Constraints from parity transformations are easy to include, see~\cite{Moult:2015aoa}.} For SCET operators this approach was formalized in~\cite{Moult:2015aoa} by defining helicity building block fields for the construction of leading power operators for jet processes. It takes advantage of the fact that collinear SCET fields are themselves collinear gauge invariant, and are each associated with a fixed external label direction with respect to which helicities can naturally be defined. We will follow the notation and conventions of~\cite{Moult:2015aoa}. We first define  collinear gluon and quark fields of definite helicity as
\begin{subequations}
	\label{eq:cBpm_quarkhel_def}
	\begin{align} 
	\label{eq:cBpm_def}
	\cB^a_{i\pm} &= -\ve_{\mp\mu}(n_i, \bn_i)\,\cB^{a\mu}_{n_i\perp,\w_i}
	\,, \\
	\label{eq:quarkhel_def}
	\chi_{i \pm}^\alpha &= \frac{1\,\pm\, \gamma_5}{2} \chi_{n_i, - \omega_i}^\alpha
	\,,\qquad\quad
	\bar{\chi}_{i \pm}^\balpha =  \bar{\chi}_{n_i, - \omega_i}^\balpha \frac{1\,\mp\, \gamma_5}{2}\,,
	\end{align}
\end{subequations}
where $a$, $\alpha$, and $\balpha$ are adjoint, $3$, and $\bar 3$ color indices respectively, and by convention the $\omega_i$ labels on both the gluon and quark building block are taken to be outgoing.

 The lowest order Feynman rules for these fields are simple. For example, for an outgoing gluon with polarization $\pm$, momentum $p$ ($p^0>0$), and color $a$ we have
 \\ ${\l g_\pm^a(p) | \cB_{i \pm}^b |0 \r = \delta^{ab} \ldel (\lp_i -p)}$, while for an incoming quark ($p^0<0$) with helicity $\pm$ and color $\alpha$ we have
$\Mae{0}{\chi^\beta_{i\pm}}{q_\pm^\balpha(-p)} = \delta^{\beta\balpha}\, \ldel(\lp_i - p)\, \ket{(-p_i)\pm}_{n_i}$.   Here we define the spinors with an SCET projection operator by $|p \pm \r_{n_i}\equiv \frac{\slashed{n}_i \slashed{\bar{n}}_i}{4} | p \pm \r$ and the $\tilde\delta(\tilde p_i-p)$ indicate that the momentum label in the building block field matches that of the state. The full set of Feynman rules are given in~\cite{Moult:2015aoa}.

To take advantage of the fact that fermions always come in pairs, Ref.~\cite{Moult:2015aoa} defined the currents
\begin{align} \label{eq:jpm_def}
J_{ij\pm}^{\balpha\beta}
& = \mp\, \sqrt{\frac{2}{\omega_i\, \omega_j}}\, \frac{   \ve_\mp^\mu(n_i, n_j) }{\langle n_j\mp | n_i\pm\rangle}   \, \bar{\chi}^\balpha_{i\pm}\, \gamma_\mu \chi^\beta_{j\pm}
\,, \\
J_{ij0}^{\balpha\beta}
& =\frac{2}{\sqrt{\vphantom{2} \omega_i \,\omega_j}\,  [n_i n_j] } \bar \chi^\balpha_{i+}\chi^\beta_{j-}
\,, \qquad
(J^\dagger)_{ij0}^{\balpha\beta}=\frac{2}{\sqrt{ \vphantom{2} \omega_i \, \omega_j}  \langle n_i  n_j \rangle  } \bar \chi^\balpha_{i-}\chi^\beta_{j+}
. \nn
\end{align}
These currents are manifestly invariant under the RPI-III symmetry of SCET, which takes $n_i^\mu \to e^\alpha n_i^\mu$ and $\bn_i^\mu \to e^\alpha \bn_i^\mu$, since $\omega_i\sim\bn_i$ and the $| n_i \rangle\sim\sqrt{n_i}$.  In general these currents consist of two spin-$1/2$ objects whose spin quantum numbers are specified along different axes, $\hat n_i$ and $\hat n_j$.  If we consider back-to-back collinear directions $n$ and $\bn$, then the two axes are the same, and these currents have definite helicity, given by
\begin{align} \label{eq:jpm_back_to_bacjdef}
& h=\pm 1:
& J_{n \bn \pm}^{\balpha\beta}
& = \mp\, \sqrt{\frac{2}{\omega_n\, \omega_\bn}}\, \frac{   \ve_\mp^\mu(n, \bn) }{\langle \bn \mp | n \pm\rangle}   \, \bar{\chi}^\balpha_{n\pm}\, \gamma_\mu \chi^\beta_{\bn \pm}
\,, \\
& h=0:
& J_{n \bn 0}^{\balpha\beta}
& =\frac{2}{\sqrt{\vphantom{2} \omega_n \,\omega_\bn}\,  [n \bn] } \bar \chi^\balpha_{n+}\chi^\beta_{\bn-}
\,, \qquad
(J^\dagger)_{n \bn 0}^{\balpha\beta}=\frac{2}{\sqrt{ \vphantom{2} \omega_n \, \omega_\bn}  \langle n  \bn \rangle  } \bar \chi^\balpha_{n-}\chi^\beta_{\bn+}
. \nn
\end{align}
The currents $J_{n \bn \pm}^{\balpha \beta}$ have helicity $h=\pm1$ along $\hat n$ respectively. The current $J_{n \bn 0}^{\balpha \beta} + (J^\dagger)_{n \bn 0}^{\balpha \beta}$ transforms as a scalar under rotations about the $n$ axis, i.e. has helicity zero (while the current $J_{n \bn 0}^{\balpha \beta} - (J^\dagger)_{n \bn 0}^{\balpha \beta}$ transforms as a pseudoscalar). We choose to use the $0$ subscript in both the back-to-back and non-back-to-back cases, to emphasize the helicity for the former case and conform with our notation for subleading currents below.

Together, the gluon building blocks $\cB^a_{i\pm}$ and the current building blocks $J_{ij\pm}^{\bar\alpha\beta}$, $J_{ij\,0}^{\bar\alpha\beta}$, and $(J^\dagger)_{ij\, 0}^{\bar\alpha\beta}$ suffice for the construction of leading power operators for all hard processes. (The only exceptions are hard processes that start at a power suppressed order.) All these objects behave like scalars under the Lorentz group, and can trivially be combined to form hard scattering operators by simple multiplication. The construction of leading power operators of this type was the focus of~\cite{Moult:2015aoa}.  We review below the organization of color structures in the leading power hard scattering operators and the decoupling of soft and collinear degrees of freedom using the BPS field redefinition. Then, in the next section we will extend this basis of building block objects to account for new structures that can appear at subleading power. 

The effective Lagrangian for hard scattering operators at any given order in the power counting, $ \cL^{(j)}_{\text{hard}}$, can be separated into a convolution between Wilson coefficients $\vec C$ encoding hard physics with $p^2\sim Q^2$, and on-shell physics encoded in SCET operators $\vec O$. In the hard scattering Lagrangian, the structure of SCET only allows convolutions between $\vec C$ and $\vec O$ in the collinear gauge invariant ${\cal O}(\lambda^0)$ momenta $\omega_i$, 
\begin{align} \label{eq:Leff_sub_explicit}
\cL^{(j)}_{\text{hard}} = \sum_{\{n_i\}} \sum_{A,\{\lambda_j\}} 
\bigg[ \prod_{i=1}^{\ell_{A}}\int\!\!\df \omega_i \bigg] \,
& \vO^{(j)\dagger}_{A\{\lambda_j\}}\big(\{n_i\};
\omega_1,\ldots,\omega_{\ell_A}\big) \,
\vC^{(j)}_{A\{\lambda_j\}}\big(\{n_i\};\omega_1,\ldots,\omega_{\ell_A} \big)
\,.
\end{align}
The operators $\vec{O}_A^{(j)}$ are traditionally constructed from the SCET building blocks in Tab. \ref{tab:PC}, whereas here we will use helicity building blocks. The hard process being considered determines the appropriate collinear sectors $\{n_i\}$, and the relevant helicity combinations $\{\lambda_j\}$, which are a series of $\pm$s and $0$s, $\{\lambda_j\}=+-0+0+\cdots$. Different classes of operators are distinguished by the additional subscript $A$. which encodes all relevant information that is not distinguished by the helicity labels, such as particle content. This $A$ is also used to label the number of convolution variables $\ell_A$. The number of $\omega_i$'s depends on the specific operator we are considering since at subleading power multiple collinear fields can appear in the same collinear sector and we must consider the inclusion of ultrasoft building blocks with no $\omega_i$ labels. At leading power the operators $\vO^\dagger_{A\{\lambda_j\}}$ are given by products of the gluon and quark helicity building block operators in \eqs{cBpm_def}{jpm_def}.

The Wilson coefficients $\vC^{(j)}_{A\{\lambda_j\}}$ appearing in \Eq{eq:Leff_sub_explicit} are ${\cal O}(\lambda^0)$, and can be determined by a matching calculation. They are vectors in an appropriate color subspace. Since we will use building blocks that are simultaneously gauge invariant under collinear and ultrasoft transformations, the constraints of SCET gauge invariance are reduced to that of global color, making it simple to construct a color basis for these objects. Decomposing both the coefficients and operators in terms of color indices following the notation of~\cite{Moult:2015aoa}, we have
\begin{align} \label{eq:Cpm_Opm_color}
C_{A\{\lambda_j\}}^{a_1\dotsb\alpha_n}
& = \sum_k C_{A\{\lambda_j\}}^k T_k^{a_1\dotsb\alpha_n}
\equiv \vT^{ a_1\dotsb\alpha_n} \vC_{A\{\lambda_j\}}
\,, \nn\\
\vO^\dagger_{A\{\lambda_j\}} &= \widetilde O_{A\{\lambda_j\}}^{a_1\dotsb \alpha_n}\, \vT^{\, a_1\dotsb \alpha_n}
\,,
\end{align}
and the color space contraction in \eq{Leff_sub_explicit} becomes explicit, $\vec O_{A\{\lambda_j\}}^\dagger \, \vec C_{A\{\lambda_j\}} = \widetilde O_{A\{\lambda_j\}}^{a_1\dotsb \alpha_n}  C_{A\{\lambda_j\}}^{a_1\dotsb\alpha_n}$.
In \eq{Cpm_Opm_color} $\vT^{\, a_1\dotsb\alpha_n}$ is a row vector of color structures that spans the color conserving subspace. The $a_i$ are adjoint indices and the $\alpha_i$ are fundamental indices.  The color structures do not necessarily have to be independent, but must be complete. This issue is discussed in detail in~\cite{Moult:2015aoa}. Color structures which do not appear in the matching at a particular order will be generated by renormalization group evolution. (For a pedagogical review of the color decomposition of QCD amplitudes see~\cite{Dixon:1996wi,Dixon:2013uaa}.) 

As discussed in \ref{sec:SCET}, in \SCETi, the leading power interactions between the soft and collinear degrees of freedom, described by $\cL^{(0)}$, can be decoupled using the BPS field redefinition of \eq{BPSfieldredefinition}. The BPS field redefinition generates ultrasoft interactions through the Wilson lines $Y_n^{(r)}$ which appear in the hard scattering operators~\cite{Bauer:2002nz}. When this is done consistently for S-matrix elements it accounts for the full physical path of ultrasoft Wilson lines~\cite{Chay:2004zn,Arnesen:2005nk}, so that some ultrasoft Wilson lines instead run over $(-\infty,0)$. We can organize the result of this field redefinition by grouping the Wilson lines $Y_n^{(r)}$ together with elements in our color structure basis $\vT^{\, a_1\dotsb\alpha_n}$. We will denote the result of this by $\vT_{\rm BPS}^{\, a_1\dotsb\alpha_n}$.  As a simple leading power example of this, consider the operators
\be \label{eq:349}
O^{a\bar \alpha \beta}_{+(\pm)}= \cB_{1 +}^a\,J_{23\pm}^{\balpha\bt}\,,     \qquad     O^{a\bar \alpha \beta}_{-(\pm)}= \cB_{1 -}^a\,J_{23\pm}^{\balpha\bt} \,.
\ee
In this case there is a unique color structure before the BPS field redefinition, namely
\be \label{eq:gqqcolor}
\vT^{ a \al\bbeta} = \left ( T^a \right )_{\alpha \bbeta}\,.
\ee
After BPS field redefinition, we find the Wilson line structure,
\be \label{eq:example_BPS}
\vT_{\BPS}^{ a \al\bbeta} = Y^{\dagger \alpha \bar{\gamma}} _{n_2} T^b_{\gamma \bar{\sigma}} \cY_{n_1}^{ba}   Y_{n_3}^{\sigma \bbeta} \,.
\ee
The non-local structure encoded in these ultrasoft Wilson lines is entirely determined by the form of the operator in \Eq{eq:349}, and the definition of the BPS field redefinition in \Eq{eq:BPSfieldredefinition}.  After the BPS field redefinition, the building block fields are ultrasoft gauge invariant, but still carry global color indices. This will play an important role in defining gauge invariant helicity building blocks at subleading power, when ultrasoft fields appear in the hard scattering operators. In general we will use the notation
\begin{equation} \label{eq:Opm_BPScolor}
\vO^\dagger_{\{\lambda_j\}}
= O_{\{\lambda_j\}}^{a_1\dotsb \alpha_n}\, \vT_{\BPS}^{\, a_1\dotsb \alpha_n}
\,,
\end{equation}
for the operators with definite color indices that are obtained after the BPS field redefinition. After BPS field redefinition, $\vT_{\BPS}$ contains both color generators and ultrasoft Wilson lines, as in \Eq{eq:example_BPS}. This generalizes the vector of color structures used in the decomposition of the pre-BPS hard scattering operators in \Eq{eq:Cpm_Opm_color}, where to distinguish we included an extra tilde on the operators with specified color indices. More examples will be given in \Sec{sec:ang_cons}.

\section{Complete Set of Helicity Building Blocks}\label{sec:helops}

We now carry out the main goal of this chapter, namely the extension of the scalar building blocks of \Eqs{eq:cBpm_def}{eq:jpm_def} to include all objects that are needed to describe subleading power interactions in the hard scattering Lagrangian. This will include defining operator building blocks involving multiple collinear fields in the same collinear sector, $\cP_\perp$ insertions, and explicit ultrasoft derivatives and fields.  We will continue to exploit the conservation of fermion number by organizing the fermions into bilinear currents.

\begin{table}
	\begin{center}
		\begin{tabular}{|c|c|cc|ccc|c|ccc|}
			\hline \phantom{x} & \phantom{x} & \phantom{x} 
			& \phantom{x} & \phantom{x} & \phantom{x} & \phantom{x} 
			& \phantom{x} & \phantom{x} & \phantom{x} & \phantom{x} 
			\\[-13pt]                      
			Field: & 
			$\cB_{i\pm}^a$ & $J_{ij\pm}^{\balpha\beta}$ & $J_{ij0}^{\balpha\beta}$ 
			& $J_{i\pm}^{\balpha \beta}$ 
			& $J_{i0}^{\balpha \beta}$ & $J_{i\bar 0}^{\balpha \beta}$  
			& $\cP^{\perp}_{\pm}$ 
			& $\partial_{us(i)\pm}$ & $\partial_{us(i)0}$ & $\partial_{us(i)\bar{0}}$
			\\[3pt] 
			Power counting: &	
			$\lambda$ &  $\lambda^2$ &  $\lambda^2$
			& $\lambda$ & $\lambda^2$& $\lambda^2$ & $\lambda^2$ 
			& $\lambda^2$ & $\lambda^2$  & $\lambda^2$
			\\
			Equation: & 
			(\ref{eq:cBpm_def}) & \multicolumn{2}{c|}{(\ref{eq:jpm_def})} 
			& \multicolumn{3}{c|}{(\ref{eq:coll_subl})} & (\ref{eq:Pperppm}) 
			& \multicolumn{3}{c|}{(\ref{eq:partialus})}
			\\
			\hline  
		\end{tabular}\\
		\vspace{.3cm} \hspace{2.85cm}
		\begin{tabular}{|c|cc|cccc|cc|}
			\hline  \phantom{x} &  \phantom{x} & \phantom{x} & \phantom{x} 
			& \phantom{x} & \phantom{x} & \phantom{x} & \phantom{x} & \phantom{x}
			\\[-13pt]                        
			Field: & 
			$\cB^a_{us(i)\pm}$ & 
			\!\!$\cB^a_{us(i)0}$ & 
			$J_{i(us)\pm}^{\balpha\beta}$  &
			$J_{i(\overline{us})\pm}^{\balpha\beta}$ &
			$J_{i(us)0}^{\balpha\beta}$ &
			$J_{i(\overline{us})0}^{\balpha\beta}$ &
			$J_{(us)^2ij\pm}$ & $J_{(us)^2ij0}$ 
			\\[3pt] 
			Power counting: &
			$\lambda^2$ & $\lambda^2$ & $\lambda^4$ &  $\lambda^4$ &  $\lambda^4$
			& $\lambda^4$ & $\lambda^6$& $\lambda^6$
			\\ 
			Equation: & 
			\multicolumn{2}{|c|}{(\ref{eq:Bus})}  & \multicolumn{4}{c|}{(\ref{eq:Jus})} 
			& \multicolumn{2}{c|}{(\ref{eq:Jus2})} 
			\\
			\hline
		\end{tabular}
	\end{center}
	\vspace{-0.3cm}
	\caption[The complete set of helicity building blocks in $\text{SCET}_\text{I}$]{The complete set of helicity building blocks in $\text{SCET}_\text{I}$, together with their power counting order in the $\lambda$-expansion, and the equation numbers where their definitions may be found. The building blocks also include the conjugate currents $J^\dagger$ in cases where they are distinct from the ones shown.
	} 
	\label{tab:helicityBB}
\end{table}

A summary of our final results for the complete set of scalar building blocks valid to all orders in the \SCETi power expansion is shown in \Tab{tab:helicityBB}, along with the power counting of each building block and the equation number where it is defined. The building blocks that appeared already at leading power~\cite{Moult:2015aoa}, were given above in \eqs{cBpm_def}{jpm_def}. We will discuss each of the additional operators in turn.

For collinear gluons, the fields $\cB^a_{i\pm}$ suffice even at subleading power. An operator with an arbitrary number of collinear gluons in the same sector with arbitrary helicity and color indices can be formed by simply multiplying the $\cB^a_{i\pm}$ building blocks with the same collinear sector index $i$, such as $\cB^a_{i+}\cB^b_{i+}$.  On the other hand, for a quark-antiquark pair in the same collinear sector, the bilinear current building blocks of \eq{jpm_def} are not suitable. Indeed, the SCET projection relations
\begin{align} \label{eq:proj}
\frac{\Sl n_i \Sl {\bar n}_i}{4}  \chi_{n_i}=\chi_{n_i}, \qquad \Sl n_i \chi_{n_i}=0 
\,,
\end{align} 
enforce that the scalar current $\bar \chi_{n_i} \chi_{n_i}=0$, vanishes, as do the plus and minus helicity components of the vector current $\bar \chi_{n_i} \gamma_\perp^{\pm} \chi_{n_i}=0$.
In other words, the SCET projection relations enforce that a quark-antiquark pair in the same sector must have zero helicity if they are of the same chirality. Similarly, a quark-antiquark pair in the same sector with opposite chirality must have helicity $\pm 1$.
We therefore define the helicity currents 
\begin{align}\label{eq:coll_subl}
& h=0:
& J_{i0}^{\balpha \beta} 
&= \frac{1}{2 \sqrt{\vphantom{2} \omega_{\bar \chi} \, \omega_\chi}}
\: \bar \chi^\balpha_{i+}\, \Sl{\bar n}_i\, \chi^\beta_{i+}
\,,\qquad
J_{i\bar 0}^{\balpha \beta} 
= \frac{1}{2 \sqrt{\vphantom{2} \omega_{\bar \chi} \, \omega_\chi}}
\: \bar \chi^\balpha_{i-}\, \Sl {\bar n}_i\, \chi^\beta_{i-}
\,, \\[5pt]
& h=\pm 1:
& J_{i\pm}^{\balpha \beta}
&= \mp  \sqrt{\frac{2}{ \omega_{\bar \chi} \, \omega_\chi}}  \frac{\epsilon_{\mp}^{\mu}(n_i,\bar n_i)}{ \big(\l n_i \mp | \bar{n}_i \pm \r \big)^2}\: 
\bar \chi_{i\pm}^\balpha\, \gamma_\mu \Sl{\bar n}_i\, \chi_{i\mp}^\beta
\,. \nn
\end{align}
Because of the SCET projection relations of \Eq{eq:proj}, this set of currents, when combined with those of \Eq{eq:jpm_back_to_bacjdef} provides a complete set of building blocks for constructing hard scattering operators involving collinear fermions at all powers in the SCET expansion. Hard scattering operators involving arbitrary numbers of collinear quarks in different sectors, with arbitrary helicity and color indices, can be formed from products of these building blocks. The  $J_{i0}^{\balpha \beta}$ and $J_{i\bar 0}^{\balpha \beta}$ transform together as a scalar/pseudoscalar under rotations about the $\hat n_i$ axis, i.e. have helicity $h=0$. Similarly, the operators $J_{i\pm}^{\balpha \beta}$ have helicity $h=\pm 1$.  These four currents with quarks in the same collinear direction are shown in the second category in \Tab{tab:helicityBB}. These currents are again RPI-III invariant and our choice of prefactors is made to simplify their Feynman rules. The Feynman rules are simple to obtain, but we do not give them explicitly here. The Feynman rules for all currents in \SCETi and \SCETii will be given in~\cite{subhel:long}.  

Subleading power operators can also involve explicit insertions of the $\cP_{i\perp}^\mu$ operator.  Since the $\cP_{i\perp}^\mu$ operator acts on the perpendicular subspace defined by the vectors $n_i, \bar n_i$, which is spanned by the polarization vectors $\epsilon^{\pm}(n_i, \bar n_i)$, it naturally decomposes as 
\begin{align} \label{eq:Pperppm}
\cP_{i+}^{\perp}(n_i,\bar n_i)=-\epsilon^-(n_i,\bar n_i) \cdot \cP_{i\perp}\,, \qquad \cP_{i-}^{\perp}(n_i,\bar n_i)=-\epsilon^+(n_i,\bar n_i) \cdot \cP_{i\perp}\,.
\end{align} 
This decomposition is performed for the $\cP_{i\perp}$ operator in each sector. 
As we mentioned earlier, power counting ensures that the sector on which $\cP_{i\perp}$ acts is unambiguous. Hence we can simply drop the subscript $i$ and use $\cP^\perp_\pm$ as building blocks, as shown in \Tab{tab:helicityBB}. 

To see how this decomposition applies to operators written in more familiar notation, we consider the example operator $ \cP_\perp \cdot \cB_{i\perp}$. Using the completeness relation
\begin{align}\label{eq:completeness}
\sum\limits_{\lambda=\pm} \epsilon^\lambda_\mu(n_i,\bar n_i) \big[ \epsilon^\lambda_\nu(n_i,\bar n_i)  \big]^* = - g^\perp_{\mu \nu} ( n_i, \bn_i)\,,
\end{align}
the decomposition into our basis is given by
\begin{align}\label{eq:pdotbexample}
\cP_\perp \cdot \cB_{i\perp}=-\cP^{\perp}_{+} \cB_{i-}-\cP^{\perp}_{-} \cB_{i+}\,.
\end{align}
When acting within an operator containing multiple fields, square brackets are used to denote which fields are acted upon by the $\cP^{\perp}_{\pm}$ operator. For example
$\cB_{i+} \left [ \cP^{\perp}_{+}  \cB_{i-}  \right]  \cB_{i-}$,
indicates that the $\cP^{\perp}_{+}$ operator acts only on the middle field. Note that $\cP^\perp_\pm$ carry helicity $h=\pm 1$, and that the products in \eq{pdotbexample} behave like scalars.

To denote insertions of the $\cP^{\perp}_{\pm}$ operator into the currents of \Eq{eq:coll_subl} we establish a notation where the $\cP^{\perp}_{\pm}$ operator acts on only one of the two quark building block fields, by writing it either to the left or right of the current, and enclosing it in curly brackets. For example,
\begin{align}\label{eq:p_perp_notation}
\big\{ \cP^{\perp}_\lambda J_{i 0 }^{\balpha \beta} \big\}  
& = \frac{1}{2 \sqrt{\vphantom{2}\omega_{\bar \chi} \, \omega_\chi }} \:
\Big[  \cP^{\perp}_{\lambda}  \bar \chi^\balpha_{i +}\Big] \Sl {\bar n}_i \chi^\beta_{i+}
\,, \\
\big\{ J_{i0 }^{\balpha \beta} (\cP^{\perp}_{\lambda})^\dagger \big\}
&=  \frac{1}{2\sqrt{\vphantom{2}\omega_{\bar \chi} \, \omega_\chi}} \:
\bar \chi^\balpha_{i+} \Sl {\bar n}_i \Big[   \chi^\beta_{i+} (\cP^{\perp}_{\lambda})^\dagger \Big]
\,. \nn
\end{align}
If we wish to instead indicate a $\cP^\perp_\pm$ operator that acts on both building blocks in a current then we use the notation $\big[\cP^\perp_\lambda J_{i 0 }^{\balpha \beta}\big]$. The extension to multiple insertions of the $\cP^\perp_\pm$ operators should be clear.   Since the $\cP^\perp_\pm$ operators commute with ultrasoft Wilson lines, they do not modify the construction of the color bases either before or after the BPS field redefinition.

The operators defined in \Eq{eq:jpm_def}, \Eq{eq:coll_subl}, and \Eq{eq:Pperppm} form a complete basis of building blocks from which to construct hard scattering operators involving only collinear fields. As with the leading power operators, each of these subleading power operators is collinear gauge invariant, and therefore the treatment of color degrees of freedom proceeds as in \Eq{eq:Cpm_Opm_color}. Subleading hard scattering operators appearing in the ${\cal L}_{\rm hard}$ part of the SCET Lagrangian of \Eq{eq:SCETLagExpand} can be constructed simply by taking products of the scalar building blocks. Examples demonstrating the ease of this approach will be given in \Sec{sec:example}.

We now consider the remaining building blocks listed in \Tab{tab:helicityBB}, which all involve ultrasoft gluon fields, ultrasoft quark fields or the ultrasoft derivative operator $\partial_{us}$.  The simplicity of the collinear building blocks does not trivially extend to ultrasoft fields, since prior to the BPS field redefinition all collinear and ultrasoft objects transform under ultrasoft gauge transformations. This implies that constraints from ultrasoft gauge invariance must be imposed when forming an operator basis, and that the color organization of \Sec{sec:scethelicity} cannot be trivially applied to operators involving ultrasoft fields.  To overcome this issue, we can work with the hard scattering operators after performing the BPS field redefinition of \Eq{eq:BPSfieldredefinition}. The BPS field redefinition introduces ultrasoft Wilson lines, in different representations $r$, $Y^{(r)}_n(x)$, into the hard scattering operators. These Wilson lines can be arranged with the ultrasoft fields to define ultrasoft gauge invariant building blocks. The Wilson lines which remain after this procedure can be absorbed into the generalized color structure, $\vT_{\BPS}$, as was done at leading power in \Eq{eq:Opm_BPScolor}.

We begin by defining a gauge invariant ultrasoft quark field
\begin{align} \label{eq:usgaugeinvdef}
\psi_{us(i)}=Y^\dagger_{n_i} q_{us}\,,
\end{align}
where the direction of the Wilson line $n_i$ is a label for a collinear sector. Since the ultrasoft quarks themselves are not naturally associated with an external label direction, $n_i$ can be chosen arbitrarily, though there is often a convenient or obvious choice. This choice does not affect the result, but modifies the structure of the Wilson lines appearing in the hard scattering operators at intermediate stages of the calculation.  We also perform the following decomposition of the gauge covariant derivative in an arbitrary representation, $r$,
\begin{align}\label{eq:soft_gluon}
Y^{(r)\,\dagger}_{n_i} i D^{(r)\,\mu}_{us} Y^{(r)}_{n_i }=i \partial^\mu_{us} + [Y_{n_i}^{(r)\,\dagger} i D^{(r)\,\mu}_{us} Y^{(r)}_{n_i}]=i\partial^\mu_{us}+T_{(r)}^{a} g \cB^{a\mu}_{us(i)}\,,
\end{align}
where we have defined the ultrasoft gauge invariant gluon field by
\begin{align} \label{eq:softgluondef}
g \cB^{a\mu}_{us(i)}= \left [   \frac{1}{in_i\cdot \partial_{us}} n_{i\nu} i G_{us}^{b\nu \mu} \cY^{ba}_{n_i}  \right] \,.
\end{align}
In the above equations the derivatives act only within the square brackets. Again, the choice of collinear sector label $n_i$ here is arbitrary. This is the ultrasoft analogue of the gauge invariant collinear gluon field, which can be written in the similar form
\begin{align}  
g\cB_{n_i\perp}^{A\mu} =\left [ \frac{1}{\bar \cP}    \bar n_{i\nu} i G_{n_i}^{B\nu \mu \perp} \cW^{BA}_{n_i}         \right]\,.
\end{align}
From the expression for the gauge invariant ultrasoft quark and gluon fields of \Eqs{eq:usgaugeinvdef}{eq:softgluondef} we see that unlike the ultrasoft fields, the operator $\cB^{A\mu}_{us(i)}$ is non-local at the scale $\lambda^2$, and depends on the choice of a collinear direction $n_i$. However the non-locality in our construction is entirely determined by the BPS field redefinition, and we can not simply insert arbitrary powers of dimensionless Wilson line products like $(Y_{n_1}^\dagger Y_{n_2})^k$ into the hard scattering operators. In practice this means that we can simply pick some $n_i$ for the Wilson lines in the building blocks in \eqs{usgaugeinvdef}{softgluondef} and then the BPS field redefinition determines the unique structure of remaining ultrasoft Wilson lines that are grouped with the color structure into  $\vT_{\BPS}^{a \al\bbeta}$. Determining a complete basis of color structures is straightforward. Detailed examples will be given in~\cite{subhel:long}, where the hard scattering operators for $e^+e^-\to$ dijets  involving ultrasoft fields will be constructed.

With the ultrasoft gauge invariant operators defined, we can now introduce ultrasoft fields and currents of definite helicity, which follow the structure of their collinear counterparts. Note from \eq{softgluondef}, that  $n_i\cdot \cB^{a}_{us(i)}= 0$.   For the ultrasoft gluon helicity fields we define the three building blocks
\begin{equation} \label{eq:Bus}
\cB^a_{us(i)\pm} = -\ve_{\mp\mu}(n_i, \bn_i)\,\cB^{a\mu}_{us(i)},\qquad  \cB^a_{us(i)0} =\bar n_\mu  \cB^{a \mu}_{us(i)}   
\,.\end{equation}
This differs from the situation for the collinear gluon building block in \eq{cBpm_def}, where only two building block fields were required, corresponding to the two physical helicities. For the ultrasoft gauge invariant gluon field we use three building block fields to describe the two physical degrees of freedom because the ultrasoft gluons are not fundamentally associated with any direction. Without making a further gauge choice, their polarization vectors do not lie in the perpendicular space of any fixed external reference vector.  If we use the ultrasoft gauge freedom to choose $\cB^a_{us(j)0}=0$, then we will still have $\cB^a_{us(i)0} \ne 0$ and $\cB^a_{us(i)\pm}\ne 0$ for $i\ne j$. We could instead remove $\cB^a_{us(j)0}$ for every $j$ using the ultrasoft gluon equation of motion, in a manner analogous to how $[W_{n_j}^\dagger i n_j\cdot D_{n_j} W_{n_j}]$ is removed for the collinear building blocks. However this would come at the expense of allowing inverse ultrasoft derivatives, $1/(in_j \cdot\partial_{us})$, to appear explicitly when building operators. While in the collinear case the analogous $1/\bnP$ factors are ${\cal O}(\lambda^0)$ and can be absorbed into the Wilson coefficients, this absorption would not be not possible for the ultrasoft case. Therefore, for our \SCETi construction we choose to forbid explicit inverse ultrasoft derivatives that can not be moved into Wilson lines, and allow $\cB^a_{us(i)0}$ to appear. An example of a case where the non-locality can be absorbed is given in \eq{softgluondef}, where the $1/(in\cdot\partial_{us})$ is absorbed into ultrasoft Wilson lines according to \eq{soft_gluon}. Thus the only ultrasoft non-locality that appears in the basis is connected to the BPS field redefinition.

We also decompose the ultrasoft partial derivative operator $\partial_{us}^\mu$ into lightcone components,
\begin{equation}  \label{eq:partialus}
\partial_{us(i)\pm} = -\ve_{\mp\mu}(n_i, \bn_i)\,\partial^{\mu}_{us},\qquad   \partial_{us(i)0} =\bar n_{i\mu} \partial^{\mu}_{us}, \qquad \partial_{us(i)\bar 0} = n_{i \mu} \partial^{\mu}_{us}
\,.\end{equation}
In contrast with the collinear case, we cannot always eliminate the $n_i \cdot \partial_{us}$ using the equations of motion without introducing inverse ultrasoft derivatives (e.g. $1/(\bn_i \cdot \partial_{us})$) that are unconnected to ultrasoft Wilson lines. When inserting ultrasoft derivatives into operators we will use the same curly bracket notation defined for the $\cP_\perp$ operators in \Eq{eq:p_perp_notation}. In other words, $\{i \partial_{us(i)\lambda} J\}$ indicates that the ultrasoft derivative acts from the left on the first field in $J$ and $\{  J (i\partial_{us(i)\lambda})^\dagger\}$ indicates that it acts from the right on the second field in $J$.

Gauge invariant ultrasoft quark fields also appear explicitly in the operator basis at subleading powers. Due to fermion number conservation they are conveniently organized into scalar currents. From \tab{PC}, we see that ultrasoft quark fields power count like $\lambda^3$. However, for factorization theorems involving a single collinear sector, as arise when describing a variety of inclusive and exclusive $B$ decays (see e.g.  ~\cite{Manohar:1993qn,Beneke:2000ry,Beneke:2000wa,Bauer:2000yr,Beneke:2001at,Bauer:2001ct,Bauer:2001cu,Bosch:2001gv,Bauer:2001yt,Bauer:2002aj,Beneke:2003pa,Beneke:2004dp,Arnesen:2005ez,Lee:2005pwa,Lee:2006gs}), operators involving ultrasoft quarks appear at leading power. The currents involving both collinear and ultrasoft quarks that are necessary to define subleading power operators at any desired order are 
\begin{align} \label{eq:Jus}
J_{i(us)\pm}^{\balpha\beta}
&= \mp  \:
\frac{\ve_\mp^\mu(n_i, \bar n_i)}{ \l \bn_i \mp | n_i \pm \r}\: 
\bar{\chi}^\balpha_{i\pm}\,  \gamma_\mu \psi^\beta_{us(i)\pm}
\,,  \\
J_{i(\overline{us})\pm}^{\balpha\beta}
&=\mp  
\frac{\ve_\mp^\mu(\bar{n}_i, n_i)}{ \l n_i \mp | \bar{n}_i \pm \r}\: \bar{\psi}_{us(i) \pm}^\balpha\, \gamma_\mu \chi^\beta_{i\pm} 
\, , \nn \\
J_{i(us)0}^{\balpha\beta}
&=   \bar \chi^\balpha_{i+}\psi^\beta_{us(i)-}
\,, \qquad\qquad\qquad
(J^\dagger)_{i(us)0}^{\balpha\beta}
=   \bar \psi^\balpha_{us(i)-} \chi^\beta_{i+}
\,, \nn\\
J_{i(\overline{us})0}^{\balpha\beta}
&=  \bar\psi^\balpha_{us(i)+} \chi^\beta_{i-}
\,, \qquad\qquad\qquad
(J^\dagger)_{i(\overline{us})0}^{\balpha\beta}
= \bar \chi^\balpha_{i-}\psi^\beta_{us(i)+}
\,, \nn
\end{align}
For these mixed collinear-ultrasoft currents we choose to use the collinear sector label $i$ in order to specify the ultrasoft quark building block field. In addition, we need currents that are purely built from ultrasoft fields,
\begin{align}  \label{eq:Jus2}
J_{(us)^2 ij\pm}^{\balpha\beta}
&= \mp\, \frac{\ve_\mp^\mu(n_i, n_j)}{\langle n_j\mp | n_i\pm\rangle}\: 
\bar{\psi}^\balpha_{us(i)\pm} \gamma_\mu \psi^\beta_{us(j)\pm} 
\,, \\
J_{(us)^2 ij 0}^{\balpha\beta}
&=
\bar \psi^\balpha_{us(i)+}\psi^\beta_{us(j)-}
\,,\qquad\qquad\quad
(J^\dagger)_{(us)^2 ij 0}^{\balpha\beta}
=
\bar \psi^\balpha_{us(i)-}\psi^\beta_{us(j)+}
\,. \nn
\end{align}
To specify the building blocks in these ultrasoft-ultrasoft currents we use two generic choices, $i$ and $j$, with $n_i \neq n_j$ so as to make the polarization vector well defined. Although the ultrasoft quark carries these labels, they are only associated with the Wilson line structure and, for example, the ultrasoft quark building block fields do not satisfy the projection relations of \eq{proj}.  

The ultrasoft currents in \eq{Jus} complete our construction of the complete set of scalar building blocks given in \Tab{tab:helicityBB}.  The objects in this table can be used to construct bases of hard scattering operators at any order in the power counting parameter $\lambda$, by simply taking products of the scalar building blocks.


There are several extensions to this construction that should be considered. One is the extension to \SCETii with collinear and soft fields, rather than collinear and ultrasoft fields. A table of scalar building block operators for \SCETii that is analogous to \Tab{tab:helicityBB} will be given in~\cite{subhel:long}.  Also, the completeness of the set of helicity building blocks relies on massless quarks and gluons having two helicities, which is specific to $d=4$ dimensions. Depending on the regularization scheme, this may or may not be true when dimensional regularization with $d=4-2\epsilon$ dimensions is used, and evanescent operators \cite{Buras:1989xd,Dugan:1990df,Herrlich:1994kh}, beyond those given in \Tab{tab:helicityBB} can appear. While evanescent operators are not required at leading power, (see~\cite{Moult:2015aoa} for a detailed discussion), this need no longer be the case at subleading power, and will be discussed further in~\cite{subhel:long}.

\section{Constraints from Angular Momentum Conservation}\label{sec:ang_cons}

If we include the spin of objects that are not strongly interacting, such as electrons and photons, then the overall hard scattering operators in \eq{Leff_sub_explicit} are scalars under the Lorentz group. In this section we will show that this constraint on the total angular momentum gives restrictions on the angular momentum that is allowed in individual collinear sectors. These restrictions become nontrivial beyond leading power, when multiple operators appear in the same collinear sector.

If we consider a leading power hard scattering process where two gluons collide to produce two well separated quark jets plus an $e^+e^-$ pair, then this is described by a leading power operator with each field sitting alone in a well separated collinear direction, such as
\begin{align} \label{eq:BBJJee}
\cB_{1\lambda_1}^a \cB_{2\lambda_2}^b J_{34\lambda_q}^{\balpha\beta} J_{e56\lambda_e} \,.
\end{align}
Here, the leading power electron current is defined in a similar way as the quark current, but without gluon Wilson lines,
\begin{equation}
J_{e \pm } \equiv 
J_{e ij\pm }
= \mp \sqrt{\frac{2}{\omega_i\, \omega_j}}\, \ve_\mp^\mu(n_i, n_j)\,    \frac{\bar{e}_{i\pm} \gamma_\mu e_{j \pm}} {\langle n_j\mp | n_i\pm\rangle}
\,.
\end{equation}
For notational convenience we will drop the explicit $ij$ label on the electron current, denoting it simply by $J_{e\pm}$. Although the operator in \eq{BBJJee} has to be a scalar, there are still no constraints on the individual values of the $\lambda_i$. Each building block has spin components that are defined with respect to a distinct axis $\hat n_i$, and yields a linear combination of spin components when projected onto a different axis. Thus, projecting all helicities onto a common axis we only find the trivial constraint that the angular momenta factors of $1$ or $1/2$ from each sector must together add to zero.\footnote{There are of course simple examples where this constraint reduces the basis of operators. For example, for gluon fusion Higgs production, angular momentum conservation implies that only two operators are required in the basis
	\begin{align}
	O_{++}^{ab}
	= \frac{1}{2}\, \cB_{1+}^a\, \cB_{2+}^b\,  H_3
	\,, \qquad
	O_{--}^{ab}
	= \frac{1}{2}\, \cB_{1-}^a\, \cB_{2-}^b\, H_3\nn
	\,,\end{align}
	where $H_3$ is the scalar Higgs field.
}
In the example of \eq{BBJJee}, this is $1\oplus 1\oplus\frac12\oplus\frac12\oplus\frac12\oplus\frac12 =0$ for a generic kinematic configuration.\footnote{If we were in a frame where the gluons were back-to-back, there spins would be combined along a single axis. In this example, this would still not give us any additional restrictions.} Note that for the quark and electron currents here, we have individual spin-$1/2$ fermions in different directions, so $\lambda_q$ and $\lambda_e$ do not correspond to helicities. As another example, consider $4$-gluon scattering, with all gluon momenta well separated and thus in their own collinear sectors, we have the operators 
\begin{align}
\cB_{1\lambda_1}^{a_1}\cB_{2\lambda_2}^{a_2}\cB_{3\lambda_3}^{a_3}\cB_{4\lambda_4}^{a_4}\,. 
\end{align}
Here we can again specify the helicities $\lambda_i=\pm$ independently, because each of these helicities is specified about a different quantization axis.  Each carries helicity $h=\pm1$, and angular momentum is conserved because these four spin-$1$ objects can add to spin-$0$.  Therefore all helicity combinations must be included.

To understand the constraints imposed by angular momentum conservation at subleading power, it is interesting to consider a specific example in more detail. As a simple example, consider an $e^+e^-$ collision in the center of mass frame producing two back-to-back jets, where we label the associated jet directions as $n$ and $\bar n$. The leading power operators are
\begin{align}\label{eq:LP_basis_angsec}
O_{(+;+)}^{(0)\balpha\bt}
=J^{\balpha\bt}_{n \bar n+}J_{e +}\,, \qquad
O_{(+;-)}^{(0)\balpha\bt}
=J^{\balpha\bt}_{n \bar n+}J_{e -}\,,  \\
O_{(-;+)}^{(0)\balpha\bt}
=J^{\balpha\bt}_{n \bar n-}J_{e +}\,, \qquad
O_{(-;-)}^{(0)\balpha\bt}
=J^{\balpha\bt}_{n \bar n-}J_{e -}\,,\nn
\end{align}
where $J_{n\bn\pm}^{\balpha\bt}$ were defined in \eq{jpm_back_to_bacjdef}.
Here, we can view $J^{\balpha\bt}_{n \bar n \pm}$ as creating or destroying a state of helicity $h=\pm1$ about the $n$ axis, and $J_{e\pm}$ as creating or destroying a state of helicity $h=\pm 1$ about the electron beam axis. Defining $\theta$ as the angle between the quark and electron and taking all of the particles to be outgoing, the spin projection implies that the Wilson coefficients are proportional to the Wigner $d$ functions,
\begin{alignat}{2}
&C_{(+;+)}^{(0)\balpha\bt}
\propto 1+\cos \theta\,, \qquad & &
C_{(+;-)}^{(0)\balpha\bt}
\propto 1-\cos \theta   \,,  \\
&C_{(-;+)}^{(0)\balpha\bt}
\propto 1-\cos \theta  \,, \qquad &&
C_{(-;-)}^{(0)\balpha\bt}
\propto 1+\cos \theta \,.\nn
\end{alignat}
As expected, all helicity combinations are non-vanishing (except when evaluated at special kinematic configurations).

\begin{figure}[t!]
	\begin{center}
		\includegraphics[width=0.35\columnwidth]{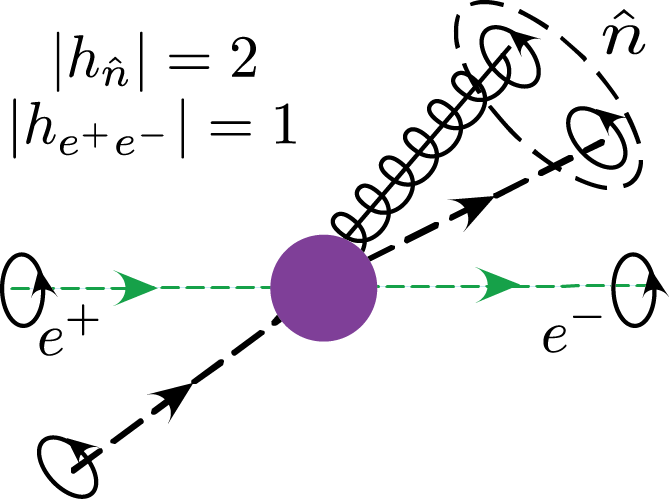} 
		\hspace{1.4cm}
		\includegraphics[width=0.35\columnwidth]{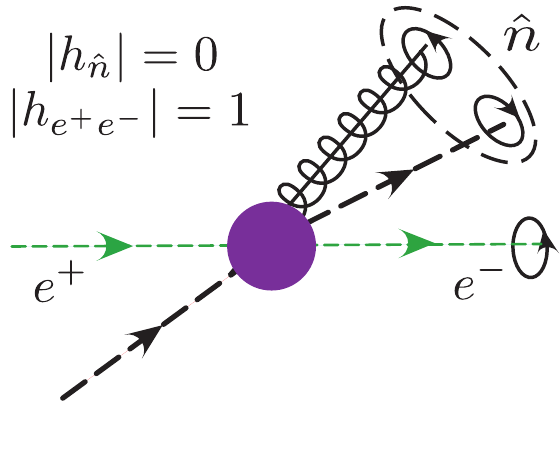} 
		\raisebox{0cm}{ \hspace{-3.0cm} 
			$a$)\hspace{6.6cm}
			$b$)\hspace{2.4cm} }
		\\[-25pt]
	\end{center}
	\vspace{-0.4cm}
	\caption[Illustration of helicity selection rule with two axes]{ 
		A schematic illustration of the helicity selection rule with two axes, as relevant for the case of $e^+e^-\to $ dijets. In a) the $n$-collinear sector carries $|h|=2$, and therefore has a vanishing projection onto the $J_{e\pm}$ current. In b), the collinear sector carries $|h|=0$ and has a non-vanishing projection onto the $J_{e\pm}$ current. } 
	\label{fig:helicity_constraint}
\end{figure}

Considering this same example at subleading power, the analysis of angular momentum becomes more interesting, since multiple fields are present in a single collinear sector. For the subleading $e^+e^-\to $ dijet operators with only $n$-collinear and $\bn$-collinear fields, we only have a single axis $\hat n$ for all strongly interacting operators, and can simply add up their helicities to determine the helicity $h_{\hat n}$ in this direction.  Since the operator in the only other direction, $J_{e\pm }$, has spin-1, this implies that the total helicity for the $n$-$\bar n$ sector must be $h_{\hat n}=0,1,-1$ for the operator to have a non-vanishing contribution. Any operator with $|h_{\hat n}|>1$ must belong to a representation of spin $J>1$, and is ruled out because we can not form a scalar when combining it with the spin-$1$ electron current. An example of this is shown in \Fig{fig:helicity_constraint}.

As an explicit example of the constraints that this places on the subleading power helicity operators, consider the $\mathcal{O}(\lambda)$ back-to-back collinear operators  involving two collinear quark fields and a single collinear gluon field, which appears at $\cO(\lambda)$. For the case that the quarks are in different collinear sectors we can start by considering the operator list
\begin{alignat}{2} \label{eq:Z1_basis_cons}
&O_{+(+;\pm)}^{(1)a\,\balpha\bt}
=\cB_{n+}^a \, J_{n\bar n\,+}^{\balpha\bt}\,  J_{e\pm }
\,,\qquad &
&O_{+(-;\pm)}^{(1)a\,\balpha\bt}
= \cB_{n+}^a\, J_{n\bar n\,-}^{\balpha\bt}\, J_{e\pm }
\,,  \\
&O_{-(+;\pm)}^{(1)a\,\balpha\bt}
= \cB_{n-}^a\, J_{n\bar n\,+}^{\balpha\bt}\, J_{e\pm }
\,,\qquad &
&O_{-(-;\pm)}^{(1)a\,\balpha\bt}
= \cB_{n-}^a \, J_{n\bar n\, -}^{\balpha\bt}\,J_{e\pm }
\,, \nn
\end{alignat}
while for the case that the quarks are in the same collinear sector we consider
\begin{alignat}{2} \label{eq:Z1_basis_diff_cons}
&O_{\bar{n}+(0;\pm)}^{(1)a\,\balpha\bt}
= \cB_{n+}^a\, J_{\bar n0\, }^{\balpha\bt}\,J_{e\pm }
\,,\qquad &
&O_{\bar{n}+(\bar 0;\pm)}^{(1)a\,\balpha\bt}
=  \cB_{n+}^a\, J_{\bar n\bar 0\, }^{\balpha\bt}\,J_{e\pm }
\,,\\
&O_{\bar{n}-(0;\pm)}^{(1)a\,\balpha\bt}
= \cB_{n-}^a\,  J_{\bar n0\, }^{\balpha\bt}\,J_{e\pm }
\,,\qquad &
&O_{\bar{n}-(\bar 0;\pm)}^{(1)a\,\balpha\bt}
=\cB_{n-}^a\, J_{\bar n\bar 0\, }^{\balpha\bt} \,J_{e\pm }
\,.\nn
\end{alignat}
We have used the fact that chirality is conserved in massless QCD, eliminating the need to consider $J_{n\bn 0}^{\balpha\bt}$ or $J_{\bn \pm}^{\balpha\bt}$ for the process being considered here. There are also operators with $\cB_{\bn\pm}^a$ that are obtained from those in \eqs{Z1_basis_cons}{Z1_basis_diff_cons} by taking $n\leftrightarrow \bn$. Furthermore, we do not consider the color structure, as it is irrelevant for the current discussion.  (Also note that we are not attempting to enumerate all ${\cal O}(\lambda)$ operators here. This is done in~\cite{subhel:long}.)

The constraint from conservation of angular momentum gives further restrictions, implying that only a subset of the eight operators in \eqs{Z1_basis_cons}{Z1_basis_diff_cons} are non-vanishing. In \eq{Z1_basis_cons} the strongly interacting operators have $h_{\hat n}=0$ or $h_{\hat n}=\pm2$,  and only those with $h_{\hat n}=0$ can contribute to the $J=0$ hard scattering Lagrangian, leaving only
\begin{alignat}{2} \label{eq:Z1_basis_cons_actuallyconserved}
&O_{+(-;\pm)}^{(1)a\,\balpha\bt}
= \cB_{n+}^a\, J_{n\bar n\,-}^{\balpha\bt}\,  J_{e\pm }
\,,\qquad &
&O_{-(+;\pm)}^{(1)a\,\balpha\bt}
= \cB_{n-}^a\, J_{n\bar n\,+}^{\balpha\bt}\,J_{e\pm} \,.
\end{alignat}
Thus angular momentum reduces the number of hard scattering operators by a factor of two in this case. On the other hand, for the case with both quarks in the same collinear sector in \eq{Z1_basis_diff_cons}, the operators all have $h_{\hat n}=\pm1$, and therefore all of them are allowed. 

Having understood how the angular momentum conservation constraint appears in the helicity operator language, it is interesting to examine how it appears if we instead work with the traditional operators of \tab{PC}. 
Here we must construct the SCET currents $\mathcal{J}^\mu$ at $\mathcal{O}(\lambda)$ involving two collinear quarks and a collinear gluon. The Lorentz index on $\mathcal{J}^\mu$ is contracted with the leptonic tensor to give an overall scalar, and thus preserve angular momentum. The operators in a basis for $\mathcal{J}^\mu$ can be formed from Lorentz and Dirac structures, as well as the external vectors, $n^\mu$ and $\bar n^\mu$. When the collinear quarks are each in a distinct collinear sector, the SCET projection relations of \Eq{eq:proj} imply that $\bar \chi_{\bar n} {\Sl {n}} \chi_n =  \bar \chi_{\bar n} {\Sl {\bar n}} \chi_n =0$. To conserve chirality we must have a $\gamma^\perp_\nu$ between the quark building blocks, and this index must be contracted with the other free $\perp$-index, $\nu$, in the collinear gluon building block $\cB_{\perp n}^\nu$ (which we again choose to be in the $n$ direction). Therefore an $n$ or $\bar n$ must carry the $\mu$ Lorentz index. After the BPS field redefinition it can be shown\footnote{Note that in constructing a complete basis of Lorentz and Dirac structures for \eqs{dirac_op_1}{dirac_op_2}, that all other operators can be eliminated using symmetry properties and the conservation of the current,  $q_\mu \mathcal{J}^{(1)\mu}_i=0$. Eliminating operators here is tedious compared to the helicity operator approach.} that for photon exchange the unique $\mathcal{O}(\lambda)$ operator with collinear quark fields in distinct collinear sectors is 
\begin{align}\label{eq:dirac_op_1}
\mathcal{J}^{(1)\mu}_1 =r^\mu_-\bar \chi_{\bar n}Y^\dagger_{\bar n}Y_n \Sl\cB_{\perp n} \chi_{n}\,,
\end{align}
where, defining $q^\mu$ as the sum of the momenta of the colliding leptons, we have
\begin{align}
r^\mu_-= \frac{n \cdot q}{2} \bar n^\mu -\frac{\bar n \cdot q}{2} n^\mu\,.
\end{align}
In the case that both collinear quark fields are in the same collinear sector, similar arguments using the SCET projection relations can be used to show that the collinear gluon field must carry the Lorentz index, and that the unique operator is 
\begin{align} \label{eq:dirac_op_2}
\mathcal{J}^{(1)\mu}_2 =\bar \chi_{\bar n}Y^\dagger_{\bar n}Y_n\cB_{ \perp n}^{\mu}Y^\dagger_{n} Y_{\bar n}\Sl r_{- } \chi_{\bar n} \,.
\end{align}

We see a direct correspondence between \Eqs{eq:Z1_basis_diff_cons}{eq:dirac_op_2}. In both equations the collinear quark fields have $h=0$ and thus form a scalar, and the collinear gluon field carries the spin that is combined with the leptonic current. For photon exchange, all of the Wilson coefficients of the operators in \eq{Z1_basis_diff_cons} are related by CP properties and angular momentum constraints, so there is only one combination of the four operators that appears with a nontrivial Wilson coefficient. This combination maps exactly to the single operator in \eq{dirac_op_2}.  We also see a correspondence between \Eqs{eq:Z1_basis_cons_actuallyconserved}{eq:dirac_op_1}, where both collinear quarks are contracted with the collinear gluon to form a $h=0$ combination. Indeed, using the completeness relation of \Eq{eq:completeness} for $g^\perp_{\mu \nu} ( n_i, \bn_i)$, the operators of \Eqs{eq:dirac_op_1}{eq:dirac_op_2} can straightforwardly be converted to the helicity operators of \Eqs{eq:Z1_basis_diff_cons}{eq:Z1_basis_cons_actuallyconserved}. 

It is interesting to note that when working in terms of building blocks involving Lorentz and Dirac structures, the SCET projection relations, which were ultimately what allowed us to define helicity fields along given axes, played a central role in reducing the basis. One is also forced to incorporate the constraints from the total angular momentum as part of the analysis, by the need to keep track of the contraction of Lorentz indices. In the helicity operator basis the same constraints appear as simple elimination rules on the allowed helicities when taking products of building blocks in the same collinear sector (and any back-to-back sector if one is present). These products can be classified by the minimal total angular momentum object for which they are a component, and eliminated if this value is too large.

We can now specify the general constraint from angular momentum on the helicities of an operator basis. The operator basis must be formed such that $J^{(i)}_{\rm min}$, the minimal angular momentum carried by the $n_i$-collinear sector,  satisfies 
\begin{align} \label{eq:Jmin}
J^{(i)}_\text{min}\leq \sum_{j \text{ with }  \hat n_j\neq \hat n_i} J_{\rm min}^{(j)} \,.
\end{align}
If the helicities in the $n_i$-collinear sector of some operator add up to $h_{n_i}^{\rm tot}$, then the minimum angular momentum for that sector is $J^{(i)}_\text{min}=|h_{n_i}^{\rm tot}|$. Therefore we can write \eq{Jmin} in a form that is useful for constraining the helicity of operators,
\begin{align} \label{eq:hmin}
|h_{n_i}^\text{tot}| \leq \sum_{j \text{ with }  \hat n_j\neq \hat n_i}  | h_{n_j}^\text{tot}|  \,.
\end{align}
In cases where two of our light-like vectors are back-to-back, $n_i\cdot n_{k} =2 +{\cal O}(\lambda^2)$, then the operators in both the $n_i$ and $n_k$ collinear directions are considered simultaneously when calculating the value of $h_{n_i}^\text{tot}$ (where $\pm$ for $n_k$ count as $\mp$ for $n_i$), and not as distinct terms in the sum. This includes the case where $n_k=\bn_i$.  \eq{hmin} prevents subleading power operators from having exceedingly large angular momenta about any particular collinear direction.

This constraint of angular momentum conservation of the hard scattering process shows that when writing down a basis of helicity operators, not all helicity combinations should be included in the basis. Especially when working at higher powers, this places considerable constraints on the basis, and supplements additional constraints from parity and charge conjugation invariance (see~\cite{Moult:2015aoa}). This reduction can be contrasted with the leading power operators explored in~\cite{Moult:2015aoa}, where most often all possible different helicity combinations had to be included in the basis of hard scattering operators.

\section{Example: $q\bar qgg$ Operators for $n$-$\bn$ Directions}\label{sec:example}

To demonstrate the simplicity of the helicity operator approach, in this section we  will explicitly construct a basis of hard scattering operators with two back-to-back collinear sectors, $n$ and $\bn$. For simplicity, we will restrict ourselves to the channel involving two collinear gluons, a collinear quark and a collinear antiquark.  The operators to be discussed in this section are suppressed by $\cO(\lambda^2)$ compared to the leading power operator, which involves a quark and antiquark field in opposite collinear sectors, and contribute at subleading power to $e^+ e^- \to$ dijet event shapes, Drell-Yan, or DIS with one jet. They do not in themselves constitute a complete basis of ${\cal O}(\lambda^2)$ operators, but do make up a unique subset which we can use to illustrate the power of our approach. The complete ${\cal O}(\lambda^2)$ basis of operators will be presented and analyzed in~\cite{subhel:long}.

The angular momentum arguments of \sec{ang_cons} enforce that the helicity along the single jet axis satisfies $|h_{\hat n}^{\text{tot}}| \le 1$. Additionally, for the particular process $e^+e^-\to$ dijets the quark and antiquark have the same chirality, which provides further restrictions on the allowed operators that we will enumerate below. Using the notation of \eq{Cpm_Opm_color} we write the three-dimensional color basis for the $q\bar q gg$ channels as
\begin{equation} \label{eq:ggqqll_color}
\vT^{\, ab \alpha\bbeta}
= \Bigl(
(T^a T^b)_{\alpha\bbeta}\,,\, (T^b T^a)_{\alpha\bbeta} \,,\, \tr[T^a T^b]\, \delta_{\alpha\bbeta}
\Bigr)
\,.\end{equation} 
The color basis after BPS field redefinition will be given separately for each distinct partonic configuration, each of which will be discussed in turn.

We begin by considering operators where the quark and antiquark fields have distinct collinear sector labels, and the gluon fields are in the same collinear sector. In this case, a basis of helicity operators is
\begin{align}
\boldsymbol{(gg q)_n (\bar q)_{\bn}:}     {\vcenter{\includegraphics[width=0.18\columnwidth]{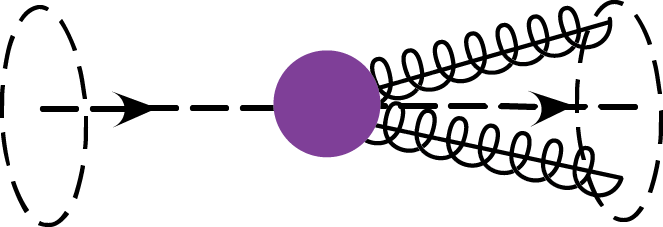}}}
\nn
\end{align}
\vspace{-0.4cm}
\begin{alignat}{2} \label{eq:eeqqgg_basis1}
&O_{\cB1++(-;\pm)}^{(2)ab\, \balpha\bt}
= \frac{1}{2}   \cB_{n+}^a\, \cB_{n+}^b\, J_{n\bar n\, - }^{\balpha\bt}   J_{e\pm }
\, , \qquad 
&&O_{\cB1--(+;\pm)}^{(2)ab\, \balpha\bt}
= \frac{1}{2}  \cB_{ n-}^a\, \cB_{ n-}^b \, J_{n\bar n\, + }^{\balpha\bt}   J_{e\pm }
\, ,  \nn\\
&O_{\cB1+-(+;\pm)}^{(2)ab\, \balpha\bt}
=  \cB_{n+}^a\, \cB_{n-}^b  \, J_{n\bar n\, +}^{\balpha\bt}   J_{e\pm }
\, , \qquad 
&&O_{\cB1+-(-;\pm)}^{(2)ab\, \balpha\bt}
=  \, \cB_{n+}^a\, \cB_{ n-}^b J_{n\bar n\, -}^{\balpha\bt}    J_{e\pm }
\, , 
\end{alignat}

\begin{align}
\boldsymbol{(gg \bar q)_n (q)_{\bn}:}   {\vcenter{\includegraphics[width=0.18\columnwidth]{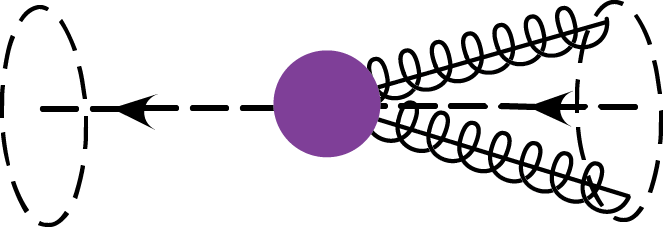}}}
\nn
\end{align} 
\vspace{-0.4cm} 
\begin{alignat}{2} \label{eq:eeqqgg_basis1a}
&O_{\cB2++(-;\pm)}^{(2)ab\, \balpha\bt}
= \frac{1}{2}   \cB_{n+}^a\, \cB_{n+}^b\, J_{\bar n n\, + }^{\balpha\bt}   J_{e\pm }
\, , \qquad 
&&O_{\cB2--(+;\pm)}^{(2)ab\, \balpha\bt}
= \frac{1}{2}  \cB_{ n-}^a\, \cB_{ n-}^b \, J_{\bar n n\, - }^{\balpha\bt}   J_{e\pm }
\, , \nn \\
&O_{\cB2+-(+;\pm)}^{(2)ab\, \balpha\bt}
=  \cB_{n+}^a\, \cB_{n-}^b  \, J_{\bar n n\, +}^{\balpha\bt}   J_{e\pm }
\, , \qquad 
&&O_{\cB2+-(-;\pm)}^{(2)ab\, \balpha\bt}
=  \, \cB_{n+}^a\, \cB_{ n-}^b J_{\bar n n\, -}^{\balpha\bt}    J_{e\pm }
\, . 
\end{alignat}
Here we have used constraints from angular momentum conservation to eliminate operators whose non-leptonic component do not have $h=0,\pm1$ along the $\hat n$ axis. For example, we have not allowed the operators $ \cB_{n+}^a\, \cB_{n+}^b\, J_{n\bar n\, + }^{\balpha\bt}   J_{e\pm }$ which have $h=+3$ along the $n$ axis and could not be created from the intermediate vector boson. Also, we have used the $n \leftrightarrow \bn$ symmetry to only write operators with both gluons in the $n$-collinear sector, a simplification that we will make repeatedly in this section. Operators with $\bn$-collinear gluons are obtained by simply taking $n \leftrightarrow \bn$. The color basis for the operators in \eqs{eeqqgg_basis1}{eeqqgg_basis1a} after the BPS field redefinition is
\begin{equation} \label{eq:BPS1}
\vT_{\BPS}^{\, ab \alpha\bbeta}
= \Bigl(
(T^a T^bY^\dagger_n Y_{\bar n})_{\alpha\bbeta}\,,\, (T^b T^aY^\dagger_n Y_{\bar n})_{\alpha\bbeta} \,,\, \tr[T^a T^b]\, [Y^\dagger_n Y_{\bar n}]_{\alpha\bbeta}
\Bigr)
\,.
\end{equation}
In order to see how this is derived, we will go through the algebra explicitly for the first color structure. Using the result for the transformations in \eq{BPSfieldredefinition}, we see that each gluon field from (\ref{eq:eeqqgg_basis1}) or (\ref{eq:eeqqgg_basis1a}) contributes an adjoint Wilson line while each fermion contributes a fundamental Wilson line. So, our color structure becomes
\begin{align} \label{eq:color-BPS-explicit}
(T^aT^b)_{\alpha \bbeta} &\to (Y_n^\dagger T^{a'} \cY_n^{a' a} T^{b'} \cY_n^{b' b} Y_{\bn})_{\alpha \bbeta} = (Y_n^\dagger Y_n T^{a} Y_n^{\dagger} Y_n T^{b} Y_n^{\dagger} Y_{\bn})_{\alpha \bbeta} \nn \\
&= (T^a T^bY^\dagger_n Y_{\bar n})_{\alpha\bbeta}\,,
\end{align}
where we have used $T^{a'} \cY_i^{a' a} = Y_i T^a Y^\dagger_i$. Similar manipulations give the other Wilson line structures in \eq{BPS1}.

Next we consider the operators where the quark and antiquark fields have distinct collinear sector labels, as do the gluons. In this case, the basis of helicity operators is
\begin{align}
\boldsymbol{(g q)_n (g \bar q)_{\bn}:}   {\vcenter{\includegraphics[width=0.18\columnwidth]{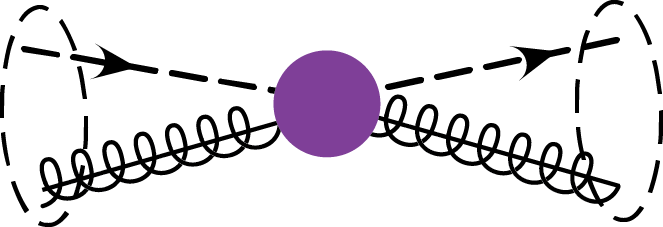}}}  \nn
\end{align}
\vspace{-0.45cm}
\begin{alignat}{2} \label{eq:eeqqgg_basis2}
&O_{\cB3++(+;\pm)}^{(2)ab\, \balpha\bt}
= \cB_{n+}^a\, \cB_{\bar n+}^b \, J_{n\bar n\, + }^{\balpha\bt}   J_{e\pm } \, , \qquad 
&&O_{\cB3--(-;\pm)}^{(2)ab\, \balpha\bt}
=    \cB_{n-}^a\, \cB_{\bar n-}^b \, J_{n\bar n\, -}^{\balpha\bt} J_{e\pm }\, ,  
\nn \\
&O_{\cB3++(-;\pm)}^{(2)ab\, \balpha\bt}
= \cB_{n+}^a\, \cB_{\bar n+}^b \, J_{n\bar n\, - }^{\balpha\bt}   J_{e\pm } \, , \qquad 
&&O_{\cB3--(+;\pm)}^{(2)ab\, \balpha\bt}
=    \cB_{n-}^a\, \cB_{\bar n-}^b \, J_{n\bar n\, +}^{\balpha\bt} J_{e\pm }\, , 
\\
&O_{\cB3+-(-;\pm)}^{(2)ab\, \balpha\bt}
=    \cB_{n+}^a\, \cB_{\bar n-}^b \, J_{n\bar n\, - }^{\balpha\bt}  J_{e\pm }\, , \qquad 
&&O_{\cB3-+(+;\pm)}^{(2)ab\, \balpha\bt}
=\cB_{n-}^a\, \cB_{\bar n+}^b  \, J_{n\bar n\, + }^{\balpha\bt} J_{e\pm }  \, , \nn
\end{alignat}
where we have used angular momentum to eliminate operators such as $\cB_{n+}^a\, \cB_{\bar n-}^b \, J_{n\bar n\, + }^{\balpha\bt}   J_{e\pm }$ and $\cB_{n-}^a\, \cB_{\bar n+}^b \, J_{n\bar n\, - }^{\balpha\bt}   J_{e\pm }$. Here the post-BPS color basis is given by
\begin{equation}
\vT_{\BPS}^{\, ab \alpha\bbeta}
= \Bigl(
(T^a  Y^\dagger_n Y_{\bar n} T^b)_{\alpha\bbeta}\,,\, (Y^\dagger_n T^d \cY_{\bn}^{db}  T^c \cY^{ca}_n Y_{\bar n})_{\alpha\bbeta} \,,\, \tr[ T^c \cY^{ca}_n T^d \cY_{\bn}^{db} ]\, [Y^\dagger_n Y_{\bar n}]_{\alpha\bbeta}
\Bigr)
.
\end{equation}
This is easily obtained following the steps described below \Eq{eq:BPS1}.

The next relevant case is when the gluons are in distinct collinear sectors and the quarks are in the same collinear sector. Here, the basis of helicity operators is
\begin{align}
\boldsymbol{(g q\bar q)_n (g)_{\bn}:}   {\vcenter{\includegraphics[width=0.18\columnwidth]{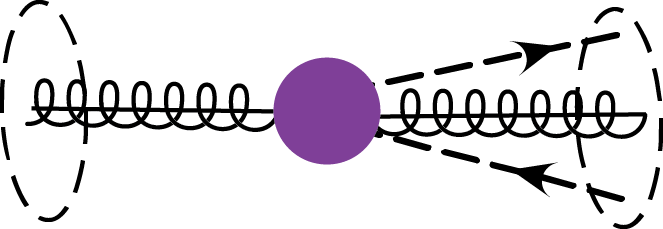}}} \nn
\end{align}
\vspace{-0.4cm}
\begin{alignat}{2} \label{eq:eeqqgg_basis3}
&O_{\cB4++(0:\pm)}^{(2)ab\, \balpha\bt}
=   \cB_{n+}^a\, \cB_{\bar n+}^b \, J_{n\,{0} }^{\balpha\bt}  J_{e\pm }\, , \qquad 
&&O_{\cB4++(\bar 0:\pm)}^{(2)ab\, \balpha\bt}
=\cB_{n+}^a\, \cB_{\bar n+}^b  \, J_{n\,{\bar 0} }^{\balpha\bt}   J_{e\pm }\, ,  \\
&O_{\cB4--(0:\pm)}^{(2)ab\, \balpha\bt}
=  \cB_{n-}^a\, \cB_{\bar n-}^b \, J_{n\,{0} }^{\balpha\bt}    J_{e\pm }\, , \qquad 
&&O_{\cB4--(\bar 0:\pm)}^{(2)ab\, \balpha\bt}
= \cB_{ n-}^a\, \cB_{\bar n-}^b  \, J_{n\,{\bar 0} }^{\balpha\bt}   J_{e\pm }\, . \nn
\end{alignat}
In writing \eq{eeqqgg_basis3} we have again used constraints of angular momentum conservation to restrict the allowed operators in the basis (e.g. we have eliminated $\cB_{n+}^a\, \cB_{\bar n-}^b \, J_{n\,{0} }^{\balpha\bt}  J_{e\pm }$). The color basis after BPS field redefinition in this case is
\begin{equation}
\vT_{\BPS}^{\, ab \alpha\bbeta}
= \Bigl(
(T^a Y^\dagger_n Y_{\bar n} T^b Y^\dagger_{\bar n} Y_n)_{\alpha\bbeta}\,,\, (Y^\dagger_n Y_{\bar n} T^bY^\dagger_{\bar n} Y_n T^a)_{\alpha\bbeta} \,,\, \tr[T^c \cY_{n}^{ca} T^d \cY_{\bn}^{db}]\, \delta_{\alpha\bbeta}
\Bigr)
\,.
\end{equation}

Finally, we consider the basis of operators with both quarks in the same collinear sector, and both gluons in the other collinear sector. Imposing angular momentum conservation reduces the basis from four to two distinct operators
\begin{align}
& \boldsymbol{(q \bar q)_n (gg)_{\bn}:}{\vcenter{\includegraphics[width=0.18\columnwidth]{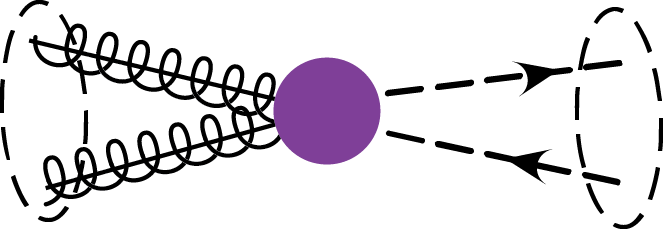}}}  \nn
\end{align}
\vspace{-0.4cm}
\begin{align}\label{eq:eeqqgg_basis4}
&O_{\cB5+-(0:\pm)}^{(2)ab\, \balpha\bt}
= \cB_{\bar n+}^a\, \cB_{ \bar n-}^b  \, J_{n\,{0} }^{\balpha\bt}   J_{e\pm }\,, \qquad 
&O_{\cB5+-(\bar 0:\pm)}^{(2)ab\, \balpha\bt}
= \cB_{\bar n+}^a\, \cB_{ \bar n-}^b  \, J_{n\,{\bar 0} }^{\balpha\bt}   J_{e\pm }\,.
\end{align}
Here, the color basis after BPS field redefinition is
\begin{equation}
\vT_{\BPS}^{\, ab \alpha\bbeta}
= \Bigl(
(Y_n^\dagger Y_\bn T^a T^b Y^\dagger_{\bar n} Y_n )_{\alpha\bbeta}
\,,\, 
(Y_n^\dagger Y_\bn T^b T^a Y^\dagger_{\bar n} Y_n )_{\alpha\bbeta}
\,,\, 
\tr[T^a T^b]\, \delta_{\alpha\bbeta} \Bigr)
\,.
\end{equation}

These operators, provide a complete basis of hard scattering operators with two back to back collinear sectors in the $q\bar qgg$ channel. This example illustrates several key aspects of using the subleading helicity operators: imposing the angular momentum constraints has helped reduce the number of distinct helicity labels that we must consider, the structure of the ultrasoft Wilson lines is determined by the BPS field redefinition and the enumeration of a complete basis is as simple as writing down all allowed helicity choices. The analysis of this channel only gives partial results for the ${\cal O}(\lambda^2)$ operator basis. The full basis of subleading operators for the back-to-back case at $\cO(\lambda)$ and $\cO(\lambda^2)$ will be discussed in detail in~\cite{subhel:long}, including an analysis of relations that occur from parity and charge conjugation.


\chapter{Conclusions and Outlook}
\label{ch:conclusion}

Effective theories (SCET in particular) give an organized framework for carrying out QCD calculations to high precision. Factorization in SCET provides a powerful tool for resumming large logs and calculating cross sections for event shapes at particle colliders.

A large portion of our work has focused on extracting the value of the strong coupling constant, $\alpha_s(m_Z)$ from the cross section for the C-parameter event shape. In order to do this, we first computed the resummed cross section to N${}^3$LL$'$ order, resumming the most important large logs. Additionally, we extracted the nonsingular pieces of the cross section to $\cO(\alpha_s^3)$ and the two-loop constants for the soft function. We treated the hadronic effects analytically, using a shape function whose first moment was given by $\Omega_1$. Further, we switched to a short-distance scheme for this leading power correction, which removed sensitivity from an $\cO(\lqcd)$ renormalon. We improved the calculation by including hadron mass effects, which allowed us to test universality between various event shapes. The final components of the theoretical calculation were profile functions that give the appropriate physical development of the energy scales $\mu_S$, $\mu_J$ and $\mu_H$ across the C-parameter spectrum. By including all of these effects, we saw very good pertubative convergence, with an uncertainty of, on average, 2.5 \% in the tail of the distribution for $Q=m_Z$.

Using this convergent cross section, we performed a fit in the tail region of the distribution for $\alpha_s(m_Z$) and $\Omega_1$. This fit was done at multiple $Q$ values to break the degeneracy between the fit parameters. As our final result from the \mbox{\it C-parameter global fit} we obtained,
\begin{align} \label{eq:allerror}
\alpha_s(m_Z) &  = \, 0.1123 \pm 0.0015\,,\\
\Omega_1(R_\Delta,\mu_\Delta) &  = \, 0.421 \pm 0.063\,\mbox{GeV},\nn
\end{align}
where $\alpha_s$ is defined in the $\msbar$ scheme, and $\Omega_1$ in the Rgap scheme (without hadron-mass effects) at the reference scales $R_\Delta=\mu_\Delta=2$\,GeV. In order to test universality, we compared this with a {\it global fit for thrust}, which gives
\begin{align} \label{eq:allerrortau}
\alpha_s(m_Z) & \, = \, 0.1128 \,\pm\, 0.0012\,,\\
\Omega_1(R_\Delta,\mu_\Delta) & \, = \, 0.322 \,\pm\, 0.068\,\mbox{GeV}.\nn
\end{align}
From these, we conclude that our results support universality between event shapes. Our result is significantly lower than the world average value for $\alpha_s(m_Z)$ and the source of this disagreement is an important open question.

The remainder of our work focused on the construction of a complete set of helicity operator building blocks which can be used to construct operators at any order in the SCET power expansion. We have shown that using the helicity building blocks simplifies the construction of bases of hard-scattering operators, focusing on the case of two collinear directions as a basic example. Additionally, we developed useful selection rules from the conservation of angular momentum that allow us to eliminate some operators from a given basis based on their helicity field content. We expect that further exploration of these subleading helicity methods will provide simpler approaches to factorization beyond leading power in SCET. In particular, considering threshold resummation in DIS or Drell-Yan and the calculation of the subleading thrust cross section are the next steps towards controlling SCET beyond leading power. These studies will allow for higher precision understanding of important physical processes.


\chapter{Acknowledgments}
\label{ch:acknowledgments}

First thanks to Iain Stewart, for basically teaching me everything. To my postdocs: Vicent Mateu, thanks for helping out when I was young and didn't know anything; Piotr Pietrulewicz, thanks for helping out when I was old and somehow still didn't know anything. I'm also grateful to Ian Moult, for his unique perspective on everything from neutrinos \cite{Retiere:2009zz} to effective theories \cite{Moult:2016aah}. A big shout out to Saif Rayyan for his support.

I have truly appreciated all the CTP folks who could always be counted on for an interesting conversation about the physics behind building an effective Death Star or the proper weight advantage in martial arts. Especially thanks to Cathy Modica, Joyce Berggren, Charles Suggs and Scott Morley who are the cream of the administrative crop. My gratitude to my committee members Jesse Thaler and Mike Williams, who were always available when I needed them.

Of course, a million thanks to all of the people outside of work who made me \textit{me}. Mom and Dad, for the genes and such; Mike and Hiva, for a  championship and Persian food; all the SP people, for keeping me mostly sane over the past five years; and the Part I study crowd, for proving something good can come out of quals. Everyone else: you're so vain, you probably think these acknowledgments are about you (and they might be).

This work was supported in part by the Office of Nuclear Physics of the U.S. Department of Energy under Grants No. DE-SCD011090 and Contract No.\ DE-SC0011090.  Additional support came from MIT MISTI global seed funds
and the ESI summer program on ``Jets and Quantum Fields for the LHC and Future Colliders''. Special thanks to Simons for our weekly Supersymmecheese, which greatly bolstered journal club discussions.

\appendix

\chapter{Formulae required for C-parameter Cross Section}
\label{ap:Cformula}

In this appendix we collect all formulae used in our analysis of the C-parameter cross section. In \ref{ap:formulae}, we lay out the formulae required for massless quarks. Section \ref{ap:MC-comparison} shows how to do the Monte Carlo comparison to extract the two loop soft constants. In \ref{ap:Gijcoefficients}, we give the fixed-order coefficients for the C-parameter cross section. In \ref{ap:hadronmassR}, we show the formulae required to do R-evolution with hadron mass effects and in \ref{ap:ArcTan} we discuss the how we treat the gap scheme in the fixed-order region. Finally, \ref{ap:subtractionchoice} is devoted to explaining the Rgap scheme based on the C-parameter soft function.

\section{Formulae} \label{ap:formulae}
Here we focus on our analysis for the case of massless quarks. Since we want to compare to experimental data, which are normalized to the total number of events, we need to calculate $(1/\sigma) \df \sigma / \df C$, our normalized cross section. To do this we can either self-normalize our results by integrating over $C$ or use the fixed-order result for the total hadronic cross section, which, at three loops for massless quarks at \mbox{$\mu=Q$}, is (see Ref.~\cite{Chetyrkin:1996ia} for further discussion)
\begin{align} \label{eq:Rhad}
R_{\rm had}\,&=\,1 + 0.3183099\,\alpha_s(Q) + 0.1427849\,\alpha_s^2(Q)
\ -\,0.411757\,\alpha_s^3(Q)\,.
\end{align}

Throughout our analysis, we use $m_{Z}=91.187\,\mathrm{GeV}$, and all numerical results quoted below are for SU(3) color with $n_f=5$ active light flavors, for simplicity.

\head{Singular Cross Section Formula}\vspace*{1pt}

\noindent
For the singular part of the differential cross section given in \Eq{eq:singular-resummation}, we simplify the numerical evaluation using our freedom to take $\mu=\mu_J$. This means
$U_J^\tau(s-s',\mu_J,\mu_J)=\delta(s-s')$, and we can write
\begin{align} \label{eq:dswithP}
&\frac{1}{\sigma_0}\!\int\! {\rm d}k\, \frac{{\rm d}\hat\sigma_s}{ {\rm d}C}\Big(C-\frac{k}{Q}\Big) 
F_{C} \Big(k- 3 \pi \bar\Delta(R,\mu_S)\Big) \nn\\
&= \frac{Q}{6}
H(Q,\mi_H)\,U_H\big(Q,\mi_H,\mi_J\big)\,
\int {\rm d}k\,P\big(Q,Q C/6-k/6,\mi_J\big)\,\nn\\
&\qquad \times
e^{-{3 \pi\, \delta(R,\mu_s)}\frac{{\rm d}}{{\rm d}k}}\, F_{C} \Big(k-3 \pi\, \bar\Delta(R,\mu_S)\Big),
\end{align}
Here $\sigma_0$ is the tree-level (Born) cross section for \mbox{$e^+e^-\to q\bar q$}. 
Here we have combined the perturbative corrections from the partonic soft function, jet
function, and soft evolution factor into a single function, 
\begin{align}
P(Q,k,\mu_J) & =\int \!{\rm d}s\!\int\!
{\rm d}k' J_\tau(s,\mu_J)\, U_S^{\tau}(k',\mu_J,\mu_S)
\hat{S}_{\widetilde{C}}(k-k'-s/Q,\mu_S) \
\,.
\end{align}
The large logarithms of $C/6$ are summed up in the evolution factors $U_H$ and $U_S^{\tau}$. We can carry out the integrals in $P$ exactly, and the results are enumerated below. The shape function \mbox{$F_{C}(k-3\pi\,\bar{\Delta})$} is discussed in Sec.~\ref{sec:power}, and we have used integration by parts in \Eq{eq:dswithP} to have the derivative in the exponential act on $F_C$, which is simpler than acting the derivative on the perturbative soft function. For our numerical calculation, we expand $H$, $J_{\tau}$, and $\hat{S}_{\widetilde{C}}$ order by order as a series in $\alpha_s(\mu_H)$, $\alpha_s(\mu_J)$, and $\alpha_s(\mu_S)$ respectively, with no large logs. Additionally, we expand $\exp(-\,3 \pi\, \delta(R,\mu_S) {\rm d}/{\rm d}k)$ [see \eq{deltaseries}] as a series in $\alpha_s(\mu_S)$, which must be done consistently to cancel the renormalon present in $\hat{S}_{\widetilde{C}}$.

The hard function to ${\cal O}(\alpha_s^3)$ with $n_f=5$ is~\cite{Matsuura:1987wt,Matsuura:1988sm,Gehrmann:2005pd,Moch:2005id,Lee:2010cg,Baikov:2009bg,Gehrmann:2010ue,Abbate:2010xh}
\begin{align} \label{eq:Hardnumeric}
& H(Q,\mu_H) \nn\\
& =
1+\alpha_{s}(\mu_H)\Big(\!0.745808\!-\!1.27324 L_{Q}\!-\!0.848826L_{Q}^{2}\Big)
\notag\\&
+\alpha_{s}^{2}(\mu_H)\Big(2.27587- 0.0251035\, L_{Q}- 1.06592\, L_{Q}^{2}
+0.735517L_{Q}^{3}+0.360253L_{Q}^{4}\Big)
\notag\\&
+\alpha_{s}^{3}(\mu_H)\Big(0.00050393 \,h_{3}+2.78092 L_{Q}-2.85654 L_{Q}^{2}
-0.147051 L_{Q}^{3}
\notag\\& \qquad \qquad +0.865045L_{Q}^{4}-0.165638 L_{Q}^{5}
-0.101931\, L_{Q}^{6}\Big)\,,
\end{align}
where $L_Q=\ln\frac{\mi_H}{Q}$ and $h_3=8998.080$ from Ref.~\cite{Baikov:2009bg}.

The resummation of large logs between $\mu_H$ and $\mu_J$ is given by
$U_H(Q,\mi_H,\mi_J)$, the solution of the RGE for the
hard function,  which can be written as \cite{Bauer:2000yr}
\begin{align} \label{appendix:U_H}
U_H(Q,\mi_H,\mi)=
e^{2 K(\Gamma_H,\gamma_H,\mi,\mi_H)}
\bigg(\frac{\mi_H^2}{Q^2}\bigg)^{\!\!\omega(\Gamma_H\!,\,\mi,\mi_H)},
\end{align}
where $\omega$ and $K$ are given in \eqs{w}{K} below.

In momentum space, we can use the results from Ref.~\cite{Ligeti:2008ac} to calculate the convolution of the plus-functions in $P$ to give the form
\begin{align} \label{appendix:P}
&P\big(Q,k,\mi_J\big)=
\frac{1}{\xi}\,E_S\big(\xi,\mi_J,\mi_S\big)\nonumber\\
&\times\sum_{\substack{n,m,k,l=-1\\m+n+1\geq k\\k+1\geq l}}^{\infty}
V_{k}^{mn}\,\,J_m\Big[\as(\mi_J),\frac{\xi\, Q}{\mi^2_{J}}\Big]
\,S_{n}^{\widetilde C}\Big[\as(\mi_S),\frac{\xi}{\mi_S}\Big]\,
\notag\\
&\times
V_l^k\big[-2\,\omega(\Gamma_S,\mi_J,\mi_S)\big]\,\mathcal
L_l^{-2\omega(\Gamma_S,\mi_J,\mi_S)}\Big(\frac{k}{\xi}\Big) \,.
\end{align}
Here $\xi$ is a dummy variable that does not affect the value of the result. \footnote{When convolved
	with $F_C$ we use our freedom in choosing $\xi$ to simplify
	the final numerical integration, picking \mbox{$\xi = QC/6 - 3 \pi\, \bar{\Delta}(R,\mu_S)$}.} $E_S(\xi,\mu_J,\mu_S)$ is given by~\cite{Balzereit:1998yf,Neubert:2004dd}
\begin{align} \label{appendix:E}
&E_{S}(\xi,\mi_J,\mi_S)=
\exp\!\big[2 K(\Gamma_S,\gamma_S,\mi_J,\mi_S)\big]
\\&\times
\Big(\frac{\xi}{\mi_S}\Big)^{-2\omega(\Gamma_S,\mi_J,\mi_S)}\ 
\frac{\exp\!\big[2\gamma_E\, \omega(\Gamma_S,\mi_J,\mi_S) \big]}
{\Gamma\big[1-2\,\omega(\Gamma_S,\mi_J,\mi_S)\big]} \,,
\notag
\end{align}
and encodes part of the running between $\mu_S$ and $\mu_J$. The rest of the running is included in the $V$ coefficients and the plus-functions, $\mathcal L_l$.

The $J_m$ and $S_n$ in Eq.~(\ref{appendix:P}) are the coefficients of the momentum-space soft and jet functions, given by
\begin{align}
J_\tau(p^-k,\mu_J) &= \frac{1}{p^-\xi} \sum_{m=-1}^\infty
J_m\Big[\alpha_s(\mu_J),\frac{p^-\xi}{\mu_J^2}\Big] {\cal L}_m\Big(\frac{k}{\xi}\Big) , \nn\\
\hat S_{\widetilde C}(k,\mu_S) &= \frac{1}{\xi} \sum_{n=-1}^\infty
S_n^{\widetilde C}\Big[\alpha_s(\mu_S),\frac{\xi}{\mu_S}\Big] {\cal
	L}_n\Big(\frac{k}{\xi}\Big) .
\end{align}
Here the C-parameter soft function coefficients are
\begin{align}
S_{-1}^{\widetilde C}[\alpha_s,x] 
&= S_{-1}^{\widetilde C}(\alpha_s) + \sum_{n=0}^\infty S_n^{\widetilde C}(\alpha_s)
\frac{\ln^{n+1} x}{n+1} \,,\nn \\
S_n^{\widetilde C}[\alpha_s,x] 
&= \sum_{k=0}^\infty \frac{(n+k)!}{n!\, k!} S_{n+k}^{\widetilde C}(\alpha_s) \ln^k x
\,,
\end{align}
which for $n_f=5$ can be written as
\begin{align} \label{eq:Sncoeff}
S_{-1}^{\widetilde C}(\alpha_s) &= 1 + 1.0472 \alpha _s +
(1.75598 + 0.012666\, s_2^{\widetilde{C}}) \alpha _s^2 
\notag\\
&+ \left (
2.59883 + 0.0132629 \, s_2^{\widetilde{C}} 
+ 0.00100786\, s_3^{\widetilde C} \right) \alpha_s^3 
\nn \,,\\
S_0^{\widetilde C}(\alpha_s) &= 1.22136 \alpha _s^2 + 
(2.63481- 0.0309077\, s_2^{\widetilde{C}}) \alpha _s^3 
\nn\,,\\
S_1^{\widetilde C}(\alpha_s) &= -1.69765\, \alpha_s - 7.45178\, \alpha_s^2
\notag\\
&
-(19.1773+0.021501\, s_2^{\widetilde{C}})\, \alpha _s^3 \,,\nn \\
S_2^{\widetilde C}(\alpha_s) &= 1.03573\, \alpha_s^2 + 2.3245\, \alpha_s^3 \,,\nn\\
S_3^{\widetilde C}(\alpha_s) &= 1.44101\, \alpha_s^2 + 10.299\, \alpha_s^3 \,,\nn\\
S_4^{\widetilde C}(\alpha_s) &= -\,1.46525\, \alpha_s^3 \,,\nn\\
S_5^{\widetilde C}(\alpha_s) &= -\,0.611585\, \alpha_s^3 \,.
\end{align}
Note that $s_2^{\widetilde{C}}$ and $s_3^{\widetilde C}$ are the ${\cal O}(\alpha_s^{2,3})$ coefficients of the
non-logarithmic terms in the series expansion of the logarithm of the position
space C-parameter soft function. The coefficients for the jet function are
\begin{align}
J_{-1}[\as,x]=&
J_{-1}(\as)+\sum_{n=0}^{\infty}\,J_{n}(\as)\,\frac{\ln^{n+1}x}{n+1} 
\,, \nn\\
J_{n}[\as,x]=& \sum_{k=0}^{\infty}\,\frac{(n+k)!}{n!\, k!}
\,J_{n+k}(\as)\,\ln^{k}x\,,
\end{align}
and are known up to ${\cal O}(\alpha_s^3)$ except for the constant $j_3$
term~\cite{Lunghi:2002ju,Bauer:2003pi,Bosch:2004th,Becher:2006qw,Moch:2004pa,Becher:2008cf}. With $n_f=5$ we have
\begin{align} \label{eq:Jncoeff}
J_{-1}(\as) &= 1 - 0.608949\, \as - 2.26795\, \as^2
\notag\\&\qquad + (2.21087 + 0.00100786\, j_3)\,\as^3  \,,\nn \\
J_{0}(\as) &= -\,0.63662\, \as + 3.00401\, \as^2 + 4.45566\, \as^3 \,,\nn\\
J_{1}(\as) &= 0.848826\, \as - 0.441765\, \as^2 - 11.905\, \as^3\,,\nn \\
J_{2}(\as) &= -\,1.0695\, \as^2 + 5.36297\, \as^3\,,\nn\\
J_{3}(\as) &= 0.360253\, \as^2 + 0.169497\, \as^3\,,\nn\\
J_{4}(\as) &= -\,0.469837\, \as^3\,,\nn \\
J_{5}(\as) &= 0.0764481\,\as^3.
\end{align}
The plus-distributions, denoted by $\cL(x)$, are given by
\begin{equation} \label{eq:cLna_def}
\cL_n^a(x) = \biggl[\frac{\theta(x)\ln^n x}{x^{1-a}}\biggr]_+
= \frac{\df^n}{\df a^n}\, \cL^{a}(x) \,\,\, [n \ge 0]\,,
\end{equation}
${\cal L}_{-1}^a(x) = {\cal L}_{-1}(x) = \delta(x)$, and for $a> -1$
\begin{equation}
\label{eq:cLa_def}
\cL^a(x) = \biggl[\frac{\theta(x)}{x^{1-a}} \biggr]_+
= \lim_{\e\to 0}\, \frac{\df}{\df x}
\biggl[ \theta(x - \e)\, \frac{x^a - 1}{a} \biggr] \,.
\end{equation}
	In Eq.~(\ref{appendix:P}) we also take advantage of the shorthand for the $V$ coefficients presented in Ref.~\cite{Ligeti:2008ac},
	\begin{align} \label{eq:Vkna_def}
	\V_k^n(a) &= \begin{cases}
	\displaystyle a\, \frac{\df^n}{\df b^n}\,\frac{\V(a,b)}{a+b}\bigg\vert_{b = 0}\,,
	&   k=-1\,, \\[10pt]
	\displaystyle  a\, \binom{n}{k}   \frac{\df^{n-k}}{\df b^{n-k}}\, \V(a,b)
	\bigg\vert_{b = 0} + \delta_{kn} \,, \quad
	& 0\le k\le n \,,  \\[10pt]
	\displaystyle  \frac{a}{n+1} \,,
	& k=n+1  \,,
	\end{cases}
	\end{align}
	and the coefficients
	\begin{align} \label{eq:Vkmn_def}
	\V_k^{mn} &= \begin{cases}
	\displaystyle \frac{\df^m}{\df a^m}\, \frac{\df^n}{\df b^n}\,\frac{\V(a,b)}{a+b}\bigg\vert_{a = b = 0} \,,
	& k=-1\,, \\[10pt]
	\displaystyle  \sum_{p=0}^m\sum_{q=0}^n\delta_{p+q,k}\,\binom{m}{p} \binom{n}{q}
	\frac{\df^{m-p}}{\df a^{m-p}}\, \frac{\df^{n-q}}{\df b^{n-q}} \ \V(a,b)
	\bigg\vert_{a = b = 0}\,, \quad
	& 0\le k \le m+n \,,\\[15pt]
	\displaystyle  \frac{1}{m+1} + \frac{1}{n+1}\,,
	& k=m+n+1 \,,
	\end{cases}
	\end{align}
	for
	\begin{equation}
	\V(a,b) = \frac{\Gamma(a)\,\Gamma(b)}{\Gamma(a+b)} - \frac{1}{a} - \frac{1}{b}
	\,.
	\end{equation}
	We also need the special cases
	\begin{align}
	&\V_{-1}^{-1}(a) = 1
	\,,
	&\V_0^{-1}(a) &= a
	\,,
	&\V_{k \geq 1}^{-1}(a) &= 0
	\,,\quad
	&\V^{-1,n}_k &= \V^{n,-1}_k = \delta_{nk}
	\,.\end{align}

	\head{Evolution factors and Anomalous Dimensions}\vspace*{1pt}
	
	The running between scales is encoded in just a few functions. In Eqs.~(\ref{appendix:U_H}),
	(\ref{appendix:P}), and (\ref{appendix:E}), we use
	\begin{align} \label{eq:w}
	&\omega(\Gamma,\mi,\mi_0) =2\!\int_{\alpha_s(\mu_0)}^{\alpha_s(\mu)}\frac{{\rm
			d}\,\alpha}{\beta(\alpha)}\,\Gamma(\alpha) \notag\\
	&=  -\,\frac{\Gamma_0}{\beta_0}\bigg\{\!\ln \kappa
	+\frac{\alpha_s(\mu_0)}{4\pi}\Big(\frac{\Gamma_1}{\Gamma_0}
	-\frac{\beta_1}{\beta_0}\Big)(\kappa-1) 
	+\frac{1}{2}
	\frac{\alpha_s^2(\mu_0)}{(4\pi)^2}\Big(\frac{\beta_1^2}{\beta_0^2}
	-\frac{\beta_2}{\beta_0}+\frac{\Gamma_2}{\Gamma_0}
	-\frac{\Gamma_1\beta_1}{\Gamma_0\beta_0}\Big)(\kappa^2-1) \notag\\& +\frac{1}{3}
	\frac{\alpha_s^3(\mu_0)}{(4\pi)^3}
	\bigg[\frac{\Gamma_3}{\Gamma_0}-\frac{\beta_3}{\beta_0}
	+\frac{\Gamma_1}{\Gamma_0}\Big(\frac{\beta_1^2}{\beta_0^2}-\frac{\beta_2}{\beta_0}\Big)
	-\frac{\beta_1}{\beta_0}\Big(\frac{\beta_1^2}{\beta_0^2}-
	2\,\frac{\beta_2}{\beta_0}+\frac{\Gamma_2}{\Gamma_0}\Big) \bigg](\kappa^3-1)\!\bigg\},
	\end{align}
	and
	\begin{align} \label{eq:K}
	&K(\Gamma,\gamma,\mi,\mi_0)-\omega\Big(\frac{\gamma}{2},\mi,\mi_0\Big)
	= 2\!\int_{\alpha_s(\mu_0)}^{\alpha_s(\mu)}
	\frac{{\rm d}\,\alpha}{\beta(\alpha)}\,\Gamma(\alpha)\!
	\int_{\alpha_s(\mu_0)}^{\alpha}\frac{{\rm d}\alpha'}{\beta(\alpha')}
	\notag\\[3pt]
	 =&\frac{\Gamma_0}{2\beta_0^2}\Bigg\{\frac{4\pi}{\alpha_s(\mu_0)}
	\Big(\ln \kappa+\frac{1}{\kappa}-1\Big) +
	\Big(\frac{\Gamma_1}{\Gamma_0}-\frac{\beta_1}{\beta_0}\Big)(\kappa-1-\ln \kappa)
	-\frac{\beta_1}{2\beta_0}\ln^2 \kappa
	\nn \\
	&+\frac{\alpha_s(\mu_0)}{4\pi}\bigg[
	\Big(\frac{\Gamma_1\beta_1}{\Gamma_0\beta_0}-\frac{\beta_1^2}{\beta_0^2}\Big)
	(\kappa-1-\kappa \ln \kappa) -B_2 \ln \kappa
	+\Big( \frac{\Gamma_2}{\Gamma_0}-\frac{\Gamma_1\beta_1}{\Gamma_0\beta_0} +
	B_2 \Big)\frac{(\kappa^2\!-\!1)}{2}
	\nn \\
	&+\Big(\frac{\Gamma_2}{\Gamma_0}-\frac{\Gamma_1\beta_1}{\Gamma_0\beta_0}\Big) 
	(1\!-\!\kappa)\bigg]
	\nn \\
	&+\frac{\alpha_s^2(\mu_0)}{(4\pi)^2}
	\bigg[ \Big[\Big(\frac{\Gamma_1}{\Gamma_0}-\frac{\beta_1}{\beta_0} \Big)B_2
	+\frac{B_3}{2}\Big]\frac{(\kappa^2\!-\!1)}{2} + \Big(\frac{\Gamma_3}{\Gamma_0} -
	\frac{\Gamma_2\beta_1}{\Gamma_0\beta_0}
	+\frac{B_2\Gamma_1}{\Gamma_0}+B_3\Big) \Big(\frac{\kappa^3-1}{3}-\frac{\kappa^2-1}{2}\Big)
	\nn \\
	&-\frac{\beta_1}{2\beta_0}
	\Big(\frac{\Gamma_2}{\Gamma_0}-\frac{\Gamma_1\beta_1}{\Gamma_0\beta_0}+B_2\Big)
	\Big(\kappa^2\ln \kappa-\frac{\kappa^2-1}{2}\Big) -\frac{B_3}{2}\ln \kappa
	-B_2\Big(\frac{\Gamma_1}{\Gamma_0}-\frac{\beta_1}{\beta_0} \Big)(\kappa-1)
	\bigg]
	\Bigg\},
	\end{align}
	where here $\kappa=\alpha_s(\mu)/\alpha_s(\mu_0)$ requires the known 4-loop running couplings, and
	we have defined $B_2 \equiv \beta_1^2/\beta_0^2-\beta_2/\beta_0$ and $B_3 \equiv -\beta_1^3/\beta_0^3+2\beta_1\beta_2/\beta_0^2-\beta_3/\beta_0$.
\begin{figure*}[t!]
	\subfigure[]
	{
		\includegraphics[width=0.3\textwidth]{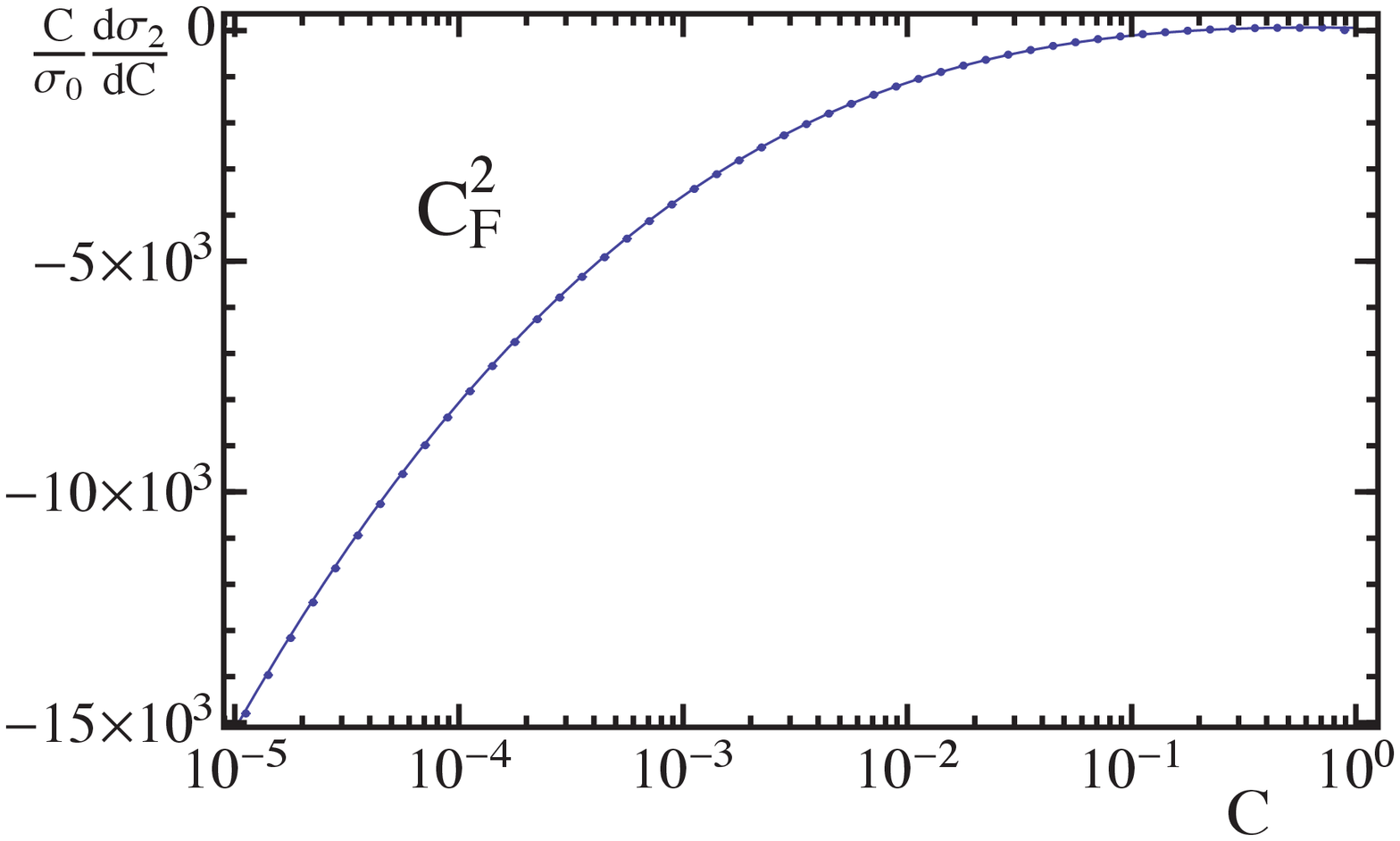}
		\label{fig:EVENT2-CF}
	}
	\subfigure[]{
		\includegraphics[width=0.3\textwidth]{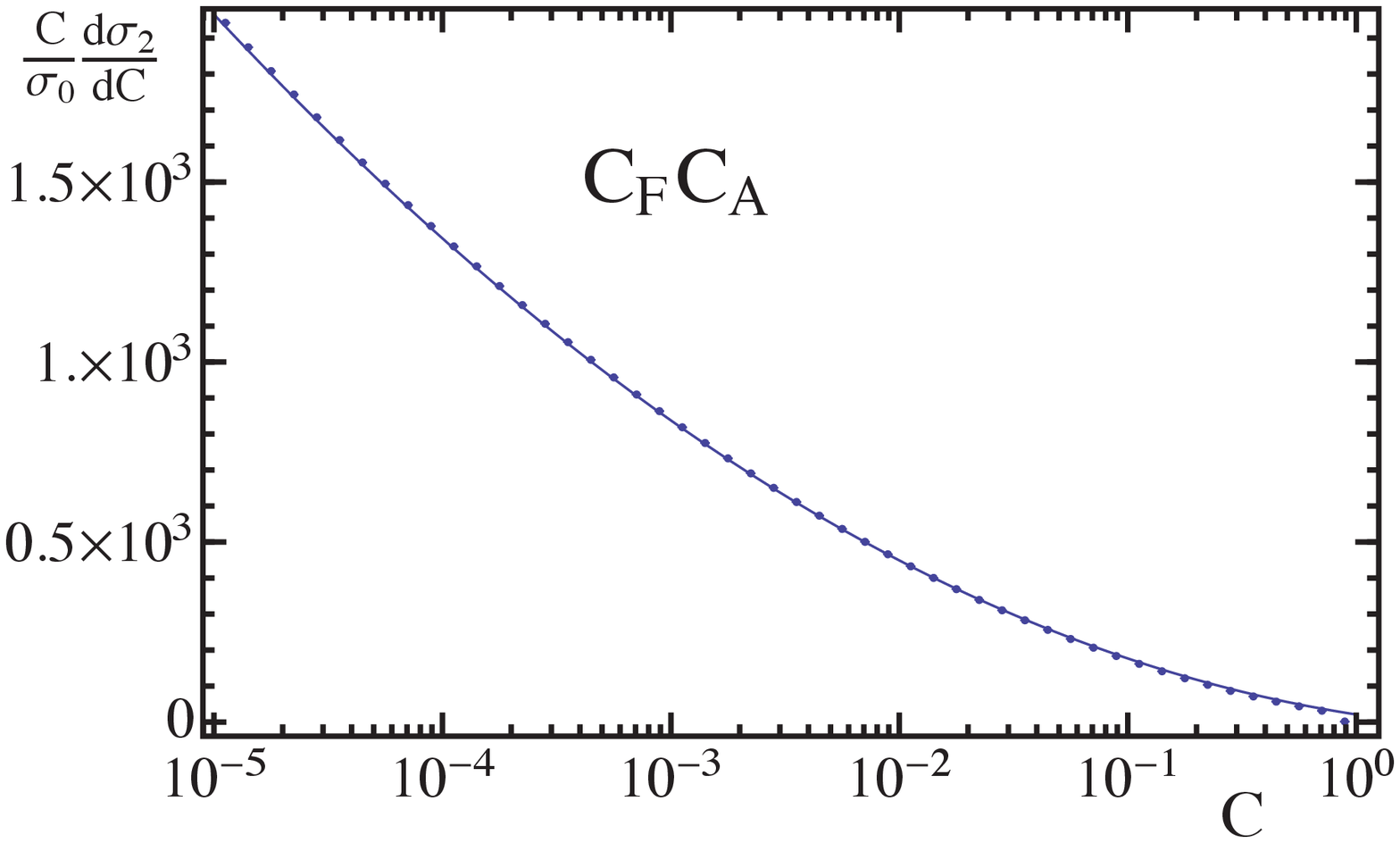}
		\label{fig:EVENT2-CA}
	}
	\subfigure[]{
		\includegraphics[width=0.31\textwidth]{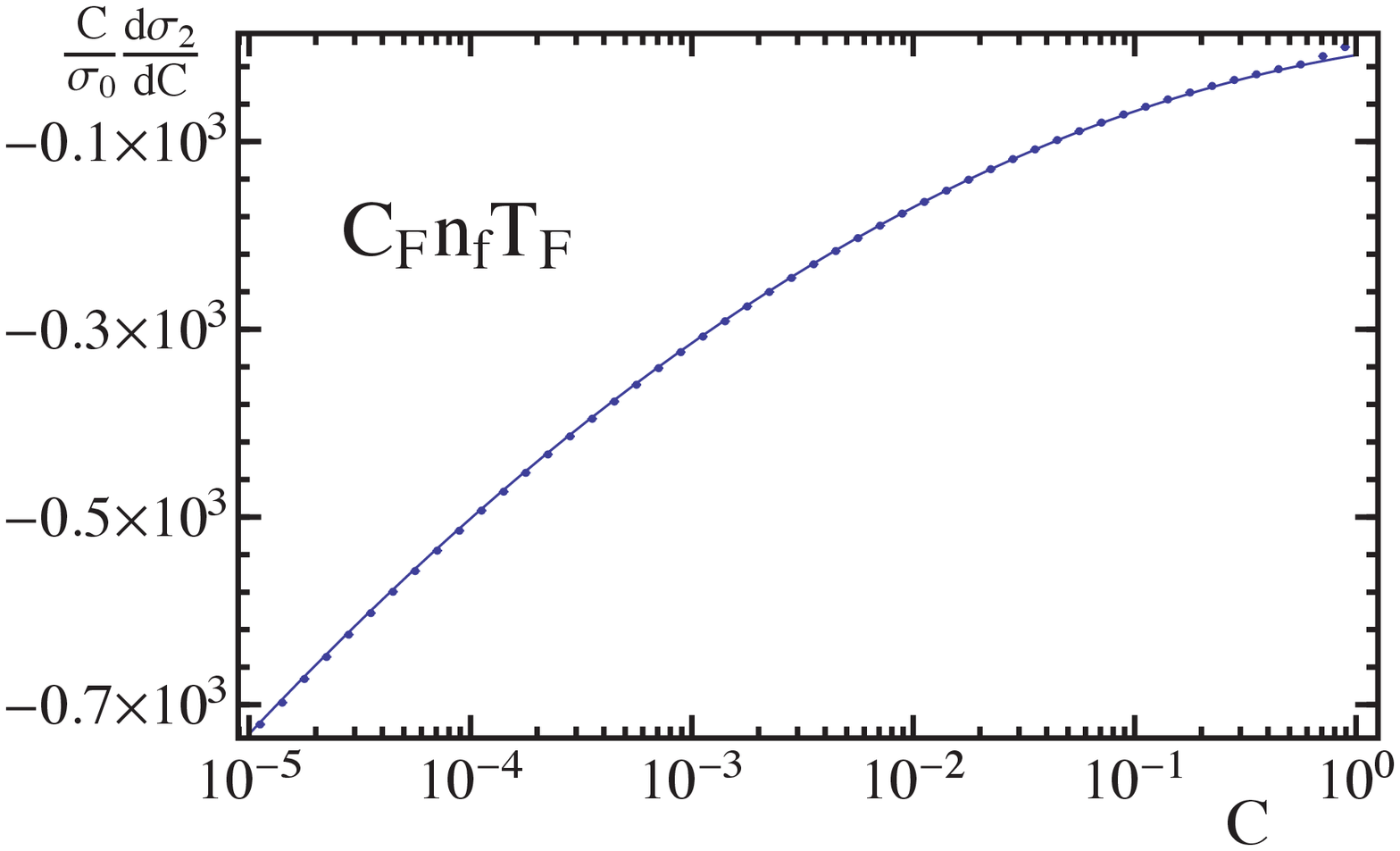}
		\label{fig:EVENT2-NF}
	}
	\caption[Comparison of FO analytic SCET prediction for $\mathcal{O}(\alpha_s^2)$ piece of cross section with EVENT2]{ \label{fig:1loop-comparison} Comparison of the fixed-order analytic SCET prediction for the $\mathcal{O}(\alpha_s^2)$ piece with the parton level Monte Carlo EVENT2. The decomposition in the three color structures $C_F^2$, $C_F C_A$, and $C_F n_f T_f$ is shown in panels (a), (b), and (c), respectively. The factor $\alpha_s^2/(2\pi)^2$ has been divided out. We use a log binning in the horizontal axis to emphasize the dijet region.}
	\label{fig:EVENT2}%
\end{figure*}
\noindent
These results are expressed in terms of the coefficients of
series expansions of the QCD $\beta$ function $\beta[\alpha_s]$,  $\Gamma[\alpha_s]$ (which is given
by a constant of proportionality times the QCD cusp anomalous dimension) and of
a non-cusp anomalous dimension $\gamma[\alpha_s]$. These expansion coefficients are defined by the equations
\begin{align}
&\beta(\as)=-\,2\,\as\,\sum_{n=0}^{\infty}\beta_n\Big(\frac{\as}{4\pi}\Big)^{\!\!n+1}\!\!,\;\\
&\Gamma(\as)=\sum_{n=0}^{\infty}\Gamma_n\Big(\frac{\as}{4\pi}\Big)^{\!\!n+1}\!\!,\;
\gamma(\as)=\sum_{n=0}^{\infty}\gamma_n\Big(\frac{\as}{4\pi}\Big)^{\!\!n+1}\!\!.\notag
\end{align}
For $n_f=5$, the relevant coefficients are~\cite{Tarasov:1980au, Larin:1993tp,
	vanRitbergen:1997va, Korchemsky:1987wg, Moch:2004pa,Czakon:2004bu}
\begin{align}\label{eq:betaCusp}
\beta_0 &= 23/3\,,\quad
\beta_1 = 116/3\,,\quad
\beta_2 = 180.907,
\\
\beta_3 &= 4826.16,
\nn \\
\Gamma^{\rm{cusp}}_0 &= 16/3,\quad
\Gamma^{\rm{cusp}}_1 = 36.8436,\quad
\Gamma^{\rm{cusp}}_2 = 239.208
\,.
\nn
\end{align}
As was mentioned in the text, we use a Pad\'e
approximation for the unknown four-loop cusp anomalous dimension, assigning a large uncertainty to this estimate:
\begin{align}
\Gamma_{3}^{\rm cusp} = (1 \pm 2) \frac{(\Gamma _2^{\rm
		cusp}){}^2}{\Gamma_{1}^{\rm cusp}}.
\end{align}
For the hard, jet, and soft functions, the anomalous dimensions are the same as in the thrust case and are given by~\cite{vanNeerven:1985xr,Matsuura:1988sm,Catani:1992ua,Vogt:2000ci,Moch:2004pa,Neubert:2004dd,Moch:2005id,Idilbi:2006dg,Becher:2006mr}
\begin{align}
\Gamma^H_n &= -\,\Gamma^{\rm{cusp}}_n,\quad
& \Gamma^J_n &= 2\,\Gamma^{\rm{cusp}}_n,\quad
& \Gamma^S_n &= -\,\Gamma^{\rm{cusp}}_n,
\nn \\[4pt]
\gamma^H_0 &= -\,8,\quad
&\gamma^H_1 &= 1.14194,\quad
&\gamma^H_2 &= -\,249.388,
\nn \\[4pt]
\gamma^J_0 &= 8,\quad
&\gamma^J_1 &= -\,77.3527,\quad
&\gamma^J_2 &= -\,409.631,
\nn \\
\gamma^S_n &= -\,\gamma^H_n-\gamma^J_n.
\end{align}

For the 4-loop running of the strong coupling constant, we use a form that agrees very well numerically with the solution to the beta function. For $n_f=5$, the value of the coupling is given by
\begin{align} \label{alphas}
\frac{1}{\alpha_s(\mu)} &= \frac{X}{\alpha_s(m_Z)} + 0.401347248\, \ln X 
\\
&+ \frac{\alpha_s(m_Z)}{X} \big[ 0.01165228\, (1-X) + 0.16107961\,
{\ln X}\big]
\nn\\
&+ \frac{\alpha_s^2(m_Z)}{X^2}
\big[ 0.1586117\, (X^2-1) +0.0599722\,(X
\nn\\
&+\ln X-X^2)
+0.0323244\, \{(1-X)^2-\ln^2 X\} \big] ,
\nn
\end{align}
where we have used the values from \eq{betaCusp} for the $\beta_i$ and $X=1+\alpha_s(m_Z) \ln(\mu/m_Z) \beta_0/(2\pi)$.

For the singular cross section, we have implemented the formulas described in this appendix into a Mathematica~\cite{mathematica} code. Additionally, we have created an independent Fortran~\cite{gfortran} code based on a Fourier space implementation (the nonsingular distributions have also been implemented into the two codes independently). These two codes agree with each other at $10^{-6}$ or better.

\section{Comparison to Parton Level Monte Carlos}
\label{ap:MC-comparison}
A useful way to validate our results is to compare the SCET prediction, computed using the needed formulas above, with fixed scales $\mu_H=\mu_J=\mu_S=Q$ to
the fixed-order prediction, approaching small values of $C$. This serves as an important test on the
accuracy of parton level Monte Carlos such as EVENT2 and EERAD3 where they should agree with SCET.
Beginning at $\mathcal{O}(\alpha_s^2)$ we compare our theory runs to the parton level Monte Carlo given by EVENT2, splitting the output
in the various color structures. This comparison used an approach with logarithmically-binned EVENT2 distributions across the entire
spectrum. Details on the run parameters (number of events and cutoff parameter) have
been given in Sec.~\ref{sec:nonsingular}.

Additionally, Fig.~\ref{fig:EVENT2} clearly shows that for all color structures the
agreement is excellent all the way to $C\sim 10^{-5}$. The very large number of events used in our runs ($3\times 10^{11}$) makes the error bars here essentially invisible. Also note that the cross section shoulder at $C=0.75$ is all in the largest $C$ bin and hence not visible in these plots.

\begin{figure*}[t!]
	\subfigure[]
	{
		\includegraphics[width=0.3\textwidth]{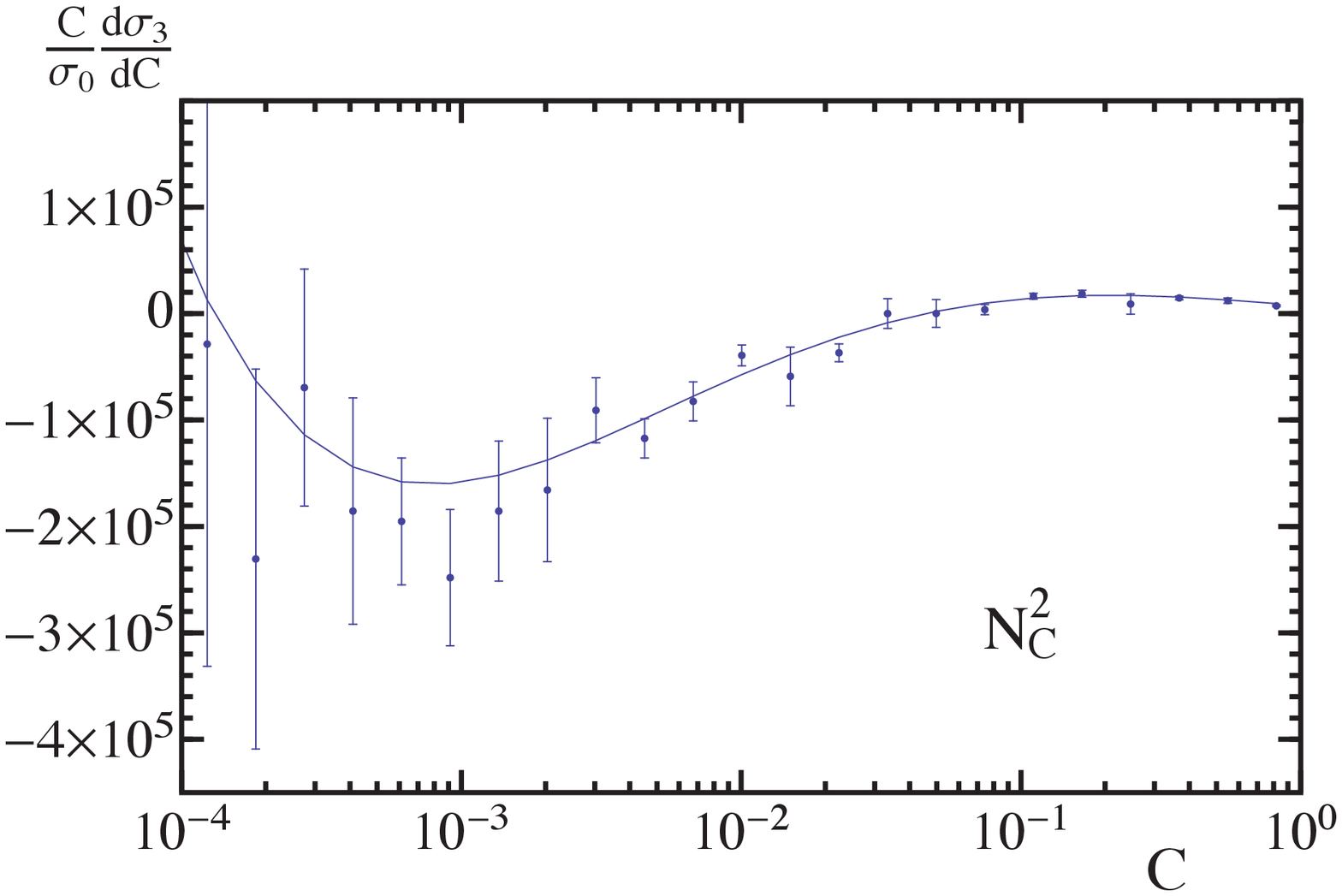}
		\label{fig:EERAD-NC2}
	}
	\subfigure[]{
		\includegraphics[width=0.3\textwidth]{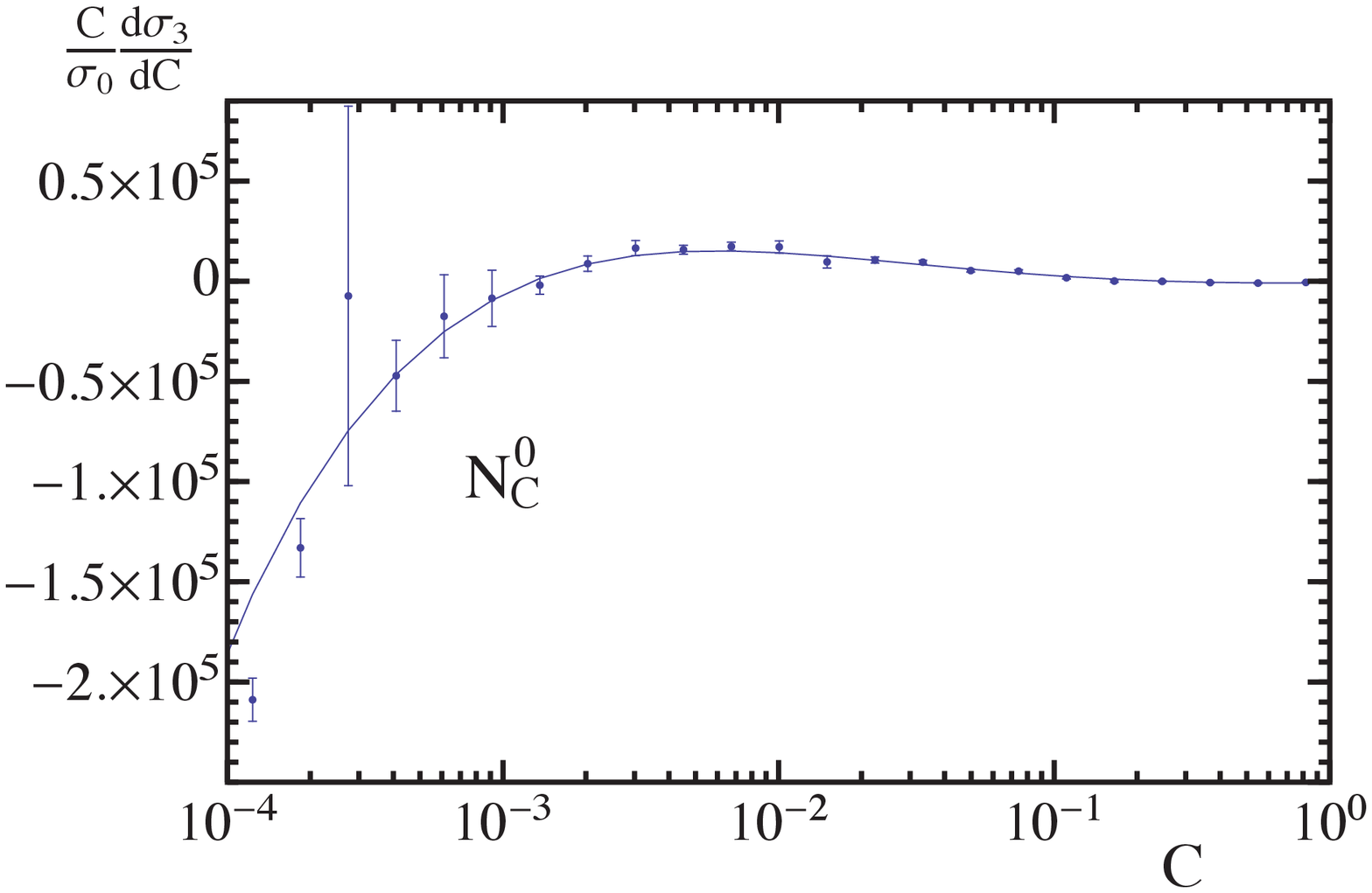}
		\label{fig:EERAD-NC0}
	}
	\subfigure[]{
		\includegraphics[width=0.3\textwidth]{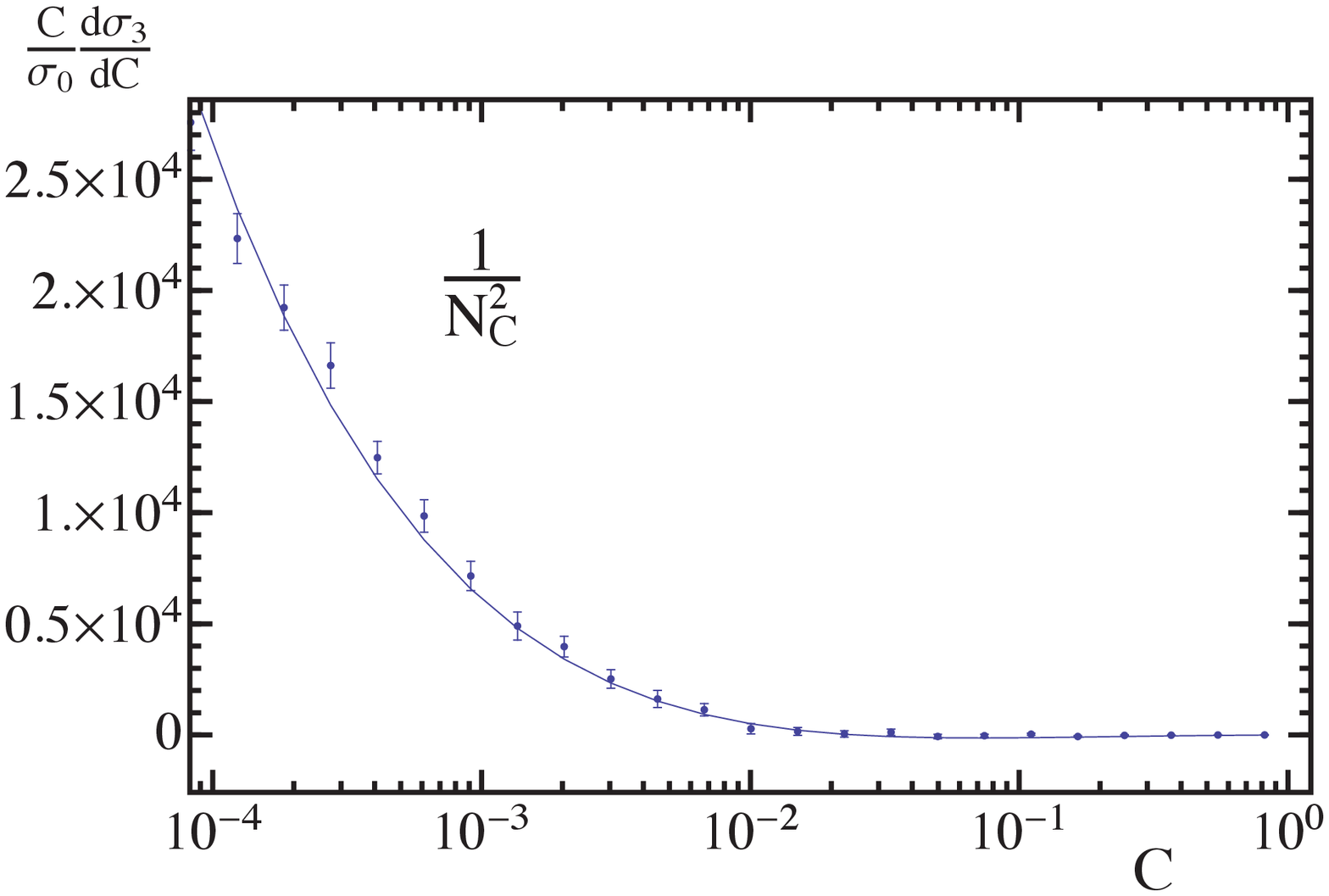}
		\label{fig:EERAD-NC-2}
	}
	\subfigure[]
	{
		\includegraphics[width=0.3\textwidth]{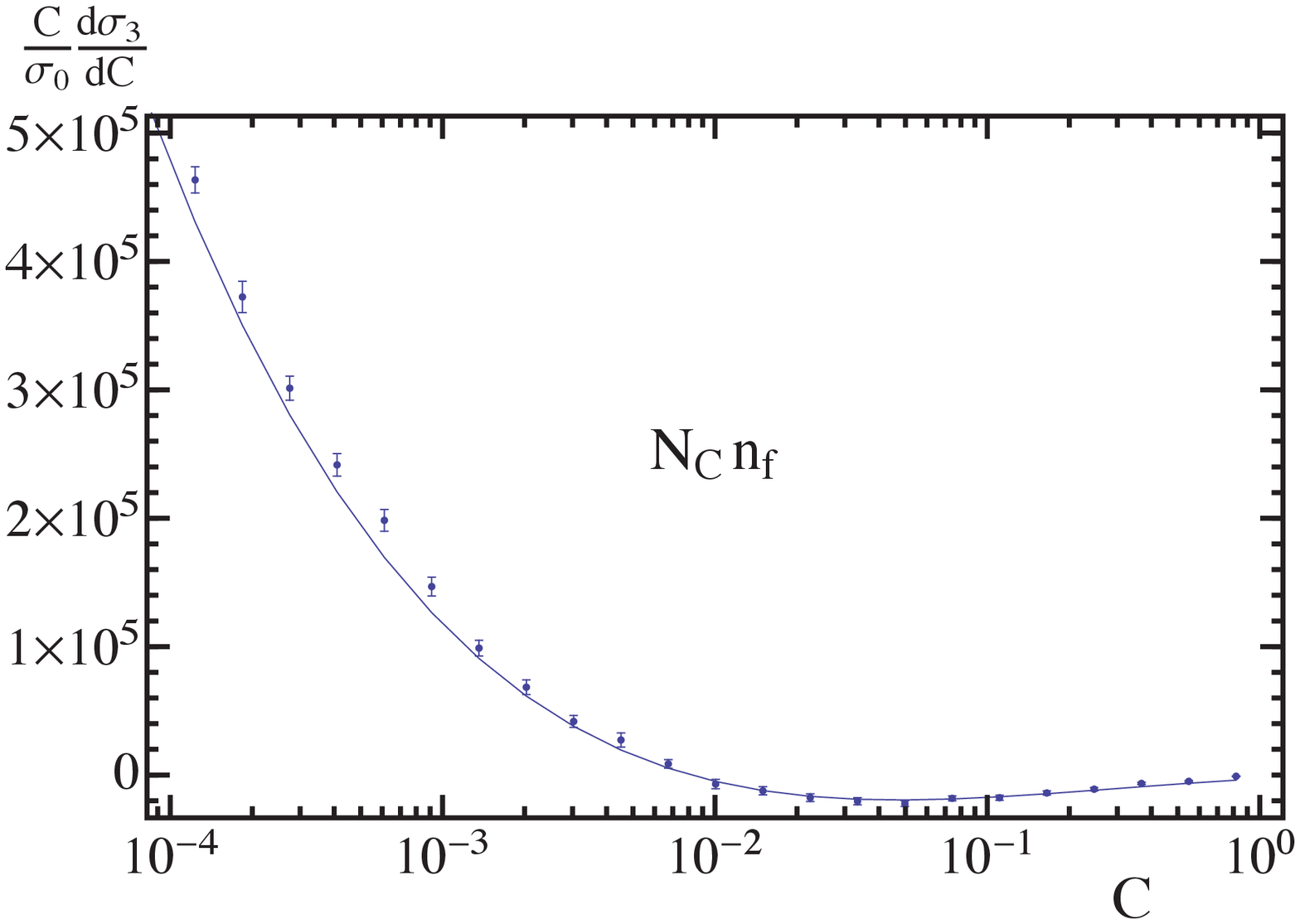}
		\label{fig:EERAD-NC-nf}
	}
	\subfigure[]{
		\includegraphics[width=0.3\textwidth]{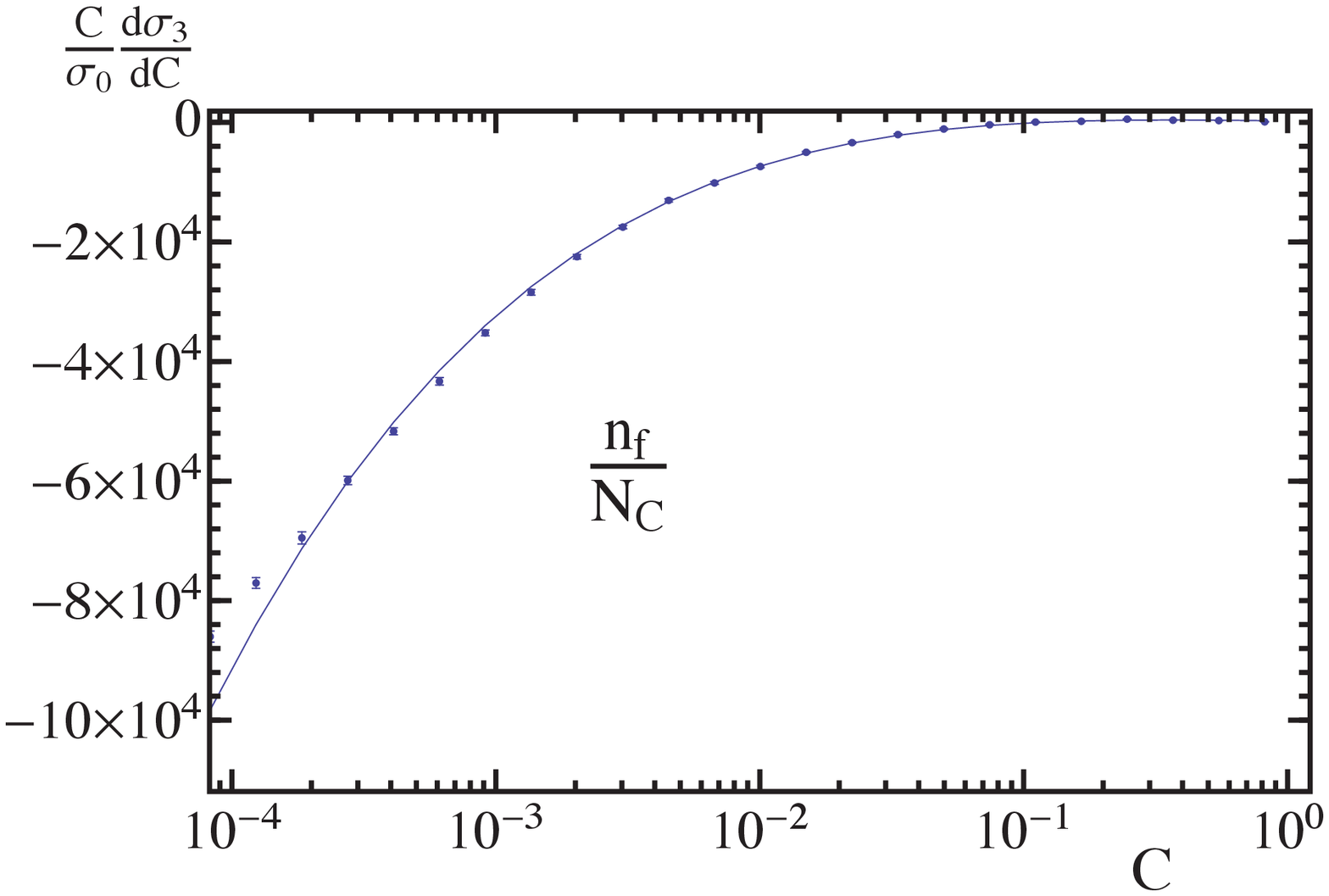}
		\label{fig:EERAD-NC-1_nf}
	}
	\subfigure[]{
		\includegraphics[width=0.3\textwidth]{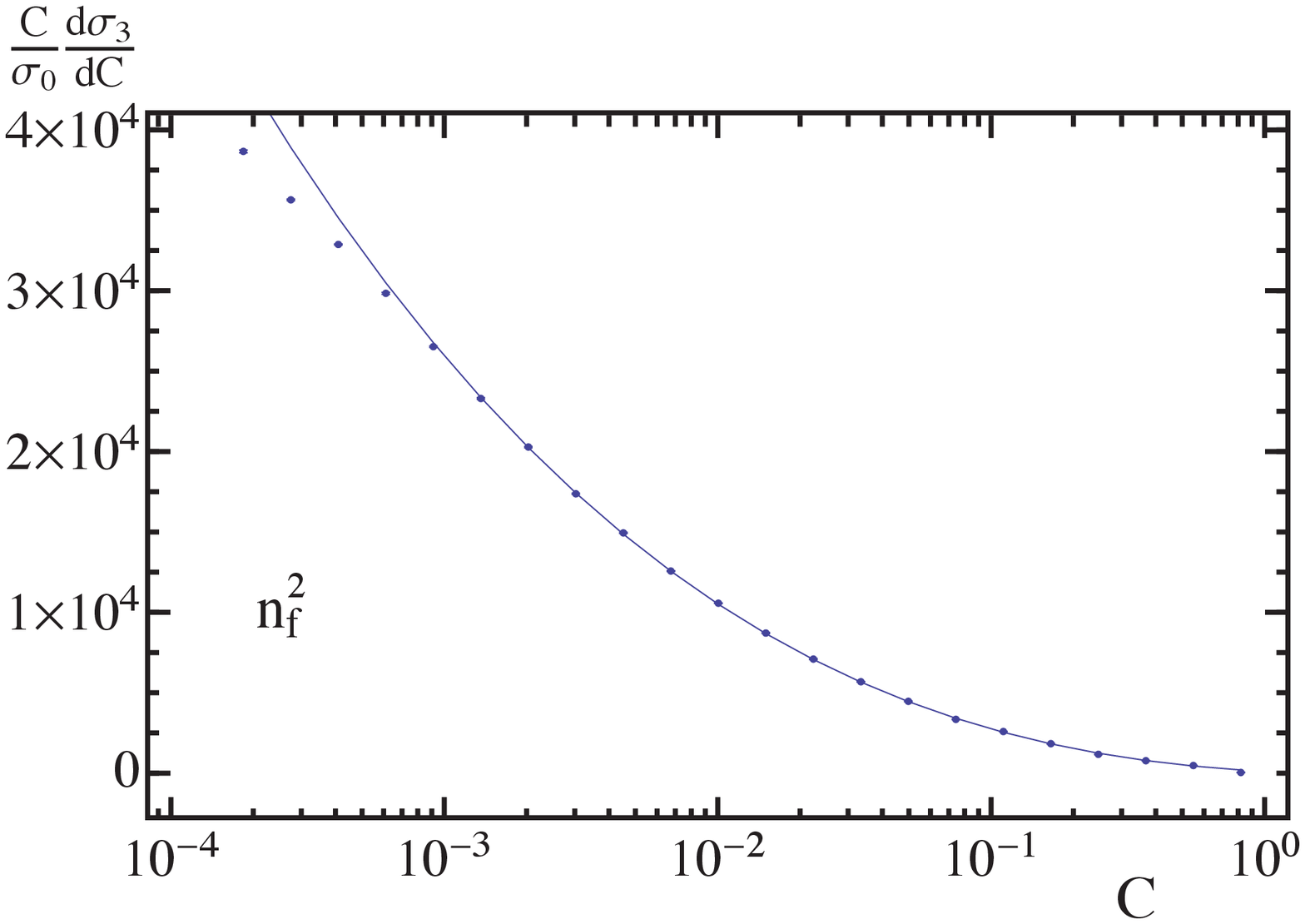}
		\label{fig:EERAD-nf-2}
	}
	\caption[Comparison of FO analytic SCET prediction for $\mathcal{O}(\alpha_s^3)$ piece of cross section with EERAD3]{Comparison of the fixed-order analytic SCET prediction for the $\mathcal{O}(\alpha_s^3)$ piece with the parton
		level Monte Carlo EERAD3. The decomposition in the three color structures $N_C^2$, $1$, and $N_C^{-2}$ are shown in panels (a), (b), and (c), of the first row, respectively, and $N_C\,n_f$, $n_f/N_C$ and $n_f^2$ are shown in panels (d), (e) and (f),
		respectively, on the second row. The factor $\alpha_s^3/(2\pi)^3$ has been divided out. We use a log binning in the 
		horizontal axis to emphasize the dijet region.}
	\label{fig:EERAD3}%
\end{figure*}
\begin{figure*}[t!]
	\includegraphics[width=0.45\textwidth]{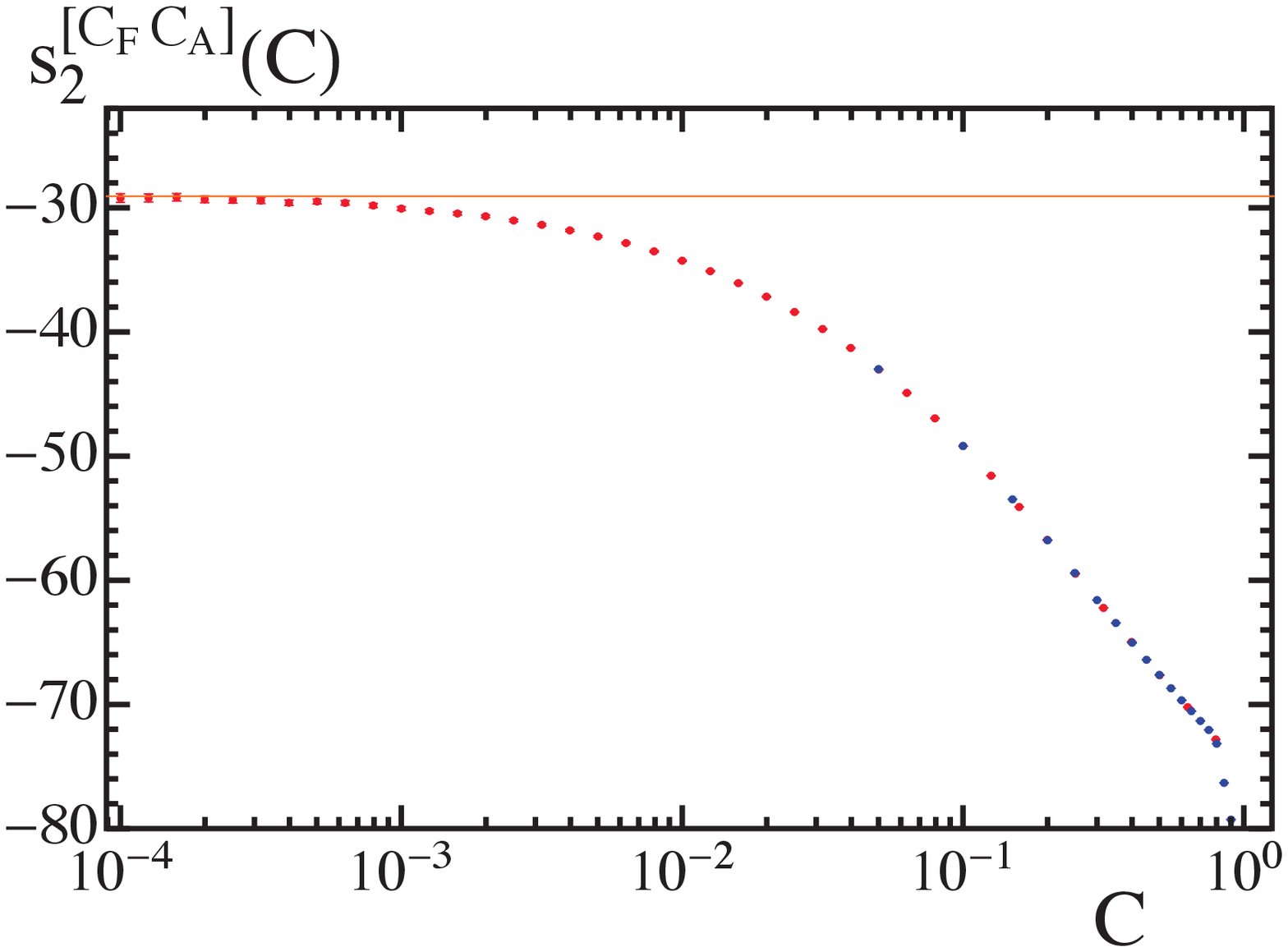}~~~~~~~~
	\includegraphics[width=0.45\textwidth]{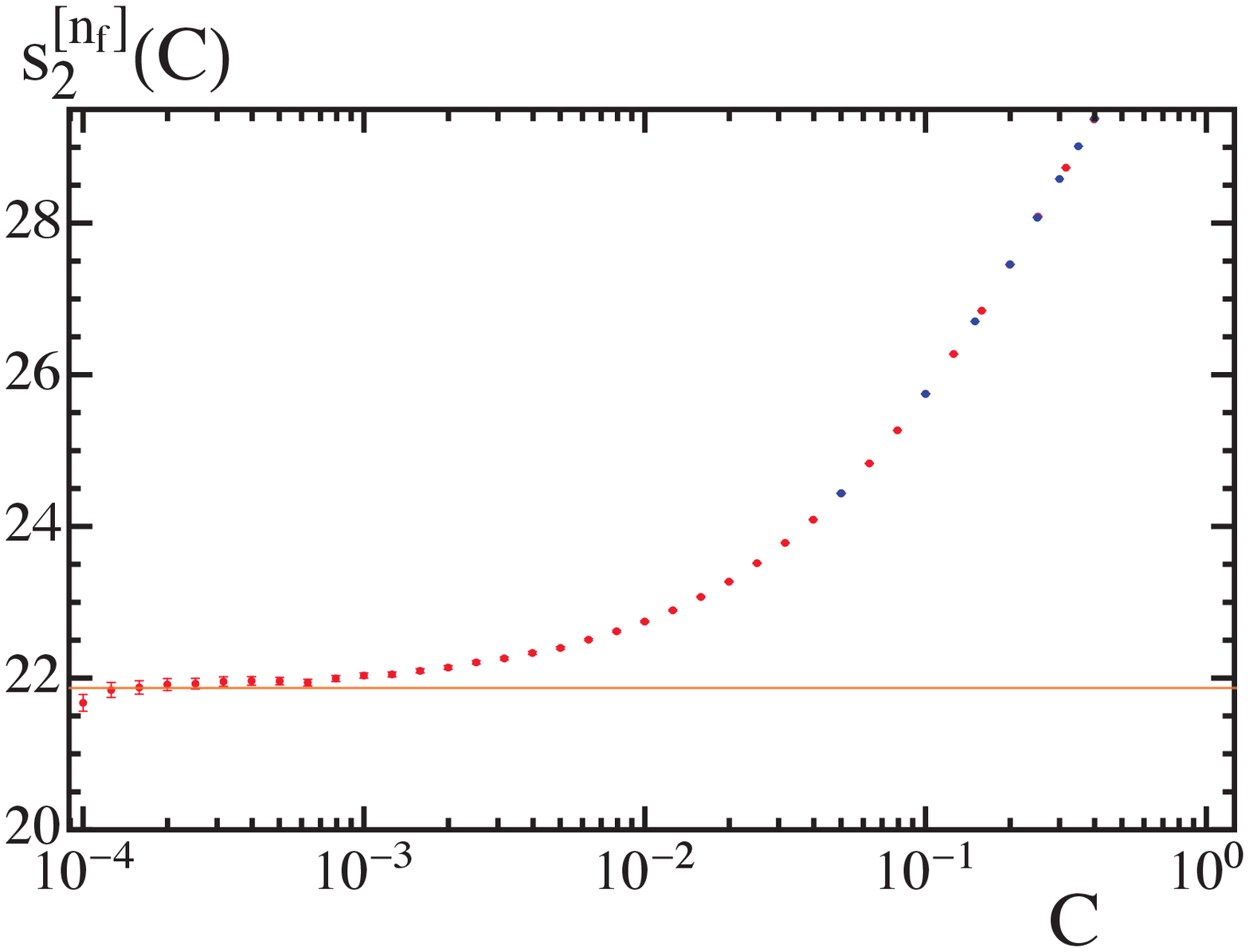}
	\caption[Comparison of methods for determining two loop soft function constant]{Comparison of the determination of the soft function non-logarithmic constants at ${\cal O}(\alpha_s^2)$ as explained in Sec.~\ref{sec:softtwoloop} (shown as horizontal lines), with a determination employing
		the method used in Ref.~\cite{Hoang:2008fs} (plateau at $10^{-4}\lesssim C \lesssim 10^{-3}$). The function $s_2^{\widetilde C}(C)$ is defined in Eq.~(\ref{eq:s2-lim}).
		The $C_F C_A$ and $C_F n_f T_f$ color structures are shown in left and right panels, respectively.
		The logarithmic horizontal axis emphasizes the $C\to 0$ extrapolation. Error bars are included and the blue and red points correspond to data from a different binning.}
	\label{fig:s2}%
\end{figure*}

At $\mathcal{O}(\alpha_s^3)$ we compare to the EERAD3 parton level Monte Carlo output, again splitting the results
in the various color structures. Once again we use logarithmically binned distributions for this exercise. The results are shown in \Fig{fig:EERAD3}. The comparison looks very good for the numerically most relevant color structures (which also have the biggest uncertainties) and quite good for other structures. Slight deviations are observed in some cases for $C < 10^{-3}$, presumably indicating systematic uncertainties due the numerical infrared cutoff. The dominant color structures do not have this problem and have larger uncertainties, so one can still use the distribution even for $C\sim 10^{-4}$ as long as all color structures are summed up.

In Fig.~\ref{fig:s2} we compare our numerical determination of $s_2^{\widetilde C}$, that was described above in Sec.~\ref{sec:softtwoloop}, to the alternate method used in Ref.~\cite{Hoang:2008fs} and show that both procedures yield very similar results. The result from Sec.~\ref{sec:softtwoloop} is shown as an orange line (whose width is its uncertainty). The method of Ref.~\cite{Hoang:2008fs} gives the points. In Ref.~\cite{Hoang:2008fs} $s_2^\tau$ is computed from a relation very similar to Eq.~(\ref{eq:s2extractionfinalintegral}), written in the following form for \mbox{C-parameter}:
\begin{align}\label{eq:s2-lim}
s_2^{\widetilde C} &= \dfrac{1}{\sigma_0}\bigg\{\sigma_{\rm had}^{(2)} -
\Sigma_{\rm s}^{(2)}(1)\Big|_{s_2^{\widetilde C} = 0}-\lim_{C\to 0}\Big[\Sigma_{\rm ns}^{(2)}(1)-\Sigma_{\rm ns}^{(2)}(C)\Big]\bigg\} 
\equiv \lim_{C\to 0} s_2^{\widetilde C}(C)\,,
\end{align}
where we have implicitly defined the function $s_2^{\widetilde C}(C)$, and used
\begin{align}
\Sigma_{\rm s,ns}^{(2)}(C)  &= \int_0^C \df C\,\dfrac{\df \sigma_{\rm s,ns}^{(2)}}{\df C}\,.
\end{align}
Eq.~(\ref{eq:s2-lim}) can be broken down into various color factors. The limit in Eq.~(\ref{eq:s2-lim}) has to be taken numerically from the output of EVENT2. This is best achieved if events are distributed in logarithmic bins, such that the $C\to 0$ region is enhanced, as can be seen in Fig.~\ref{fig:s2}. The limit can be identified as the value at which the log-binned distribution reaches a plateau, which in the case of \mbox{C-parameter} happens for $10^{-4}\lesssim C \lesssim 10^{-3}$. Figure~\ref{fig:s2} shows that our determination of $s_2^{\widetilde C}$ as described in Sec.~\ref{sec:softtwoloop}, represented by an orange line, agrees with the plateau for the two nontrivial color structures.

\section{$\mathbf{B_i, G_{ij}}$ coefficients}
\label{ap:Gijcoefficients}
The resummed cross section at N$^3$LL$^\prime$ can be used to compute various fixed-order coefficients, as in \Eqs{eq:BG1}{eq:BG2}.  The results for coefficients up to ${\cal O}(\alpha_s^3)$ in perturbation theory are summarized here.

The $B_i$ coefficients read
\begin{align}
B_{1}^{[0]} &=C_F \!\left(\dfrac{2 \pi^2}{3}-1 \right), \\[1.5mm]
B_{2}^{[0]} &=C_F^2\! \left(-\,6 \,\zeta
_3+1-\frac{17 \pi ^2}{24}+\frac{11 \pi ^4}{36}\right) \nn \\
&+C_A C_F \!\left(\frac{s_2^{[C_F C_A]}}{4}+\frac{283 \zeta
	_3}{18}-\frac{73 \pi ^4}{360}+\frac{85 \pi
	^2}{24}+\frac{493}{324}\right) \nn \\
&+C_F
n_f T_F\! \left(\frac{s_2^{[n_f]}}{4}-\frac{22 \zeta
	_3}{9}-\frac{7 \pi ^2}{6}+\frac{7}{81}\right), \nn \\[1.5mm]
B_{3}^{[0]} &=C_A^2 C_F \!\left(\frac{620179 \zeta _3}{1944}-\frac{41 \pi
	^2 \zeta _3}{2}-\frac{284 \zeta _3^2}{9}-\frac{217 \zeta
	_5}{18} \right.  \left. -\,\frac{51082685}{209952} \right. 
\nn \\
&
\left.
 \qquad \qquad \qquad +\frac{1294933 \pi
	^2}{34992}-\frac{3641 \pi ^4}{7776}+\frac{4471 \pi
	^6}{102060}\right) \nn \\
&+C_A C_F^2\! \left(\frac{5 \pi ^2
	s_2^{[C_F C_A]}}{24}-\frac{s_2^{[C_F C_A]}}{4}-\frac{2273 \zeta
	_5}{9}+\frac{2 \zeta _3^2}{3} \right.  \left. +\,\frac{248 \pi ^2 \zeta
	_3}{9}-\frac{89 \zeta _3}{27}
 \right. 
 \nn \\
 &
 \left.
  \qquad \qquad \qquad-\frac{14887 \pi
	^6}{68040}+\frac{23093 \pi ^4}{19440}+\frac{172585 \pi
	^2}{3888} -\,\frac{185039}{1296}\right)\nn \\
&+C_A C_F n_f
T_F \!\left(-\frac{352 \zeta _3}{3}+\frac{13 \pi ^2 \zeta
	_3}{3}-\frac{2 \zeta _5}{3} \right.  \left. +\,\frac{1700171}{13122}-\frac{103903 \pi
	^2}{4374}+\frac{227 \pi ^4}{4860}\right)\nn \\
&+C_F^3 \bigg(-\,167\, \zeta
_3\left. +\,\frac{38 \pi ^2 \zeta _3}{3}-\frac{4 \zeta _3^2}{3}+22 \zeta
_5-\frac{4679}{96}+\frac{139 \pi ^2}{18}-\frac{109 \pi
	^4}{40} \right.  \left. +\,\frac{42757 \pi ^6}{136080}\right)
\nn \\
&+C_F^2 n_f
T_F\! \left(\frac{5 \pi ^2
	s_2^{[n_f]}}{24}-\frac{s_2^{[n_f]}}{4}+\frac{368 \zeta
	_5}{9} \right.  \left. -\,\frac{94 \pi ^2 \zeta _3}{9}+\frac{4324 \zeta
	_3}{81}-\frac{497 \pi ^4}{2430}-\frac{35503 \pi
	^2}{1944}+\frac{112073}{972}\right) \nn \\
&+C_F n_f^2 T_F^2\!
\left(\frac{808 \zeta _3}{243}-\frac{190931}{13122}+\frac{257 \pi
	^2}{81}+\frac{52 \pi
	^4}{1215}\right) \nn \\
&+\,\frac{j_3}{4}+\frac{s_3^{\widetilde C}}{8}, \nn \\[1.5mm]
B_1 &= C_F \left(\frac{2 \pi ^2}{3}-\frac{5}{2}\right), \\
B_2&=C_F^2\! \left(-\,6\, \zeta
_3+\frac{41}{8}-\frac{41 \pi ^2}{24}+\frac{11 \pi
	^4}{36}\right)\nn \\
&+C_A C_F \!\left(\frac{s_2^{[C_F C_A]}}{4}+\frac{481 \zeta
	_3}{18}-\frac{73 \pi ^4}{360}+\frac{85 \pi
	^2}{24}-\frac{8977}{648}\right)\nn \\
&+C_F n_f T_F\!
\left(\frac{s_2^{[n_f]}}{4}-\frac{58 \zeta _3}{9}-\frac{7 \pi
	^2}{6}+\frac{905}{162}\right), \nn \\[1.5mm]
B_3&=C_A^2 C_F \!\left(\frac{915775 \zeta _3}{1944}-\frac{41 \pi
	^2 \zeta _3}{2}-\frac{284 \zeta _3^2}{9}+\frac{113 \zeta
	_5}{18}   -\,\frac{95038955}{209952}+\frac{1353739 \pi
	^2}{34992}
\right. \nn \\
& \left.
\qquad \qquad \qquad -\frac{3641 \pi ^4}{7776}+\frac{4471 \pi
	^6}{102060}\right) \nn \\
&+C_A C_F^2\! \left(\frac{5 \pi ^2
	s_2^{[C_F C_A]}}{24}-\frac{5 s_2^{[C_F C_A]}}{8}-\frac{3263 \zeta
	_5}{9}+\frac{2 \zeta _3^2}{3} +\,\frac{314 \pi ^2 \zeta
	_3}{9}+\frac{67 \zeta _3}{108}
\right. \nn \\
& \left.
\qquad \qquad \qquad -\frac{14887 \pi
	^6}{68040}+\frac{14503 \pi ^4}{9720}+\frac{56039 \pi
	^2}{1944}  -\,\frac{87719}{1296}\right)
\nn \\
&+C_A C_F n_f
T_F \!\left(-\,\frac{1952 \zeta _3}{9}+\frac{13 \pi ^2 \zeta
	_3}{3}-\frac{22 \zeta _5}{3} +\,\frac{3585851}{13122}-\frac{109249 \pi
	^2}{4374}+\frac{227 \pi ^4}{4860}\right)
\nn \\
&+C_F^3 \bigg(-\,158\, \zeta
_3 \left. +\,\frac{38 \pi ^2 \zeta _3}{3}-\frac{4 \zeta _3^2}{3}+22\, \zeta
_5-\frac{5093}{96}+\frac{1517 \pi ^2}{144}-\frac{191 \pi
	^4}{60}  +\,\frac{42757 \pi ^6}{136080}\right)
\nn \\
&+C_F^2 n_f
T_F \!\left(\frac{5 \pi ^2 s_2^{[n_f]}}{24}-\frac{5
	s_2^{[n_f]}}{8}+\frac{728 \zeta _5}{9} -\,\frac{118 \pi ^2 \zeta
	_3}{9}
\right. \nn \\
& \left.
\qquad \qquad \qquad+\frac{2839 \zeta _3}{81} -\frac{497 \pi ^4}{2430}-\frac{24973
	\pi ^2}{1944}+\frac{188173}{1944}\right) 
\nn \\
&+C_F n_f^2
T_F^2 \!\left(\frac{4912 \zeta
	_3}{243}-\frac{484475}{13122}
+\frac{275 \pi ^2}{81}+\frac{52 \pi
	^4}{1215}\right) +\,\frac{j_3}{4}+\frac{s_3^{\widetilde C}}{8}. \nn
\end{align}
The results for the first few $G_{ij}$ coefficients read
\begin{align}
G_{12}=& -2\, C_F, \:\:\:\: G_{11}= 3\, C_F,\\[1.5mm]
G_{23}=&\,C_F \!\left(\frac{4 n_f T_F}{3}-\frac{11 C_A
}{3}\right), \nn \\[1.5mm]
G_{22}=&C_F \!\left[ C_A\!
\left(\frac{\pi ^2}{3}- \frac{169}{36}\right)-\frac{4}{3} \pi ^2 C_F+\frac{11 n_f
	T_F}{9}\right], \nn \\[1.5mm]
G_{21}=&\, C_F\!\left[ C_A\! \left(-\,6\, \zeta _3+\frac{57}{4}+\frac{11 \pi
	^2}{9}\right) +\,C_F\! \left(-\,4\, \zeta _3+\frac{3}{4}+\pi
^2\right) +n_f
T_F \!\left(-\,5-\frac{4 \pi ^2}{9}\right) \right],  \nn \\[1.5mm]
G_{34}=&\,C_F \!\left(-\,\frac{847 C_A^2}{108}+\frac{154}{27} C_A
n_f T_F-\frac{28}{27} n_f^2
T_F^2 \right), \nn \\[1.5mm]
G_{33}=&\,C_F\! \left[ C_A^2\!
\left(\frac{11 \pi ^2}{9}-\frac{3197}{108}\right) -\frac{22\pi^2}{3} C_A
C_F +C_A
n_f T_F\! \left(\frac{512}{27}-\frac{4 \pi ^2}{9}\right) +
\right. \nn \\
&\left.
\frac{64 C_F^2 \zeta
	_3}{3}+\,C_F n_f T_F\! \left(2+\frac{8 \pi ^2}{3}\right) -\frac{68}{27} n_f^2 T_F^2 \right], \nn \\
&\nn \\
G_{32}=&\,C_F\! \left[ C_A^2\! \left(11\, \zeta _3-\frac{11323}{648}+\frac{497 \pi
	^2}{54}-\frac{11 \pi ^4}{90}\right) \right. + C_A C_F \!\left(-\,110\,
\zeta _3+\frac{11}{8}-\frac{70 \pi ^2}{27}+\frac{4 \pi^4}{9}\right) \nn \\
&+C_A n_f T_F \!\left(4\, \zeta
_3+\frac{673}{162}-\frac{152 \pi ^2}{27}\right) +C_F^2\!
\left(\frac{8 \pi ^4}{45}-48 \zeta _3\right)
\nn \\
&+C_F n_f
T_F \!\left(32\, \zeta _3+\frac{43}{6}+\frac{8 \pi
	^2}{27}\right)\left.+\,n_f^2 T_F^2\! \left(\frac{70}{81}+\frac{8 \pi ^2}{9}\right) \right]
, \nn \\[1.5mm]
G_{31}=&\,C_A^2 C_F\! \left(\frac{11 s_2^{[C_F C_A]}}{6}+10\, \zeta
_5-\frac{361 \zeta _3}{27}-\frac{541 \pi ^4}{540} +\frac{892 \pi
	^2}{81} \right. \nn \\
& \left. +\,\frac{77099}{486}\right)+C_A C_F^2\!
\left(\frac{452 \zeta _3}{9}+2 \pi^2 \zeta _3+30\, \zeta
_5  +\frac{23}{2} \right. \nn \\
& \left. +\,\frac{161 \pi^2}{72}-\frac{49 \pi
	^4}{135}\right)+C_A C_F n_f T_F\!
\left(-\,\frac{2 s_2^{[C_F C_A]}}{3} \right. \nn \\
& \left. +\,\frac{11 s_2^{[n_f]}}{6} -\frac{608
	\zeta _3}{27}+\frac{10 \pi ^4}{27}-\frac{520 \pi
	^2}{81}-\frac{24844}{243}\right) \nn \\
&+C_F^3 \!\left(53\, \zeta
_3-\frac{44 \pi^2 \zeta _3}{3}+132\zeta _5+\frac{29}{8}+\frac{5
	\pi ^2}{4}-\frac{8 \pi ^4}{15}\right) \nn \\
&+C_F^2 n_f T_F\!
\left(-\,\frac{208 \zeta _3}{9}-\frac{77}{4}-\frac{31 \pi
	^2}{18}+\frac{8 \pi ^4}{135}\right) \nn \\
&+C_F n_f^2 T_F^2\!
\left(\!-\,\frac{2 s_2^{[n_f]}}{3}+\frac{176 \zeta _3}{27}+\frac{64 \pi
	^2}{81}+\frac{3598}{243}\right). \nn
\end{align}
Note that the entire infinite series of $G_{ij}$ coefficients listed in \Tab{tab:Gijorders} is determined by our resummation results.

\section{R-evolution with and without Hadron Mass Effects}
\label{ap:hadronmassR}
The result for R-evolution in the case of no hadron masses is given in \Eq{eq:DeltaRevolution}.
The resummed $\omega$ appearing in this equation was given in App.~\ref{ap:formulae}. The remaining coefficients and variables that appear in this equation are
\begin{align} \label{eq:Sjnor}
S_0 &= \frac{\gamma_0^R}{2 \beta_0} \,,\,\,
S_1  = \frac{\gamma_1^R}{(2 \beta_0)^2} - (\hat{b}_1 + \hat{b}_2) \frac{\gamma_0^R}{2 \beta_0}\,, \nn \\
S_2 &=  \frac{\gamma_2^R}{(2 \beta_0)^3} - (\hat{b}_1 + \hat{b}_2) \frac{\gamma_1^R}{(2 \beta_0)^2}\,,  \nn \\
\hat{b}_1 &= \frac{\beta_1}{2 \beta_0^2}\,, \;\; \hat{b}_2 = \frac{\beta_1^2 - \beta_0 \beta_2}{4 \beta_0^4}\,,\\
\hat{b}_3 &= \frac{\beta_1^3-2 \beta_0 \beta_1 \beta_2 + \beta_0^2 \beta_3}{8 \beta_0^6}\,,  \nn \\
t_1 &= -\,\frac{2 \pi}{\beta_0 \alpha_s(R)},\,t_0 = -\,\frac{2 \pi}{\beta_0 \alpha_s(R_\Delta)}\,. \nn
\end{align}
\begin{figure}[t!]
	\begin{center}
		\includegraphics[width=0.5\columnwidth]{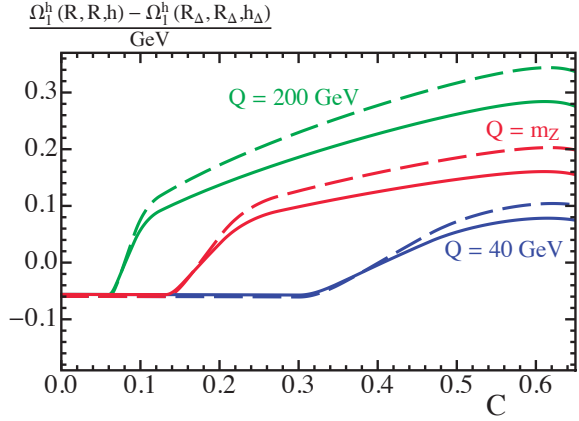}
		\caption[Running of the short-distance power correction $\Omega_1^h(R,R,h)$]{Running of the short-distance power correction $\Omega_1^h(R,R,h)$ with respect to the reference value $\Omega_1^h(R_\Delta,R_\Delta,1)$. The scale $R$ is set to the default profile, and the scheme parameter $h$ is set to the function displayed in Eq.~(\ref{eq:arctan}). Red , blue,  and green correspond to center-of-mass energies of $91.2$, $40$, and $200$\,GeV, respectively. The solid lines do not include hadron-mass effects, whereas the dashed ones do.}
		\label{fig:Omega-run}
	\end{center}
\end{figure}

The impact of R-evolution on the value of the hadronic parameter $\Omega_1(R,R)$ is shown in \Fig{fig:Omega-run} (with and without hadron-mass effects). Here we have set $\mu = R$ and used the default profile function for $R(C)$. We actually plot the hybrid scheme $\Omega_1^h(R,R,h)$ which accounts for a reasonable treatment of threshold effects at the shoulder $C=0.75$ and which is discussed in detail below in \App{ap:ArcTan}. The  effect of using the hybrid scheme rather than  the Rgap scheme is quite small in the region of this figure but does cause the bending over of the curves that is visible at $C=0.6$. Above the shoulder the hybrid scheme uses the $\msbar$ power correction, which has a flat behavior. 

When we introduce hadron-mass effects, it is necessary to extent these results to account for the value of $r$,
the transverse velocity. Due to the running in \Eq{eq:omega1-r-murunning}, the scheme change result in \Eq{eq:omegaschemechange} now becomes
\begin{align} \label{eq:rrunningschemechange}
g_C(r)\,\Omega_1^C(R,\mu,r) &= g_C(r)\,\overline{\Omega}_1(\mu,r) - \delta (R,\mu,r)\,,
\end{align}
where there is additional $\mu$ dependence from the hadron-mass induced running. Our scheme for $\delta$ now becomes
\begin{align} \label{eq:delta-scheme-littler}
\delta(R,\mu,r) &= \left[ \frac{\alpha_s(\mu)}{\alpha_s(R)}\right]^{\hat{\gamma}_1(r)}\!\delta(R,\mu)\,, 
\end{align}
and the $r$ dependence is encoded in the known one-loop anomalous dimension~\cite{Mateu:2012nk}
\begin{align} \label{eq:gamma1hatdef}
\hat{\gamma}_1(r) &= \frac{2\, C_{\!A}}{\beta_0} \ln (1 - r^2)\,.
\end{align}
With this $r$-dependent scheme change, we can once again derive the equations that govern the $R$ evolution and $\mu$
running for $\Omega_1^C$, which following Ref.~\cite{Hoang:2015zz} become
\begin{align} \label{eq:RandmuRGElittler}
&  R \frac{\df}{\df R} \Big\{ \Big[ g_C(r)\, \Omega_1^C (R,\mu,r) 
\big( \alpha_s(\mu) \big)^{-\hat\gamma_1(r)} \Big]_{\mu=R} \Big\}
\\[4pt]
&\quad
= -\, R\, \big(\alpha_s(R)\big)^{-\hat{\gamma}_1(r)} \sum_{n=0}^\infty \gamma_{n}^{R\,r}(r) \left( \frac{\alpha_s(R)}{4 \pi} \right)^{\!\!n+1}
\, , \nn \\
& \mu \frac{\df}{\df\mu} \Big\{  g_C(r)\,\Omega_1^C (R,\mu,r) 
\big( \alpha_s(\mu) \big)^{-\hat\gamma_1(r)}  \Big\}
\nn\\
&\quad
= 2\, R\, e^{\gamma_E} \big(\alpha_s(R)\big)^{-\hat{\gamma}_1(r)}
\sum_{n=0}^\infty \Gamma_n^\text{cusp}
\left( \frac{\alpha_s(\mu)}{4 \pi} \right)^{\!\!n+1}
, \nn
\end{align}
for
\begin{align} \label{eq:gammaRlittler}
\gamma_0^{R\,r}(r) &= \gamma_0^R, \qquad 
\gamma_1^{R\,r}(r) = \gamma_1^R + 2 \beta_0 \hat{\gamma}_1(r)  \gamma_0^R  \,, \nn \\
\gamma_2^{R\,r}(r) &= \gamma_2^R + 2 \beta_0 \hat{\gamma}_1(r) (\gamma_1^R + 2 \beta_0 \gamma_0^R) \,.
\end{align}
Solving this set of equations gives \Eq{eq:omega1R-r-murunning}, with the $S_i^r$ given by
\begin{align} \label{eq:littlerSdef}
S^r_0(r) &= S_0\,,\qquad
S^r_1 (r) = S_1
+ \frac{\hat{\gamma}_1(r) \gamma_0^R}{2 \beta_0}\,, \\
S^r_2(r) & =  S_2 + (\gamma_1^R + 2 \beta_0 \gamma_0^R)\nn \\
&\ \ \times  \bigg[ (1 + \hat b_1)\, \hat b_2 + \frac{\hat b_2^2 + \hat b_3}{2} \bigg]
\frac{\hat{\gamma}_1(r)}{(2 \beta_0)^2}\,. \nn
\end{align}
Values for the $\gamma_i^R$ anomalous dimensions were given above in Eq.~(\ref{eq:gammaR}).

\section{Gap Scheme in the Fixed-Order Region}
\label{ap:ArcTan}

In the fixed-order (or far-tail) region, there is no longer a hierarchy between the hard, jet, and soft scales, and one sets them equal to reproduce the fixed-order QCD predictions exactly. This is done through our profiles.
In this region the singular and non-singular terms are of similar size, and hence the factorization of the cross section in a hard factor and a convolution of jet and soft functions is no longer relevant.
The issue is further complicated by the analytic structure of the shoulder region, which is located in the fixed-order region and contains the integrable singularity at $C = 0.75$ which has its own logarithmic series (see Fig.~\ref{fig:component-plot} and the accompanying discussion there). Thus, in the far-tail region $C\gtrsim 0.75$, the structure of nonperturbative 
corrections is likely to differ from that of the shape function $F_C$ and thus is unknown at this time. Nevertheless, we do expect a smearing by a function whose width is $\sim\Lambda_{\rm QCD}$, and hence our smearing by $F_C$ is simply a proxy for a more detailed analysis in this region.  Since fits for $\alpha_s$ can be carried out with $C<0.75$, the treatment of this region, and the discussion below, are not relevant for predicting the shape in the fit region.  The region $C\ge 0.75$ does contribute when computing the total cross section from our resummed result, and this motivates us to use a cross section formula that still obtains realistic results in this region.

Due to the shoulder at $C=0.75$, the use of the infrared subtractions $\delta$ in this region can cause an unphysical behavior of the cross section. The reason for this is that these subtractions yield derivatives acting on the partonic cross section. In the shoulder region with the singular discontinuity starting at $\mathcal{O}(\alpha_s^2)$, these derivatives can cause an artificially enhanced singular behavior if the subtraction is not carefully defined in this region. If this is not done, the convolution with the shape function $F_C$ may be insufficient to achieve a smooth cross section near \mbox{$C\simeq 0.75$}. In Ref.~\cite{Catani:1998sf} it was shown that the singularities at $C = 0.75$ can be cured by including soft gluon resummation which makes the cross section smooth. However, this treatment does not resolve the question of the proper field theoretic nonperturbative function for this region, nor any accompanying infrared subtractions due to renormalons. 

To deal with the region $C\gtrsim 0.75$ we take an alternative approach, which is to implement a smooth transition between the Rgap scheme in the dijet region to $\msbar$ in the fixed-order region. This avoids any subtractions at $C=0.75$. To that end we define a new scheme that depends on a continuous parameter $h$ which takes values between $0$ and $1$ and that smoothly switches off the gap subtractions when we get near $C=0.75$. We start by rewriting \Eq{eq:deltasplitting} as
\begin{align} \label{eq:deltasplittingH}
\Delta_C  & =  \frac{3\pi}{2}[\,\bar{\Delta}^h(R,\mu,h) + h\,\delta(R,\mu)\,]\,.
\end{align}
This defines a hybrid short-distance scheme for $\Omega_1$ which we call $\Omega_1^h$,
\begin{align} \label{eq:omegaschemechangeH}
\Omega_1^h(R,\mu,h) &= \overline{\Omega}_1^C - 3\pi\,h\,\delta (R,\mu)\,,
\end{align}
which becomes the $\msbar$ scheme for $h = 0$ and the Rgap scheme for $h = 1$. One can easily derive RGE equations in $\mu$ and $R$, for $\bar{\Delta}^h$, which we write in the convenient form\,\footnote{The right-hand sides of the running
	equations for the power correction $\Omega_1^h(R,\mu_S,h)$ are simply $3\pi$ times the right-hand sides of
	Eqs.~(\ref{eq:running-h}) and (\ref{eq:change-h}).}
\begin{align}\label{eq:running-h}
R \frac{\df}{\df R}  \bar{\Delta}^h (R,R,h) & = -\,h\,R \, \gamma^R [\alpha_s(R)] \, , \\
\mu \frac{\df}{\df\mu}  \bar{\Delta}^h (R,\mu,h) &= 2\, R\,h\, e^{\gamma_E} \Gamma_{\rm cusp}[\alpha_s(\mu)] \, ,\nn
\end{align}
and a relation to switch from different $h$ schemes:
\begin{align}\label{eq:change-h}
\bar{\Delta}^h (R,\mu,h_1) - \bar{\Delta}^h (R,\mu,h_2) = (h_2 - h_1)\,\delta(R,\mu)\,.
\end{align}
The solution to these three equations is rather simple,
\begin{align}\label{eq:sol-run-h}
\bar{\Delta}^h (R,\mu,h) & = \bar{\Delta}^h (R_\Delta,\mu_\Delta,h_\Delta) - (h - h_\Delta)
\delta(R,\mu) \nn\\
&+ h_\Delta\,\Delta^{\rm diff}(R_\Delta,R,\mu_\Delta,\mu)\,,
\end{align}
where $\Delta^{\rm diff}$ has been defined in \Eq{eq:DeltaRevolution}. We choose to evolve first in $R$
and $\mu$ in the $h_\Delta$ scheme, where above the peak region [see Eq.~(\ref{eq:muRprofile})] there is only a single evolution since $\mu_S(C)=R(C)$. Close to the shoulder region \mbox{$C\sim 0.75$}, we then smoothly transform from the $h_\Delta$ scheme to the $h$ scheme. This implements the transition from the Rgap scheme with ${\cal O}(\Lambda_{\rm QCD})$ renormalon subtraction to the  $\msbar$ scheme where this renormalon is not subtracted. The procedure entails a 
residual dependence on $R$ in the region $C\gtrsim 0.75$ even once $h=0$, which comes from the fact that we are transforming from Rgap to $\msbar$ at the scale $R>R_\Delta$. This residual dependence leads to a somewhat smaller effect of the ${\cal O}(\Lambda_{\rm QCD})$ renormalon even though one employs $\overline{\Omega}_1$ in this region.

In the hybrid scheme described above, the first moment of the shape function reads
\begin{equation}
\!\!\int\! \!\df k \,k\,F_C(k) = \Omega_1^h(R_\Delta,\mu_\Delta,h_\Delta) \,-\,
3\pi\, \bar\Delta^h(R_\Delta,\mu_\Delta,h_\Delta)\,.
\end{equation}
For the practical implementation, we choose $h_\Delta = 1$ and thus identify $\Delta^h(R_\Delta,\mu_\Delta,1) = \Delta(R_\Delta,\mu_\Delta)$ as well as $\Omega_1^{h}(R_\Delta,\mu_\Delta,1)=\Omega_1(R_\Delta,\mu_\Delta)$. In the numerical codes, this amounts to inserting a factor $h$ in front of each $\delta$ and substituting each $\bar{\Delta}(R,\mu)$ of \Eq{eq:DeltaRevolution} appearing in any of the equations shown in the main text by $\bar{\Delta}^h(R,\mu,h)$ of \Eq{eq:sol-run-h}. 
Note again that all of the changes induced by the use of the hybrid scheme rather than the Rgap scheme only influence the shape of the C-parameter cross section for $C\gtrsim 0.75$.

\begin{figure*}[t!]
	\begin{center}
		\includegraphics[width=0.33\columnwidth]{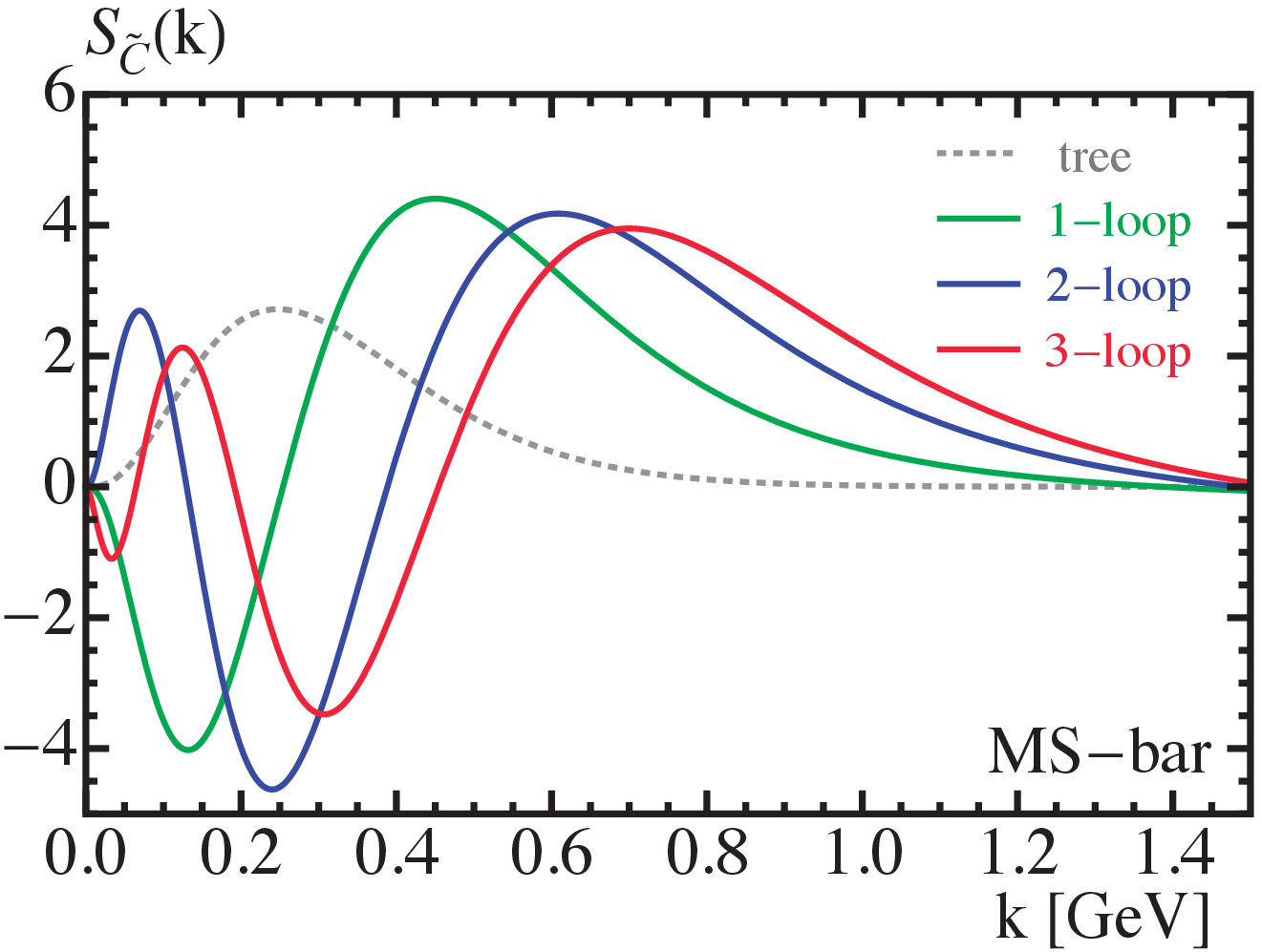}~
		\includegraphics[width=0.33\columnwidth]{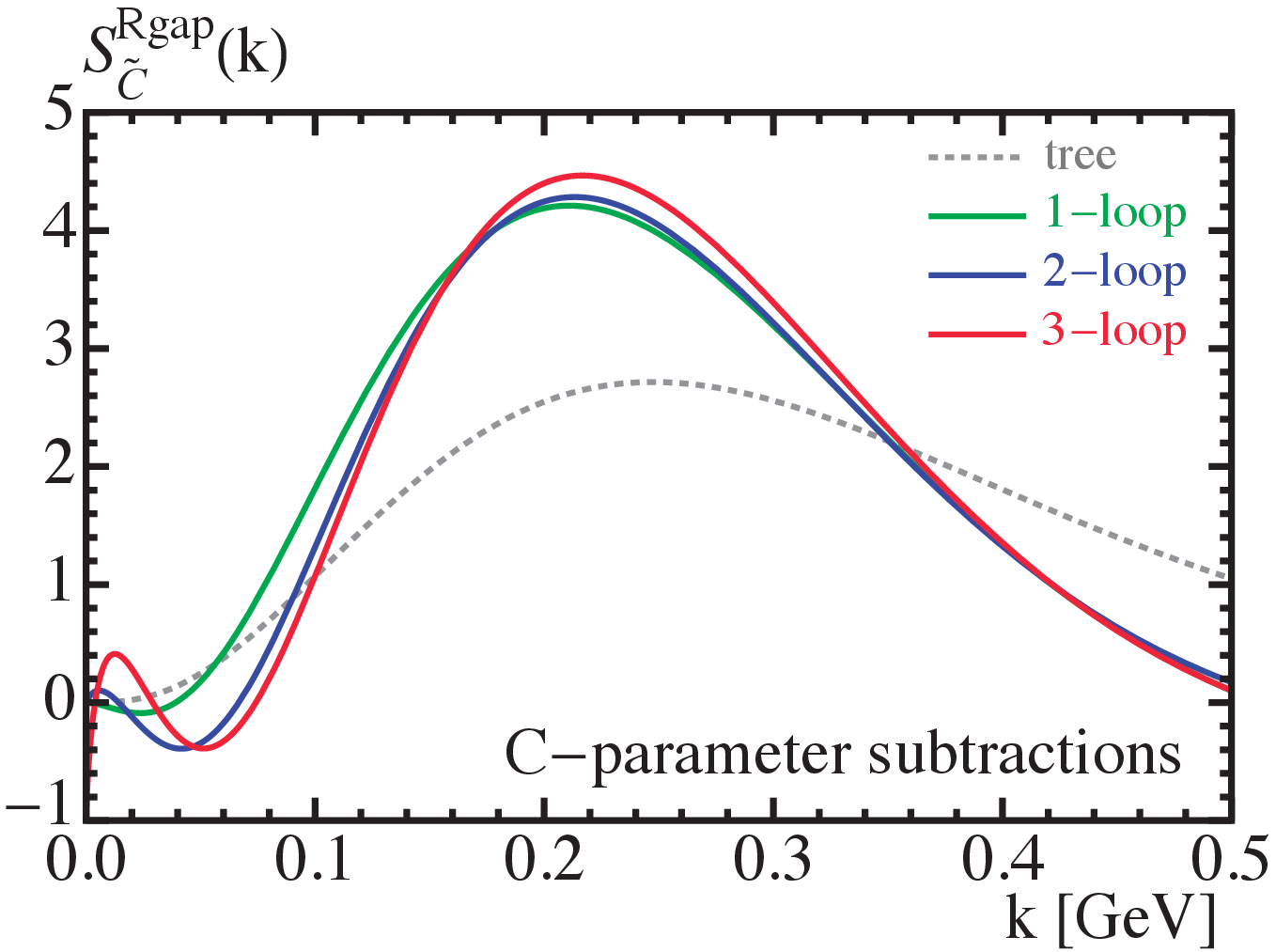}
		\includegraphics[width=0.33\columnwidth]{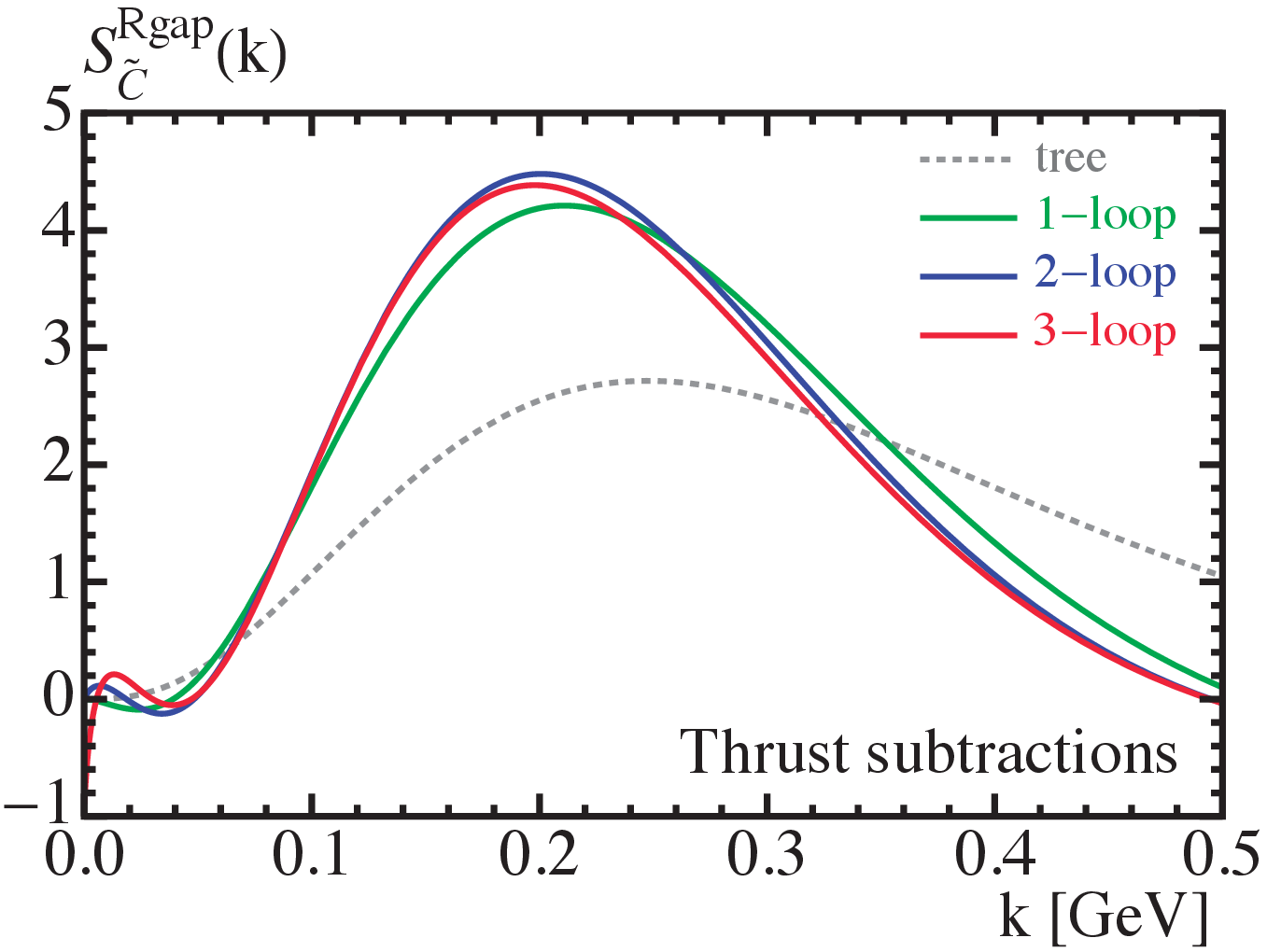}
		\caption[Convolution of perturbative and nonperturbative soft C-parameter functions in various schemes]{Convolution $S = \hat S \otimes F$ combining the perturbative and nonperturbative C-parameter soft functions in the $\overline{\rm MS}$ scheme (left), Rgap with C-parameter subtractions (center), and Rgap with thrust subtractions (right). The results at 1-, 2- and 3-loops are shown in green, blue, and red, respectively, whereas the tree level result is depicted as a gray dotted line. We use $\mu = 1\,$GeV, $R=0.8\,$GeV, $\alpha_s(m_Z) = 0.1141$, and a fixed generic shape function whose first moment is $\Omega_1 = 0.33\,$\,GeV.}
		\label{fig:gap-effects-soft}
	\end{center}
\end{figure*}

One can also easily extend the hybrid scheme to account for hadron-mass effects by defining
\begin{align} \label{eq:rrunningschemechangeh}
g_C(r)\,\Omega_1^{C,h}(R,\mu,h,r) &= g_C(r)\,\overline{\Omega}_1^C(\mu,r) - h\,\delta (R,\mu,r)\,,
\end{align}
and the evolution and scheme-transformation equations simply read
\begin{align} \label{eq:RandmuRGElittlerh}
& R \frac{\df}{\df R} \Big\{\Big[g_C(r)\big(\alpha_s(\mu)\big)^{-\hat{\gamma}_1(r)}\Omega_1^{C,h} (R,\mu,h,r)\Big]_{\mu=R}\Big\} \\
&= -\,h \,R \big(\alpha_s(R)\big)^{-\hat{\gamma}_1(r)} \gamma_r^{R}[\alpha_s(R),r] \, , \nn \\
&\mu \frac{\df}{\df\mu} \Big\{g_C(r)\big(\alpha_s(\mu)\big)^{-\hat{\gamma}_1(r)}\Omega_1^{C,h} (R,\mu,h,r)\Big\}\nn\\
&= 2 \, h\, R\, e^{\gamma_E} \big(\alpha_s(R)\big)^{-\hat{\gamma}_1(r)} \Gamma_{\rm cusp}[\alpha_s(\mu)] \, , \nn\\
&g_C(r)\big[\,\Omega_1^{C,h} (R,\mu,h_1,r) - \Omega_1^{C,h} (R,\mu,h_2,r)\big]\nn \\
& = (h_2 - h_1)\,\delta(R,\mu,r)\,.\nn
\end{align}
The solution to these equations is again simple:
\begin{align}\label{eq:sol-run-Omega-h}
&g_C(r)\,\Omega_1^{C,h} (R,\mu,h,r)  = \\
&g_C(r)\!\left[ \frac{\alpha_s(\mu)}{\alpha_s(\mu_\Delta)} \right]^{\hat{\gamma}_1(r)}\!
\Omega_1^{C,h}(R_\Delta, \mu_\Delta,h_\Delta,r)\nn \\
&\!\!\!- (h_\Delta - h)\, \delta(R_\Delta,\mu_\Delta,r)
+ h_\Delta\,\Delta^{\rm diff}(R_\Delta,R,\mu_\Delta,\mu,r)\,.\nn
\end{align}
For phenomenological analyses that consider $C\gtrsim 0.75$ or integrate over $C$ to compute a normalization, we must specify $h=h(C)$. It must be a function of $C$ which smoothly interpolates between the values
$1$ and $0$ as one transitions from the resummation to the fixed-order region near \mbox{$C\sim 0.75$}. To achieve this we use the following simple form:
\begin{align}\label{eq:arctan}
h(C) & = \frac{1}{2} - \frac{1}{\pi} \arctan\,[\,\eta\,(C - C_0)\,]\,,\\
C_0 & = 0.7\,,\; \eta = 30\,.\nn
\end{align}

\section{Rgap Scheme based on the C-Parameter Soft Function}
\label{ap:subtractionchoice}
As an alternative to the Rgap scheme subtraction function used in this work and defined in \Eq{eq:delta-scheme}, one may employ the analog relation based on the C-parameter partonic soft function:
\begin{equation} \label{eq:delta-C-scheme}
\delta_{\widetilde{C}}(R,\mu) = R\, e^{\gamma_E} \frac{\df}{\df \ln(ix)}
\big[ \ln S^{\text{part}}_{\widetilde C}(x,\mu) \big]_{x = (i R e^{\gamma_E})^{-1}}.
\end{equation}
In this scheme the analogue of \Eq{eq:soft-nonperturbative-subtract} reads
\begin{align} \label{eq:soft-nonperturbative-subtractC}
\!\!\!S_C(k,\mu) & = \!\!\int \!\df k' \,e^{-\, 6\,\delta_{\widetilde C} \frac{\partial}{\partial k}}
\hat S_C(k-k',\mu) F_C(k' - 6\bar{\Delta}_{\widetilde{C}})\,,
\end{align}
with 
\begin{align} \label{eq:deltasplittingC}
\Delta_{\widetilde C}  & =  \,\bar{\Delta}_{\widetilde{C}}(R,\mu) + \delta_{\widetilde{C}}(R,\mu)\,\,.
\end{align}
Here the subtraction function $\delta_{\widetilde{C}}$ can be written as
\begin{align} \label{eq:deltaseriesC}
\delta_{\widetilde{C}}(R,\mu) =
R \,e^{\gamma_E} \sum_{i=1}^\infty \alpha_s^i(\mu)\, \delta^i_{\widetilde{C}}(R,\mu)\,,
\end{align}
where the coefficients for five light flavors read
\begin{align} \label{eq:d123C}
\delta_{\widetilde{C}}^1(R,\mu) &= -\,1.69765\, L_R \,, \nn \\
\delta_{\widetilde{C}}^2(R,\mu) &= 0.539295 - 0.933259\, L_R - 1.03573\, L_R^2 \,, \nn \\
\delta_{\widetilde{C}}^3(R,\mu) &= 0.493255 + 0.0309077\, s_2^{\widetilde C} + 0.833905\, L_R
\nn\\&\quad -\,1.55444\, L_R^2 - 0.842522\, L_R^3 \,,
\end{align}
for $L_R = \ln(\mu/R)$.

Figure~\ref{fig:gap-effects-soft} shows the effect of the renormalon subtractions on the soft function $S_C(k,\mu)$ 
from \Eqs{eq:soft-nonperturbative-subtract}{eq:soft-nonperturbative-subtractC} [we will refer to using \Eq{eq:soft-nonperturbative-subtract} as thrust subtractions and to using \Eq{eq:soft-nonperturbative-subtractC} as C-parameter subtractions], which are compared with the result in the $\overline{\rm MS}$ scheme with any subtractions. The key thing to consider is the stability of the soft function when higher orders in perturbation theory are included, illustrated by the green, blue, and red curves at 1-, 2-, and 3-loop orders. In the leftmost panel of Fig.~\ref{fig:gap-effects-soft}, we show the C-parameter soft function in the $\overline{\rm MS}$ scheme. Here the presence of the $\Lambda_{\rm QCD}$ renormalon is apparent from the shifting of the soft function to the right as we increase the perturbative order. The $\overline{\rm MS}$ result also exhibits  a large negative dip at small momentum, which makes predictions for the cross section at small $C$ inaccurate in this scheme. With either the C-parameter subtractions (middle panel) or the thrust subtractions (rightmost panel), one achieves significantly better convergence for the soft function and alleviates most of the negative dip.  Making an even closer comparison of the C-parameter and thrust subtraction results, it becomes evident that the thrust subtractions exhibit better convergence near the peak (comparing the difference between the blue and red lines in the two panels) and also more completely remove the negative dip at small momenta. (Similar conclusions hold for the thrust soft function, where again thrust subtractions are preferred.) This improvement for the thrust subtractions can be traced back to the fact that the sign of the non-logarithmic 2-loop term in the C-parameter subtractions in \Eq{eq:d123C} is positive, which is opposite to the sign of the renormalon. In the resummation region,  $R(\tau)=\mu_S(\tau)$, so $L_R\to 0$, and numerically the subtraction goes in the opposite direction to the renormalon in this scheme at 2-loops. This term has an impact even when the logarithmic terms are active in the small $C$ nonperturbative region, which we can see by taking $R=R_0$ and $\mu_S=\mu_0$.  For the thrust subtractions, this gives $ \pi/2 \{\delta^1, \delta^2, \delta^3\} = \{-\,0.603, -\,0.743, -\,1.621\}$, whereas for the C-parameter subtractions, we have $\{\delta_{\widetilde C}^1, \delta_{\widetilde C}^2, \delta_{\widetilde C}^3\} = \{-\,0.767, -\,0.094, -\,0.861\}$. The small value for this \mbox{2-loop} C-parameter subtraction coefficient is due to the positive constant term.

\chapter{Computation of 1-loop Soft Function}
\label{ap:softoneloop}

%
In this section we present a general computation of the 1-loop soft function for any
event shape $e$ which can be expressed in the dijet limit as
\begin{align}
e = \frac{1}{Q}\sum_i p_i^\perp f_e(y_i)\,,
\end{align}
where the sum is over all particles in the final state, $p_i$ is the magnitude of the
transverse momentum and $y_i$ is the rapidity of the particle, both measured with
respect to the thrust axis.\,\footnote{For perturbative computations partons are taken as massless and
	hence rapidity $y$ and pseudorapidity $\eta$ coincide.} For thrust one has $f_\tau(y) = \exp(-|y|)$, for angularities
one has $f_{\tau^a}(y) = \exp[-(1-a)|y|\,]$, and for C-parameter one has $f_{C}(y) = 3/\cosh y$ and
$f_{\widetilde C}(y) = 1/(2\cosh y)$.

\begin{figure}
	\begin{center}
		\includegraphics[width=0.5\columnwidth]{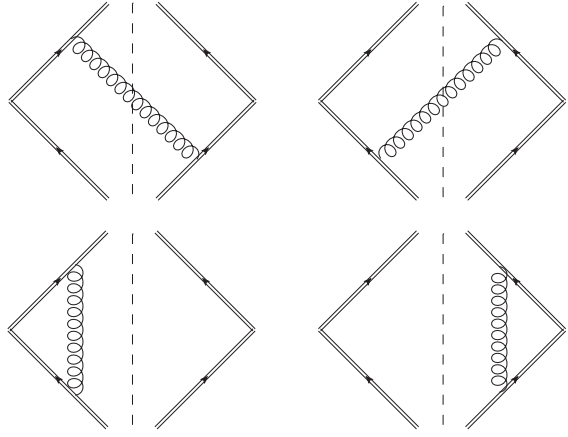}
		\caption[One loop soft function diagrams]{Diagrams contributing to the soft function at one loop.}
		\label{fig:soft-one-loop}
	\end{center}
\end{figure}
One needs to compute the four diagrams in Fig.~\ref{fig:soft-one-loop} in order to
determine the soft function. The two diagrams on the bottom are scaleless and vanish in
dimensional regularization. They actually convert the IR divergences in the two diagrams
on the top into UV divergences. We take the space-time number of dimensions to be
\mbox{$d = 4 - 2\epsilon$}. A direct computation in momentum space gives
\begin{align}\label{eq:soft-general}
\!\!\!S^{\rm 1-loop}_e(\ell) = 4\, g_s^2 C_F \!\!\int \!
\frac{\df^{3-2\epsilon}\vec{p}}{(2\pi)^{3-2\epsilon}2 |\vec{p}\,|}
\frac{\delta[\ell - p_T f_e(y)]}{p^+ p^-}\,.
\end{align}
After integrating the angular variables, it is convenient to make a change of variables
from $p^\pm$ to ($p_T$, $y$):
\begin{align}\label{eq:chage-variable}
\frac{\df^{3-2\epsilon}{\vec p}}{(2\pi)^{3-2\epsilon}2|\vec{p}\,|} = 
\frac{2}{(4\pi)^{2-\epsilon}}\frac{p_T^{1-2\epsilon}}{\Gamma(1-\epsilon)}
\,\df p_T \df y\,.
\end{align}
Using Eq.~(\ref{eq:chage-variable}) in Eq.~(\ref{eq:soft-general}) and imposing the
on-shell condition $p^+p^- = p_T^2$, where $\vec{p\,}^2 = p_z^2 + p_T^2$, we obtain
\begin{align}\label{eq:soft-momentum}
& S_e^{\rm 1-loop}(\ell)  \\
&= \frac{2\,\alpha_s(\mu) C_F e^{\epsilon \gamma_E}}{\mu\,\pi\,\Gamma(1-\epsilon)}
\!\int \! \df p_T\, \df y\, \Big(\frac{p_T}{\mu}\Big)^{-1-2\epsilon} 
\delta[\ell - p_T f_e(y)] \nonumber\\
&= \frac{2\,\alpha_s(\mu) C_F
	e^{\epsilon \gamma_E}}{\mu\,\pi\,\Gamma(1-\epsilon)}
\Big(\frac{\ell}{\mu}\Big)^{-1-2\epsilon}\,I_e(\epsilon)\,,\nonumber
\end{align}
where we have defined
\begin{align}
I_e(\epsilon) = \int_{-\infty}^\infty\! \df y\, [\,f_e(y)\,]^{2\epsilon}\,.
\end{align}
Similarly in Fourier space, one gets
\begin{align}\label{eq:soft-position}
\!\!\!\tilde{S}_e^{\rm 1-loop}(x)  = \frac{2\,\alpha_s C_F}{\pi}
\frac{\Gamma(-2\epsilon)}{\,\Gamma(1-\epsilon)}
(i\, x\, \mu)^{2\epsilon}e^{\epsilon \gamma_E}\,I_e(\epsilon)\,.
\end{align}
For thrust and angularities, one trivially obtains
\begin{align}
I_{\tau}(\epsilon) = \frac{1}{\epsilon}\,,\qquad
I_{\tau^a}(\epsilon) = \frac{1}{1-a}\frac{1}{\epsilon}\,.
\end{align}
For C-parameter it is convenient to perform a change of variables,
\begin{align}
y = \frac{1}{2}\ln\!\Big(\frac{1+x}{1-x}\Big)\,,\quad
\cosh y = \frac{1}{\sqrt{1-x^2}}\,,
\end{align}
obtaining for the integral:
\begin{align}
I_{\widetilde C}(\epsilon) & = 2^{1-2\epsilon}\!\int_0^1\!\df x\, (1-x^2)^{-1+\epsilon}
= \frac{1}{2}\frac{\Gamma(\epsilon)^2}{\Gamma(2\epsilon)}\,.
\end{align}
Expanding in $\epsilon\to 0$ and upon renormalization in ${\overline{\rm MS}}$, we find for the position space soft
function
\begin{align}
\tilde{S}^{\rm 1-loop}_{\widetilde C} & = -\frac{\alpha_s(\mu)}{4\pi}\,C_F
\Big[\frac{\pi^2}{3}+8\ln^2(ix\mu e^{\gamma_E})\Big]\,,\\
\tilde{S}^{\rm 1-loop}_{\tau^a} & = -\frac{1}{1-a}\frac{\alpha_s(\mu)}{4\pi}\,C_F
\Big[\pi^2+8\ln^2(ix\mu e^{\gamma_E})\Big]\,.\nonumber
\end{align}
Writing the logarithm of the soft function in Fourier space evaluated at the point
$x = -\,i\exp(-\,\gamma_E)/\mu$ in a generic form as
\begin{align}
\ln {\widetilde S_e} = 2\sum_{n=1}^\infty \Big(\frac{\alpha_s(\mu)}{4\pi}\Big)^n s_n^e \,.
\end{align}
we obtain
\begin{align}
s_1^{\widetilde C} = -\,\frac{\pi^2}{6}\,C_F\,,\qquad
s_1^{\tau_a} = -\, \frac{1}{1-a}\frac{\pi^2}{2}\,C_F\,.
\end{align}
Fourier transforming the result, one obtains the renormalized momentum space soft function:
\begin{align}
S^{\rm 1-loop}_{\widetilde C} & = \frac{\alpha_s(\mu)}{4\pi}\,C_F
\bigg(\!\pi^2\,\delta(\ell) -
\frac{16}{\mu}\bigg[\frac{\ln(\ell/\mu)}{\ell/\mu}\bigg]_+\bigg)
\,,\\
S^{\rm 1-loop}_{\tau_a}   & = \frac{1}{1-a}\frac{\alpha_s(\mu)}{4\pi}\,C_F
\bigg(\!\frac{\pi^2}{3}\,\delta(\ell)-
\frac{16}{\mu}\bigg[\frac{\ln(\ell/\mu)}{\ell/\mu}\bigg]_+\bigg)\,.\nn
\end{align}
%

\chapter{Comparison with other fits}
\label{app:fit-compare}
In this appendix, we compare the main results of the $\alpha_s(m_Z)$ extraction from the C-parameter cross section with other possible fits.

\section{Comparison of thrust and C-parameter subtractions}
\label{ap:subtractions}
\begin{figure}[t!]
	\begin{center}
		\includegraphics[width=0.5\columnwidth]{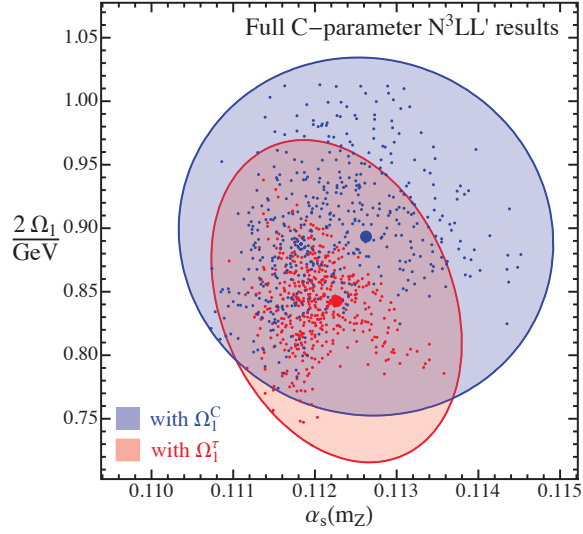}
		\caption[Comparison of $\alpha_s$ determinations from thrust Rgap scheme and C-parameter Rgap scheme ]{Comparison of $\alpha_s$ determinations from \mbox{C-parameter} tail fits in the thrust Rgap scheme (lower red ellipse) and the
			\mbox{C-parameter} Rgap scheme (upper blue ellipse). The leading power correction $\Omega_1^C$ in the \mbox{C-parameter} Rgap scheme is converted to $\Omega_1$ in the thrust Rgap scheme in order to have a meaningful comparison. Theoretical uncertainty ellipses are shown which are suitable for projection onto one dimension to obtain the $1$-$\sigma$ uncertainty, without experimental uncertainties.}
		\label{fig:Cgap}
	\end{center}
\end{figure}

In Fig.~\ref{fig:Cgap} we compare fits performed in the Rgap scheme with C-parameter gap subtractions as the upper red ellipse, and for our default fits in the Rgap scheme with thrust gap subtractions as the lower blue ellipse. At N$^3$LL$^\prime$ order with C-parameter subtractions the results are $\alpha_s(m_Z) = 0.1126 \pm 0.0002_{\rm exp} \pm 0.0007_{\rm hadr} \pm
0.0022_{\rm pert}$ and $\Omega_1(R_\Delta,\mu_\Delta) = 0.447 \pm 0.007_{\rm exp}
\pm 0.018_{\alpha_s} \pm 0.065_{\rm pert}$~GeV, with $\chi^2_{\rm min}/{\rm dof} = 0.988$.
One can see that, even though both extractions are fully compatible, the thrust subtractions lead to smaller perturbative uncertainties. This is consistent with the better perturbative behavior observed for the cross section with thrust subtractions in Ch, \ref{ch:Cparam-theory}.

\section{Comparison of thrust results with Ref.~\cite{Abbate:2010xh}}
\label{ap:thrustresults}

\begin{figure}[t!]
	\begin{center}
		\includegraphics[width=0.5\columnwidth]{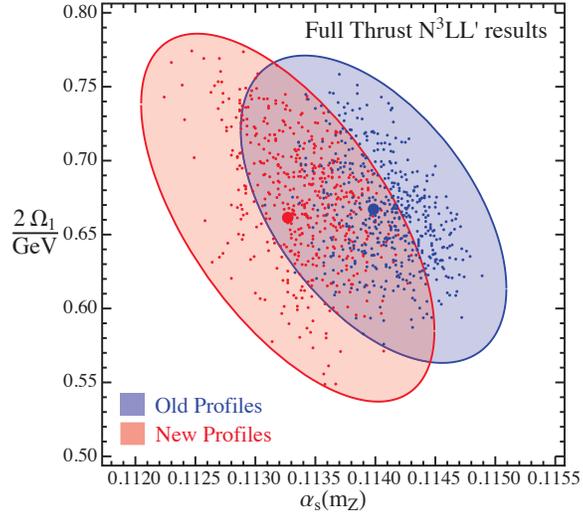}
		\caption[Comparison of thrust $\alpha_s$ determinations using updated profiles and the profiles of Ref.~\cite{Abbate:2010xh}]{Comparison of thrust $\alpha_s$ determinations using our new profiles (left red ellipse) and the profiles of Ref.~\cite{Abbate:2010xh} (right blue ellipse). Theoretical uncertainty ellipses are shown which are suitable for projection onto one dimension to obtain the $1$-$\sigma$ uncertainty, without experimental uncertainties.}
		\label{fig:new-old-profiles}
	\end{center}
\end{figure}

In Fig.~\ref{fig:new-old-profiles} we compare global fits for the thrust distribution using the profiles of Ref.~\cite{Abbate:2010xh} (shown by the right ellipse in blue) and the profiles used here (shown by the left ellipse in red). As mentioned earlier, the profiles used here have several advantages over those of Ref.~\cite{Abbate:2010xh} in terms of their ability to independently impact the different regions of the thrust distribution, and in particular do a better job in the nonperturbative region (which is outside our fit region). The two versions of the profiles are consistent within their variations, and the fit results shown for 39\% CL for two dimensions in Fig.~\ref{fig:new-old-profiles} (which is 68\% CL for each one-dimensional projection) are fully compatible.

%

\chapter{Spinor Helicity Identities and Conventions}
\label{app:helicity}

The four-component spinor $u(p)$ of a massless Dirac particle with momentum $p$, satisfies the massless Dirac equation,
\begin{equation} \label{eq:Dirac}
\Sl p\, u(p)=0
\,, \qquad
p^2 = 0
\,,\end{equation}
as does the charge conjugate (antiparticle) spinor $v(p)$. We can therefore choose a representation such that $v(p) = u(p)$. We denote the spinors and conjugate spinors for the two helicity states by
\begin{align} \label{eq:braket_def}
\ket{p\pm} = \frac{1 \pm \ga_5}{2}\, u(p)
\,,\qquad
\bra{p\pm} = \mathrm{sgn}(p^0)\, \bar{u}(p)\,\frac{1 \mp \ga_5}{2}\,.
\end{align}
Here the $\mathrm{sgn}(p^0)$ is included in the definition to simplify relations under crossing symmetry.

The spinors $\ket{p \pm}$ have an overall phase that is left free by the Dirac equation. Using the Dirac representation,
\begin{equation}
\ga^0 = \begin{pmatrix} 1 & 0 \\ 0 & -1 \end{pmatrix}
\,,\quad
\ga^i = \begin{pmatrix} 0 & \sigma^i \\ -\sigma^i & 0 \end{pmatrix}
\,,\quad
\ga_5 = \begin{pmatrix} 0 & 1 \\ 1 & 0 \end{pmatrix}
\,.\end{equation}
If we take $n_i^\mu = (1,0,0,1)$, we get the standard solutions~\cite{Dixon:1996wi}
\begin{equation} \label{eq:ket_explicit}
\ket{p+} = \frac{1}{\sqrt{2}}
\begin{pmatrix}
\sqrt{p^-} \\
\sqrt{p^+} e^{i  \phi_p} \\
\sqrt{p^-} \\
\sqrt{p^+} e^{i  \phi_p}
\end{pmatrix}
,\quad
\ket{p-} = \frac{1}{\sqrt{2}}
\begin{pmatrix}
\sqrt{p^+} e^{-i \phi_p} \\
-\sqrt{p^-} \\
-\sqrt{p^+} e^{-i \phi_p} \\
\sqrt{p^-}
\end{pmatrix}
,\end{equation}
where
\begin{equation}
p^\pm = p^0 \mp p^3
\,,\qquad
\exp(\pm i  \phi_p) = \frac{p^1 \pm i  p^2}{\sqrt{p^+ p^-}}
\,.\end{equation}
For negative $p^0$ and $p^\pm$ we use the usual branch of the square root, such that for $p^0 > 0$
\begin{equation}
\ket{(-p)\pm} = i  \ket{p\pm}
\,.\end{equation}
We also define, $\bra{p\pm}$, the conjugate spinors, as
\begin{equation}
\bra{p\pm} = \mathrm{sgn}(p^0)\, \overline{\ket{p\pm}}
\,.\end{equation}
We include the additional minus sign for negative $p^0$ as we want to use the same branch of the square root for both types of spinors. We see for $p^0 > 0$
\begin{equation}
\bra{(-p)\pm} = - \overline{\ket{(-p)\pm}} = -(-i ) \bra{p\pm} = i \bra{p\pm}
\,.\end{equation}
This makes all spinor identities correct for momenta of both signs, allowing easier utilization of crossing symmetry. These signs will appear in expressions with explicit complex conjugation, including the most important example,
\begin{equation} \label{eq:spin_conj}
\l p-|q+\r^* = \mathrm{sgn}(p^0 q^0)\, \l q+|p-\r
\,.\end{equation}
The spinor products are denoted by
\begin{equation}
\langle p q \rangle = \langle p-|q+\rangle
\,,\qquad
[p q] = \langle p+|q-\rangle
\,.\end{equation}
Several useful identities satisfied by the spinor products are
\begin{align}
&\ang{pq} = - \ang{qp}
\,,\qquad
[pq] = - [qp]
\,, \qquad [p|\ga^\mu \ket{p}=\bra{p} \ga^\mu |p] = 2p^\mu\,, \\
&\ket{p\pm}\bra{p\pm} = \frac{1 \pm \ga_5}{2}\,\pslash
\,, \qquad
\pslash=|p]\bra{p}+\ket{p}[p|,\nn\\
&\langle pq \rangle [qp] = \frac{1}{2}\,\tr\bigl\{(1 - \ga_5) \pslash \qslash \bigr\} = 2 p\cdot q \,, \qquad
|\ang{pq}| = |[pq]| = \sqrt{|2p \cdot q|}\,, \nn\\
&\bra{p}\gamma^\mu |q]  = [q|\gamma^\mu \ket{p}
\,, \qquad
[p|\gamma_\mu \ket{q} [k|\gamma^\mu \ket{l}= 2[pk]\ang{lq}
\,, \nn\\
&\ang{pq} \ang{kl} = \ang{pk} \ang{ql} + \ang{pl} \ang{kq}
\,.\nn
\end{align}
Momentum conservation $\sum_{i=1}^n p_i = 0$ also implies the relation
\begin{equation} \label{eq:spinormomcons}
\sum_{i=1}^n [ji] \ang{ik} = 0
\,.\end{equation}
Under parity the spinors transform as
\begin{align} \label{eq:spinorparity}
\ket{p^\P\pm} = \pm e^{\pm i \phi_p}\gamma^0\,\ket{p\mp}\,, \quad
\ang{p^\P q^\P} = -e^{i (\phi_p + \phi_q)} [pq]
\,, \quad
[p^\P q^\P] = -e^{-i (\phi_p + \phi_q)} \ang{pq}
\,.\end{align}
The polarization vector satisfies the completeness relation
\begin{align}
\sum\limits_{\lambda=\pm} \epsilon^\lambda_\mu(p,q) \left (\epsilon^\lambda_\nu(p,q) \right )^*=-g_{\mu \nu}+\frac{p_\mu q_\nu+p_\nu q_\mu}{p\cdot q}.
\end{align}
In SCET the projected collinear quark fields 
\begin{align}
\ket{p\pm}_{n}=\frac{\nslash \bnslash}{4}\ket{p\pm}\,,
\end{align}
satisfy the relation
\begin{equation}
\nslash\,\Bigl(\frac{\nslash\bnslash}{4}\ket{p\pm}\Bigr) = 0
\,,\end{equation}
and are therefore proportional to $\ket{n\pm}$. Working in the basis in \eq{ket_explicit}, we find
\begin{align}\label{eq:ketn}
\frac{\nslash\bnslash}{4} \ket{p}=& \sqrt{p^0}\left[ \cos\left (\frac{\theta_n}{2}\right )\cos \left(\frac{\theta_p}{2}\right )
+e^{i(\phi_p-\phi_n)} \sin \left(\frac{\theta_n}{2}\right )   \sin \left(\frac{\theta_p}{2} \right )   \right ] \ket{n}\,,  \\
\frac{\nslash\bnslash}{4} |p]=& \sqrt{p^0}\left [ e^{i \left(\phi _p-\phi _n\right)} \cos \left(\frac{\theta _n}{2}\right)  \cos \left(\frac{\theta _p}{2}\right) 
+\sin \left(\frac{\theta _n}{2}\right) \sin \left(\frac{\theta _p}{2}\right)   \right ] |n]\,.\nn
\end{align}
Here $\theta_n, \phi_n$, and $\theta_p, \phi_p$, are the polar and azimuthal angle of the $n$ and $p$ vectors, respectively.
Choosing $n_i^\mu=p_i^\mu/p_i^0$, we have
\begin{align}
\frac{\nslash\bnslash}{4} \ket{p\pm}=\sqrt{\frac{\bn_i\cdot p}{2}}\: \ket{n_i\pm}\,.
\end{align}

\begin{singlespace}
\bibliography{thesis.bib}
\bibliographystyle{jhep}
\end{singlespace}
\end{document}